\documentclass[a4paper,english,12pt]{report}

\usepackage{graphicx,epstopdf,amssymb,amsfonts,amsmath,amsthm,array,
mathrsfs,amscd,bbm,enumerate}
\usepackage{subcaption}
\usepackage{xcolor}   % Handling colors
\usepackage[top=2.5cm, bottom=3.5cm, left=2.5cm, right=2.5cm]{geometry}
\usepackage{tcolorbox}   % Environment for colored and framed boxes
\usepackage{wrapfig}
\usepackage{comment}
\usepackage{appendix}
\usepackage{mathbbol}
\usepackage{framed}
\usepackage{atbegshi, picture}
\usepackage{csquotes}   %\blockquote{}
\usepackage[backgroundcolor=white, linecolor=black]{todonotes}   % Add remarks in the margin (\todo{todo text}), add missing figures with text (\missingfigure[figwidth=6cm]{text}) and list all the todos (\listoftodos)
\usepackage{tensor}   % Bla bla bla
\usepackage{esvect} %  Vectors
\usepackage{hyperref}   % Hyperlinks for references
\hypersetup{
	colorlinks,
	citecolor=blue,
	filecolor=blue,
	linkcolor=blue,
	urlcolor=blue,			
	pdfborder = {0 0 0},   % Removes hyperlinks boxes
	linktocpage
}

\numberwithin{equation}{chapter}
\numberwithin{figure}{chapter}

\def\be{\begin{equation}}\def\ee{\end{equation}}
\def\tr{\mathop{\text{tr}}\nolimits}

\def\re{\mathop{\text{Re}}\nolimits}
\def\deg{\mathop{\text{deg}}\nolimits}
\def\ind{\mathop{\text{ind}}\nolimits}

\newcommand{\tra}[1]{#1^{\text{T}}}

\def\suN{\text{SU}(N)}
\def\uN{\text{U}(N)}

\def\oN{\text{O}(N)}
\def\oD{\text{O}(D)}

\def\cG{\mathcal G}
\def\gG{\mathcal G}
\def\cH{\mathcal H}
\def\cB{\mathcal B}
\def\gB{\mathcal B}
\def\gJ{\mathcal J}
\def\cK{\mathcal K}
\def\gK{\mathcal K}
\def\cT{\mathcal T}

\def\cN{\mathcal N}
\def\cA{\mathcal A}
\def\cV{\mathcal V}
\def\cE{\mathcal E} 

\def\cT{\mathcal T}
\def\cI{\mathcal I}
\def\cS{\mathcal S}

\newtheorem{lemma}{Lemma}[section]
\newtheorem{proposition}{Proposition}[section]

\newtheorem{theorem}{Theorem}

 \theoremstyle{remark}

\definecolor{darkblue}{HTML}{004C93} 
\tcbset{
    center title,
		top=1cm,
		right=0cm,
		left=0cm,
    colback=darkblue,
    colframe=white,
    width=15.85cm,
		height=5.62cm,
		halign=right,
		valign=top,
		sharp corners=all,
		flush right
    }
    
\newcommand{\pbs}[1]{\let\temp=\\#1\let\\=\temp}

\def\cvp{\raise 2pt\hbox{,}}

\def\imath#1#2#3{{\it Invent math }{\bf #1} (#2) #3}

\newcommand{\bee}{\begin{equation}}

\newcommand{\bea}{\begin{eqnarray}}
\newcommand{\eea}{\end{eqnarray}}

\begin{document}

% Page de garde 
\thispagestyle{empty}
\newgeometry{left=2cm, right=2cm, top=2cm, bottom=1cm}
\includegraphics[width=12.3cm, height=3.07cm]{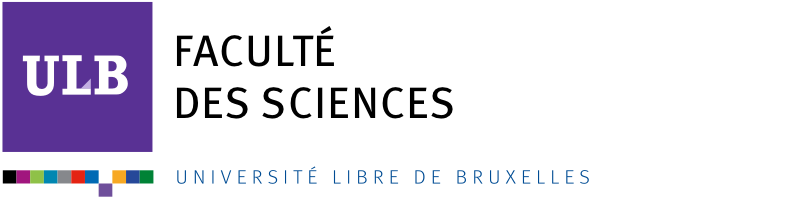} \\ \\

\begin{tcolorbox}
\color[rgb]{1,1,1}
\Large{\textbf{New Limits for Large $N$ Matrix and Tensor Models}} \\
\large{\textit{Large $D$, Melons and Applications}}
\end{tcolorbox} 

\begin{tcolorbox}[colback=white, halign=left]
\color{darkblue}
\large{\textbf{Thesis presented by Guillaume VALETTE}} \\
\color{black}
\large{with a view to obtaining the Ph.D. Degree in Sciences (``Docteur en Sciences")\\
Academical year 2019-2020}
 
\vspace{2cm}

\begin{flushright}
\color{darkblue}
Supervisor: Prof.\ Frank FERRARI \\
Co-supervisor: Prof.\ St\'ephane DETOURNAY \\
\color{black}
Service de Physique Math\'ematique des Interactions Fondamentales
\end{flushright}
\end{tcolorbox}
\vspace{4cm}

\begin{wrapfigure}{r}{0.2\textwidth}
\includegraphics[scale=0.9]{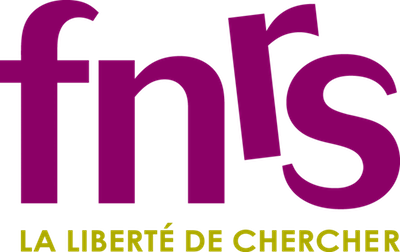}
\end{wrapfigure}
\noindent
\textbf{Thesis jury : } \\  \\
Prof.\ Riccardo ARGURIO (Universit\'e Libre de Bruxelles, Chair)  \\
Prof.\ Razvan GURAU (Ecole Polytechnique) \\
Prof.\ Vincent RIVASSEAU (Universit\'e Paris-Sud XI) \\
Ass.\ Prof.\ Reiko TORIUMI (Okinawa Institute of Science and Technology) \\

\restoregeometry

% White page
\newpage
\thispagestyle{empty}
\null
\newpage

\pagenumbering{roman}

\thispagestyle{empty}
\newgeometry{top=2.5cm, bottom=3.5cm, left=2.2cm, right=2.2cm}
% Information
\begin{center}

\LARGE{New Limits for Large $N$ Matrix and Tensor Models} \\
\bigskip
\Large{Large $D$, Melons and Applications}

\vspace{1cm}

Guillaume~Valette
 
\bigskip

{\small{\textit{Service de Physique Th\'eorique et Math\'ematique\\
Universit\'e Libre de Bruxelles\\
Campus de la Plaine, CP 231, B-1050 Bruxelles, Belgique}}}

\smallskip

{\small{\tt guillaume.valette@ulb.ac.be}}

\end{center}

\vspace{1cm}

\noindent This manuscript corresponds to the pedagogical part of the complete Ph.D.\ thesis.

\vspace{1cm}

\noindent The Ph.D. thesis was defended behind closed doors on the 12th of September 2019 and in public on the 9th of October 2019 at the Universit\'e Libre de Bruxelles.

\vspace{1cm}

\noindent The author is a Research Fellow (\textit{Aspirant}) at the Belgian F.R.S.-FNRS (2016-2020).

\vspace{1cm}

\noindent The original contributions of this thesis are based on the following papers:

\begin{itemize}

\item[\cite{Ref2}] F.~Ferrari, V.~Rivasseau and G.~Valette, \emph{A New Large $N$ Expansion for General Matrix-Tensor Models}, Commun.\ Math.\ Phys.\ \textbf{370} (2019) 403-448, arXiv:1709.07366 [hep-th].

\item[\cite{Ref1}] T.~Azeyanagi, F.~Ferrari, P.~Gregori, L.~Leduc and G.~Valette, \emph{More on the New Large $D$ Limit of Matrix Models}, Annals Phys.\ \textbf{393} (2018) 308-326, arXiv:1710.07263 [hep-th].

\item[\cite{Ref3}] S.~Dartois, O.~Evnin, L.~Lionni, V.~Rivasseau and G.~Valette, \emph{Melonic Turbulence}, to appear in Commun.\ Math.\ Phys., arXiv:1810.01848 [math-ph].

\end{itemize}

\restoregeometry

% White page
\newpage
\thispagestyle{empty}
\null
\newpage

% Abstract
\chapter*{Abstract}
\setcounter{page}{1}
\addcontentsline{toc}{chapter}{Abstract}

Large $N$ matrix models play an important role in modern theoretical physics, ranging from quantum chromodynamics to string theory and holography. However, they remain a difficult technical challenge because in most cases it is not known how to perform the sum over planar graphs, which dominate the models at large $N$. 

In this thesis, we study large $D$ matrix models as they provide a framework to build new limits for matrix models in which the sum over planar graphs simplifies when $D$ is large. The basic degrees of freedom are real matrices of size $N\times N$ with $r$ additional indices of range $D$ and with symmetry group $\oN^2\times\oD^r$. These matrices can be interpreted as a real tensor of rank $R=r+2$ with indices of different ranges, making a compelling connection with tensor models.

We define a new large $D$ limit for the sum over Feynman graphs of fixed genus in matrix models, based on an enhanced large $D$ scaling of the coupling constants. Using the combinatorial techniques developed in tensor models, we show that the resulting large $D$ expansion is well-defined and organized according to a half-integer called the index. When $N=D$, the result also provides a new large $N$ limit for general $\oN^R$ invariant tensor models.

In the large $D$ limit, the sum over planar graphs of large $N$ matrix models simplifies to a non-trivial sum over generalized melonic graphs. This class of graphs extends the one obtained in tensor models with standard scaling and allows for a wider class of interactions, including all the maximally single-trace terms.

The general classification of generalized melonic graphs remains an open problem. However, in the case of the complete interaction of order $R+1$ for $R$ a prime number, we identify them in detail and demonstrate that they exhibit the same important features as the SYK model with $q=(R+1)$-fold random interactions, including the emergent conformal symmetry in the infrared regime and maximal chaos.

The advantage of large $D$ matrix models over the SYK model and its variants is that they correspond to genuine quantum field theories. In addition, for $r=1$, they have a natural interpretation in terms of $D$-brane constructions in string theory, making a possible relation with holography clearer.

Another part of this thesis applies the tools developed in tensor models to study non-linear resonant flows in many variables. By averaging over both the tensor coupling and the initial conditions, we prove that in some regime of perturbation theory, melonic graphs dominate the dynamics and are responsible for turbulent energy cascades.

% White page
\newpage
\thispagestyle{empty}
\null
\newpage

%% R\'esum\'e
%\chapter*{R\'esum\'e}
%\setcounter{page}{3}
%\addcontentsline{toc}{chapter}{R\'esum\'e}

%Table of contents
\tableofcontents

%%%%%%%%%%%%%%%%%%%%%%%%%%%%%%%%%%%%%%%%%%%%%%%%
%%%%%%%%%%%%%%%%%%%  MAIN  %%%%%%%%%%%%%%%%%%%%%%%%%
%%%%%%%%%%%%%%%%%%%%%%%%%%%%%%%%%%%%%%%%%%%%%%%%

\chapter*{Introduction}
\pagenumbering{arabic}
\addcontentsline{toc}{chapter}{Introduction}
\label{chap:Intro}

\section*{Context and motivation}
\label{sec:Context}

During the twentieth century, theoretical physics has been marked by the development of two important pillars in the understanding of the fundamental laws of nature. On the one hand, there is the theory of General Relativity, which describes the gravitational interaction, and on the other hand, the Standard Model of particle physics, which explains the other three fundamental interactions, namely the electromagnetic, the weak and the strong interactions. During the past decades, a significant research interest has been focused on the construction of a unified theory that consistently accounts for the four types of interactions at all energy scales. However, in spite of many efforts, the basic principles underlying such a theory remain to a large extent elusive. The reason is deeply rooted in the incompatibility of General Relativity with the quantum world. 

The Standard Model is described in the realm of quantum field theory (QFT). It accounts for all the non-gravitational interactions in a fully relativistic framework which is consistent with quantum mechanics. The particle content is described in terms of dynamical fields that spread over spacetime. The Standard Model provides theoretical predictions that match with experimental observations to a high precision. In particular, the Brout-Englert-Higgs scalar particle, which is an essential ingredient of this theory, was detected in 2013 at the Large Hadron Collider (LHC).

Einstein's theory of General Relativity (GR) describes the gravitational force at the classical level. It is a theory of spacetime itself where gravitation is elegantly understood as a consequence of the curvature of the spacetime manifold. The field content includes the metric on this manifold, whose dynamics satisfies Einstein's equations. Importantly, the classical description of GR is only reliable at low energy scales, that is, at energy scales much lower than the Planck scale. Indeed, when viewed as a QFT, GR is non-renormalizable, which means that it becomes strongly coupled in the ultraviolet regime. Thus, GR cannot be a complete theory; a more fundamental one, often referred to as a theory of quantum gravity, must come into play at high energy scales. However, it does not mean that GR is wrong. Rather, it should be interpreted as a low-energy effective field theory that flows from this still unknown theory of quantum gravity. In fact, many predictions of GR have been confirmed experimentally at moderate energy scales. This includes the detection of gravitational waves emitted after the collision of two black holes by the Laser Interferometer Gravitational-Wave Observatory (LIGO) in 2016 and the first photo of a black hole by the Event Horizon Telescope (EHT) in 2019.

The incompatibility of GR as a quantum theory can also be understood from the study of black holes. These objects were predicted long ago in the context of GR. However, their classical description cannot be complete because it predicts the existence of singularities inside their horizon, where spacetime curvature is infinite, which is inconsistent. 

Other paradoxes related to black holes result from their thermodynamical description. It was realized in the 1970s that the laws of black hole mechanics are strikingly similar to the laws of thermodynamics. For instance, the second law of black hole mechanics, which states that the total area of the event horizon can never decrease, is analogous to the second law of thermodynamics, which says that the total entropy of a system can never decrease. This led Bekenstein \cite{Bekenstein} to propose that black holes should have an entropy and that it should be proportional to the horizon area, in Planck units. Further evidence for the thermodynamical interpretation of black holes was then obtained by Hawking \cite{Hawking}, who proved that at the semi-classical level, black holes evaporate by emitting thermal radiation at a constant temperature. 

Black holes thus seem to behave like thermodynamical objects with a temperature and an entropy. However, this raises important puzzles on the nature of spacetime and unitarity. For instance, the suggestion that black holes have an entropy proportional to their horizon area cannot be accounted for in classical GR. It also begs the question of what the microscopic degrees of freedom counted by this entropy are and where they come from. Besides, the thermal radiation emitted by an evaporating black hole stands in sharp contradiction with the unitarity of quantum mechanics, because any initial pure quantum state would evolve into a mixed state, destroying the initial information. These so-called black hole entropy problem \cite{BHEntropy} and information paradox \cite{InfoParadox} are key questions that cannot be answered in the realm of classical GR. They provide two challenges to any putative theory of quantum gravity.

Over the years, different approaches to define quantum gravity have been developed, the most notable being string theory. In the framework of string theory, the point particles of usual QFT are replaced by one-dimensional extended objects called strings. Interestingly, the low-energy description of strings gives rise to classical GR, together with all the non-gravitational interactions present in the Standard Model. Moreover, it is well-defined in the ultraviolet regime. Hence, string theory is a good candidate for a unified theory of all the fundamental interactions at all energy scales. However, it also has some limitations. For instance, string theory does not have a fully non-perturbative definition, unlike ordinary QFTs. In addition, it seems to lack predictive power because of the many different possibilities of compactifying the extra spatial dimensions, leading to a huge number of different possible vacuum states. Finally, solving the black hole entropy problem and information paradox in the most interesting cases remains a big challenge in the direct context of string theory.

\subsubsection{Gauge/gravity correspondence and holography}

In spite of these various drawbacks, string theory still offers a framework in which one can begin to explore issues that were not accessible before. In particular, it provides a powerful tool to study quantum gauge theories from a totally new perspective, called the gauge/gravity correspondence \cite{Malda,GaugeGravity}. This correspondence conjectures an equivalence between a quantum gravitational theory in a bulk spacetime and an ordinary quantum gauge theory, with no gravity, on the boundary of the bulk spacetime. It is compelling that two totally different theories, living in different dimensions, are claimed to be dual to each other, meaning that they describe the same physics in two different languages. 

Historically, the first setup of the gauge/gravity correspondence was proposed by Maldacena \cite{Malda}, where a type IIB superstring theory on a bulk $\text{AdS}_5\times \text{S}^5$ was related to $\cN=4$ super-Yang-Mills theory on the four-dimensional boundary. This is the celebrated Anti-de Sitter/Conformal Field Theory (AdS/CFT) correspondence. Various consistency checks of this setup have been completed; however, there is currently no complete proof that these two theories are strictly equivalent. 

A generic setup of the gauge/gravity correspondence can be summarized as follows \cite{Sonner}, where the relationships between the parameters of the two theories are indicated.
\begin{center}
\begin{tabular}{ c c c }
$\biggl(\begin{array}{c}\text{\small Quantum gravity in an asymptotically} \\ \text{\small AdS spacetime in $d$ dimensions}\end{array}\biggr)$ & $\leftrightarrow$ & $\biggl(\begin{array}{c}\text{\small Conformal quantum gauge field theory} \\ \text{\small $\text{SU}(N)$ in $d-1$ dimensions}\end{array}\biggr)$ \\ [1cm]
\( \displaystyle \biggl(\frac{\ell_{Pl}}{L}\biggr)^{d-2} \) & $\leftrightarrow$ & \( \displaystyle\frac{1}{\cN}\) \\  [1cm]
\( \displaystyle \biggl(\frac{\ell_{s}}{L}\biggr)^{2}\) & $\leftrightarrow$ & \( \displaystyle\frac{1}{\lambda}\)    
\end{tabular}
\end{center}
On the bulk side, $\ell_{Pl}$ corresponds to the Planck length, which governs the quantum effects, $\ell_{s}$ is the string length, which governs the stringy effects, and $L$ is the characteristic length associated with the geometry of the bulk, which is typically the AdS radius. On the boundary side, $N$ corresponds to the rank of the gauge group $\text{SU}(N)$, $\cN$ denotes the number of degrees of freedom, which is of order $N^2$ in typical setups, and $\lambda$ is some coupling constant.

The correspondence between the two theories is conjectured for any value of $N$ and $\lambda$. However, for moderate values of $N$ and $\lambda$, it is difficult to do any calculations on both sides: the boundary theory is complicated because the perturbative regime is not accessible whereas the bulk theory corresponds to a full string theory with important quantum corrections. Thus, it is of interest to consider specific regimes of parameters for which computations can be made on at least one side of the correspondence. One such regime is associated with classical, Einstein-like gravity in the bulk. It is obtained in the thermodynamical limit $\cN\rightarrow\infty$, which typically corresponds to the limit of a gauge group of large rank $N$, so that the bulk string theory becomes classical. Then, one needs to take the strong coupling limit $\lambda\rightarrow\infty$, so that the stringy corrections become negligible. From the bulk perspective, it allows one to access a regime where computations are feasible. At the same time, it corresponds to the most inaccessible regime on the quantum field theory side, namely the strong coupling regime. In other words, the gauge/gravity correspondence may in principle deal with difficult questions in strongly coupled field theories by answering simpler questions in classical, Einstein-like gravity. This is one of the attractive features of the gauge/gravity correspondence.

In fact, the gauge/gravity correspondence is a realization of the holographic principle, initiated by 't~Hooft \cite{tHooftHolo} and Susskind \cite{SusskingHolo}. This principle is inspired from the behavior of black hole entropy, which is proportional to the horizon area while in usual statistical systems, entropy scales like a volume. This property seems to imply that the microscopic information about black holes is actually contained on the boundary horizon and not in the interior bulk. By extension, the holographic principle states that a theory with gravity in a bulk spacetime is dual to a non-gravitational theory with less dimensions, which accounts for the microscopic degrees of freedom. From the perspective of holography, the bulk gravitational description is thus an emergent phenomenon, that is, it emerges from the strongly coupled dynamics of a fundamental quantum gauge theory with less dimensions.

%From the above discussion, the study of gauge field theories at strong coupling might reveal interesting properties about the corresponding emergent gravitational theory. However, it is in general a difficult task to analyze any quantum field theory at strong coupling. Until recently, this had been done only in very special cases, where calculations at strong coupling could be inferred from the weak coupling regime because of supersymmetry or integrabitility. A few years ago, recent developments (see below) explored an interesting class of quantum models, whose strong coupling regime can be accessed. These models are being studied actively and constitute a central piece of this thesis.

%
\subsubsection{Quantum models of black holes}

An important application of the gauge/gravity correspondence is the study of black holes. According to this correspondence, black holes in the bulk spacetime are mapped to thermal states in the dual gauge theory. This framework thus provides an interesting description of black holes, which are interpreted as emerging from some quantum states in the dual theory. In principle, this holographic description of black holes offers a solution to the entropy problem and the information paradox discussed earlier. The entropy of the black hole is interpreted as the entropy of the dual gauge theory, which counts the microscopic states in the corresponding Hilbert space. On the other hand, since any quantum field theory is consistent with unitarity, it means that the thermal properties of black holes also become manifestly consistent with unitarity in this picture. 

However, in practice, the gauge/gravity correspondence is unable to provide at the moment a satisfactory solution to these two problems. One of the difficulties comes from the necessity to tackle strongly coupled gauge theories to recover an ordinary bulk gravitation description. Secondly, holography is only reasonably well-understood when the dual gauge theory corresponds to a conformal field theory. In the emergent gravity picture, it is equivalent to the fact that the bulk spacetime is asymptotically AdS. Thirdly, supersymmetry is often assumed. In contrast, real-world black holes are not supersymmetric and they are not likely to be associated with a dual conformal field theory nor AdS. It is currently an active field of research to look for possible extensions of the original gauge/gravity correspondence to non-conformal or non-supersymmetric setups. 

Besides these limitations, not any strongly coupled gauge theory can describe a black hole. In the framework of the gauge/gravity correspondence, prime candidates are quantum gauge theories of $N\times N$ matrices in the large $N$ limit and at strong coupling. Indeed, as explained below, such models naturally arise from $Dp$-brane constructions in string theory. Explicit examples include the $\text{AdS}_5$ Schwarzschild black hole \cite{AdS5} and the $D0$-brane black hole \cite{BFFS}. A comprehensive overview of the importance of matrix models in the study of holographic quantum black holes can be found in \cite{PaoloThesis}.

We now briefly comment on these ``quantum gauge theories of $N\times N$ matrices in the large $N$ limit and at strong coupling". Firstly, the models must be studied at strong coupling. As already emphasized, this regime is difficult to analyze and inaccessible to conventional methods. There exists however a class of interesting quantum field theories that greatly simplify in the limit of a large number of degrees of freedom, which usually corresponds to a gauge group of large rank $N$. They simplify because their solution admits a perturbative expansion in powers of $1/N$. In the large $N$ limit, only a small number of terms survive in the expansion, which can then be sometimes evaluated analytically. In addition, the large $N$ solution obtained by this method is in several cases a good approximation of the original model and it allows one to access physics at strong coupling by analytical continuation. Theories that become solvable at large $N$ hold a crucial role in the context of this thesis. In particular, large $N$ matrix models belong to this class of theories. Their large $N$ solution is computed by summing over planar graphs, as first realized by 't~Hooft in the context of Quantum Chromodynamics (QCD) \cite{tHooft}.

\begin{figure}[h!]
\centerline{\includegraphics[scale=1]{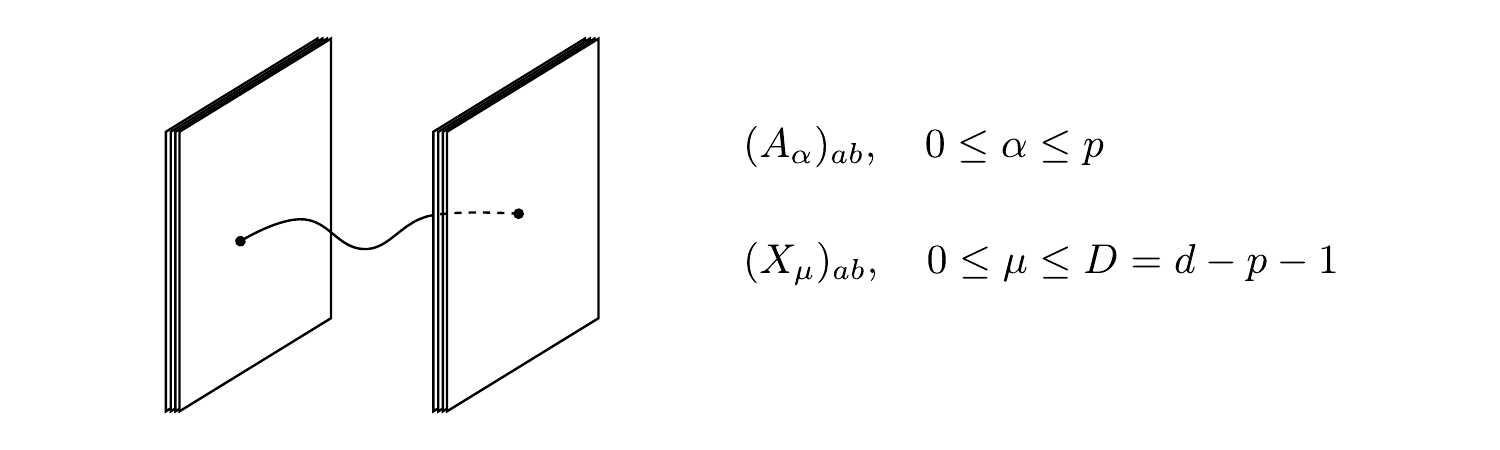}}
\caption{Two stacks of $N$ $Dp$-branes in spacetime on which open strings can attach. The excitations of the open strings are then described microscopically by $N\times N$ matrix degrees of freedom, the matrix indices corresponding to the Chan-Paton factors associated with the two end points of the open strings. The fields $A_\alpha$ correspond to longitudinal excitations whereas the fields $X_\mu$ correspond to transverse excitations.}\label{Branes}
\end{figure}

Secondly, matrix degrees of freedom naturally arise in the context of holography from $Dp$-brane constructions. The matrix indices correspond to the Chan-Paton associated with the two end points of open strings attached on stacks of $N$ $Dp$-branes. This is illustrated in Figure \ref{Branes}. The $p+1$ fields $(A_\alpha)_{ab}$ correspond to excitations of the strings parallel to the branes while the $D$ fields $(X_\mu)_{ab}$ correspond to transverse excitations. The gauge group associated with the matrix indices is typically $\uN$. As for the global symmetry transverse to the branes, it is given by the rotation group $\text{O}(D)$, under which the matrices $X_\mu$ transform like a vector.

In general, large $N$ matrix quantum gauge theories still remain a difficult challenge at the technical level, because in the most interesting cases, it is not known how to perform the sum over planar graphs. It is therefore natural to look for new strategies that can lead to amenable simplifications of the sum over planar graphs. This is one of the motivations of this thesis, as further explained below. 

\subsubsection{Discretized spacetime and random tensors}

Besides string theory and the gauge/gravity correspondence, there exist other approaches in the ongoing quest for a theory of quantum gravity. A natural, geometrical approach is based on a discretized version of the Einstein-Hilbert action. As mentioned above, GR is a classical field theory of dynamical spacetime, whose field content includes the metric $g$ and whose dynamics is governed by the Einstein-Hilbert action $S_{\text{EH}}$. On the other hand, the other three fundamental interactions are unified within the Standard Model of particle physics with action $S_{\text{SM}}$, which contains the typical matter fields $\Phi$ of our universe. Following the standard Feynman path integral quantization for field theories, a possible starting point for a theory of quantum gravity is a ``sum over histories" of the form \cite{GurauRandTens}
\be \label{EinsteinHilbert}
\int_\mathcal{M} [dg][d\Phi] e^{-S_{\text{EH}} - S_{\text{SM}}} \, .
\ee
It corresponds to a full path integral over spacetime manifolds $\mathcal{M}$, including all possible topologies and metrics, as well as over the matter fields. However, such an integral is too complicated and it cannot be mathematically defined in a rigorous way.

One way to make better sense of the above path integral is to replace the integration over continuous geometries with a summation over discretized versions thereof, that is,
\be \label{EinsteinHilbert2}
\int_\mathcal{M} [dg] \rightarrow \sum_{\text{discretized versions } of \mathcal{M}}  \, .
\ee
In this approach, the actions $S_{\text{EH}}$ and $S_{\text{SM}}$ are also replaced by their discretized versions. In fact, the idea of approximating the spacetime manifold with a discretized version, obtained by gluing together elementary building blocks such as simplices (the discretized version is then a triangulation), comes back to Regge \cite{Regge} in the context of classical GR. Once working in the discrete setup, one first needs to decide the weights that should be used in the sum over triangulations. Then, one needs to understand the combinatorial properties of these different triangulations. Finally, one needs to define a ``continuum limit" to go back to continuous geometries from the discrete ones.

The discrete approach is expected to be difficult. For instance, geometry and topology become more and more complicated as the number of dimensions increase. In particular, there is no simple classification of topological manifolds in dimensions higher than two. Nevertheless, as we now put forward, impressive results have been observed in dimension two with the help of matrix models and progress has been made in higher dimensions with tensor models.

Random matrix models offer an elegant definition of quantum gravity in dimension two within the discrete approach \cite{DiFran}. The Feynman perturbative expansion of these models can indeed be interpreted as a sum over graphs embedded on surfaces (or ribbon graphs), which precisely provide the required notion of discretized surfaces. In addition, the weight of each Feynman embedded graph is fixed by the Feynman rules, which depend on the details of the models. Matrix models therefore correspond to canonical theories of discretized surfaces. 

As usual in QFT, perturbative expansions are by nature divergent: there are too many Feynman graphs to sum over. However, matrix models simplify in the large $N$ limit. In this limit, the perturbative expansion can be rearranged as an expansion in powers of $1/N$, indexed by the genus \cite{tHooft}, which is a topological invariant for connected closed surfaces. In the large $N$ limit, the Feynman embedded graphs that dominate are the planar graphs, which have genus zero. In the context of two-dimensional quantum gravity, the family of planar graphs can be moreover studied analytically \cite{BrezAl}. The perturbative expansion restricted to planar graphs becomes convergent and one can then define a consistent continuum limit by tuning the coupling constants to some critical value.

The success of matrix models for two-dimensional quantum gravity consequently led to an extension of the results to higher dimensions, including the physically relevant case of dimension four. To this effect, random tensor models are natural candidates \cite{QG} because they generalize the Feynman perturbative expansion of matrix models to a sum over higher dimensional discretized geometries. The lack of any power counting argument similar to the $1/N$ expansion of matrix models however prevented any further progress. 

This was until new graph theoretical tools were invented, which led Gurau to discover the first $1/N$ expansion \cite{LargeNColored} for colored tensor models \cite{Colored}, which was later extended to the more general uncolored tensor models \cite{BGR}. In the case of tensor models, $N$ corresponds to the range of the indices of the tensors. The original $1/N$ expansion of tensor models is governed by a new quantity, called the Gurau degree, which plays the role of the genus in higher dimensions, although it is not a topological invariant. In the large $N$ limit, the Feynman graphs that dominate have Gurau degree zero and are called melonic graphs \cite{Melons}. The family of melonic graphs can be evaluated analytically. Just like matrix models, the perturbative expansion restricted to melonic graphs is convergent and one can therefore define a continuum limit to continuous higher-dimensional geometries. 

When going from matrix models to tensor models, one would \textit{a priori} expect the complexity of the large $N$ limit to increase. However, surprisingly, the family of melonic graphs is more restricted than the family of planar graphs. In particular, the continuous geometries generated by the melonic graphs in the continuum limit are more akin to tree-like geometries. From the point of view of quantum gravity, this does not seem satisfying. An important current research topic is to find extensions of the original $1/N$ expansion that include more Feynman graphs in the large $N$ limit, as initiated in \cite{BonzomNewLargeExp}. In this way, one can hope to describe more elaborate geometries in the continuum limit and ultimately uncover some aspects of quantum gravity; a programme which has been nicknamed the tensor track \cite{TensorTrack}.

\subsubsection{Recent developments: SYK model}

In a series of interesting recent developments, a quantum mechanical model of fermions has been shown to reproduce some important features expected for quantum models of black holes in holography. Inspired by the Sachdev-Ye model \cite{SY1}, originally introduced in the context of condensed matter physics to study spin glass phase transitions, and by related studies \cite{SY2}, Kitaev proposed a quantum mecha\-nical model of $N$ Majorana fermions coupled through quartic all-to-all random interactions \cite{KitaevSYK}. The Hamiltonian of the so-called Sachdev-Ye-Kitaev (SYK) model is given by
\begin{equation}\label{SYK}
H_{\text{SYK}} = \sum_{1\leq i<j<k<l\leq N} J_{ijkl}\,\psi_{i}\psi_{j}\psi_{k} \psi_{l} \, ,
\end{equation}
where the $\psi_i$, $i=1,\ldots,N$, are $N$ Majorana fermionic operators satisfying \smash{$\{\psi_i,\psi_j\}=\delta_{ij}$} and the interaction coupling constants are arranged in a real antisymmetric tensor $J_{ijkl}$ with indices ranging from $1$ to $N$. Moreover, the coupling constants are chosen to be random variables drawn from a Gaussian distribution with properties
\begin{equation}\label{SYKGaussian}
\mu=\bigl\langle J_{ijkl} \bigr\rangle=0 \, , \quad \quad \quad \sigma^2=\bigl\langle J_{ijkl} ^2\bigr\rangle \, = \frac{3!J^2}{N^{3}}\, ,
\end{equation}
for some parameter $J$, where $\mu$ denotes the mean value of the distribution and $\sigma^2$ its variance. We note that such a model in fact originally appeared in the context of nuclear physics under the name of fermionic Gaussian embedded ensemble \cite{SYKNuclear1,SYKNuclear2}.

The random couplings $J_{ijkl}$ are chosen to be time-independent, a case known as quenched disorder in the condensed matter literature \cite{Quench} (more generally, the couplings must be constant on the timescale over which the fields $\psi_i$ fluctuate). This has important consequences on the way one must average over the disorder, that is to say, over the random couplings with distribution \eqref{SYKGaussian}. More precisely, in models with quenched disorder, each observable \textit{a priori} depends on the random couplings, including the free energy $F$ given by
\begin{equation}\label{SYKFreeEnergy}
F(N, J_{ijkl})= -\log \int  \mathcal{D} \psi_i \, e^{- S(\psi_i, N, J_{ijkl})} \, ,
\end{equation}
where $S$ is the (Euclidean) action. However, in the large $N$ limit, which corresponds to the limit of a large system, some physical quantities, called self-averaging, do not depend on the disorder. Thus, averaging these quantities over the disorder leads to a result that agrees with their disorder-independent value. From the point of view of distributions, it is equivalent to say that in the large $N$ limit, self-averaging quantities are peaked around their disorder-averaged value with a variance tending to zero. An example of self-averaging quantity is the free energy \eqref{SYKFreeEnergy} \cite{Quench}; hence, we are interested in evaluating the following, quenched averaged quantity in the large $N$ limit
\begin{equation}\label{SYKFreeEnergyAveraged}
\bigl\langle F\bigr\rangle= \int \frac{dJ_{ijkl}}{\sqrt{2\pi\sigma^2}} \, e^{-\frac{J_{ijkl}^2}{2\sigma^2}}\, F(N,J_{ijkl}) \ .
\end{equation}
Furthermore, since the free energy is a self-averaging quantity, the same holds for all the connected correlation functions.

Before discussing the relevant properties of the SYK model, we describe two possible generalizations. A first generalization consists in promoting the Majorana fermions to complex fermions $\psi_i, \psi_i^\dagger$ satisfying the creation-annihilation anti-commutation relations $\{\psi_i,\psi_j^\dagger\}=\delta_{ij}$. The resulting model, called complex SYK \cite{ComplexSYK}, has Hamiltonian given by 
\begin{equation}\label{complexSYK}
H_{\text{SYK,complex}} = \sum_{1\leq i<j<k<l\leq N} J_{ijkl}\,\psi_{i}^\dagger\psi_{j}^\dagger\psi_{k} \psi_{l} \, ,
\end{equation}
where the real random interaction couplings $J_{ijkl}$ satisfy $J_{ijkl}=-J_{jikl}=J_{klij}$ to ensure that the Hamiltonian is Hermitian. We remark that working with complex fermions allows one to add a non-trivial mass term of the form $m\sum_{1\leq i\leq N} \psi_i^\dagger \psi_i$ in the Hamiltonian, yielding an extra parameter $m$ in the model. This approach has been studied in detail in \cite{FrankPlus}, though in a different but related framework introduced below. 

A second generalization corresponds to enlarging the quartic interactions of the SYK model to interactions between an even number $q$ of Majorana fermions at a time \cite{SYKq}. The Hamiltonian of such models writes
\begin{equation}\label{SYKq}
H_{\text{SYK},q}=i^{q/2}\sum_{1\leq i_1<i_2<\cdots<i_q\leq N} J_{i_1i_2\cdots i_q}\psi_{i_1}\psi_{i_2}\cdots \psi_{i_q} \, ,
\end{equation}
where the interaction couplings are now arranged in a real antisymmetric Gaussian random tensor $J_{i_1i_2\cdots i_q}$ with statistics
\begin{equation}\label{SYKqGaussian}
\mu=\bigl\langle J_{i_1i_2\cdots i_q} \bigr\rangle=0 \, , \quad \quad \quad \sigma^2=\bigl\langle J_{i_1i_2\cdots i_q} ^2\bigr\rangle \, = \frac{(q-1)!J^2}{N^{q-1}}\, .
\end{equation}
As for the factor $i^{q/2}$, it ensures that the Hamiltonian is Hermitian when $q/2$ is odd. In these models, the parameter $q$ allows for the study of two interesting limits where the models simplify, namely the $q=2$ and the $q\rightarrow\infty$ limits \cite{SYKq}. 

For the sake of completeness, we mention that other generalizations of the SYK model have also appeared in the literature, including two-dimensional \cite{2dSYK}, colored \cite{ColoredSYK} and supersymmetric \cite{SUSYSYK} versions.

\subsubsection{Properties of the SYK model}

One of the most important features of the SYK model and its generalizations lies in the structure of the large $N$ leading order Feynman diagrams, obtained in perturbation theory after averaging over the disorder. These leading order Feynman diagrams are similar to the family of melonic graphs found in tensor models. In particular, like in tensor models, they can be resummed explicitly and thus, the model becomes solvable at any value of the coupling, including in the strong coupling regime. It is interesting to point out that even though the fundamental degrees of freedom of the SYK model are fermionic vectors, the structure of the melonic graphs departs from the structure of the large $N$ leading order Feynman diagrams found in usual vector models, which correspond to bubble or tadpole graphs. The difference comes from the type of interactions used in the SYK model, which consist in four fermionic vectors $\psi_i$ contracted with a Gaussian random tensor $J_{ijkl}$. Then, averaging over the disorder results in a pairing of the vertices in Feynman diagrams and leads to the melonic dominance when $N\rightarrow\infty$ \cite{KitaevSYK,SYKq,BNT}.\footnote{We remark that even though the structure of the melonic graphs found in the SYK model differs from the structure of bubble graphs obtained in usual large $N$ vector models, the former can also be described effectively by bubble graphs with bi-local interaction vertices. This bi-local structure is essential for the derivation of the interesting properties of the SYK model and stands in contrast to the tadpole-like structure of the bubble graphs.}

The large $N$ melonic dominance of the SYK model is responsible for many of its remarkable properties. In particular, the summability of the melonic graphs allows one to compute, in principle, all the (disorder averaged) connected correlation functions with the help of Schwinger-Dyson equations. In the following, we briefly review those properties, how they are obtained and their interpretation in the context of holography. More details can be found in \cite{KitaevSYK,SYKq,SYK1,PaoloThesis}.

Firstly, in the case of the connected Euclidean $2$-point function, analytical progress can be made in the strong coupling regime, or equivalently, in the infrared (IR) regime $J|\tau|\gg1$, where $|\tau|$ is the (Euclidean) time interval between the two insertions. In this limit, the corresponding Schwinger-Dyson equation becomes invariant under time reparameterizations. Furthermore, it is solved by a power-law ansatz of the form
\begin{equation}\label{SYK2Point}
G(\tau)=\frac{b}{|\tau|^{2\Delta}}\,\text{sgn}(\tau)\, , \quad \quad \Delta=\frac{1}{4} \, ,
\end{equation}
for some constant $b$, where $\text{sgn}(\tau)$ denotes the sign function. It means that the SYK model flows to a conformal IR fixed point where the fermions acquire a scaling dimension \smash{$\Delta=1/4$}. It also means that the time reparameterization symmetry is spontaneously broken down to $\text{SL}(2,\mathbb{R})$ by the solution \eqref{SYK2Point}, and the associated pseudo-Goldstone modes can be shown to be governed at low energy by a Schwarzian action \cite{SYKq}. Interestingly, a similar situation arises in gravity when one considers the near-horizon geometry of near-extremal black holes. The relevant gravitational effects can be described by Jackiw-Teitelboim dilaton gravity \cite{JT,AdS2}, which displays the same symmetry breaking pattern and the same low energy effective action \cite{MaldaStanYang}. Hence, the SYK model seems to provide a setup for a holographic duality of the type $\text{NAdS}_2/\text{NCFT}_1$, where N stands for ``nearly" because of the symmetry breaking pattern \cite{SYKq}. 

Secondly, one can obtain the finite temperature IR $2$-point function from \eqref{SYK2Point} by mapping the real line to the thermal circle using the time reparameterization symmetry. In one dimension, there is however an infinite number of such maps so that a prescription needs to be found. In \cite{PaoloThesis}, it is shown that if one requires \eqref{SYK2Point} to remain invariant under time translations after a time reparameterization, as expected for a time independent Hamiltonian, then there exists a unique map from the real line to the thermal circle, namely the tangent map $\tau\rightarrow \tan \frac{\tau \pi}{\beta}$, where $\beta=1/T$ is the inverse temperature. The resulting finite temperature IR $2$-point function is then given by 
\begin{equation}\label{SYK2PointTh}
G_\beta(\tau)= b\biggl[\frac{\pi}{\beta \sin \frac{\tau \pi}{\beta}} \biggr]^{2\Delta} \text{sgn}(\tau)\, ,
\end{equation}
where $\tau \sim \tau +\beta$. One can then access the real Lorentzian version of the above $2$-point function by analytic continuation $\tau\rightarrow-it$, which turns the $\sin \frac{\tau \pi}{\beta}$ factor into $\sinh \frac{\tau \pi}{\beta}$, yielding an exponential decay at late times. This behavior for the thermal $2$-point function corresponds to a thermalization process. In the context of holography, it is the expected behavior for the quantum dual of a black hole. Indeed, any small perturbation in a black hole geometry eventually decays with damped oscillations, a phenomenon known as quasi-normal behavior \cite{QNBehavior1,QNBehavior2}. According to the gauge/gravity correspondence, a black hole in AdS corresponds to a thermal state in the dual gauge theory. Hence, the decay of a perturbation of a black hole is interpreted as a return to thermal equilibrium for the dual thermal state, which is what is observed in \eqref{SYK2PointTh} for the SYK model. 

Thirdly, one can evaluate the thermal entropy of the SYK model in the low temperature (or infrared) regime $J\beta\gg1$ by plugging the finite temperature solution \eqref{SYK2PointTh} into the free energy. This is usually done by first rewriting the free energy as an effective functional integral over bi-local fields and then by evaluating the saddle point in the large $N$ limit \cite{SYKq,EffectiveSYK}. As a result, one observes that the SYK model displays a non-zero zero-temperature entropy:
\begin{equation}\label{SYKEntropy}
\lim_{\beta\rightarrow\infty} \lim_{N\rightarrow\infty} S(\beta) \sim N \, .
\end{equation}
In particular, the entropy is of order $N$. This is rather atypical for usual quantum mechanical models, which have a small number of grounds states; hence, a small zero-temperature entropy. In contrast, the SYK model exhibits a macroscopically large ground state degeneracy, which is possible because we take the large $N$ limit first at fixed temperature \cite{SYK1}. This property is in fact in good agreement with the expected properties for the quantum dual of a black hole in holography, since extremal black holes have a large finite entropy at zero temperature, proportional to the area of their event horizon. 

Finally, an analytical treatment of the connected $4$-point function can be made in the IR or strong coupling regime. In this case, the melonic dominance implies that one has to sum over ladder diagrams \cite{SYKq,SYK1}. The careful evaluation of the $4$-point function can then be used fruitfully as a probe for ``quantum chaos", in a way that we now briefly sketch.

In classical dynamical systems, chaos is characterized by the sensitive dependence of the solutions on small changes in the initial conditions. More precisely, in classical chaotic systems, the separation $\delta x(t)$ between two trajectories, initially separated by $\delta x(0)$, grows exponentially in time \cite{ClassicalChaos}
\begin{equation}\label{ClassicalChaos}
\delta x(t) \sim \delta x(0) \, e^{\lambda_L t} \, ,
\end{equation}
where $\lambda_L>0$ is called the Lyapunov exponent and describes the exponential rate of expansion. This chaotic behavior can also be written in terms of the Poisson bracket as
\begin{equation}\label{ClassicalChaos2}
\frac{\partial x(t)}{\partial x(0)} = \{x(t), p(0) \} \sim e^{\lambda_L t} \, ,
\end{equation}
where $x$ and $p$ are the position and the momentum respectively. 

On the other hand, a similar approach fails in quantum systems because there is no well-defined notion of trajectory, due to the uncertainty principle. Hence, one is led to look for purely quantum mechanical criteria that can distinguish between two types of dynamics, regular or chaotic, and which match the classical notion of chaos as $\hbar\rightarrow 0$. Such criteria in fact exist, some of them being based on the statistics of the energy spectra and its relation with random matrix theories \cite{QuantumChaos}. In the context of the SYK model, the relevant notion of ``quantum chaos" is however different. Recently, another diagnostic for the presence of chaos in thermal quantum systems has been proposed \cite{QChaos,MSS}, based on its resemblance, in the semiclassical limit $\hbar\rightarrow0$, with the classical definition of chaos in terms of the Lyapunov exponent. This diagnostic relies on two related quantities, called the commutator squared (CS) and the out-of-time-ordered correlator (OTOC).

The CS is obtained from the quantum counterpart of \eqref{ClassicalChaos2} with the usual replacement $\{ \cdot , \cdot\} \rightarrow \frac{1}{i\hbar} [ \cdot ,\cdot ]$, then by squaring the commutator (to avoid phase cancellations) and finally by averaging over the thermal ensemble:
\begin{equation}\label{CS1}
- \bigl\langle [ x(t), p(0)]^2 \bigr\rangle_\beta \, ,
\end{equation}
where $\bigl\langle \cdot\bigr\rangle_\beta = Z^{-1} \tr (e^{-\beta H} \cdot)$ is the thermal expectation value, $Z$ the partition function and $H$ the Hamiltonian. This quantity was first introduced in \cite{OldCS}, though in a different context related to superconductivity. The discussion can be generalized to any pair of Hermitian operators $V$ and $W$, so that one is led to study the following quantity
\begin{equation}\label{CS2}
C(t) = - \bigl\langle [ W(t), V(0)]^2 \bigr\rangle_\beta \, .
\end{equation}
A quantum system is said to be chaotic if this quantity grows exponentially in time \cite{QChaos}:
\begin{equation}\label{QuantumChaos1}
C(t) \sim \frac{1}{N^2} \, e^{2\lambda_L t} \, ,
\end{equation}
where the factor $2$ comes from squaring \eqref{ClassicalChaos2} and $N$ corresponds to the number of degrees of freedom in the system. This behavior is expected for times $t_d \ll t \ll t_\star$, where $t_d\sim\beta$ is the dissipation time (i.e.\ the characteristic time of the exponential decay of the thermal $2$-point function) and $t_\star\sim \lambda_L^{-1} \log N$ is the scrambling time (i.e.\ the characteristic time of the spreading of quantum information across the whole system \cite{Scrambling}). We remark that the exponential growth of the CS is usually well-defined when there is a large separation between the dissipation time and the scrambling time. 

Besides, the OTOC is defined as
\begin{equation}\label{OTOC}
F(t) = \bigl\langle V(0)W(t)V(0)W(t)\bigr\rangle_\beta \, ,
\end{equation}
and is related to the CS \eqref{CS2} by expanding the commutator squared. For $t_d \ll t \ll t_\star$, it can be argued that the growth \eqref{QuantumChaos1} of the CS is due to a decrease of the OTOC \eqref{OTOC}. More precisely, in chaotic thermal quantum systems, one expects the OTOC to behave as \cite{QChaos}
\begin{equation}\label{QuantumChaos2}
F(t) \sim f_0 - \frac{f_1}{N^2} \, e^{2\lambda_L t} \, ,
\end{equation}
where $f_0$ and $f_1$ are constants of order one. This behavior thus provides an equivalent diagnostic for chaos in thermal quantum systems.\footnote{We note that for regularization concerns in quantum field theories, a convenient prescription is to consider the following CS, $- \bigl\langle y^2[ W(t), V(0)]y^2[ W(t), V(0)] \bigr\rangle_\beta$, where $y=Z^{-1/4}e^{-\beta H/4}$, and the following OTOC, $\bigl\langle yV(0)yW(t)yV(0)yW(t)\bigr\rangle_\beta$, which corresponds to displacing the operator insertions around the thermal circle \cite{QChaos}. We do not enter into these details as they go beyond the scope of this introduction.} 

In the context of holography, recent progress has been made in evaluating the CS and the OTOC for thermal states in large $N$ conformal gauge theories and for their holographic dual, black holes in AdS \cite{QChaosAppl}. In both cases, a chaotic behavior has been observed with a Lyapunov exponent given by $\lambda_L=2\pi/\beta$. Based on these results, it was then conjectured in \cite{MSS} that thermal states with holographic duals are the most chaotic systems. In other words, there exists a maximal bound (hereafter called MSS bound) on the Lyapunov exponent in any thermal quantum systems with many degrees of freedom:
\begin{equation}\label{MSS}
\lambda_L\leq \frac{2\pi}{\beta} \, ,
\end{equation}
which is saturated by thermal states in CFTs dual to AdS black holes. Arguments, based on reasonable physical assumptions, are provided in \cite{MSS} to establish the MSS bound. It is also argued that the saturation of the MSS bound is a strong indication of the presence of a graviational dual. 

Returning to the evaluation of the connected $4$-point function in the SYK model, the results of \cite{KitaevSYK,SYKq} reveal that in fact, the OTOC exactly behaves, at large real time, as \eqref{QuantumChaos2} with a Lyapunov exponent that saturates the MSS bound \eqref{MSS}. In this sense, the SYK model (and its generalizations) is ``maximally chaotic". This very non-trivial property, together with the previous ones, therefore suggest that the SYK model may provide a quantum description of the near-horizon, low energy limit of near-extremal black holes in the context of holography. 

The interesting gravitational features of the SYK model came however as a surprise from the point of view of usual setups in holography and string theory. Indeed, the SYK model is non-standard because it does not correspond to a genuine quantum model, but rather to a statistical ensemble of such models. For that reason, the emergent bulk gravitational theory of the SYK model, if it exists, is likely to be unconventional.

We emphasized at the beginning of the discussion that the key element responsible for the gravitational features of the SYK model is the melonic dominance at large $N$. It is quite fascinating to observe that the melonic graphs, which initially appeared in a different context with random tensor models, end up being crucial for the SYK model and holography. On the other hand, this also suggests that one can construct genuine tensor quantum mechanical models with the same holographic properties as the SYK model. We now describe this program, nicknamed ``holographic tensors" in \cite{HoloTensors}.

\subsubsection{Link between the SYK model and tensor models}

The similarity between the melonic graphs of the SYK model and the melonic graphs of tensor models was first pointed out by Witten \cite{Witten}. He proposed to eliminate the quenched disorder of the SYK model by considering a quantum mechanical model based on colored tensor models. This model is often referred to as the Gurau-Witten (GW) model. It was shown in \cite{GW}, using the graph theoretical tools of tensor models, that the GW model admits a $1/N$ expansion governed by the Gurau degree and dominated by melonic graphs in the large $N$ limit. The large $N$ solution of the GW model is then almost identical to the one obtained in the SYK model; in particular, it captures the same gravitational features. 

A version of the GW model based on uncolored tensor models was then later developed by Klebanov and Tarnopolski \cite{KT}, using results from the work of Carrozza and Tanasa \cite{CT} in the context of tensor models. The so-called CTKT model is based on a Hamiltonian of the form
\begin{equation}\label{CTKT}
H_{\text{CTKT}}=\sum_{\substack{{a_1, a_2, a_3 \, \ }\\ {b_1, b_2, b_3=1}}}^N \frac{\lambda}{4N^{3/2}}\psi_{a_1a_2a_3}\psi_{a_1b_2b_3}\psi_{b_1a_2b_3}\psi_{b_1b_2a_3} \, ,
\end{equation}
where $\psi_{a_1a_2a_3}$ is a real fermionic random tensor of rank three, with indices ranging from $1$ to $N$, and $\lambda$ is the coupling constant. 

We turn our attention to an important remark which is at the root of many results presented in this thesis. It was explained earlier that melonic graphs generically dominate the $1/N$ expansion of tensor models. However, no distinction was made at this point between the melonic graphs of colored tensor models and the ones of uncolored tensor models. The leading Feynman graphs of the SYK model and the ones of the GW model both correspond to melonic graphs of colored tensor models, but they do not correspond to melonic graphs of uncolored tensor models. A crucial difference can be seen at the level of the $2$-point function of the models: in the first case, the $2$-point function is bi-local in time while in the second case, it corresponds to a tadpole. In the CTKT model, the $1/N$ expansion of the original uncolored tensor models is in fact extended to include more Feynman graphs in the large $N$ limit. For instance, the interaction used in the model, called the tetrahedric interaction, can contribute at large $N$ in the CTKT models but it cannot in the original models. Because of this extension, the leading Feynman graphs of the CTKT model, called generalized melonic graphs, also yield the SYK physics. 

The use of random tensors to build quantum mechanical models \`a la SYK, which are referred to as SYK-like tensor models, has led to many results in the literature \cite{SYKLike,NewKleba} (see also the reviews \cite{HoloTensors,KlebaLargeN}). These models correspond to genuine quantum mechanical models, since a random coupling is not necessary anymore in order to achieve the large $N$ melonic dominance. This is an advantage over the SYK model because it may make the holographic interpretation clearer. Besides, tensor models have a much richer structure than the SYK model, for which the interaction consists in a single tensor saturated with vectors. Indeed, even though the large $N$ melonic dominance is achieved in both models, the richness lies in the structure of the subleading terms \cite{SYKCorr}. 

The connection between random tensor models and quantum models of black holes in holography opens up a promising research direction \cite{HoloTensors}. In this context, one important issue is how to relate the initial motivation for tensor models, which is concerned with random higher dimensional geometries, with holography where the bulk gravitational dual is fixed and emerges from the boundary dual theory. A step towards this direction was proposed in \cite{SYKRandomTrees} where QFTs are studied on random trees, the latter providing a natural way to randomize the fixed time of the SYK-type models. 

\subsubsection{Large $D$ limit of matrix models}

From the point of view of usual setups in string theory and holography, the SYK model with quenched disorder and SYK-like tensor models still remain rather unconventional. As mentioned earlier, quantum gauged theories based on matrices are singled out in this framework. It was then realized by Ferrari \cite{FrankLargeD} that the large $N$ melonic dominance can also be relevant in matrix quantum mechanics. 

The models introduced in \cite{FrankLargeD} are based on $D$ matrices of size $N\times N$, which can be interpreted as transverse excitations of strings $(X_\mu)_{ab}$. The symmetry group associated with the matrix indices is then $\uN$ and the one associated with the transverse directions to the branes is $\oD$. In the large $N$ limit, the models correspond to the usual planar limit of matrix models, which is too difficult in general. But now, the additional parameter $D$ in the models allows one to consider the large $D$ limit of the sum over planar graphs. Until recently, the large $D$ limit had not been able to reproduce the typical properties of black holes because it suppresses too many planar Feynman graphs, yielding a physics similar to vector models \cite{LargeDStandard}. It was then shown in \cite{FrankLargeD} that the large $D$ limit can actually be enhanced in such a way that the generalized melonic graphs of the CTKT model also dominate the sum over planar graphs. This class of Feynman graphs is much larger than the one for vector models and in particular, it yields the SYK physics, as explained above.

From the dual gravitational point of view, the large $D$ limit of the models in \cite{FrankLargeD} can be interpreted as the limit of large spacetime dimension $d$ in GR (see Figure \ref{Branes}). Such limit has been actively studied by Emparan et al. \cite{Emparan}. Interestingly, the large $d$ limit of GR greatly simplifies the analytical treatment and at the same time keeps the important features of classical black holes, in the same spirit as the large $D$ limit of matrix models. The precise connection between the two limits is however not known at present.

The study of melonic large $D$ matrix models provide an interesting framework to build new, analytically tractable limits in matrix models, which might capture the relevant physics associated with the full sum over planar graphs. Furthermore, these models make an interesting link with usual tensor models because the basic variables can be equivalently interpreted as tensors with three indices having different ranges. Also, their natural interpretation in the context of string theory and holography means that their corresponding emergent gravitational bulk theories could be in principle derived following the lines of \cite{Emergent1, Emergent2}. Finally, they display many interesting new features such as a line of first order phase transitions terminating at a critical point with asymmetric critical exponents \cite{FrankPlus}.

\subsubsection{Non-linear random flows}

As we put forward previously, large $N$ melonic dominance has recently played an important role in different approaches to quantum gravity, from random tensor models to toy models of quantum black holes in holography. Interestingly, it was also realized that melonic graphs can be useful in a totally different context, namely the study of energy cascades in classical non-linear resonant systems \cite{Ref3}. Such models often emerge from non-linear wave equations with weak non-linearities and with highly resonant energy spectra of linearized perturbations. Examples include non-linear Schr\"odinger equations in harmonic potentials \cite{GPRes} and non-linear dynamics in AdS spacetime \cite{AdSRes}. 

A generic Hamiltonian for resonant systems with cubic non-linearities can be written as
\begin{equation}\label{NLResSyst}
H  = \frac{1}{2} \sum_{\substack{{j,j',k,k'=0}\\ {j+j'= k+k'}}}^\infty C_{jj'kk'} \bar \alpha_{j}(t) \bar \alpha_{j'}  (t)\alpha_k (t) \alpha_{k'} (t)  \, .
\end{equation}
In this expression, the $\alpha_j (t)$, with $j\geq 0$, correspond to an infinite set of complex-value functions of time. They can be interpreted as the Fourier amplitudes of linearized normal modes of a weakly non-linear system, which possesses a (linearized) energy spectrum given in terms of non-negative integers, $E_j\sim j$. Such an energy spectrum is called highly resonant in the sense that the difference between any two energies is an integer. Besides, the condition $j+j'=k+k'$ corresponds to a resonance condition and results from a time-averaging procedure to discard fast oscillatory terms \cite{TimeAverage}. Finally, the interaction coefficients are encoded in a real tensor with four indices and with symmetries $C_{jj'kk'}=C_{j'jkk'}=C_{kk'jj'}$. These coefficients encode the physics of the original problem. In particular, they depend on the structure of the linearized normal modes and the specific form of the non-linearity. 

The classical dynamics within the class of systems described by \eqref{NLResSyst} can be very rich in general, ranging from integrability to chaotic behavior depending on the value of the interaction coefficients \cite{QRes,CubicRes}. From a statistical point of view, it is also instructive to consider a generic system of the form \eqref{NLResSyst} whose interaction coefficients and initial conditions are drawn from some random distribution, as it would provide a picture of the expected properties that hold on average for realizations of this generic system.

An interesting connection can be made between the non-linear resonant classical system \eqref{NLResSyst} and the complex SYK quantum mechanical model \eqref{complexSYK} (or more generally fermionic embedded Gaussian ensembles). When quantizing \eqref{NLResSyst}, the complex modes $\bar \alpha_j, \alpha_j$ become creation and annihilation operators $a^\dagger_j, a_j$, which can be chosen as fermionic operators satisfying the standard commutation relation $\{ a_i, a_j^\dagger\}=\delta_{ij}$. Then, by discarding the resonance condition, one obtains a quantum Hamiltonian similar to \eqref{NLResSyst}, where the random interaction coefficients in \eqref{NLResSyst} translate into the quenched disorder of \eqref{complexSYK}. Note also that since the indices of \eqref{NLResSyst} range from $0$ to $\infty$, the quantized version is related to the complex SYK model with $N\rightarrow\infty$. A study of quantized versions of \eqref{NLResSyst}, albeit with the resonance condition, has been carried out in \cite{QRes}. In particular, it is shown that the corresponding quantum resonant systems are solvable in terms of diagonalizing finite-sized numerical matrices, which is then used to study numerically the spectral statistics of the Hamiltonian and its connections to integrability and chaos. 

Returning to the non-linear resonant system \eqref{NLResSyst}, a natural question to investigate is the presence of energy transfers between modes with different energy scales, especially energy transfers from modes of low energies to modes of higher energies. The latter phenomenon is called forward or direct cascade and is directly related to the notion of weak turbulence \cite{WeakTurb}, which often takes place in non-linear wave equation driven by weak non-linearities. In particular, this notion is typical of studies of conservative deterministic systems and has to be contrasted with the distinct notion of dissipative hydrodynamic turbulence. In this setting, weak turbulence is quantified by the growth of Sobolev norms, defined as
\begin{equation}\label{Sobolev}
S_\gamma(t) = \sum_{r=0}^{\infty} r^\gamma \bar\alpha_r (t) \alpha_r(t)\, ,
\end{equation}
where $\gamma\in\mathbb{N}$. In the cases of $\gamma=0$ and $\gamma=1$, the Sobolev norms $S_0(t)$ and $S_1(t)$ correspond to known conserved quantities, which can be interpreted as the particle number and the total energy of the linearized normal modes respectively. On the other hand, the Sobolev norms $S_\gamma(t)$ for $\gamma>1$ are generically not conserved and can be used to quantify energy transfers between modes. In particular, the growth of these quantities provides evidence of direct energy cascade and thus weak turbulence.

Recent progress has been made in evaluating the Sobolev norms \eqref{Sobolev} for perturbative solutions of the non-linear resonant system \eqref{NLResSyst}, when averaged over the random interaction coefficients and the random initial conditions \cite{Ref3}. More precisely, the averaged Sobolev norms admit a perturbative series which can represented in terms of Feynman-like diagrams. Then, in the limit of many initially excited modes of low energy, which plays the role of the large $N$ limit in usual QFTs, this perturbative series is dominated by a specific class of melonic graphs similar to the ones in random tensor models and SYK-type models. In particular, by restricting the perturbative series to these melonic graphs only, one is able to show that the averaged Sobolev norms grow at least within a certain initial time interval, proving the onset of weak turbulence. Once again, it is fascinating to discover that melonic graphs are present in a different context, where they can be used to investigate important physical phenomena.

\section*{About the thesis}
\label{sec:Contribution}

The bulk of this thesis lies at the cross section of random tensor models and the large $D$ limit of matrix models. It focuses on finding new limits for both types of models, which extend the ones already existing in the literature and which yield interesting results from the point of view of physics, in the context presented above, and also from the point of view of combinatorics. In particular, it investigates in detail the melonic dominance observed in both types of models. The principal techniques used throughout this thesis are large $N$ techniques, based on a resummation of the Feynman graphs in perturbation theory, and graph theory, which is crucial in order to study the structure of Feynman graphs. The outline of this thesis is as follows:

\

Chapter \ref{chap:Vector} is devoted to reviewing one of the simplest models that simplify in the large $N$ limit, namely the vector models. These models are a perfect playground to introduce many of the necessary tools for the rest of the thesis. In particular, we explain the basic notions underlying a $1/N$ expansion.

In Chapter \ref{chap:Matrix}, we describe large $N$ matrix models, as they hold a prominent role in the context of this thesis. In addition, working out their $1/N$ expansion and the related combinatorial tools proves to be important in the following chapters. 

We then move on to large $N$ tensor models in Chapter \ref{chap:Tensor}, which are the natural generalization of vector and matrix models. This chapter is mainly based on Ref.\ \cite{Ref2} and corresponds to an important original contribution of this thesis. We first review the original $1/N$ expansion of uncolored tensor models, suitably generalized to the symmetry group $\oN^R$ and to multiply-connected interactions. This requires to generalize the notions introduced in the original models based on the symmetry group $\uN^R$ and on connected interactions. Then, we extend this $1/N$ expansion in a non-trivial way by enhancing the large $N$ scaling of the coupling constants, so that more Feynman graphs are included in the large $N$ limit. This extension generalizes the works of \cite{CT} and \cite{FrankLargeD} for tensors of any rank $R$ and for any interaction term. Moreover, it necessitates new graph-theoretical tools that make an interesting link between matrix models and tensor models. 

Next, we introduce in Chapter \ref{chap:MatrixTensor} matrix-tensor models, which naturally extend the models of \cite{FrankLargeD} to matrices $(X_{\mu_1\cdots\mu_r})_{ab}$ with $r$ indices having the same range $D$. This chapter is also based on Ref.\ \cite{Ref2}. It constitutes an important piece of research from the point of view of the large $D$ limit of matrix models, with interesting applications to quantum models that display the SYK/black hole physics. In these models, the two parameters $N$ and $D$ allow one to consider the large $N$ and the large $D$ limits. When one uses an equivalent of the enhanced scaling for tensor models, the two limits no longer commute and one must take the large $N$ limit first, which corresponds to the usual planar limit of matrix models. Then, the large $D$ limits provides a way to simplify the sum over planar graphs while keeping some of its non-trivial features.

In the complete version of the Ph.D. thesis manuscript, two additional chapters are included, which correspond to the published version of Ref.\ \cite{Ref1} and to the accepted version of Ref.\ \cite{Ref3}. In Ref.\ \cite{Ref1}, the models of \cite{FrankLargeD} are extended to multi-trace interactions terms and to arbitrary correlation functions. In addition, the case of reducing the symmetry of the matrices is discussed. Then, melonic large $D$ matrix models based on bosons and on supersymmetry are considered. In Ref.\ \cite{Ref3}, the question of weak turbulence in non-linear resonant systems, averaged over the interaction coefficients and the initial conditions, is investigated. In particular, the graph-theoretical tools of tensor models are used in order to deduce information about direct energy cascades via the growth of Sobolev norms. Then, it is proved that in the limit of many initially low-lying modes, the melonic dominance implies that at least during a finite time interval, there is a turbulent cascade of energy.

The work in this thesis requires some knowledge on graph theory. A self-contained account of the notions used throughout the thesis can be found in Appendix \ref{app:AppA}, which provides general definitions for abstract graphs and reviews the basic concepts of graphs embedded on surfaces. Besides, Appendix \ref{app:AppC} provides the technical details for the characterization of the leading order graphs in models based on the prime-complete interaction (see Chapter \ref{chap:Tensor}), and Appendix \ref{app:AppD} gives additional details on maximally-single trace interactions (see Chapter \ref{chap:Tensor} as well).

Concluding remarks and research perspectives are provided at the end of Chapters \ref{chap:Tensor} and \ref{chap:MatrixTensor}.

\chapter{Large $N$ vector models}
\label{chap:Vector}

In this chapter, we describe a first class of models that become solvable in the limit of a large number of degrees of freedom: the $\oN$ vector models. As the name suggests, these models are based on a field that transforms in the vector or fundamental representation of the symmetry group $\oN$. Hence, the number of degrees of freedom $\cN$ corresponds to the number of components of the field: $\cN=N$. 

Historically, vector models were the first models for which a large $N$ limit was applied. It was in the context of statistical mechanics for the study of a $N$-component spin-model known as the spherical model \cite{Stan}, and it was later extended to quantum field theory with the important idea of $1/N$ expansion \cite{Wil}. 

The $1/N$ expansion of vector models can be understood as a resummation of the Feynman graphs that appear in perturbation theory. By choosing appropriately the coupling constants in the action, there is a simple class of graphs that dominate when $N$ is large, which can be evaluated analytically. In this sense, the large $N$ limit of vector models is solvable. Equivalently, the large $N$ solution corresponds to a saddle-point approximation in the path integral formalism by introducing an auxiliary field, which is possible because $N$ is large and $\cN=N$.

In this chapter, we focus on the first approach based on Feynman graphs. In other words, we mainly use combinatorial techniques to study the $1/N$ expansion. This requires some general knowledge of graph theory, which can be found in Appendix \ref{app:AppA}.\footnote{The tools reviewed in Appendix \ref{app:AppA} are self-contained. They are used to some extent in this chapter, but they really come into play from Chapter \ref{chap:Matrix} onwards.} Standard references on vector models are, for instance, \cite{Cole, Zi-Ji, Mak}. In particular, the second approach based on the saddle-point approximation is studied in details in these references. The general structure of the $1/N$ expansion is the same for all vector models. It does not depend on the dimension $d$ of spacetime nor on the statistics of the fields, i.e.\ whether they are bosonic or fermionic. Unless otherwise stated, we therefore deal for simplicity with a zero-dimensional QFT based on a real, bosonic vector field. A similar analysis could be applied for, e.g., Dirac fields (such as the Gross-Neveu model in $d=2$), non-linear sigma models and $\uN$ vector models, see \cite{Cole, Zi-Ji, Mak}.

We begin this chapter with a general description of the $\oN$ vector models. Then, we move on to the study of the Feynman graphs obtained in perturbation theory. In particular, we explain the useful stranded representation to depict the Feynman graphs. Afterwards, we introduce the concept of $1/N$ expansion. A pedagogical effort is being made to present the related large $N$ techniques because they are of prime importance in this thesis and they are used repeatedly in the upcoming chapters. The leading sector, which dominates the $1/N$ expansion when $N$ is large, is then studied and the corresponding large $N$ solution is computed explicitly. Finally, we close this chapter with further remarks regarding the $\oN$ vector models, in particular in relation with physics.

%%%%%%%%%%%%%%%%%%%%%%%%%%%%%%%%%%%%%%%%%%%%%%%%

\section{Definition of the models}
\label{sec:VecDefModels}

The basic degrees of freedom of the $\oN$ vector models are the $N$ components of a real vector $\vec{\phi}=(\phi_i)_{1\leq i \leq N}$, which transforms in the fundamental representation of the symmetry group $\oN$. The transformation law in terms of the field components $\phi_i$ is given by
\be \label{VecTransfLaw}
\phi_i \rightarrow \phi_i' = O_{ii'}\phi_{i'} \, ,
\ee
where $O\in\oN$ is a $N\times N$ orthogonal matrix and the sum over repeated indices is implied (which is always the case throughout this thesis).

An invariant action $S(\vec{\phi})$ for the $\oN$ vector models is naturally constructed from powers of $\vec{\phi}^{\, 2}=\phi_i \phi_i$, that is, 
\be\label{VecAction}
S(\vec{\phi})= \frac{1}{2}\vec{\phi}^{\, 2} + \sum_{p\geq2}g_p \bigl(\vec{\phi}^{\, 2}\bigr)^p = \frac{1}{2}\vec{\phi}^{\, 2} + V(\vec{\phi}^{\, 2}) \, ,
\ee
where $\{g_p\}_{p\geq2}$ are the coupling constants. The first term in this power series corresponds to the usual quadratic mass term (the mass can be set to one by a rescaling of the fields) whereas higher order terms correspond to interaction terms, which we regroup in the interaction potential $V(\vec{\phi}^{\, 2})$. For the sake of clarity, we restrict ourselves to a model based on the $\bigl(\vec{\phi}^{\, 2}\bigr)^2$ interaction only, that is, we set $g_p=0$ for $p\geq3$ and we write $g_2=g$. Note however that the logic and many results obtained in this chapter hold in the general case.

The partition function $Z$ and the free energy $F$ associated with the action \eqref{VecAction} are given by the usual expressions
\be \label{VecPartitionFunctionFull}
Z = \exp (-F) = \int [d\vec{\phi}] \, e^{-S(\vec{\phi})}=\int [d\vec{\phi}] \, e^{-\frac{1}{2}\vec{\phi}^{\, 2} - g \bigl(\vec{\phi}^{\, 2}\bigr)^2} \, ,
\ee
where $[d\vec{\phi}]$ is the appropriate path integral measure with respect to $\vec{\phi}$, which reduces in zero dimension to the product of $N$ simple integral measures on $\mathbb{R}$: $[d\vec{\phi}] = \prod_{i=1}^N \frac{1}{\sqrt{2\pi}}d\phi_i$. Similarly, general $n$-point functions are evaluated by the integral
\be \label{VecNPtFunctions}
\bigl\langle\phi_{i_1} \ldots \phi_{i_n}\bigr\rangle = \int [d\vec{\phi}] \, \phi_{i_1} \ldots \phi_{i_n} \, e^{-\frac{1}{2}\vec{\phi}^{\, 2} - g \bigl(\vec{\phi}^{\, 2}\bigr)^2}\, ,
\ee
and the free $n$-point functions $\bigl\langle\phi_{i_1} \ldots \phi_{i_n}\bigr\rangle_0$ are obtained for $g=0$. In particular, the free $2$-point function or free propagator is given by the inverse of the quadratic part, that is, 
\be \label{VecFree2PtFunction}
\bigl\langle\phi_i\phi_j\bigr\rangle_0\,=\delta_{ij} \, .
\ee

In order to define a $1/N$ expansion for vector models, we work in perturbation theory. Hence, we are interested in the perturbative expansion of the partition function (or the free energy) in powers of the coupling constant $g$ (which is assumed to be small). As usual, it is obtained by first Taylor expanding the interaction part of the integrand in \eqref{VecPartitionFunctionFull}, then by commuting the sum with the integral\footnote{This step is actually delicate because the original integral diverges for $g<0$ and we Taylor expand around $g=0$, which belongs to the boundary of the analyticity domain. However, we do not attend to cope with this (important) feature here, see for instance \cite{Renorm, QFT}.}, that is,
\be \label{VecPertExp}
Z=\exp (-F) = \sum_{k\geq0} \frac{(-g)^k}{k!} \int [d\vec{\phi}] \, e^{-\frac{1}{2}\vec{\phi}^{\, 2}} \Bigl(\bigl(\vec{\phi}^{\, 2}\bigr)^2 \Bigr)^k \, ,
\ee
and finally by evaluating the Gaussian integral using Wick's theorem.

\section{Feynman graphs}
\label{sec:VecFeynmanGraphs}

\begin{figure}[]
\centerline{\includegraphics[scale=1]{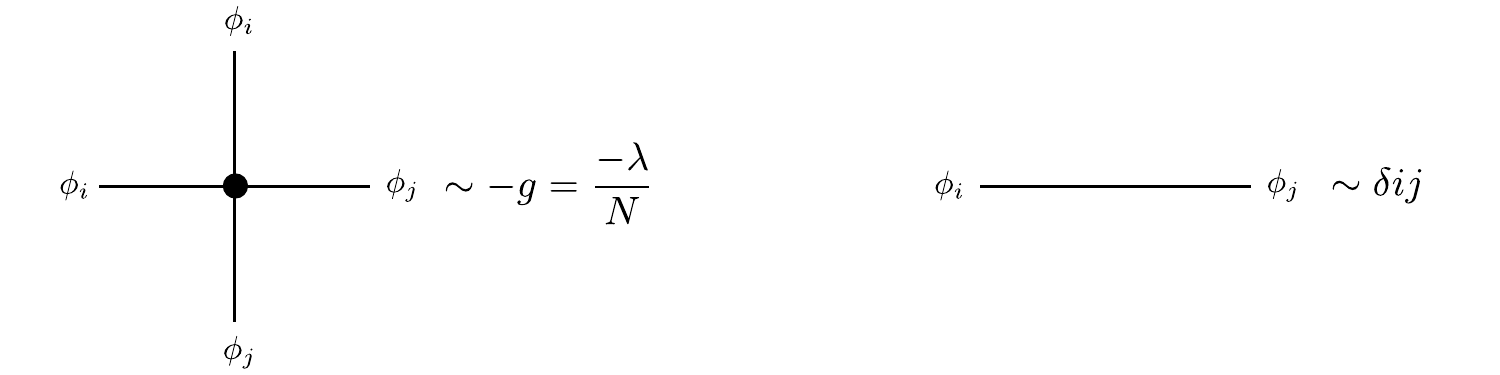}}
\caption{Feynman rules for the vertex and the propagator of the $\oN$ vector models. The coupling parameter $\lambda$ is defined in Eq.\ \eqref{VecOptScaling}.}\label{VecFeynRules}
\end{figure}

The perturbative expansion \eqref{VecPertExp} admits a graphical representation in terms of Feynman graphs. We briefly recall how it is obtained. From \eqref{VecPertExp}, we observe that at fixed order $k$, we need to evaluate the product of $k$ factors $\bigl(\vec{\phi}^{\, 2}\bigr)^2$ via a Gaussian integral. According to Wick's theorem, this is given by the sum over all pairings (or Wick's contractions) of the $4k$ copies of the field components $\phi_i$. We then represent the $k$ copies of $\bigl(\vec{\phi}^{\, 2}\bigr)^2$ in \eqref{VecPertExp} as $k$ vertices of degree four and each pairing between $\phi_i$ and $\phi_j$ as an edge that connects the two vertices associated with the entries $\phi_i$ and $\phi_j$. Finally, to each vertex we assign a factor $-g$ and to each edge (or propagator) between $\phi_i$ and $\phi_j$ we assign a factor $\delta_{ij}$, which corresponds to the free 2-point function or free propagator \eqref{VecFree2PtFunction}. These Feynman rules are illustrated in Figure \ref{VecFeynRules}. As a result, the perturbative expansion \eqref{VecPertExp} can be rewritten as an expansion onto Feynman graphs.\footnote{We remark that the terminology ``Feynman graph" is actually misleading because Feynman graphs are not abstract graphs but in fact embedded graphs \cite{Renorm}. Indeed, when we use Wick's theorem, the field components $\phi_i$ in \eqref{VecPertExp} are distinguished, i.e.\ we can tell them apart. Thus, when we perform the Wick's contractions, the cyclic ordering around each vertex matters. This consideration becomes important when one is involved with the combinatorial problem of counting the number of Feynman graphs. However, for studying the $1/N$ expansion of vector models, it has no relevance (but it does for matrix models, see Section \ref{sec:MatFeynmanGraphs}).} If we further restrict ourselves to the study of the free energy $F$, the perturbative expansion is onto connected Feynman graphs. 

\begin{figure}[]
\centerline{\includegraphics[scale=1]{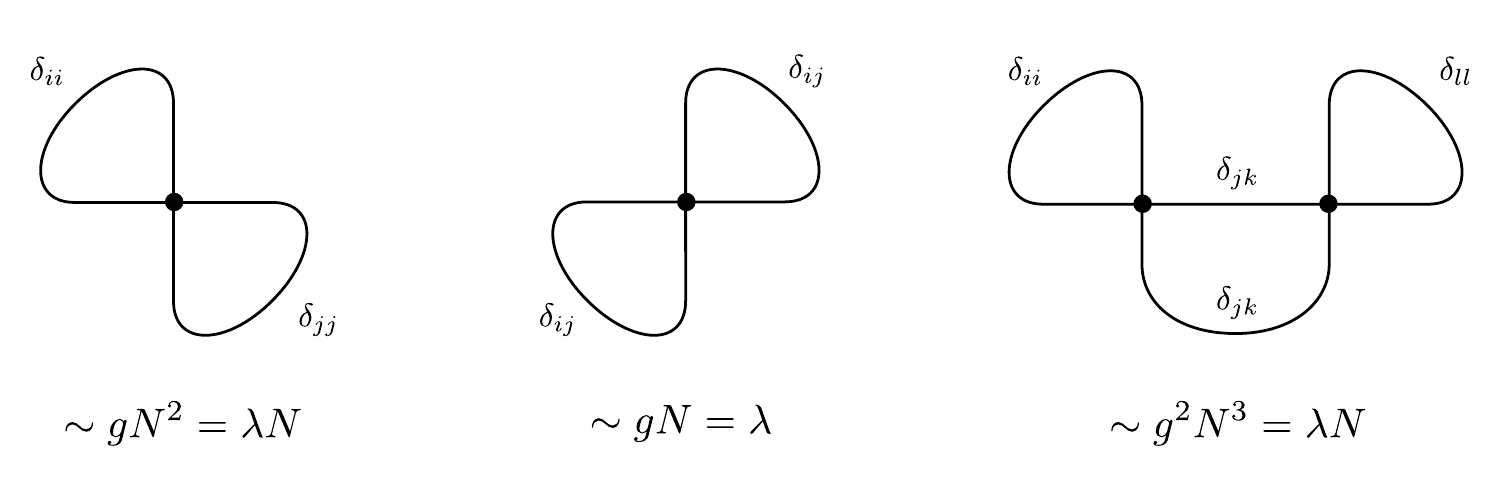}}
\caption{Examples of Feynman graphs with their factor of $g$ and $N$ in the perturbative expansion \eqref{VecPertExpFeyn}. Their scaling is also given in terms of the coupling parameter $\lambda$ defined in Eq.\ \eqref{VecOptScaling}.}\label{VecFeynGraphs}
\end{figure}

We denote by $\cG$ a connected Feynman graph and by $\{\cG_k\}$ the set of connected Feynman graphs with $k$ vertices. Then, the perturbative expansion of $F$ can be written as
\be \label{VecPertExpFeyn}
F=\sum_{k\geq0} g^k F_k\, ,
\ee
where $F_k$ corresponds to a sum over connected Feynman graphs $\cG$ with $k$ vertices (i.e.\ $\cG\in\{\cG_k\}$) weighted by some amplitude $\cA(\cG)$ that includes the contracted deltas, the sign and the combinatorial factors:
\be \label{VecPertExpFeynCoeff}
F_k= \sum_{\cG\in\{\cG_k\}} \cA(\cG) \, .
\ee
Note that the amplitude $\cA(\cG)$ of a Feynman graph $\cG$ necessarily depends on $N$ to some positive integer power because it involves products of contracted deltas. Thus, $F_k$ also depends on $N$. Examples of Feynman graphs are given in Figure \ref{VecFeynGraphs} together with their factor of $g$ and $N$ in the perturbative expansion.

\begin{figure}[]
\centerline{\includegraphics[scale=1]{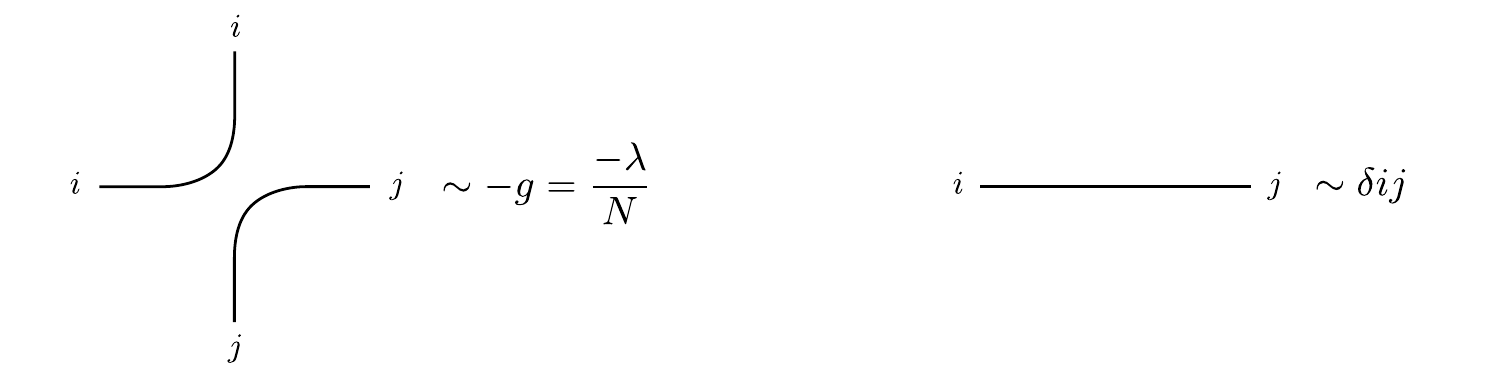}}
\caption{Feynman rules for the vertex and the propagator in the stranded representation. A corner in a vertex corresponds to the link between two paired strands.}\label{VecFeynRulesStrand}
\end{figure}

\begin{figure}[]
\centerline{\includegraphics[scale=1]{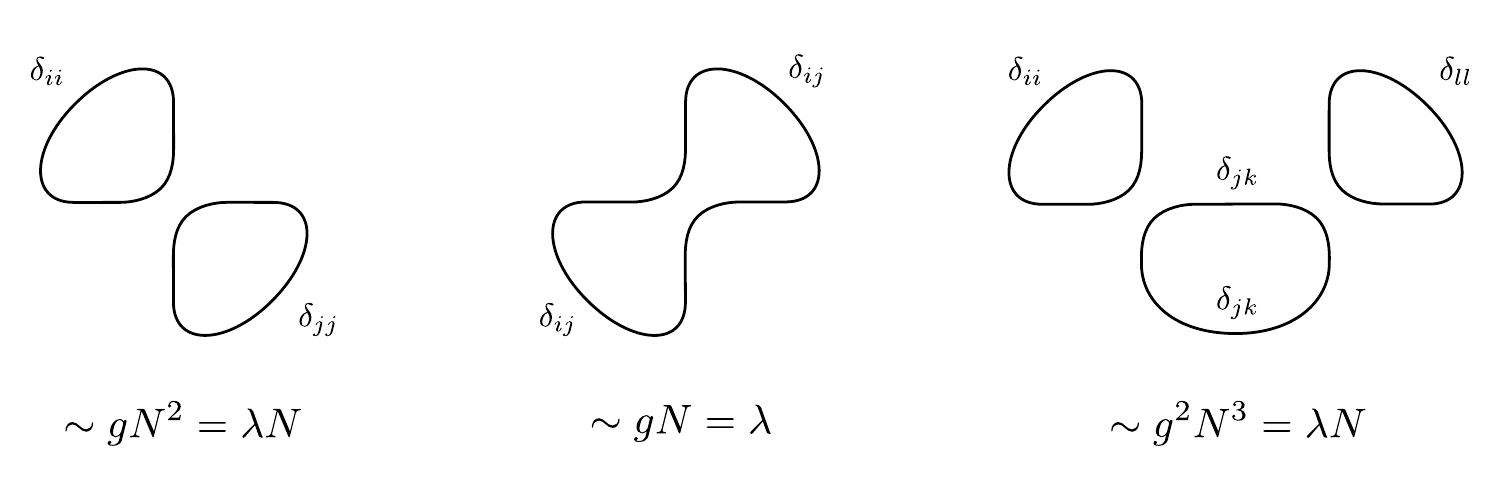}}
\caption{Feynman graphs of Figure \ref{VecFeynGraphs} in the stranded representation. When the coupling constant $g$ is used, the power of $N$ corresponds to the number of closed stranded loops.}\label{VecFeynGraphsStrand}
\end{figure}

There is actually an equivalent way of representing the Feynman graphs $\cG$, which facilitates the study of the dependence on $N$ of their amplitude. In the standard representation, vertices and edges carry information about the contraction pattern of the field components $\phi_i$. A more convenient representation consists in drawing vertices and edges according to the contraction pattern of the corresponding $\oN$ indices. This is called the stranded representation. In this representation, a vertex consists in four strands, one for each field component, such that each strand is assigned the $\oN$ index of the corresponding field component and the two strands carrying the same $\oN$ index are connected together. The link between two paired strands is called a corner. Thus, a vertex in the stranded representation is of degree four and is made of four strands and two corners. As for the edges, they each identify the $\oN$ indices $i$ and $j$ of the strands they connect with a factor $\delta_{ij}$. The stranded representation is illustrated in Figure \ref{VecFeynRulesStrand}. The Feynman graphs of Figure \ref{VecFeynGraphs} are also given in this representation in Figure \ref{VecFeynGraphsStrand}. We stress that even though the vertices now look ``disconnected", they each correspond to a single vertex in $\cG$; in particular, $\cG$ is still connected.

One can check that the two representations are equivalent. From now on, we will essentially work in the stranded representation and we will keep the notation $\cG$ for the connected Feynman graphs. The advantage of working in the stranded representation is that the power of $N$ in the amplitude of $\cG$ corresponds to the number of closed stranded loops (or just loops) in $\cG$, see Figure \ref{VecFeynGraphsStrand}.

Given a connected Feynman graph $\cG$ in the stranded representation, we denote by $p$ its number of edges or propagators, by $v$ its number of interaction vertices and by $\varphi$ its number of loops. The amplitude $\cA(\cG)$ of $\cG$ thus scales with respect to $N$ as $N^\varphi$.

\section{Large $N$ expansion}
\label{sec:VecLargeNExp}

The perturbative expansion \eqref{VecPertExpFeyn} of the free energy $F$ organizes the Feynman graphs according to their power of $g$. As we have seen, these Feynman graphs also depend on $N$. The purpose of the $1/N$ expansion is then to rearrange them according to their power of $N$ rather than their power of $g$. The new expansion parameter is in this case $1/N$. We say that the $1/N$ expansion resums the perturbative expansion. Because the end goal is to take the large $N$ limit, i.e.\ $N\rightarrow\infty$, we need to make sure that the $1/N$ expansion makes sense in this limit. As of now, we noted that the amplitude of a Feynman graph scales like $N^\varphi$, where $\varphi$ is the number of loops. Clearly, this cannot lead to a well-defined result when $N\rightarrow\infty$ since $\varphi>0$. We can however find a way out if we use the other parameter of the model, namely the coupling constant $g$, and scale it appropriately with $N$.

Hence, the first step in constructing a $1/N$ expansion is to decide how the coupling constant $g$ in the action \eqref{VecAction} scales with $N$ when $N\rightarrow\infty$. Equivalently, we need to decide which coupling parameter we keep fixed in this limit. In general, this is a fine-tuning problem: if the coupling constant is not scaled enough, the perturbative becomes trivial when $N\rightarrow\infty$; whereas if it is scaled too strongly, the perturbative expansion blows up when $N\rightarrow\infty$, as in the case of $g$. 

The first requirement for an appropriate scaling of a coupling constant is that it should lead to a well-defined $1/N$ expansion, in the sense that there exists an upper bound on the power of $N$ associated with any Feynman graph. The second requirement is that it should yield a non-trivial $1/N$ expansion, meaning that there is a sufficient number of leading order Feynman graphs, i.e.\ the ones with the highest power of $N$. Typically, we wish to obtain an infinite family of such graphs. A scaling that fulfils these two requirements is called optimal. 

The optimal scaling associated with a given coupling constant is unique. Indeed, since an optimal scaling yields a non-trivial $1/N$ expansion, the corresponding interaction vertex can appear an arbitrary number of times in leading order Feynman graphs. But then, any further enhancement of this scaling would produce Feynman graphs with amplitude proportional to an arbitrarily high power of $N$. As a result, it is impossible to enhance an optimal scaling and still have a well-defined $1/N$ expansion. This ultimately implies that an optimal scaling is unique. 

In vector models, the optimal scaling of $g$ turns out to be $g\sim1/N$.\footnote{In fact, if we had kept all the coupling constants $\{g_p\}_{p\geq2}$ in the action, the optimal scaling would have been the same for all of them: $g_p\sim 1/N$. This is a special feature of vector models.} We thus define a new coupling parameter $\lambda$ in the action \eqref{VecAction} by
\be\label{VecOptScaling}
g\equiv \frac{\lambda}{N}\, ,
\ee 
and we decide to keep $\lambda$ fixed when $N\rightarrow\infty$. The action rewrites in terms of $\lambda$ as
\be\label{VecActionScaled}
S(\vec{\phi})= \frac{1}{2}\vec{\phi}^{\, 2} + \frac{\lambda}{N} \bigl(\vec{\phi}^{\, 2}\bigr)^2\, .
\ee

Regarding a Feynman graph $\cG$, there is now a factor $-\lambda/N$ assigned to each vertex. As a result, the amplitude $\cA(\cG)$ of $\cG$ scales like $N^{-v+\varphi}$ (see Figures\ \ref{VecFeynRulesStrand} and \ref{VecFeynGraphsStrand}). The following theorem confirms that the new scaling \eqref{VecOptScaling} yields a well-defined $1/N$ expansion by analyzing the Feynman graphs.

\begin{theorem}\label{VecTheorem}
Let $\cG$ be a connected Feynman graph of the $\oN$ vector models. Then, $-v+\varphi \leq 1$. In other words, the power of $N$ associated with any connected Feynman graph $\cG$ is bounded above by $1$.
\end{theorem}

We provide two versions of the proof for this theorem. The first one is based on connectivity arguments, which will be useful in later chapters. The second one relies on a redrawing of the Feynman graphs and provides a nice combinatorial interpretation for the $1/N$ expansion.

\proof (First version) $\cG$ is connected. If we break down the degree-four vertices of $\cG$ into their two corners, this yields an effective graph with exactly $\varphi$ connected components. In this process, breaking down a given degree-four vertex into two corners can increase the number of connected components by at most one. Hence, the total number of connected component generated after the whole process is at most $v$, that is, $\varphi-1\leq v$.
\qed

\begin{figure}[]
\centerline{\includegraphics[scale=1]{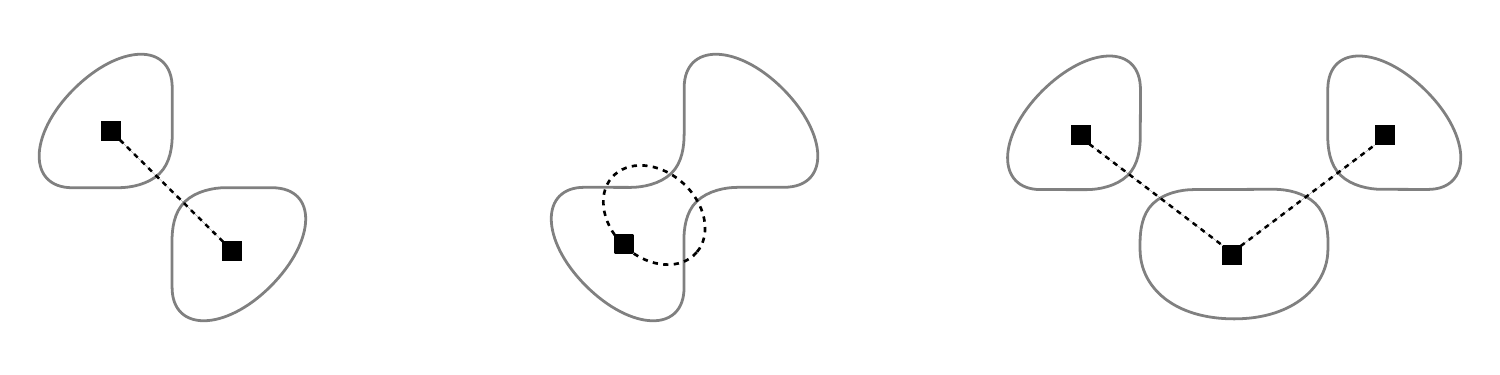}}
\caption{Redrawing of the Feynman graphs $\cG$ of Figure \ref{VecFeynGraphsStrand} into equivalent graphs $\tilde{\cG}$, whose vertices (represented by boxes) correspond to the loops in $\cG$ and whose edges (represented by dashed edges) correspond to the vertices of $\cG$.}\label{VecFeynGraphsRedrawing}
\end{figure}

\proof (Second version) We redraw $\cG$ as follows. First, at each vertex of $\cG$, we draw a (dashed) edge between the two corners. Then, we replace each loop passing through $c$ corners in $\cG$ by a new vertex of degree $c$ whose external (dashed) edges attach to these corners (one external (dashed) edge per corner). This gives rise to a new connected graph $\tilde{\cG}$ made of $V(\tilde{\cG})=\varphi$ vertices of arbitrary degree and $E(\tilde{\cG})=v$ (dashed) edges. The redrawing is illustrated in Figure \ref{VecFeynGraphsRedrawing} for the Feynman graphs of Figure \ref{VecFeynGraphsStrand}. It is clear that the two ways of drawing the Feynman graphs are in bijection. In the new description, the number of loops (more precisely independent cycles) in any Feynman graph $\tilde{\cG}$ is given by the relation $L(\tilde{\cG})=E(\tilde{\cG})-V(\tilde{\cG})+1$. In particular, $L(\tilde{\cG})$ is non-negative. Translated into the original description in terms of the corresponding $\cG$, it implies that $v-\varphi+1\geq0$.
\qed
\\

We first remark that the two versions of the proof are purely combinatorial. Actually, the second version can also be understood from a path integral point of view. It relies on the introduction of an auxiliary field $\sigma\sim\phi_i\phi_i$ that allows one to integrate out the $\vec{\phi}$ field in the path integral and obtain an effective action for $\sigma$ only. The redrawing used in the proof corresponds to the Feynman rules associated with this effective action. For clear reasons, we thus refer to the redrawing of the Feynman graphs as their effective description. Further details regarding the path integral approach can be found in \cite{Cole, Zi-Ji, Mak}.

Second, we emphasize that the number of loops $L$ in the effective description of the Feynman graphs should not be confused with the number of loops $\varphi$ in the original one. Since the two descriptions are in bijection, we will however use the slight abuse of notation $L(\cG)$.

We denote by $\{\cG_L\}$ the set of connected Feynman graph with $L$ loops (in the effective description). Then, the $1/N$ expansion of the $\oN$ vector models with scaling \eqref{VecOptScaling} can be written as
\be \label{VecLargeNExpFeyn}
F=\sum_{L\geq0} N^{1-L} F_L(\lambda)\, ,
\ee
where $F_L(\lambda)$ corresponds to a sum over connected Feynman graphs $\cG$ in $\{\cG_L\}$ weighted by some amplitude $\tilde{\cA}(\cG)$ that includes the factors $\lambda$, the combinatorial factors and the sign:
\be \label{VecLargeNExpFeynCoeff}
F_L(\lambda)= \sum_{\cG\in\{\cG_L\}} \tilde{\cA}(\cG) \, .
\ee
In particular, the $1/N$ expansion is well-defined in an explicit way.\footnote{Formally, for the $1/N$ expansion to be well-defined, each coefficient $F_L(\lambda)$ for fixed $L$ should also be a convergent perturbative series in $\lambda$ with a finite radius of convergence. It is fortunatly the case for vector models. It will be made explicit for $F_0$ in the next section.} Also, the $1/N$ expansion \eqref{VecLargeNExpFeyn} can be nicely interpreted in the effective description as a standard QFT loop expansion. In other words, the $N\rightarrow\infty$ can be interpreted as a classical limit $\hbar\rightarrow0$ that isolates graphs with $L=0$, that is, trees. Finally, it remains to show that the $1/N$ expansion is non-trivial for the scaling \eqref{VecOptScaling} to be optimal. This is a straightforward consequence of the next section.

\section{Leading sector: bubble graphs}
\label{sec:VecLO}

In the previous section, we established the $1/N$ expansion for the $\oN$ vector models. In the large $N$ limit, the leading order Feynman graphs dominate the $1/N$ expansion. The large $N$ solution of the vector models is then obtained by restricting the perturbative expansion to the leading sector only, which is the object of study of this section.

From Eq.\ \eqref{VecLargeNExpFeyn}, we deduce that the leading order (LO) Feynman graphs are characterized by $L=0$ in the effective description, that is, they correspond to trees. In the original description, the LO Feynman graphs are called bubble graphs\footnote{A recursive definition of bubble graphs is given in the next paragraph. One way of interpreting bubble graphs is as follows. According to Theorem \ref{VecTheorem}, leading order Feynman graphs $\cG$ are such that $v=\varphi-1$. Using the argument in the first version of the proof, it means that the two corners of each vertex in $\cG$ belong to two distinct loops in $\cG$. We say that $\cG$ is effectively ``maximally disconnected" with respect to its vertices.}. They form a simple class of graphs, which is tractable analytically in the sense that it can be evaluated explicitly. Examples of LO Feynman graphs include the graph on the left and the one on the right of Figure \ref{VecFeynGraphsRedrawing}. We emphasize that this is a strong result, as very few solvable models are known in QFT. In the following, we describe more precisely the large $N$ solution of the $\oN$ vector models. Unless otherwise stated, we remain in zero dimension and we work with bosonic variables.

\begin{figure}[]
\centerline{\includegraphics[scale=1]{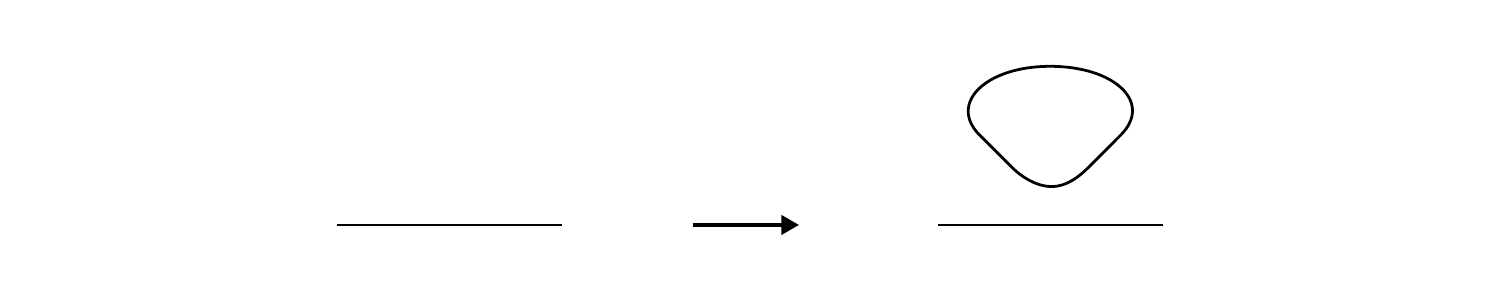}}
\caption{Bubble insertion, which consists in the replacement of an edge in a Feynman graph by an elementary 2-point bubble graph. This operation leaves the power of $N$ unchanged. Bubble graphs can be obtained by recursively applying this operation starting from the Feynman graph with no vertices, which corresponds to a single stranded loop.}\label{VecBubbleInsertion}
\end{figure}

First of all, it is important to realize that the bubble graphs can all be constructed recursively. To see how, we define the elementary bubble graph to be the bubble graph with one vertex, i.e.\ the graph on the left of Figure \ref{VecFeynGraphsStrand}. If we cut one of the two edges in the elementary bubble graph, we obtain a graph with two external (half-)edges. We call the two possible graphs obtained this way elementary 2-point bubble graphs. Then, we define a bubble insertion in a Feynman graph $\cG$ as the replacement of an edge in $\cG$ by one of the two elementary 2-point bubble graphs. This is illustrated in Figure \ref{VecBubbleInsertion}. Inserting a bubble in a Feynman graph does not change its power of $N$. Indeed, it increases the number $v$ of vertices by one and the number $\varphi$ of loops by one. Now, one can check that any bubble graph can be obtained by starting with the Feynman graph with no vertices and then inserting recursively an arbitrary number of bubbles. In the effective description, the equivalent statement is that any tree can be obtained by starting with the tree with no edge and then connecting recursively via an edge an arbitrary number of leaves.\footnote{The leaves of a tree correspond to its vertices of degree one.}

We now describe the large $N$ solution of the $\oN$ vector models. The full connected $2$-point function can be written as
\be \label{VecConn2PtFunction}
\bigl\langle\phi_i\phi_j\bigr\rangle_c \,= \delta_{ij} G(N,\lambda)\, ,
\ee
which follows from the conservation of the $\oN$ indices along the strands of the Feynman graphs. We explain below that $G(N,\lambda)$ is directly related to the free energy of the models, see Eq.\ \eqref{VecConn2PtFunctionFreeEnergy}. In particular, it admits a $1/N$ perturbative expansion onto connected Feynman graphs with two external edges, which can be obtained by cutting one edge in a connected Feynman graph that contributes to the free energy. On the other hand, $G(N,\lambda)$ is related to the connected one-particle irreducible (1PI) $2$-point function $\Sigma(N,\lambda)$ (or self-energy) by the usual relation
\be \label{Vec2PtFunction}
G(N,\lambda) = \frac{1}{1-\Sigma(N,\lambda)}\, ,
\ee
where the free propagator is $1$ in the models at hand. We write the full connected 2-point function and the self-energy restricted to the leading sector as $G_{\text{LO}}(\lambda)$ and $\Sigma_{\text{LO}}(\lambda)$ respectively.

\begin{figure}[]
\centerline{\includegraphics[scale=1]{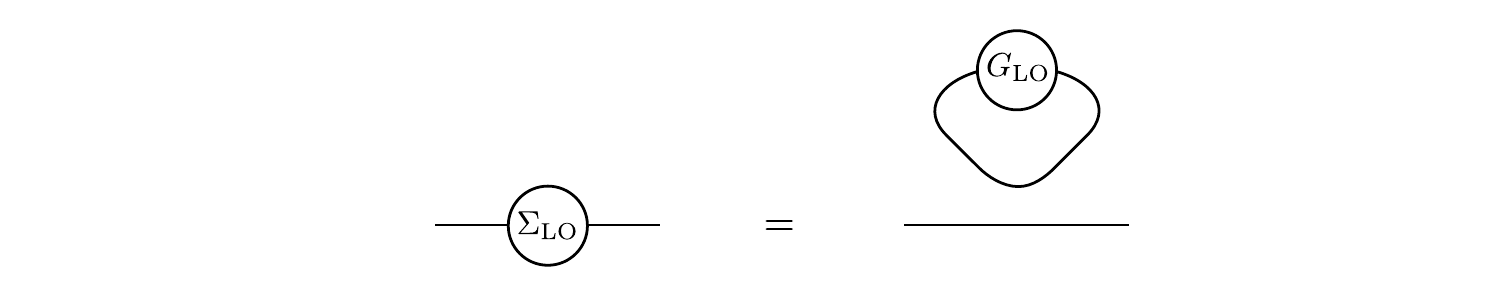}}
\caption{Diagrammatic equation for the LO self-energy $\Sigma_{\text{LO}}$ in terms of the LO full connected 2-point function $G_{\text{LO}}$. It is equivalent to Eq.\ \eqref{VecSDEq}.}\label{VecFreeEnergy}
\end{figure}

Thanks to the recursive structure of the bubble graphs, one can check that $\Sigma_{\text{LO}}(\lambda)$ necessarily has the structure represented in Figure \ref{VecFreeEnergy}. Equivalently, by taking into account the combinatorial factors, it is given by
\be \label{VecSDEq}
\Sigma_{\text{LO}}(\lambda) = -4\lambda G_{\text{LO}}(\lambda)\, .
\ee
The above relation is often referred to as a Schwinger-Dyson equation. As a result, Eq.~\eqref{Vec2PtFunction} restricted to the leading sector together with Eq.\ \eqref{VecSDEq} provide a closed equation for $G_{\text{LO}}(\lambda)$, which can be written as
\be \label{VecClosedEq}
G_{\text{LO}}(\lambda)=1-4\lambda G_{\text{LO}}(\lambda)^2\, .
\ee

This type of equation is often encountered in combinatorics (see, for instance, \cite{BookComb}). The ``physical" solution, which goes to one when $\lambda$ goes to zero (see Eq.\ \eqref{VecClosedEq}), can be written as a power series in $\lambda$ with the Catalan numbers \cite{Catalan} as coefficients:
\be \label{VecLOSolution}
G_{\text{LO}}(\lambda)=\sum_{k=0}^\infty C_k (-4\lambda)^k \quad \text{with} \quad C_k=\frac{1}{2k+1} \begin{pmatrix} 
2k+1 \\
k 
\end{pmatrix} \, .
\ee
In other words, $G_{\text{LO}}(\lambda)$ is a generating function for the Catalan numbers. In fact, this was expected. Indeed, Catalan numbers appear in many enumeration problems related to trees and we know that $G_{\text{LO}}(\lambda)$ is also related to trees because it corresponds to a sum over connected Feynman graphs with two external edges, obtained by cutting one edge in bubble graphs, which are themselves in bijection with trees. 

Since Eq.\ \eqref{VecClosedEq} is quadratic in $G_{\text{LO}}(\lambda)$, two explicit solutions can actually be found. The physical solution with behavior $G_{\text{LO}}(\lambda)\rightarrow1$ when $\lambda\rightarrow0$ is given by 
\be \label{VecLOSolutionExpl}
G_{\text{LO}}(\lambda)=\frac{2}{1+\sqrt{1+16\lambda}} =\frac{\sqrt{1+16\lambda}-1}{8\lambda}\, .
\ee

Remark that the large $N$ solution $G_{\text{LO}}(\lambda)$ makes sense even for negative $\lambda$, as long as $\lambda > \lambda_c=-1/16$. It can be shown (see Eq.\ \eqref{VecConn2PtFunctionFreeEnergy}) that the same holds for $F_0(\lambda)$, which is thus a convergent series in $\lambda$ with a non-zero radius of convergence $|\lambda_c|$. This remarkable fact stands in contrast with the initial path integral, which is divergent for $\lambda<0$. It is one of the nice features of the large $N$ limit. In particular, it means that the models exhibit a critical behavior when the coupling constant $\lambda$ approaches the critical value $\lambda_c$. 

We now give more details on the critical behavior of the LO free energy $F_0(\lambda)$ when $\lambda\rightarrow \lambda_c$. First, $G(N,\lambda)$ can be related to the free energy $F(N,\lambda)$ as follows
\be \label{VecConn2PtFunctionFreeEnergy}
G(N,\lambda) = 1- 4 \lambda \frac{d}{d\lambda} \frac{F(N,\lambda)}{N} \, ,
\ee
which is obtained by starting from the trivial identity
\begin{equation*}
\frac{1}{Z} \int [d\vec{\phi}] \, \frac{1}{N} \frac{d}{d\phi_i} \Bigl[ \phi_i \, e^{-S(\vec{\phi})} \Bigr] = 0 \, ,
\end{equation*}
and then by computing each term and using Eq.\ \eqref{VecConn2PtFunction} and $F = - \log Z$. Second, from Eq.\ \eqref{VecLOSolutionExpl}, we observe that when $\lambda\rightarrow \lambda_c$, $G_{\text{LO}}$ exhibits the critical behavior
\be \label{VecCritBehaviorG}
G_{\text{LO,sing}}(\lambda)\sim (\lambda_c-\lambda)^{\frac{1}{2}} \, ,
\ee
where $G_{\text{LO,sing}}$ denotes the leading singular part of $G_{\text{LO}}$. Finally, using Eq.\ \eqref{VecConn2PtFunctionFreeEnergy}, we deduce that the critical behavior of the LO free energy is
\be \label{VecCritBehaviorF}
F_{0,\text{sing}}(\lambda) \sim (\lambda_c-\lambda)^{2-\gamma} \quad \text{with} \ \gamma=\frac{1}{2} \, ,
\ee
where $F_{0,\text{sing}}$ corresponds to the leading singular part of $F_{0}$. The exponent $\gamma$ is known as the susceptibility critical exponent or entropy exponent. The critical behavior of the LO free energy can be interpreted as follows. When $\lambda$ approaches its critical value, $F_0(\lambda)$ loses its summability due to the Feynman graphs with a large number of vertices, that is, with a large power of $\lambda$. These Feynman graphs then dominate the behavior of $F_0(\lambda)$ close to criticality and the susceptibility critical exponent $\gamma$ characterizes this behavior. 

The regime where Feynman graphs with a large number of vertices dominate is commonly called the continuum limit. It provides an interesting way to define continuous random geometries from discrete ones, whose size can become arbitrarily large. In an appropriate scaling limit, discrete objects can indeed converge, in the sense of Gromov-Hausdorff-Prokhorov (GHP) \cite{GHP}, to a continuous space of a certain universality class. The various universality classes that can be achieved are then classified by the value of the susceptibly critical exponent.

In the case of vector models, the leading sector consists in trees in the effective description. Such discrete objects admit a GHP limit to the continuum random tree \cite{CRT}, also known as the universality class of branched polymers \cite{BranchedPolymer}. The typical value of the susceptibility critical exponent associated with this universality class is $\gamma=1/2$, which is consistent with the result \eqref{VecCritBehaviorF}. In the next chapters for matrix and tensor models, we also study the corresponding critical behaviors and continuum limits.

\section{Discussion}
\label{sec:VecDiscussion}

The study of $\oN$ vector models presented in this chapter is far for complete. We focused on the study of the $1/N$ expansion from a combinatorial perspective using Feynman graphs and we briefly examined the large $N$ solution. One could go further and study for instance phase transition phenomena, beta functions, renormalization in higher dimensions, $1/N$ corrections, etc. We refer the interested reader to \cite{Cole, Zi-Ji, Mak} for more details on these topics.

Finally, we briefly mention where vector models stand in the context of the gauge/gravity correspondence. There is evidence that quantum gauge theories based on vector models are holographically dual to higher spin ``quantum gravity" theories in AdS \cite{VecHolo}. However, it is unlikely that standard vector models capture the SYK/black hole physics. This can be seen from the Schwinger-Dyson equation in Figure \ref{VecFreeEnergy}, which corresponds to a tadpole-like behavior. Such behavior is typical of mass renormalization and it does not lead to black holes features such as a continuous spectrum.

% To illustrate this point, let us consider a quantum mechanical model based one a real bosonic vector $\phi_i(t)$ (note that in the case of real Majorana fermions, the interaction term vanishes). Then, the large $N$ Schwinger-Dyson equation \eqref{VecSDEq} can be generalized to  
%%
%\be \label{VecSDEqQM}
%\Sigma_{\text{LO}}(t-t') = -4\lambda G_{\text{LO}}(0) \delta(t-t')\, .
%\ee
%In other words, the self-energy is given by a simple delta function. This ultimately generates a tadpole in the two-point function at leading order, which can not lead to black holes-like features such as a continuous spectrum.

%%%%%%%%%%%%%%%%%%%%%%%%%%%%%%%%%%%%%%%%%%%%%%%%
%%%%%%%%%%%%%%%%%%%  Chapter 3  %%%%%%%%%%%%%%%%%%%%%%%%%
%%%%%%%%%%%%%%%%%%%%%%%%%%%%%%%%%%%%%%%%%%%%%%%%

\chapter{Large $N$ matrix models}
\label{chap:Matrix}

In this chapter, we continue on the study of models that become solvable in the limit of a large number of degrees of freedom with matrix models. In these models, the field usually corresponds to a $N\times N$ matrix so that the number of degrees of freedom is given by $\cN\sim N^2$. 

Matrix models appear in countless areas of physics and also hold a prominent role in mathematics, especially in probability theory where they correspond to random distributions for ensembles whose elements are matrices. The first appearance of so-called random matrices is due to Wishart \cite{Wish}, who studied random distributions for rectangular matrices. They were then used in the context of nuclear physics by Wigner \cite{Wig}, who was interested in the energy spectrums of heavy nuclei in the realm of quantum mechanics. Wigner realized that complicated Hamiltonians could in fact be accurately modelled by random Hermitian matrices. An important development was then achieved by 't~Hooft \cite{tHooft}, who constructed the $1/N$ expansion of interacting matrix models for the generalization of quantum chromodynamics (QCD) from gauge group $\text{SU}(3)$ to $\suN$. In this theory, the gauge fields, which describe gluons, transform in the adjoint representation of the gauge group; hence, they correspond to traceless Hermitian matrices of size $N\times N$. 't~Hooft realized that in the large $N$ limit where an appropriate coupling parameter is hold fixed, the perturbative expansion is dominated by the planar graphs. Even though his approach didn't solve QCD, it was an inspiration for the study of the resulting connection between matrix models and discretized surfaces. This was initiated by Br\'ezin, Itzykson, Parisi and Zuber \cite{BrezAl}, who computed explicitly the large $N$ solution of matrix models for cubic and quartic interactions by counting the dominant planar graphs.\footnote{Previous work on the enumeration of planar graphs includes \cite{Tutte}.} Later, Di Francesco, Ginsparg and Zinn-Justin \cite{DiFran} built a crucial bridge between two-dimensional quantum gravity and random matrices. Finally, the large $N$ limit \`a la 't~ Hooft has been of prime importance for the discovery and exploration of the gauge/gravity correspondence \cite{Malda,GaugeGravity} in the context of string theory. In particular, as emphasized in the introduction, quantum gauge theories of $N\times N$ matrices in the large $N$ limit are good candidates to describe quantum black holes in this framework.

In this chapter, we modestly focus on the study of the $1/N$ expansion of matrix models in the case of a real $N\times N$ matrix that transforms in the bi-fundamental representation of the symmetry group $\oN\times\oN$. The reason is twofold. On the one hand, the $1/N$ expansions of most matrix models have the same generic structure; for instance, planar graphs generically dominate in the large $N$ limit. Hence, studying this particular class of matrix models is sufficient to describe this structure. On the other hand, these models play an important role in the next chapters so they are a perfect playground to introduce notions and notations that are useful later. 

Like in the previous chapter for vector models, we use a Feynman graph approach to establish the $1/N$ expansion of matrix models; the steps are thus identical. Notions on embedded graphs are handy, see Section \ref{sec:App2}. We stop the discussion after the construction of the $1/N$ expansion and the description of the leading sector but the full story of matrix models is far from ending there. A pedagogical overview of matrix models can be found in \cite{Eyn}, which mainly studies matrix models from the point of view of random distributions. Descriptions of matrix models from the point of view of physics can be found, for instance, in \cite{Cole,Mak,Marino,KlebaLargeN}. Unless otherwise stated, we deal with zero-dimensional models based on real bosonic variables. 

We begin this chapter with a general description of the $\oN\times\oN$ matrix models and the construction of a general invariant action. We then present the corresponding Feynman graphs obtained in perturbation theory using two useful representations, namely the stranded and the colored representations. Next, we construct the $1/N$ expansion by scaling appropriately the coupling constants in the action. We then briefly study the leading order Feynman graphs when $N$ is large. Finally, we close the chapter with further comments on other types of matrix models, including models with complex matrices and models with Hermitian or symmetric matrices.

\section{Definition of the models}
\label{sec:MatDefModels}

We focus on matrix models whose basic degrees of freedom are the components of a single, real matrix $X=(X_{a_1a_2})_{1\leq a_1, a_2 \leq N}$, which transforms in the bi-fundamental representation of the symmetry group $\oN\times\oN$. The number of degrees of freedom is thus $\cN=N^2$. As for the transformation law, it is given in terms of the matrix components $X_{a_1a_2}$ by
\be \label{MatTransfLaw}
X_{a_1a_2} \rightarrow X'_{a_1a_2} = O^{(1)}_{a_1a_1'}O^{(2)}_{a_2a_2'}X_{a_1'a_2'} \, ,
\ee
where $O^{(1)}$ and $O^{(2)}$ are independent $N\times N$ orthogonal matrices, each in a different $\oN$ group. The two matrix indices are thus distinguishable because they transform with respect to two distinct $\oN$ groups, say, the first index with the first group in $\oN\times\oN$ and the second index with the second group. This explains the notation $a_1$ and $a_2$ for the matrix indices.
 
An invariant action $S(X)$ for these matrix models is constructed from trace invariants of the form $\tr (X\tra{X})^p$, for some $p\geq1$, and products thereof; where $\tra{X}$ denotes the transpose matrix: $\tra{X}_{a_2a_1}=X_{a_1a_2}$. This is a direct consequence of the $\oN\times\oN$ invariance. Invariants that only involve one trace are called single-trace whereas the others are called multi-trace. Here, we allow for multi-trace invariants in full generality. They only slightly complicate the combinatorics but it is important to take them into account with regard to next chapters. 

We denote by $\cI_a(X)$ an invariant, where $a$ is some label in a discrete set $\cS$, by $t_a$ the number of traces in $\cI_a(X)$ and by $s_a$ the number of entries $X$ in $\cI_a(X)$, which is necessarily even. Thus, a generic invariant $\cI_a(X)$ writes 
\be\label{MatInteraction}
\cI_a(X) = \prod_{i=1}^{t_a} \tr (X\tra{X})^{p_i(a)}\, ,
\ee
for some positive integers $\{p_i(a)\}_{i=1,\ldots,t_a}$ satisfying $\sum_{i=1}^{t_a} 2p_i(a)=s_a$. Examples of invariants are 
\begin{itemize}
\item $\tr (X\tra{X}) = X_{a_1a_2}\tra{X}_{a_2a_1}= X_{a_1a_2}X_{a_1a_2}$ ($t=1$, $s=2$);
\item $\tr (X\tra{X})^2 = X_{a_1a_2}\tra{X}_{a_2b_1}X_{b_1b_2}\tra{X}_{b_2a_1}= X_{a_1a_2}X_{b_1a_2}X_{b_1b_2}X_{a_1b_2}$ ($t=1$, $s=4$);
\item $\tr (X\tra{X}) \tr (X\tra{X}) = X_{a_1a_2}\tra{X}_{a_2a_1}X_{b_1b_2}\tra{X}_{b_2b_1} = X_{a_1a_2}X_{a_1a_2}X_{b_1b_2}X_{b_1b_2}$ ($t=2$, $s=4$);
\item $\tr (X\tra{X})^3 = X_{a_1a_2}\tra{X}_{a_2b_1}X_{b_1b_2}\tra{X}_{b_2c_1}X_{c_1c_2}\tra{X}_{c_2a_1} = X_{a_1a_2}X_{b_1a_2}X_{b_1b_2}X_{c_1b_2}X_{c_1c_2}X_{a_1c_2}$ ($\smash{t=1}$, $s=6$);
\item etc.
\end{itemize}
This concise notation allows us to write an invariant action $S(X)$ as
\be\label{MatAction}
S(X)= \frac{N}{2}\tr (X\tra{X}) + \sum_{a\in\cS} g_a \, \cI_a(X) = \frac{N}{2}\tr (X\tra{X}) + V(X) \, ,
\ee
where $g_a$ is the coupling constant associated with $\cI_a(X)$. As usual, we isolated the quadratic mass term, we set the mass to one and we regrouped higher order interaction terms in the interaction potential $V(X)$. Besides, the factor of $N$ in front of the mass term is set for further convenience.

The partition function $Z$ and the free energy $F$ associated with the action \eqref{MatAction} are given by
\be \label{MatPartitionFunctionFull}
Z = \exp (-F) = \int [dX] \, e^{-S(X)}=\int [dX] \, e^{-\frac{N}{2}\tr (X\tra{X}) - \sum_{a\in\cS} g_a \, \cI_a(X)} \, ,
\ee
where the path integral measure $[dX]$ reduces in zero dimension to the product of $N^2$ simple integral measures on $\mathbb{R}$: $[dX] = \prod_{a_1,a_2=1}^N \sqrt{\frac{N}{2\pi}}dX_{a_1a_2}$. Similar expressions hold for general $n$-point functions. In particular, the free $2$-point function or free propagator reads
\be \label{MatFree2PtFunction}
\bigl\langle X_{a_1a_2}X_{b_1b_2}\bigr\rangle_0\,=\frac{1}{N}\delta_{a_1b_1}\delta_{a_2b_2} \, .
\ee

As for vector models, we want to define a $1/N$ expansion for matrix models. To do so, we go to perturbation theory and expand the partition function (or the free energy) in powers of the coupling constants $g_a$ in the action (which are assumed to be small) and we make contact with Feynman graphs.

\section{Feynman graphs}
\label{sec:MatFeynmanGraphs}

\begin{figure}[]
\centerline{\includegraphics[scale=1]{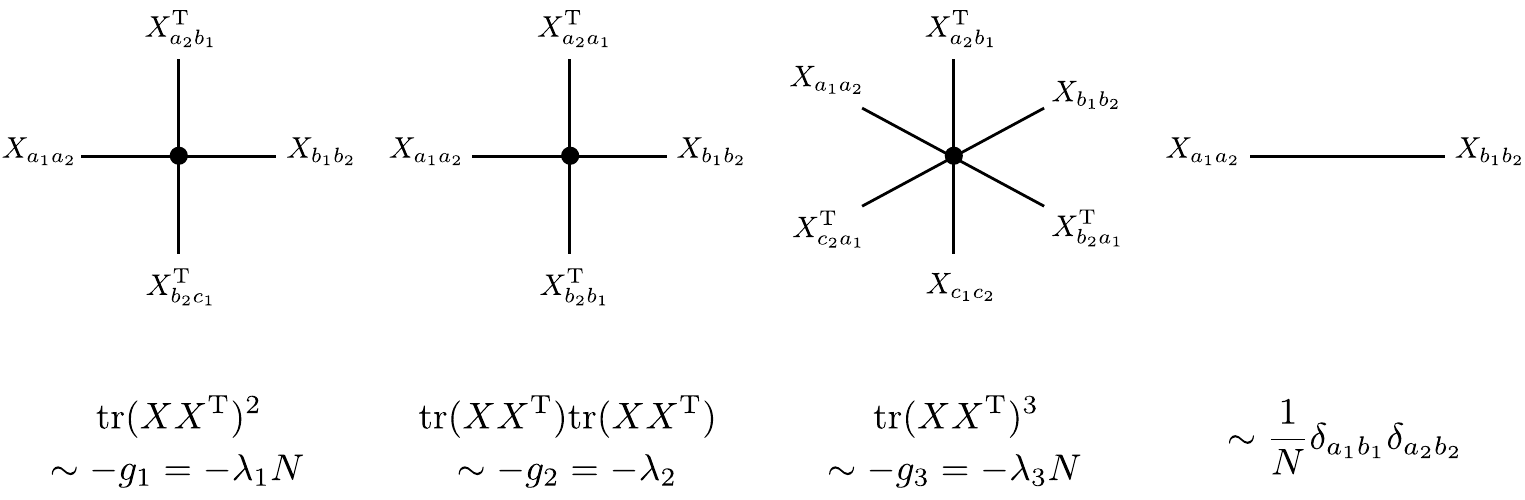}}
\caption{Feynman rules for some vertices and the propagator of the $\oN\times\oN$ matrix models. The coupling parameters $\lambda_a$ are defined in Eq.\ \eqref{MatOptScaling}.}\label{MatFeynRules}
\end{figure}

We focus on the perturbative expansion of the free energy $F$. By a similar argument to the one of Section \ref{sec:VecFeynmanGraphs}, it can be written in the form of an expansion onto connected Feynman graphs constructed according to the Feynman rules illustrated in Figure \ref{MatFeynRules}. An interaction term $\cI_a(X)$ is represented by a vertex of degree $s_a$. In Figure \ref{MatFeynRules}, we show the two vertices of degree four and the vertex of degree six mentioned above. To each interaction vertex of type $\cI_a(X)$ in a Feynman graph, we assign a factor $-g_a$ and to each edge (or propagator) that connects the field components $X_{a_1a_2}$ and $X_{b_1b_2}$, we assign a factor $\frac{1}{N}\delta_{a_1b_1}\delta_{a_2b_2}$, which corresponds to \eqref{MatFree2PtFunction}. 

In the perturbative expansion of $F$, each connected Feynman graph $\cG$ is weighted by some amplitude $\cA(\cG)$ that depends on $N$. The power of $N$ associated with a Feynman graph is however not easy to interpret in the current representation so that it is useful to introduce equivalent representations. 

\subsection{Stranded representation}
\label{sec:MatStrandRep}

As for vector models, there is an equivalent stranded representation that accounts for the contraction pattern of the $\oN$ indices associated with the field components $X_{a_1a_2}$, with the crucial difference that there are now two indices for each field component; hence, two strands. Furthermore, because the matrix indices are distinguishable, we can assign two distinct colors to them, say, color $1$ for the first index and color $2$ for the second index, and we do the same for the corresponding strands. 

\begin{figure}[]
\centerline{\includegraphics[scale=1]{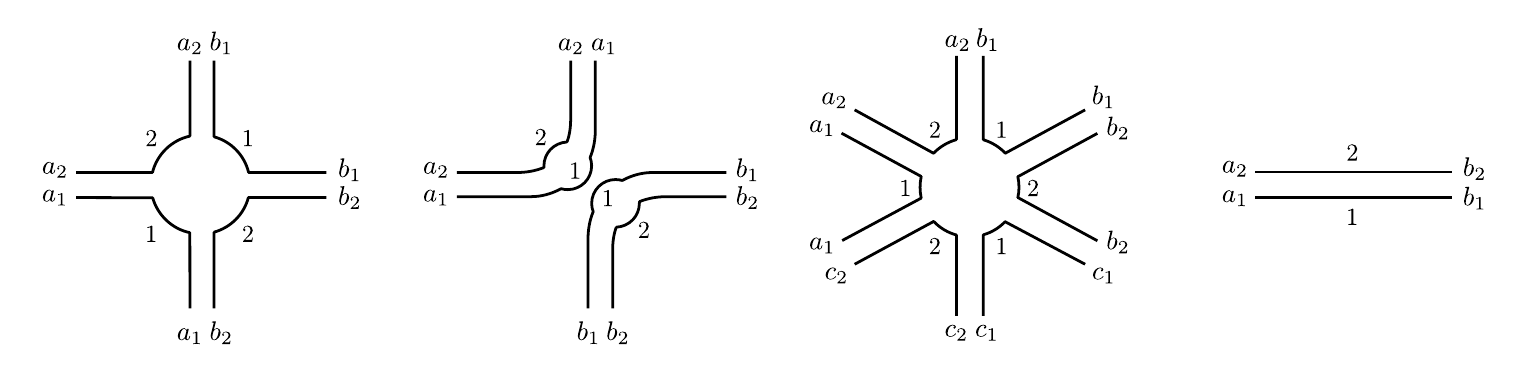}}
\caption{Stranded representation for the vertices and the propagator of Figure \ref{MatFeynRules}. Strands associated with the first (resp.\ second) matrix index are labeled with the color $1$ (resp.\ $2$). Besides, the propagator must link strands in a consistent way with color.}\label{MatFeynRulesStrand}
\end{figure}

The Feynman rules in the stranded representation are presented in Figure \ref{MatFeynRulesStrand} for the vertices and the propagator of Figure \ref{MatFeynRules}. The interaction vertex of type $\cI_a(X)$ consists in $s_a$ pairs of strands, one pair with two distinct colors for each field component, such that the strands are connected two by two via a corner if they carry the same $\oN$ index. Note that corners necessarily connect strands with the same color. Besides, an edge or propagator consists in two strands of distinct colors $1$ and $2$ representing the two $\delta$'s coming from the corresponding Wick's contraction. It is important that when a propagator connects two vertices, the strands are linked in agreement with their color. This can result in a twist in the propagator. More precisely, if we keep the distinction between $X$ and $\tra{X}$ in the interaction vertices, a propagator is untwisted when it connects a $X$ and a $\tra{X}$ while it is twisted when it connects two $X$'s or two $\tra{X}$'s.

\begin{figure}[]
\centerline{\includegraphics[scale=1]{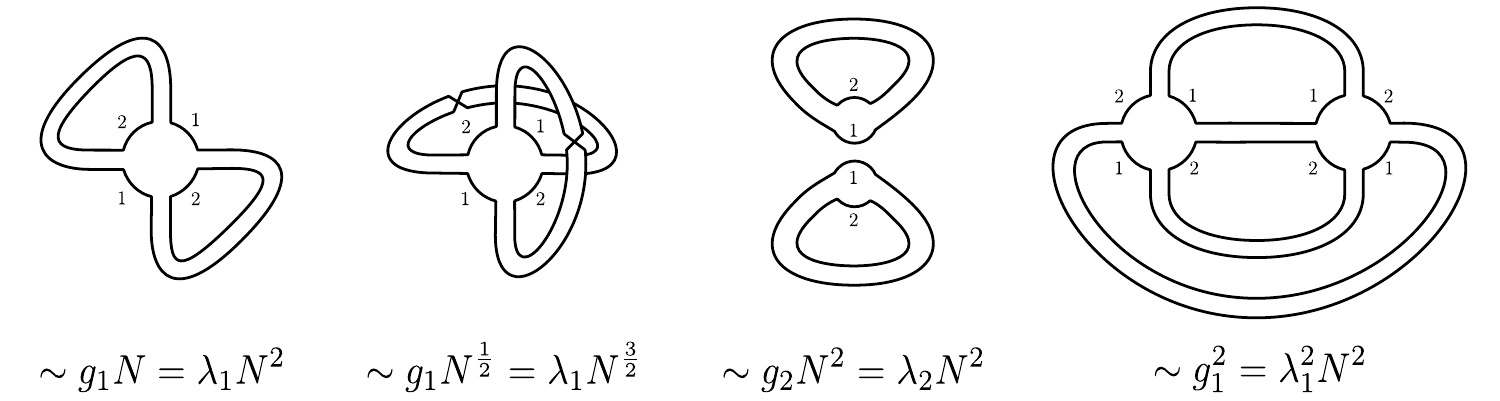}}
\caption{Examples of Feynman graphs in the stranded representation. They correspond to orientable or non-orientable embedded graphs in the ribbon graph representation. Their scaling in $g_a$ and $N$ in the perturbative expansion are indicated, as well as their scaling in terms of the coupling parameter $\lambda_a$ defined in Eq.\ \eqref{MatOptScaling}.}\label{MatFeynGraphsStrand}
\end{figure}

Examples of connected Feynman graphs in the stranded representation are given in Figure \ref{MatFeynGraphsStrand}. As one can notice, they look the same as embedded graphs in the ribbon graph representation (without the internal structure), see Figure \ref{fig:EmbGraphsc}. In fact, the Feynman graphs of matrix models correspond to embedded graphs. They can be orientable or non-orientable because of the possibility of a twist in a propagator, which corresponds to a twisted ribbon edge. As we shall see, the embedded graph structure of the Feynman graphs plays an important role in defining the $1/N$ expansion of matrix models. Also, it makes the important connection between matrix models and discretized surfaces, as advertized at the beginning of the chapter. 

%The embedded graph structure of the Feynman graphs plays an important role for defining the $1/N$ expansion of matrix models. This can be understood as follows. As explained in the footnote $3$ in Section \ref{sec:VecFeynmanGraphs}, Feynman graphs are by essence embedded graphs since different Wick's contractions lead to different cyclic orderings around each vertex. In the case of vector models, this structure has no repercussion on the $1/N$ expansion because of the (partial) symmetry of the interaction term $\bigl(\vec{\phi}^{\, 2}\bigr)^2=\phi_i\phi_i\phi_j\phi_j$. Indeed, contracting the first (resp.\ the third) field component yields the same result as contracting the second (resp.\ the fourth) field component. In contrast, there is in general no such symmetry for the interaction terms of matrix models. For instance, contracting different field components in $\tr (X\tra{X})^2 = X_{a_1a_2}X_{b_1a_2}X_{b_1b_2}X_{a_1b_2}$ gives different results. Hence, the cyclic ordering around the interaction vertices carries important information about the Feynman graphs, which in turn implies that the embedded graph structure must be taken into account in the $1/N$ expansion of matrix models.

Given a connected Feynman graph $\cG$ in the stranded representation, we denote by $p$ its number of edges or propagators, by $v$ its number of interaction vertices and by $f$ its number of closed stranded loops, which correspond to the boundary components or faces of the corresponding ribbon graph (see Appendix \ref{app:AppA}).

We remark that there are different types of interaction vertices in $\cG$, one type for each interaction term $\cI_a$ in the action. To account for this, we introduce a discrete set $\cS_\cG$ associated with $\cG$ such that each element $a\in\cS_\cG$ is also an element of $\cS$ and the number of times an element $a$ appears in $\cS_\cG$ corresponds to the number of interaction vertices of type $\cI_a$ in $\cG$. Hence, $v=|\cS_\cG|$. 

Secondly, because we deal with multi-trace interaction terms, $\cG$ may correspond to a disconnected ribbon graph, which we denote by $\tilde{\cG}$ (see for example the third graph of Figure \ref{MatFeynGraphsStrand}). Each multi-trace vertex of type $\cI_a$ in $\cG$ effectively yields $t_a$ distinct (single-trace) disk vertices in $\tilde{\cG}$. The total number of effective disk vertices in $\tilde{\cG}$ is thus $\tilde{v}=\sum_{a\in\cS_\cG} t_a$. As for the number $\tilde{p}$ of edges and the number $\tilde{f}$ of faces in $\tilde{\cG}$, they are the same as in $\cG$, i.e.\ $\tilde{p}=p$ and $\tilde{f}=f$. 

Finally, the ribbon graph $\tilde{\cG}$ corresponding to $\cG$ is characterized by a genus $g(\tilde{\cG})\in\frac{1}{2}\mathbb{N}$, which can be computed using Euler's formula \eqref{eq:EulerFormulaEmbGraphs}:
\be\label{EulerFormFeynGraphStrand}
2c(\tilde{\cG})-2g(\tilde{\cG})= \tilde{v}-\tilde{p}+\tilde{f}=\sum_{a\in\cS_\cG} t_a - p + f \, ,
\ee
where $c(\tilde{\cG})$ is the number of connected components of $\tilde{\cG}$. Note that the RHS of this equation is expressed only in terms of quantities defined for $\cG$. In the following, we use the notation $g$ to denote the genus $g(\tilde{\cG})$ of the ribbon graph associated with $\cG$.

\subsection{Colored representation}
\label{sec:MatColorRep}

Even though the stranded representation already makes things transparent to define a $1/N$ expansion, we introduce yet another equivalent representation that plays an important role in the next chapters, namely the colored representation. This representation is based on a one-to-one correspondence between the Feynman graphs and a certain class of colored graphs, called $3$-bubbles (see Appendix \ref{sec:AppABubbles} for the definition and the properties of bubbles).

\begin{figure}[]
\centerline{\includegraphics[scale=1]{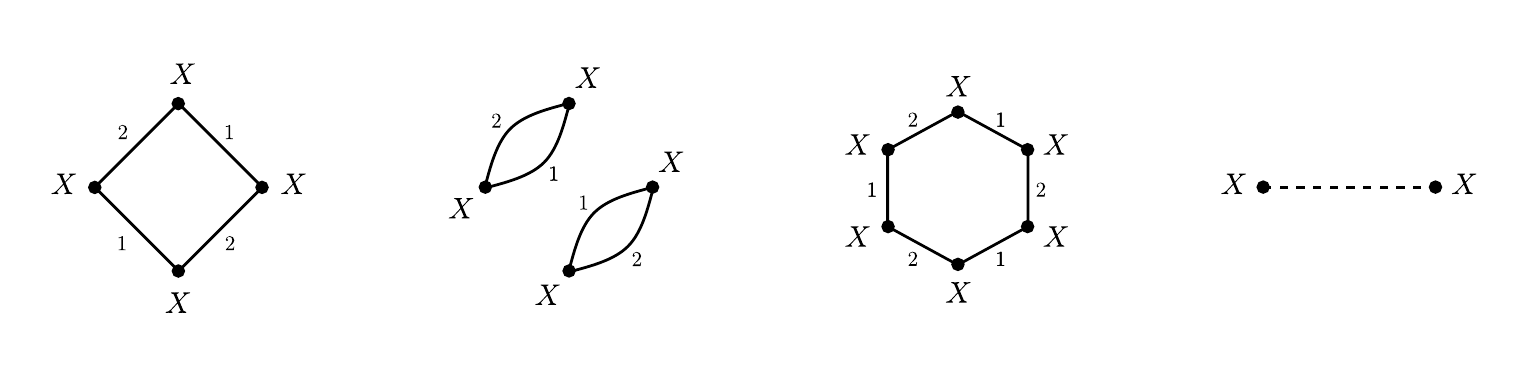}}
\caption{$2$-bubbles corresponding to the vertices of Figure \ref{MatFeynRulesStrand} and representation of the propagator in the colored representation.}\label{MatFeynRulesColor}
\end{figure}

On the one hand, there is a bijection between the interaction terms $\cI_a$ entering the action \eqref{MatAction} and $2$-bubbles $\cB_a$, which is constructed as follows. To each matrix entry $X$ in an interaction term $\cI_a(X)$, we associate a vertex in $\cB_a$. Then, we use the color $1$ for the first matrix index and the color $2$ for the second one, like in the stranded representation. Finally, we draw an edge of color $i\in\{1,2\}$ between two vertices of $\cB_a$ if the corresponding matrix entries $X$ have their $i^{\text{th}}$ index contracted together in $\cI_a$. The fact that $\cB_a$ corresponds to a $2$-bubble follows from the $\oN\times\oN$ invariance. This is illustrated in Figure \ref{MatFeynRulesColor} for the vertices of Figure \ref{MatFeynRulesStrand}.

One can check that the above construction defines a bijection between the set of interaction terms $\cI_a$ and the set of $2$-bubbles $\cB_a$ (with $V(\cB_a)>2$), which explains the use of the same label $a\in\cS$. To stick with the notations of \cite{Ref2,Ref1}, we label the interaction terms $\cI_a$ by their corresponding $2$-bubble $\cB_a$, that is, $\cI_a\equiv\cI_{\cB_a}$, and we use the notations $\cI_{\cB_a}$ and $\cB_a$ interchangeably, referring to them as interaction terms or interaction bubbles. Note that an interaction bubble $\cB_a$ may be disconnected. The number $c(\cB_a)$ of connected components of $\cB_a$ is given by the number $t_a$ of traces in $\cI_a$ (written as $t(\cB_a)$ in \cite{Ref2, Ref1}), which also corresponds to the number $F_{12}(\cB_a)$ of $(12)$-faces in $\cB_a$: $c(\cB_a)=t(\cB_a)=F_{12}(\cB_a)$. Besides, the number $V(\cB_a)$ of vertices in $\cB_a$ corresponds to the number $s_a$ of entries $X$ in $\cI_{\cB_a}$.

\begin{figure}[]
\centerline{\includegraphics[scale=1]{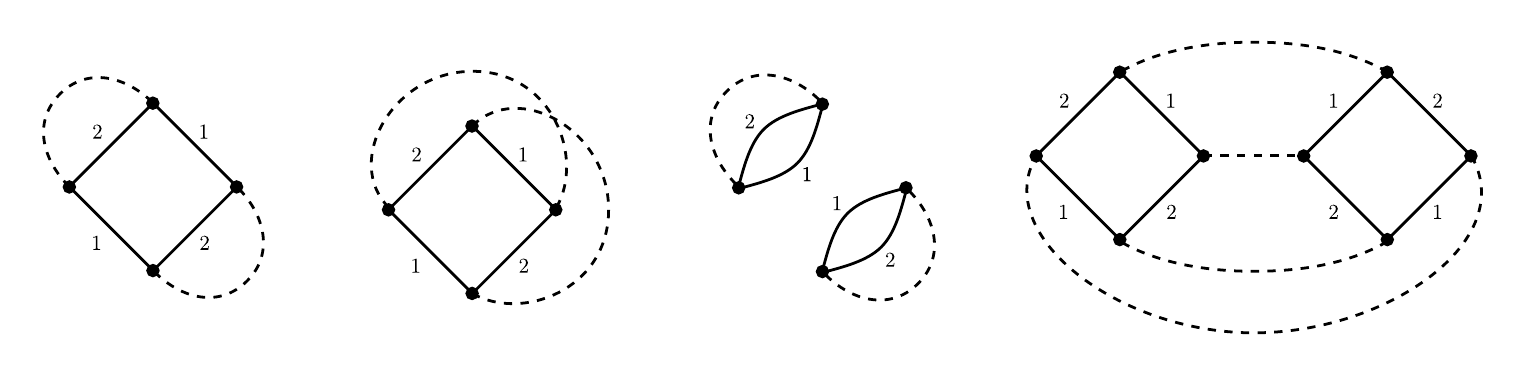}}
\caption{Feynman graphs of Figure \ref{MatFeynGraphsStrand} in the colored representation. The interaction vertices are now interaction bubbles and the faces correspond to $(0i)$-faces for $i\in\{1,2\}$.}\label{MatFeynGraphsColor}
\end{figure}

On the other hand, a moment of reflection reveals that the Feynman graphs $\cG$ can themselves be represented as $3$-bubbles. Indeed, we can represent the interaction vertices of $\cG$ as interaction bubbles, whose vertices correspond to entries $X$. Then, we can represent the propagators, which correspond to Wick's contractions between two distinct entries $X$, as a new set of edges, each connecting two distinct vertices in interaction bubbles. To distinguish the propagators from the edges in interaction bubbles, we label them with the fictitious color $0$ and we draw them as dashed edges (see Figure \ref{MatFeynRulesColor}). The Feynman rules in the colored representation directly follow from the above construction: to each interaction bubble $\cB_a$ in $\cG$ is assigned a factor $-g_a$ and to each propagator in $\cG$ is assigned a factor $\frac{1}{N}\delta_{a_1b_1}\delta_{a_2b_2}$, that it, each propagator identifies all the indices of the two vertices it connects. The Feynman graphs of Figure \ref{MatFeynGraphsStrand} are given in the colored representation in Figure \ref{MatFeynGraphsColor}.

We use the notation $\cB$ to denote a given connected Feynman graph $\cG$ in the colored representation \cite{Ref2, Ref1}. The data associated with each representation is related as follows. First, the number $p$ of propagators in $\cG$ corresponds to the number $E_0(\cB)$ of edges of color $0$ in $\cB$, which is equal to half the number $V(\cB)$ of vertices in $\cB$ (cf.\ Eq.\ \eqref{EVrel}). Second, the number $v$ of interaction vertices in $\cG$ corresponds to the number $v(\cB)$ of interaction bubbles in $\cB$. Third, each face in $\cG$ correspond to a $(0i)$-faces in $\cB$ for $i\in\{1,2\}$. Fourth, the number $\tilde{v}$ of effective disk vertices in the ribbon graph $\tilde{\cG}$ corresponds to the number $F_{12}(\cB)$ of $(12)$-faces in $\cB$, which is equal to the number $c(\cB^{(0)})\equiv B^{(0)}$ of connected components of $\cB_{(ij)}=\cB^{(0)}$ (see Section \ref{sec:AppABubbles}). Finally, the number $c(\tilde{\cG})$ of connected components of $\tilde{\cG}$ matches the number $c(\cB)\equiv B$ of connected components of $\cB$.\footnote{We stress that $\cG$ corresponds to a connected Feynman graph in the perturbative expansion; the fact that it looks disconnected in the stranded and colored representations is an artefact.} The relations between the data of the two representations is summarized below.
\be\label{MatFeynGraphData}
p=E_0(\cB)=\frac{1}{2}V(\cB)\, , \quad f=F_{01}(\cB)+F_{02}(\cB)\, , \quad  \tilde{v}=\sum_{a\in \cS_\cG} t(\cB_a)=F_{12}(\cB)=B^{(0)}\, , \quad c(\tilde{\cG})=B \, .
\ee

From the above, we see that $3$-bubbles are equivalent to ribbon graphs. It is thus natural to assign to each $3$-bubble $\cB$ a genus $g(\cB)$ through an equivalent of Euler's formula \eqref{eq:EulerFormulaEmbGraphs}:
\be\label{EulerFormFeynGraphColor}
\begin{split}
2B-2g(\cB)&= V(\cB) - E(\cB) + F(\cB) \\
	&= -\frac{1}{2}V(\cB) + F_{01}(\cB) + F_{02}(\cB) + F_{12}(\cB) \, ,
\end{split}
\ee
where we used the properties of $3$-bubbles (see Section \ref{sec:AppABubbles}). By using the identities \eqref{MatFeynGraphData} between the data of the stranded and the colored representations and comparing with Euler's formula \eqref{EulerFormFeynGraphStrand} for the ribbon graphs, it is straightforward to check that the genus of a $3$-bubble naturally coincides with the genus of the corresponding ribbon graph, i.e.\
\be\label{MatGenera}
g(\cB)=g \, .
\ee
In particular, it implies that $g(\cB)\in\frac{1}{2}\mathbb{N}$.

\section{Large $N$ expansion}
\label{sec:MatLargeNExp}

As explained in Section \ref{sec:VecLargeNExp}, the first step in constructing a $1/N$ expansion for matrix models is to define how the coupling constants $g_a$ in the action \eqref{MatAction} scale with $N$ when $N\rightarrow\infty$. In matrix models, it turns out that the optimal scaling of $g_a$ is $g_a\sim N^{2-t(\cB_a)}$, which is commonly called 't Hooft scaling for historical reasons. As a result, we define a new coupling parameter $\lambda_a$ for each interaction term $\cI_{\cB_a}$ by
\be\label{MatOptScaling}
g_a\equiv N^{2-t(\cB_a)}\lambda_a\, ,
\ee 
and we decide to keep the $\lambda_a$'s fixed when $N\rightarrow\infty$. The action rewrites in terms of the 't Hooft scalings as
\be\label{MatActionScaled}
S(X)= N\biggl(\frac{1}{2}\tr (X\tra{X}) + \sum_{a\in\cS} N^{1-t(\cB_a)}\lambda_a \, \cI_a(X)\biggr) \, .
\ee
Note that in matrix models, there exists an optimal scaling for all the coupling constants in the action, similarly to vector models. In particular, the optimal scaling is the same for the couplings of all interaction terms with the same number of traces. This is a special feature of matrix models. As a consequence, the interaction vertex associated with any interaction term contributes to the leading sector, by definition of optimal scaling.

Let us compute the power of $N$ associated with a given connected Feynman graph $\cG$. From the action \eqref{MatActionScaled}, we deduce that there is a factor $1/N$ associated to each propagator in $\cG$, a factor $N^{2-t(\cB_a)}$ to each interaction vertex of type $\cI_a$ in $\cG$ and a factor $N$ to each face (or closed stranded loop) in $\cG$. Overall, the amplitude $\cA(\cG)$ of $\cG$ thus scales with $N$ like 
\be\label{MatFeynGraphNScaling}
\cA(\cG) \sim N^{-p+\sum_{a\in\cS_\cG}(2-t(\cB_a))+f} \equiv N^{2-h} \, ,
\ee
where we introduced the parameter $h$ defined as
\be\label{MatHParam1}
h = 2+p-\sum_{a\in\cS_\cG}\bigl(2-t(\cB_a)\bigr)-f \, .
\ee
The fact that the 't Hooft scalings \eqref{MatOptScaling} yield a well-defined $1/N$ expansion is ensured by the following theorem.

\begin{theorem}\label{MatTheorem}
Let $\cG$ be a connected Feynman graph of the $\oN\times\oN$ matrix model. Then, $h\geq 0$. In other words, the power of $N$ associated with any connected Feynman graph $\cG$ is bounded above by $2$.
\end{theorem}

\proof Using Eq.\ \eqref{EulerFormFeynGraphStrand}, the parameter $h$ can be rewritten as 
\be\label{MatHParam2}
h = 2g + 2 \Bigl[1+ \sum_{a\in\cS_\cG} \bigl(t(\cB_a)-1\bigr) - c(\tilde{\cG}) \Bigr] \, .
\ee
The first term on the RHS of this expression is non-negative since $g\in\frac{1}{2}\mathbb{N}$. Let us show that the second term in brackets is also non-negative. The argument is similar to the one used in the first version of the proof for Theorem \ref{VecTheorem} in the case of vector models.\footnote{It can also be seen as a special case of the connectivity inequality \eqref{inequal2}, see Ref.\ \cite{Ref2}. Moreover, it can be interpreted as the number of loops (in the sense of independent cycles) in a connected abstract graph whose vertices are the connected components of the ribbon graph $\tilde{\cG}$ and the multi-trace interaction vertices in $\cG$, and whose edges join each multi-trace interaction vertex to all the connected components it belongs to.} Breaking down the multi-trace interaction vertices of type $\cB_a$ in $\cG$ into their $t(\cB_a)$ single-trace parts yields the ribbon graph $\tilde{\cG}$, which has $c(\tilde{\cG})$ connected components. In this process, breaking down a given interaction vertex with $t(\cB_a)$ traces can increase the number of connected components by at most $t(\cB_a)-1$. The total number of connected components generated after the whole process is thus at most $\sum_{a\in\cS_\cG} (t(\cB_a)-1)$. Since $\cG$ is connected, it implies $c(\tilde{\cG})-1\leq \sum_{a\in\cS_\cG} (t(\cB_a)-1)$. Finally, since $h$ corresponds to a sum of two non-negative terms, it is also non-negative.
\qed

\

For later purposes, it is useful to write the parameter $h$ given in Eq.\ \eqref{MatHParam2} in terms of the Feynman data in the colored representation. Using the identities \eqref{MatFeynGraphData}, this yields
\be\label{MatHParam3}
h = 2g(\cB) + 2 \Bigl[1+ F_{12}(\cB) - v(\cB) - B \Bigr] \, ,
\ee
where $v(\cB)$ corresponds to the number of interaction bubbles in $\cB$. From Eq.\ \eqref{MatHParam2} or \eqref{MatHParam3}, one can remark that $h$ is a non-negative integer, i.e.\ $h\in\mathbb{N}$.

If we denote by $\{\cG_h\}$ the set of connected Feynman graph of fixed $h$, then the $1/N$ expansion of the $\oN\times\oN$ matrix models with scaling \eqref{MatOptScaling} can be written as
\be \label{MatLargeNExpFeyn}
F=\sum_{h\in\mathbb{N}} N^{2-h} F_h\, ,
\ee
where $F_h$ corresponds to a sum over connected Feynman graphs $\cG$ in $\{\cG_h\}$ weighted by some amplitude $\tilde{\cA}(\cG)$ that includes the factors $\lambda_a$ for $a\in\cS_\cG$ and the combinatorial factors:
\be \label{MatLargeNExpFeynCoeff}
F_h= \sum_{\cG\in\{\cG_h\}} \tilde{\cA}(\cG) \, .
\ee
In particular, the $1/N$ expansion is well-defined. Similar results also hold for expectation values of invariants, which are obtained from the free energy by taking partial derivatives with respect to the coupling constants $\lambda_a$. 

\

\noindent\emph{Case of single-trace interactions}

A particular case of interest is obtained by including only single-trace interaction terms in the action \eqref{MatAction}, that is, $t(\cB_a)=1 \ \forall a\in\cS$. As a result, $t(\cB_a)=1 \ \forall a\in\cS_\cG$ and $c(\tilde{\cG})=1$ for any connected Feynman graph $\cG$ that appears in the perturbative expansion of the free energy, so that Eq.\ \eqref{MatHParam2} reduces to 
\be\label{MatHSingleTrace}
h = 2g \, ,
\ee
and the $1/N$ expansion \eqref{MatLargeNExpFeyn} becomes
\be \label{MatLargeNExpFeynSingle}
F=\sum_{g\in\frac{1}{2}\mathbb{N}} N^{2-2g} F_g\, .
\ee
The $1/N$ expansion for single-trace interactions can be nicely interpreted as a genus expansion. Since the genus is a topological invariant for connected surfaces, it is therefore a topological expansion. In particular, $F_g$ corresponds to a weighted sum over embedded graphs of genus $g$, or equivalently, over discretized surfaces of genus $g$. In many cases, it can be shown that for any fixed $g$, $F_g$ is a convergent series in $\lambda_a$ with a finite radius of convergence \cite{Mapg}.

\

For general multi-trace interactions, the discretized surfaces associated with the Feynman graphs can have several connected components. Hence, the parameter $h$ combines topological information as well as connectivity information. This is further clarified in the next section for the leading sector.

\section{Leading sector: planar graphs}
\label{sec:MatLO}

According to Eq.\ \eqref{MatLargeNExpFeyn}, the leading order Feynman graphs satisfy $h=0$. Using Eq.~\eqref{MatHParam2}, it is equivalent to the conditions
\be \label{MatLOGraph1}
\sum_{a\in\cS_\cG} \bigl(t(\cB_a)-1\bigr) = c(\tilde{\cG})-1 \, ,
\ee
\be \label{MatLOGraph2}
g=0 \, .
\ee
The first condition is similar to the interpretation of bubble graphs for vector models, as explained in Section \ref{sec:VecLO}. It implies that a leading order Feynman graph $\cG$ is effectively ``maximally disconnected" with respect to its interaction vertices. More precisely, using the argument in the proof of Theorem \ref{MatTheorem}, the single-trace parts of each multi-trace vertex must belong to distinct connected components of the ribbon graph $\tilde{\cG}$ associated with $\cG$. Note that this is automatically satisfied for single-trace interaction bubbles. As for the second condition, it says that each connected component of $\tilde{\cG}$ must be planar, that is, they must correspond to graphs embedded on the sphere $\mathbb{S}^2$. Examples of leading order Feynman graphs are the graph on the left and the two on the right of Figure \ref{MatFeynGraphsStrand} or \ref{MatFeynGraphsColor}.

A geometrical interpretation of the LO Feynman can also be given using the footnote $2$ in the proof of Theorem \ref{MatTheorem}. In terms of the connected abstract graph described in the footnote, being maximally disconnected means that it corresponds to a tree. Then, since planar graphs correspond to discretized spheres, a LO Feynman graph can thus be interpreted geometrically as a tree of discretized spheres. 

If we restrict the action to single-trace interactions only, the large $N$ limit is then governed by planar graphs. We already stressed the importance played by planar graphs in theoretical physics, from QCD to holography. Many nice properties of matrix models result from planar graphs. For instance, they can be precisely counted through algebraic equations and their number is exponentially bounded at fixed number of vertices \cite{BrezAl}.\footnote{We note that contrary to the bubble graphs of vector models, planar graphs cannot be obtained through a recursive procedure of $2$-point graph insertions.} 

The last point implies that the LO free energy $F_0$ is convergent with a non-zero radius of convergence and therefore exhibits a critical behavior. There are different methods to compute the corresponding susceptibility critical exponent, including purely combinatorial approaches and the saddle-point approximation approach. The interested reader can find more details on these types of computations in \cite{BrezAl,KlebaLargeN,Eyn,Marino}. Here, we only describe the critical properties.

In the case of matrix models based on a single generic single-trace interaction with coupling constant $\lambda$ and with 't Hooft scaling, the LO free energy exhibits a critical behavior of the form
\be \label{MatCritBehavior}
F_{0,\text{sing}}(\lambda) \sim (\lambda_c-\lambda)^{2-\gamma} \quad \text{with} \ \gamma=-\frac{1}{2} \, ,
\ee
where $\lambda_c$ is some critical value. The value $\gamma=-\frac{1}{2}$ of the susceptibility critical exponent means that generically, matrix models fall into the universality class of pure two-dimensional gravity \cite{DiFran}. As random objects, planar graphs converge in the continuum limit to a random continuous space called the Brownian map \cite{BM}. This is a strong evidence that matrix models are intimately connected to two-dimensional quantum gravity. More information on this connection can be found in \cite{DiFran} and references therein.

Results for multi-trace matrix models can also be found in the literature \cite{MTMM}. In this case, several critical behaviors can be achieved depending on the values of the different coupling constants. As explained above, the LO Feynman graphs correspond to trees of planar graphs. Thus, there is a competition between the tree-like structure of the graphs and their surface-like structure. When the tree-like structure dominates over the surface-like structure, the continuum limit corresponds to the universality class of branched polymers ($\gamma=\frac{1}{2}$); while in the reverse situation, it falls into the universality class of pure two-dimensional gravity ($\gamma=-\frac{1}{2}$). Interestingly, when the two structures are both competitive, a new continuous phase can be reached with $\gamma=\frac{1}{3}$, which describes baby universes \cite{BU}. Finally, tuning more couplings to criticality leads to more phases with various (positive) susceptibility critical exponents.

The above critical behaviors are all obtained for large $N$ matrix models in zero dimension. In this special case, the sum over planar graphs can be thoroughly analyzed with various techniques. In contrast, if one is interested in large $N$ matrix models in one dimension, i.e.\ matrix quantum mechanics, the sum over planar graphs becomes much more involved and it is not known how to evaluate it analytically. This situation happens for instance in the case of holographic quantum models of black holes. Of course, the same observation holds in higher dimensions, for instance in QCD. It is therefore natural to look for strategies that simplify the sum over planar graphs but also captures its physics. One possibility, initiated in \cite{FrankLargeD} and further discussed in Chapter \ref{chap:MatrixTensor}, is to introduce a new parameter $D$ in the models that allows for the study of the large $D$ limit of the planar graphs. For an appropriate large $D$ scaling of the coupling constants, a subset of the planar graphs dominates when $D\rightarrow\infty$, which is simple enough to be summed explicitly and studied analytically and yet relevant because it reproduces some important aspects of the physics associated with the full sum over planar graphs. In particular, these models display in the large $D$ limit the same properties observed in the SYK model and SYK-like tensor models.

\section{Discussion}
\label{sec:MatDiscussion}

In this section, we regroup further comments on matrix models. We first discuss the case of matrix models based on a complex matrix with symmetry group $\uN\times\uN$. We then describe what happens when we reduce the symmetry to a single group $\oN$ or $\uN$ for a real symmetric matrix or a complex Hermitian matrix respectively. 

\subsection{Case of complex matrices}
\label{sec:MatComplMat}

\begin{figure}[]
\centerline{\includegraphics[scale=1]{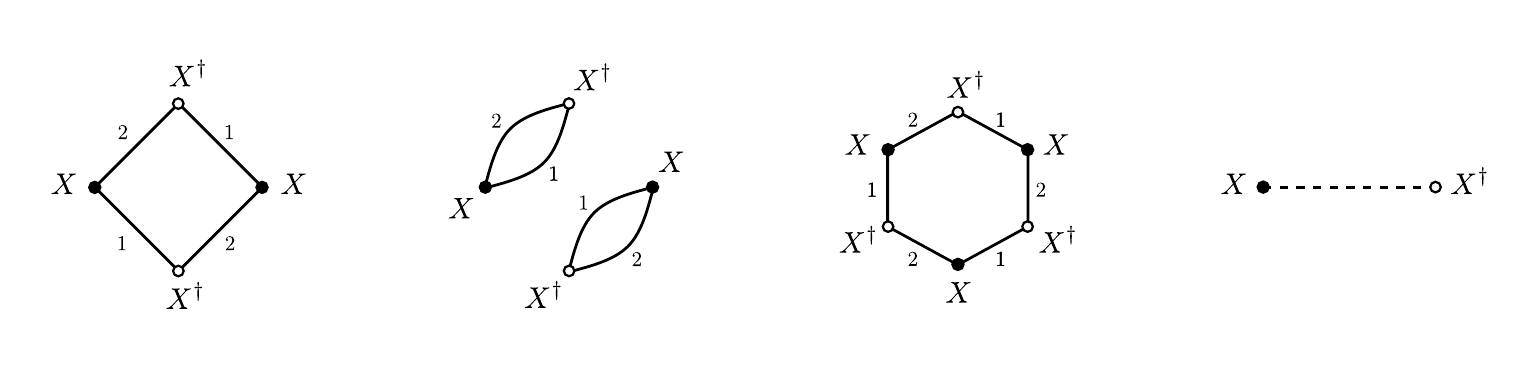}}
\caption{Examples of interaction bubbles and propagator of the $\uN\times\uN$ matrix models in the colored representation. In particular, the interaction bubbles are manifestly bipartite and the propagator always connects a black vertex with a white vertex in a Feynman graph.}\label{MatFeynRulesColorComp}
\end{figure}

The $\oN\times\oN$ matrix models can be straightforwardly modified to matrix models with symmetry $\uN\times\uN$ based on a complex matrix $X=(X_{a_1a_2})$ and its Hermitian conjugate $X^\dagger=(X^\dagger_{a_2a_1})=(\bar{X}_{a_1a_2})$, where $\bar{X}$ denotes the complex conjugate matrix. The transformation laws in terms of the matrix components are  
\be \label{MatTransfLawComp}
X_{a_1a_2} \rightarrow X'_{a_1a_2} = U^{(1)}_{a_1a_1'}U^{(2)}_{a_2a_2'}X_{a_1'a_2'} \, , \quad X^\dagger_{a_2a_1} \rightarrow X^{\dagger \, '}_{a_2a_1} = \bar{U}^{(2)}_{a_2a_2'}\bar{U}^{(1)}_{a_1a_1'}X^\dagger_{a_2'a_1'} \, ,
\ee
where $U^{(1)}$ and $U^{(2)}$ are independent $N\times N$ unitary matrices. In particular, the two matrix indices are again distinguishable; hence, we can assign two distinct colors to them. Invariant interaction terms are now of the form
\be\label{MatInteractionComp}
\cI_a(X,X^\dagger) = \prod_{i=1}^{t_a} \tr (XX^\dagger)^{p_i(a)}\, ,
\ee
and the action can be written in terms of the 't Hooft scalings as
\be\label{MatActionScaledComp}
S(X,X^\dagger)= N\biggl(\tr (XX^\dagger) + \sum_{a\in\cS} N^{1-t(\cB_a)}\lambda_a \, \cI_a(X,X^\dagger)\biggr) \, .
\ee
Note that the t'Hooft scalings are the same as earlier. From there, a reasoning similar to the one used it the previous sections can be applied. We point out a few interesting remarks. First, there are two distinct variables $X$ and $X^\dagger$ and to be consistent with the $\uN\times\uN$ symmetry, an index of $X$ is always contracted with an index of $X^\dagger$ in an invariant interaction term $\cI_a$. As a consequence, the interaction bubbles $\cB_a$ are manifestly bipartite.\footnote{Note that the interaction bubbles in the $\oN\times\oN$ models are also bipartite because of the symmetry group, even though there is a single real matrix $X$.} A few examples are illustrated in Figure \ref{MatFeynRulesColorComp}, where the vertices associated with a $X$ are drawn as black vertices and the ones associated with a $X^\dagger$ as white vertices. 

Second, the free propagator deduced from the action \eqref{MatActionScaledComp} is given by 
\be \label{MatFree2PtFunctionComp}
\bigl\langle X_{a_1a_2}X^\dagger_{b_2b_1}\bigr\rangle_0\,=\frac{1}{N}\delta_{a_1b_1}\delta_{a_2b_2} \, .
\ee
It implies that the propagators in the stranded representation always correspond to untwisted ribbon edges. The Feynman graphs of the $\uN\times\uN$ matrix models thus correspond to orientable embedded graphs and form a strict subset of the ones for the $\oN\times\oN$ matrix models. This observation follows a general trend: larger the symmetry group, smaller the set of Feynman graphs. In the colored representation, the equivalent statement is that dashed edges of color $0$ only connect a black vertex with a white vertex (see Figure \ref{MatFeynRulesColorComp}); hence, the Feynman graphs are bipartite.

%An important consequence of the previous remark is that the $1/N$ expansion of the $\uN\times\uN$ matrix models can be directly obtained by restricting the perturbative expansion \eqref{MatLargeNExpFeyn} to orientable embedded graphs only. This follows from the fact that orientable embedded graphs can always be represented with untwisted edge ribbons only (see Appendix \ref{sec:App2}). In particular, in the expression \eqref{MatHParam2} for the parameter $h$ that governs the $1/N$ expansion, $g$ is now in $\mathbb{N}$ so that $h\in2\mathbb{N}$.

Finally, in the large $N$ limit, the leading sector of $\uN\times\uN$ matrix models is the same as the one for $\oN\times\oN$ matrix models (up to combinatorial factors), that is, it contains all the ``maximally disconnected" planar graphs. 

\subsection{Case of reduced symmetry}
\label{sec:MatRedSym}

A natural question to ask is whether similar results can be obtained if we reduce the symmetry group down to a single group $\oN$ or $\uN$. For instance, the case of $\oN$ would correspond to models based on a real symmetric matrix $X=\tra{X}$ with transformation law $X\rightarrow X'=OX\tra{O}$, where $O\in\oN$. In the terminology of random matrices, $X$ would be in the Gaussian Orthogonal Ensemble (GOE) \cite{Eyn}. As for the case of $\uN$, it would correspond to models based on a complex Hermitian matrix $X=X^\dagger$ with transformation law $X\rightarrow X'=UXU^\dagger$, where $U\in\uN$; $X$ would then be in the Gaussian Unitary Ensemble (GUE) \cite{Eyn}. 

Such models differ from the ones studied above in an important aspect: the matrix indices transform with respect to the same symmetry group and thus cannot be distinguished, that is, they cannot be assigned distinct colors. The colored representation therefore breaks down because there is no well-defined notion of bubble anymore. However, the stranded representation can still be used with the difference that there are no colors assigned to the strands. As a consequence, there are many more Feynman graphs that appear in perturbation theory, which is consistent with the general trend mentioned above because the size of the symmetry group has been reduced. From the point of view of embedded graphs in the ribbon graph representation, not having colors is not a problem. The Feynman graphs of the models with reduced symmetry still correspond to embedded graphs and the free energy still admits a well-defined $1/N$ expansion onto embedded graphs, governed by the genus. More details can be found in \cite{Eyn,Marino,KlebaLargeN}. We emphasize that this result is rather specific to matrix models. In particular, it does not directly hold for tensor models (see next chapter).

%%%%%%%%%%%%%%%%%%%%%%%%%%%%%%%%%%%%%%%%%%%%%%%%
%%%%%%%%%%%%%%%%%%%  Chapter 4  %%%%%%%%%%%%%%%%%%%%%%%%%
%%%%%%%%%%%%%%%%%%%%%%%%%%%%%%%%%%%%%%%%%%%%%%%%

\chapter{Large $N$ tensor models}
\label{chap:Tensor}

After vector and matrix models, it is natural to study theories based on tensor degrees of freedom. In this case, the field corresponds to a tensor of rank $R$ whose $R$ indices take values from $1$ to $N$. The number of degrees of freedom is thus $\cN\sim N^R$. In the previous chapters, we observed that for vectors and matrices, which can be viewed as tensors of rank $1$ and $2$ respectively, the models become solvable in the limit of a large number of degrees of freedom. This was made possible by constructing a well-defined and non-trivial $1/N$ expansion for the models and by restricting the expansion to the Feynman graphs that dominate when $N$ is large. The question for tensor models is therefore whether one can define a $1/N$ expansion in the same spirit and whether the set of Feynman graphs that dominate in the large $N$ limit is analytically tractable. Because the complexity of taking the large $N$ limit increases from $R=1$ to $R=2$, one would \textit{a priori} expect that tensor models for $R>2$ are more difficult than matrix models. As we shall see, it is nevertheless possible to find tensor models that become solvable and tractable in the large $N$ limit.

Initially, tensor models were introduced as a generalization of random matrix models to higher dimensions. More precisely, we remarked in Chapter \ref{chap:Matrix} that the perturbative expansion of matrix models performs a sum over discretized surfaces. By extension, tensor models are the natural generalization that implement in a consistent way a sum over higher dimensional discretized geometries. This idea can be related to quantum gravity above two dimensions \cite{QG} in the same way as for matrix models in two dimensions. As explained in the introduction, this approach is based on a discretized version of the path integral for gravity, which can be described in $R$ dimension by a weighted sum over discretized $R$-dimensional geometries and topologies. As a result, having complete control over tensor models may in principle lead to advances in the understanding of quantum gravity. However, no progress could be made at first because one could not construct a $1/N$ expansion that simplifies the models when $N$ is large. 

The situation changed with the discovery of the $1/N$ expansion \cite{LargeNColored} for colored tensor models \cite{Colored}. These models are based on $R+1$ different complex tensors of rank $R$, $T^j_{a_1a_2\cdots a_R}$ with $j=0,1,\ldots,R$ and $a_i=1,2,\ldots,N$ for $i=1,2,\ldots,R$. Their symmetry group is $\uN^{R(R+1)/2}$ and they include a single type of interaction term. The perturbation expansion of these models supports a well-defined and non-trivial $1/N$ expansion indexed by a new integer, the Gurau degree, hereafter called degree, which plays in higher dimensions the role of the genus. However, it is no longer a topological invariant for $R\geq3$. The Feynman graphs that dominate in the large $N$ limit have degree zero and are called melons. Their structure was first identified in \cite{Melons} and it was shown that they correspond to discretized spheres in $R$ dimensions \cite{LargeNColored}. Surprisingly, for $R\geq3$, melons are more restricted than planar graphs. In particular, the leading sector of the colored tensor models is less complex that the one for matrix models. It is more similar to vector models in the sense that it is characterized by a recursive structure, it can be summed explicitly by mapping the melons to (decorated) trees \cite{Melons,Vira} and it behaves like branched polymers \cite{BranchedPolymer}. Colored tensor models provided the first analytically tractable theory of random discretized geometries in dimensions higher than two. It opened interesting perspectives on quantum gravity in arbitrary dimensions, nicknamed the tensor track \cite{TensorTrack}. An exhaustive review of colored tensor models can be found in \cite{GurauBook}.

It was later discovered that a second class of tensor models admits $1/N$ expansions, the so-called uncolored tensor models \cite{BGR}. These models differ from their colored counterpart in the number of tensor fields, in the symmetry group and in the allowed interactions.\footnote{In fact, uncolored tensor models can be obtained from colored ones by integrating out all the colored tensors but one, which leads to an effective action for a single uncolored tensor, see \cite{Vira}.} More precisely, they are based on a single, uncolored complex tensor of rank $R$, $T_{a_1a_2\cdots a_R}$ with $a_i=1,2,\ldots,N$ for $i=1,2\ldots,R$, the symmetry group is reduced to $\uN^R$ and any interaction invariant under the action of this symmetry group can be considered (see below for more details). For $R\geq 3$, there is a much larger class of possible interactions than in the case of vector, matrix or colored tensor models. To define a large $N$ expansion for uncolored tensor models, Bonzom, Gurau and Rivasseau (BGR) introduce in \cite{BGR} a particular scaling of the coupling constants that yields a well-defined and non-trivial $1/N$ expansion. Remarkably, this expansion is indexed by the degree of the Feynman graphs, as in colored tensor models. 

In this chapter, we focus on the study of uncolored tensor models or tensor models for short. The first objective is to generalize the analysis of \cite{BGR} to the symmetry group $\oN^R$ based on a single real tensor, which was first done in Ref.\ \cite{Ref2}. Note that the first instance of models based on a real tensor was \cite{CT} for $R=3$. However, the authors do not use the BGR scaling in the large $N$ limit and the general approach is distinct from the one in \cite{BGR} (see below). Here, we follow the approach of \cite{BGR} and we generalize the associated tools to establish the $1/N$ expansion, which includes the jackets and the degree. 

The second objective of this chapter concerns enhanced scalings. As explained in Section \ref{sec:VecLargeNExp}, a large $N$ scaling for a particular coupling constant is optimal if it yields a well-defined and non-trivial $1/N$ expansion. In particular, it means that it is impossible to enhance an optimal scaling and when an optimal scaling is found for a given coupling constant, the corresponding interaction vertex can appear an arbitrary number of times in leading order Feynman graphs. As we shall review, the BGR scaling is optimal for melonic interactions, but is not optimal in general. For a given interaction, the corresponding optimal scaling, if it exists at all, can be very complicated to compute in tensor models. In fact, optimal scalings are understood only for a small subset of all possible interactions. This is drastically different from what happens in vector and matrix models, where the optimal scaling is known for any interaction.

Finding optimal scalings for non-melonic interactions is an important and active research topic, first initiated in \cite{BonzomNewLargeExp}. It allows one to define new large $N$ expansions for tensor models for which the class of leading order Feynman graphs is larger than the class of melons. In terms of physics, a larger leading sector means that more physical phenomena can be \textit{a priori} described when $N$ is large. Up to now, the study of optimal scaling has been mainly conducted on a case-by-case basis for specific interactions or specific ranks. An interesting instance of non-BGR scaling was introduced in \cite{nonBGR1} for a mixture of melonic and non-melonic quartic interactions at rank four. The leading sector interpolates between the usual expansion of tensor models dominated by melonic graphs and the topological expansion of matrix models dominated by planar graphs. Other examples treated in the literature can be found in \cite{nonBGR2}. 

An important example of non-BGR scaling was studied in \cite{CT} for the $\oN^3$ tensor models. The relevance of the non-BGR scaling can be understood at the level of the two quartic interactions, called the pillow and the tetrahedric interactions. If the BGR scaling is used, the pillow interaction, which is melonic, always dominates the tetrahedric interaction, which is non-melonic. Instead, the authors use the optimal scaling for the tetrahedric interaction obtained by enhancing the BGR scaling. As a result, both interactions contribute to the leading sector, which is therefore much larger than for the BGR scaling. The remarkable point is that in spite of the non-trivial enhancement, the large $N$ limit still exists and the leading order Feynman graphs can be identified for the quartic interactions. In particular, they are no longer melonic in the traditional BGR sense.

The approach of \cite{CT} is inspired from another class of tensor models, called multi-orientable (MO) \cite{MOTensor}. These models are based on a mixed symmetry group $\uN^2\times\oN$ and a single interaction term identical to the tetrahedric interaction. Because of the mixed symmetry, they yield in perturbation theory a class of Feynman graphs strictly larger than the class of Feynman graphs of standard models. It was proven in \cite{MOLargeN} that MO tensor models admit a well-defined $1/N$ expansion and that in the large $N$ limit, melonic graphs (in the sense of colored models) dominate the expansion. More details on MO tensor models can be found in the review \cite{MOReview} and references therein. 

The non-BGR scaling used in \cite{CT} was an important inspiration for the new scaling introduced in \cite{FrankLargeD} and extended in Ref.\ \cite{Ref1}, both in the context of $\oD$-invariant matrix models. These models correspond to matrix-tensor models with $R=3$ (see Chapter \ref{chap:MatrixTensor}) and for $N=D$, they reduce to usual tensor models with symmetry group $\uN^2\times\oN$. In \cite{FrankLargeD}, the new scaling enhances the BGR scaling for all non-planar interactions. It also enhances the scaling of \cite{CT} for infinitely many interactions, except when it is already optimal, as for the tetrahedric interaction. Then, it is shown that the new scaling leads to a well-defined large $N$ limit. We will come back to \cite{FrankLargeD,Ref1} in the next chapters. 

In the three types of models described above--namely, the $\oN^3$ tensor models, MO tensor models and $\oD$-invariant matrix models--the approach seems to rely heavily on the particular case of rank three. In Ref.\ \cite{Ref2}, it was generalized to $\oN^R$ tensor models of any rank $R$ and to all interactions. This is an important part of this thesis. In this reference, a new large $N$ scaling is defined and studied, which enhances the BGR scaling for all non-melonic interactions and which coincides with the scaling of \cite{FrankLargeD} for $R=3$. Tensor models with $\uN^R$ symmetry group can be treated as special cases. This enhanced scaling yields a well-defined and non-trivial large $N$ expansion with expansion parameter $1/N^{\frac{1}{R-1}}$. This is in contrast with the BGR scaling, for which the expansion parameter is $1/N$. Moreover, the large $N$ expansion is indexed by a new quantity called the index of the Feynman graphs, which plays in higher dimensions the role of the genus but is not a topological invariant for $R\geq 3$, similarly to the degree. This index has a very natural interpretation in terms of all the possible matrix models one can embed in the tensor model. The leading Feynman graphs, called generalized melons, have index zero and form a larger class than the standard melons, which have degree zero. Their general classification remains a difficult open problem. Interestingly, the construction of Ref.\ \cite{Ref2} singles out a new interesting family of models based on maximally single-trace (MST) interactions, which generalize to tensor models the single-trace interactions of matrix models. The enhanced scaling can be shown to be optimal for all MST interactions. Finally, another important result of Ref.\ \cite{Ref2} is the classification of the leading order Feynman graphs in the particular case of the complete interaction of order $R+1$, when $R$ is a prime number.

We emphasize that in Ref.\ \cite{Ref2}, most of the results are obtained in the context of matrix-tensor models, which are introduced in Chapter \ref{chap:MatrixTensor}. Tensor models can be straightforwardly obtained from matrix-tensor models by setting $N=D$. For the sake of clarity, we decided to present in this chapter the results of Ref.\ \cite{Ref2} in the language of tensor models so that they are easily accessible. In Chapter \ref{chap:MatrixTensor}, we will give the results a second time for matrix-tensor models, which means that there are necessarily many overlaps between this chapter and the chapter on matrix-tensor models. Also, many parts of this chapter are taken from Ref.\ \cite{Ref2}, with slight adaptations and additional details. 

The structure and the approach used in this chapter are the same as in the previous chapters. First, we start with a general description of the $\oN^R$ tensor models and we define the invariant interactions. We then describe the Feynman graphs obtained in perturbation theory using the stranded and the colored representations. In the case of tensor models, the colored representation is particularly well-suited. In Section \ref{sec:DBubbles}, we introduce important graph-theoretical tools in this representation, which include $d$-bubbles, jackets, degree, $3$-bubble subgraphs and index. In Section \ref{sec:TensLargeNExp1}, we review the BGR scaling and the corresponding large $N$ expansion suitably generalized to $\oN^R$ tensor models with multiply-connected interaction bubbles. We also describe the leading sector which involves the melonic graphs. In Section \ref{sec:TensLargeNExp2}, we introduce the new enhanced scaling, we work out the corresponding large $N$ expansion and we discuss the leading sector made of generalized melons. Moreover, we show that the enhanced scaling is optimal for MST interactions and we give a few details on the classification of the leading order Feynman graphs for models based on the complete interaction (the full proof can be found in Appendix \ref{app:AppC}).

\section{Definition of the models}
\label{sec:TensDefModels}

We consider in this chapter models based on a real tensor of rank $R$, $T=(T_{a_1a_2\cdots a_R})$ with $a_i=1,2,\ldots,N$ for $i=1,2\ldots,R$, which transforms under the external tensor product of fundamental representations of the direct product of $R$ copies of $\oN$, denoted as $\oN^R$. The number of degrees of freedom is thus $\cN=N^R$ and the transformation law in terms of the tensor components $T_{a_1\cdots a_R}$ is given by 
\be \label{TensTransfLaw}
T_{a_1a_2\cdots a_R} \rightarrow T'_{a_1a_2\cdots a_R} = O^{(1)}_{a_1a_1'}O^{(2)}_{a_2a_2'}\cdots O^{(R)}_{a_Ra_R'}T_{a_1'a_2'\cdots a_R'} \, ,
\ee
where $O^{(i)}, i=1,2,\ldots,R$, are $R$ independent $N\times N$ orthogonal matrices. In other words, each $\oN$ group of $\oN^R$ acts independently on one of the tensor index of $T$, say, the first group on the first tensor index $a_1$, the second group on the second tensor index $a_2$ and so on. The $R$ tensor indices are thus distinguishable. Remark that for $R=2$, we naturally recover the matrix models studied in Section \ref{sec:MatDefModels}.

To construct an invariant action, we need to understand the invariants that can be built from the tensor $T$. We usually deal with trace invariants \cite{BGR}, which are polynomials in the tensor components $T_{a_1\cdots a_R}$ such that the indices that transform under the same group are all contracted two by two. This ensures the $\oN^R$ invariance. We denote by $\cI_a(T)$ a trace invariant, where $a$ is some label in a discrete set $\cS$, and by $s_a$ the number of entries $T$ in $\cI_a(T)$. A generic invariant can thus be formally written as
\be \label{TensInteraction}
\cI_a(T) = T_{a_{1,1}\cdots a_{R,1}}T_{a_{1,2}\cdots a_{R,2}} \cdots T_{a_{1,s_a}\cdots a_{R,s_a}}\, ,
\ee
where the $\oN$ indices transforming under the same group are contracted pairwise and summed over. In particular, it means that the number $s_a$ of entries is necessarily even. As we shall see in Sec\ \ref{sec:TensColorRep}, these trace invariants $\cI_a(T)$ are mapped to $R$-bubbles $\cB_a$; hence, we label them as $\cI_a(T)\equiv\cI_{\cB_a}(T)$. Note that this is again consistent with $\oN^2$ matrix models (see Section \ref{sec:MatDefModels}). Examples of trace invariants are:
\begin{itemize}
\item $T \cdot T \equiv T_{a_1a_2\cdots a_R}T_{a_1a_2\cdots a_R}$ ($s=2$);
\item $\cI_{\cB_1}=T_{a_1a_2a_3}T_{a_1b_2a_3}T_{b_1b_2b_3}T_{b_1a_2b_3}$ ($R=3$, $s_1=4$);
\item $\cI_{\cB_2}=T_{a_1a_2a_3}T_{a_1b_2b_3}T_{b_1b_2a_3}T_{b_1a_2b_3}$ ($R=3$, $s_2=4$);
\item $\cI_{\cB_3}=T_{a_1a_2a_3}T_{a_1a_2a_3}T_{b_1b_2b_3}T_{b_1b_2b_3}$ ($R=3$, $s_3=4$);
\item $\cI_{\cB_4}=T_{a_1a_2a_3a_4}T_{a_1b_2a_3a_4}T_{b_1b_2b_3b_4}T_{b_1a_2b_3b_4}$ ($R=4$, $s_4=4$);
\item etc.
\end{itemize}
In these examples, we singled out the unique invariant with two entries $T$, which corresponds to the mass term in the action. Remark that since tensors have more indices than matrices, many more interaction terms can be built.

An invariant action $S(T)$ for the $\oN^R$ tensor models can thus be written as
\be\label{TensAction}
S(T)= \frac{N^{R-1}}{2} T \cdot T + \sum_{a\in\cS} g_a \, \cI_{\cB_a}(T) = \frac{N^{R-1}}{2} T \cdot T + V(T) \, ,
\ee
where $g_a$ is the coupling constant associated with $\cI_{\cB_a}(T)$. Again, we isolated the quadratic mass term, we set the mass to one and we regrouped higher order interaction terms $\cI_{\cB_a}$ in the interaction potential $V(T)$. The factor of $N^{R-1}$ in front of the mass term is set for further convenience.

The partition function $Z$ and the free energy $F$ associated with the action \eqref{TensAction} are given by
\be \label{TensPartitionFunctionFull}
Z = \exp (-F) = \int [dT] \, e^{-S(T)}=\int [dT] \, e^{-\frac{N^{R-1}}{2} T \cdot T - \sum_{a\in\cS} g_a \, \cI_{\cB_a}(T)} \, ,
\ee
where the path integral measure $[dT]$ reduces in zero dimension to the product of $N^R$ simple integral measures on $\mathbb{R}$: $[dT] = \prod_{a_1,a_2,\ldots, a_R=1}^N \sqrt{\frac{N^{R-1}}{2\pi}}dT_{a_1a_2\cdots a_R}$. Similar expressions hold for general $n$-point functions. In particular, the free $2$-point function or free propagator is given by
\be \label{TensFree2PtFunction}
\bigl\langle T_{a_1a_2\cdots a_R}T_{b_1b_2\cdots b_R}\bigr\rangle_0\,=\frac{1}{N^{R-1}}\delta_{a_1b_1}\delta_{a_2b_2}\ldots \delta_{a_R b_R} \, .
\ee

In order to define a $1/N$ expansion for tensor models, we go to perturbation theory and expand the partition function (or the free energy) in powers of the coupling constants $g_a$ in the action (which are assumed to be small) so as to make contact with Feynman graphs.

\section{Feynman graphs}
\label{sec:TensFeynmanGraphs}

As usual, we focus on the perturbative expansion of the free energy $F$, which can be written in the form of an expansion onto connected Feynman graphs. The Feynman rules for tensor models can be constructed in the same way as in Section \ref{sec:VecFeynmanGraphs} for vector models or Section \ref{sec:MatFeynmanGraphs} for matrix models. On the one hand, an interaction term $\cI_{\cB_a}$ is represented as a vertex of degree $s_a$ with factor $-g_a$ and with $s_a$ external field components $T_{a_1a_2\cdots a_R}$. On the other hand, each edge or propagator between field components $T_{a_1a_2\cdots a_R}$ and $T_{b_1b_2\cdots b_R}$ corresponds to a Wick's contraction with assignment \eqref{TensFree2PtFunction}. We denote by $\cG$ a connected Feynman graph that appears in the perturbative expansion of $F$. We write its amplitude $\cA(\cG)$, which is $N$-dependent. As for $\oN^2$ matrix models, it is useful to describe the Feynman graphs using equivalent representations.

\subsection{Stranded representation}
\label{sec:TensStrandRep}

\begin{figure}[]
\centerline{\includegraphics[scale=1]{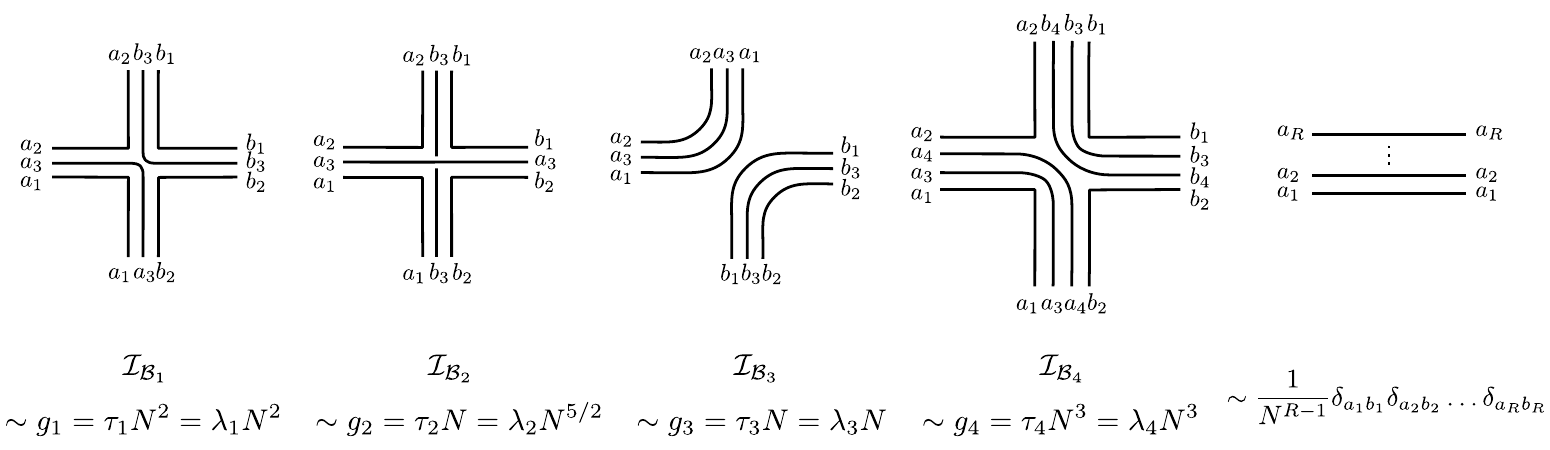}}
\caption{Feynman rules in the stranded representation for some vertices and for the propagator of the $\oN^R$ tensor models. Strands associated with the first (resp.\ second, third, etc.) tensor index are labeled with the color $1$ (resp.\ $2$, $3$, etc.). For clarity, the colors of the strands have not been indicated. The coupling parameters $\tau_a$ are defined in Eq.\ \eqref{BGRScaling} and correspond to the BGR scaling whereas the coupling parameters $\lambda_a$ are defined in Eq.\ \eqref{EnScaling} and correspond to the enhanced scaling.}\label{TensFeynRulesStrand}
\end{figure}

The stranded representation for tensor models is constructed similarly as for vector and matrix models. Since the field components $T_{a_1a_2\cdots a_R}$ have $R$ distinguished indices, we represent them as $R$ strands with $R$ distinct colors, say, color $1$ for the first index, color $2$ for the second index, etc. The Feynman rules for the vertices associated with $\cI_{\cB_1}$, $\cI_{\cB_2}$, $\cI_{\cB_3}$ and $\cI_{\cB_4}$ defined above and for the propagator are given in the stranded representation in Figure \ref{TensFeynRulesStrand}. The vertex associated with $\cI_a(T)$ consists in $s_a$ $R$-tuples of strands, one $R$-tuple with $R$ distinct colors for each field component, and strands are linked two by two via a corner if they carry the same $\oN$ index. As for the propagator, it corresponds to $R$ strands of distinct colors $i\in\{1,2,\ldots,R\}$, which connects two $R$-tuples of strands in vertices, respecting the colors. 

Given a connected Feynman graph $\cG$ in the stranded representation, we denote by $p$ its number of propagators, by $v$ its number of interaction vertices, by $f$ its number of closed stranded loops and by $\cS_\cG$ the discrete set that accounts for the different types of interaction vertices appearing in $\cG$. Note that $\cG$ may effectively correspond to a disconnected stranded graph, denoted as $\tilde{\cG}$, because nothing restricts an interaction vertex to be effectively disconnected (see Figure \ref{TensFeynRulesStrand}). We denote by $\tilde{v}$ the number of effective ``connected'' interaction vertex in $\tilde{\cG}$.

In fact, the stranded representation for tensor models is rather cumbersome to use so that we generally rely on the colored representation.

\subsection{Colored representation}
\label{sec:TensColorRep}

The usefulness of the colored representation becomes manifest for tensor models. It is constructed in a similar way as for $\oN^2$ matrix models (see Section \ref{sec:MatColorRep}). 

\begin{figure}[]
\centerline{\includegraphics[scale=1]{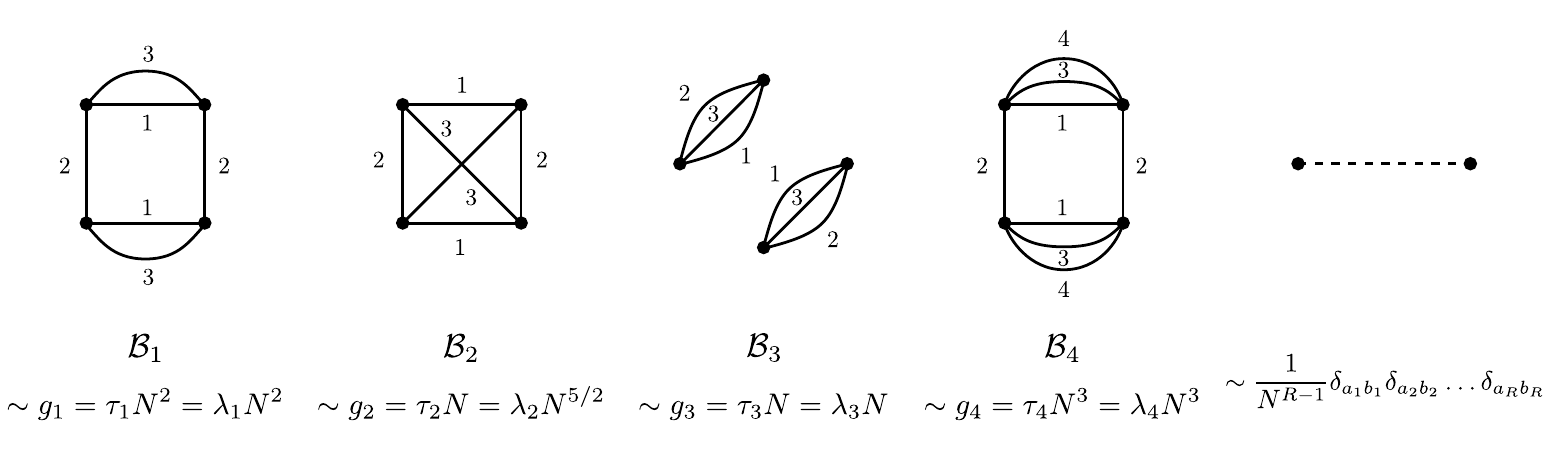}}
\caption{Interaction bubbles corresponding to the vertices of Figure \ref{TensFeynRulesStrand} and representation of the propagator in the colored representation. The first graph on the left is commonly called the pillow interaction and the second one the tetrahedric interaction. They both have four vertices and correspond to tensors of rank three. Their scaling with $N$ is the same as in Figure \ref{TensFeynRulesStrand}.}\label{TensFeynRulesColor}
\end{figure}

First, we represent the interaction terms $\cI_{\cB_a}$ entering the action \eqref{TensAction} as $R$-bubbles $\cB_a$. To each tensor field $T$ in $\cI_{\cB_a}$, we associate a vertex in $\cB_a$ and then we draw an edge of color $i\in\{1,2,\ldots,R\}$ between two vertices of $\cB_a$ if the corresponding tensor fields have their $i^{\text{th}}$ index contracted together in $\cI_{\cB_a}$. This construction yields a $R$-bubble thanks to the $\oN^R$ invariance. This is illustrated in Figure \ref{TensFeynRulesColor} for the vertices of Figure \ref{TensFeynRulesStrand}. The first interaction bubble on the left is often referred to as the pillow interaction (of order four with three colors) and the second one on the left as the tetrahedric interaction (of order four with three colors), which is based on the complete graph $\cK_4$ on four vertices. 

Similarly to $\oN^2$ matrix models, we use the notations $\cI_{\cB_a}$ and $\cB_a$ interchangeably, referring to them as interaction terms or interaction bubbles. An interaction bubble $\cB_a$ for tensor models may be disconnected; we denote by $c(\cB_a)$ the number of connected components of $\cB_a$. Unlike $\oN^2$ matrix models, the number of connected components is however not related to the number of traces as there is no notion of ``trace" for tensors of rank $R>2$.

If we restrict ourselves to bipartite interaction bubbles $\cB_a$, then we obtain interaction terms for $\uN^R$ tensor models based on a complex tensor, in the same spirit as in Section \ref{sec:MatComplMat}. For instance, the interaction bubbles presented in Figure \ref{TensFeynRulesColor} are all bipartite except the tetrahedric interaction, which therefore does not exist in $\uN^R$ tensor models.

\begin{figure}[]
\centerline{\includegraphics[scale=1]{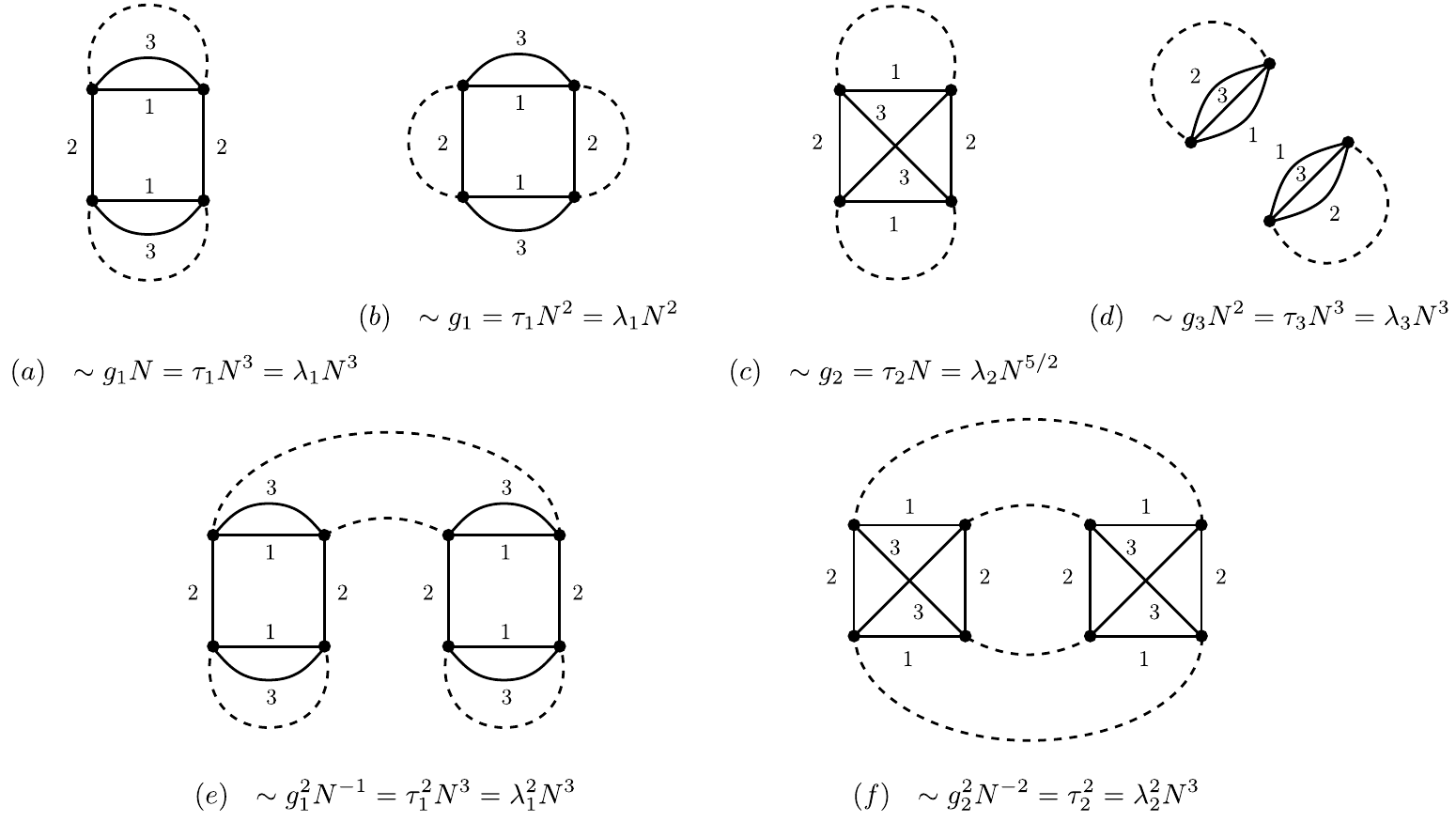}}
\caption{Example of Feynman graphs in the colored representation for the $\oN^3$ tensor models, constructed from the vertices of Figure \ref{TensFeynRulesColor}. Their scaling in $g_a$ and $N$ in the perturbative expansion are indicated, as well as their scaling in terms of the coupling parameters $\tau_a$ (see Eq.\ \eqref{BGRScaling}) and $\lambda_a$ (see Eq.\ \eqref{EnScaling}). Feynman graphs that scale like $N^3$ are part of the leading sector.}\label{TensFeynGraphsColor}
\end{figure}

On the other hand, the Feynman graphs $\cG$ in the colored representation are mapped to $(R+1)$-bubbles; the interaction vertices being represented as interaction $R$-bubbles and the propagators as dashed edges of color $0$ between vertices of interaction bubbles. Examples of Feynman graphs are shown in Figure \ref{TensFeynGraphsColor}. If we only consider bipartite $(R+1)$-bubbles, then we recover the Feynman graphs of $\uN^R$ tensor models. However, we stress that any $(R+1)$-bubble can be considered in the case of a real tensor.   

We use once more the notation $\cB$ to denote a given connected Feynman graph $\cG$ in the colored representation \cite{Ref2, Ref1}. We can relate the data associated with the stranded and colored representations in a similar way as in Section \ref{sec:MatColorRep}. The number $p$ of propagators in $\cG$ corresponds to the number $E_0(\cB)=\frac{1}{2}V(\cB)$ of edges of color $0$ in $\cB$. Then, the number $v$ of interaction vertices in $\cG$ corresponds to the number $v(\cB)$ of interaction bubbles in $\cB$. Third, each closed stranded loop in $\cG$ corresponds to a $(0i)$-face in $\cB$ for $i\in\{1,2,\ldots,R\}$. Fourth, the number $\tilde{v}$ of effective vertices in $\tilde{\cG}$ correspond to the number $B^{(0)}$ of connected components of $\cB^{(0)}$. Finally, the number $c(\tilde{\cG})$ of connected components of $\tilde{\cG}$ again matches the number $B$ of connected components of $\cB$. To sum up,
\be\label{TensFeynGraphData}
p=E_0(\cB)=\frac{1}{2}V(\cB)\, , \quad f=\sum_{i=1}^R F_{0i}(\cB)\, , \quad  \tilde{v}=\sum_{a\in \cS_\cG} c(\cB_a)=B^{(0)}\, , \quad c(\tilde{\cG})=B \, .
\ee

For $R=2$, we explained in Section \ref{sec:MatColorRep} that the Feynman graphs in the colored representation, which correspond to $3$-bubbles, are equivalently described in terms of ribbon graphs in the stranded representation. For $R\geq3$, we emphasize that the Feynman graphs in the stranded representation are not ribbon graphs. In particular, they cannot be assigned a genus. There exist however ``generalizations" of the genus for $R$-bubbles with $R\geq3$, which we introduce in the next section. 

\section{Results on $d$-bubbles}
\label{sec:DBubbles}

In this section, we define different notions associated with general $d$-bubbles (see Section \ref{sec:AppABubbles} for the definition). The basic idea is to find different types of embedded graphs associated with a given $d$-bubble. Then, we can define new quantities for $d$-bubbles constructed from quantities for these embedded graphs, such as the genus. It turns out that this plays a crucial role in establishing $1/N$ expansions for tensor models. We stress however that the notions introduced in this section hold in full generality for any $d$-bubble (with $d>2$, unless otherwise indicated). In particular, it does not depend on the models where these bubbles may appear. 

First, we review the concept of jackets associated with a $d$-bubble, which give rise to the degree. Then, we introduce another quantity called the index with respect to a color, which is based on identifying $3$-bubble subgraphs in a $d$-bubble. Finally, we prove addition formulas that hold for the degree and the index of $d$-bubbles. 

\subsection{Jackets and degree}

The jackets of a $d$-bubble $\cB$ were introduced in \cite{Jacket,LargeNColored} as a class of ribbon graphs associated with $\cB$. In the case of uncolored tensor models \cite{BGR}, they were defined for complex tensor models with symmetry $\uN^R$; hence, for bipartite $d$-bubbles. In Ref.\ \cite{Ref2}, the definition of jackets was extended to the case of $\oN^R$ tensor models, which is more general in the sense that $d$-bubbles are not necessarily bipartite. This part corresponds to Sections 2.2.2 and 2.2.3 in Ref.~\cite{Ref2}. 

As explained in Appendix \ref{sec:App2}, there are in general many ways of embedding a given abstract graph on a surface. Each such embedding can be described using different equivalent representations. Here, we mainly use the signed rotation system representation explained in Section \ref{sec:AppA23}, in which distinct embedded graphs are labeled by a choice of cyclic ordering of the edges incident to each vertex together with a choice of signature $+1$ or $-1$ on each edge. The connection with the ribbon graph representation is straightforward. We use the terms embedded graphs and ribbon graphs interchangeably in the following.

There is a natural set of embedded graphs associated to any $d$-bubble $\cB$. We pick a cycle $\sigma\in S_{d}$ and a partition $\pi$ of the set of vertices of the graph into two disjoint subsets $\mathscr V_{+}(\gB)$ and $\mathscr V_{-}(\gB)$. The vertices in $\mathscr V_{+}(\gB)$ and $\mathscr V_{-}(\gB)$ are called filled and unfilled, respectively. The embedded graph associated with the pair $(\sigma,\pi)$ is then defined as follows. The colored edges are cyclically ordered clockwise according to $\sigma$ around each unfilled vertex and anticlockwise around each filled vertex. Then, the signature of an edge joining two vertices of the same type (filled or unfilled) is chosen to be $-1$ and the signature of an edge joining two vertices of different types is chosen to be $+1$. The resulting embedded graph is called a jacket associated with $\gB$ and is denoted as $\gJ(\gB;\sigma,\pi)$. As any jacket of $\cB$ contains all the vertices and edges of $\cB$, it has the same connectivity as $\cB$. In addition, since any jacket of $\gB$ corresponds to an embedded graph, it is characterized by a genus $g(\gJ(\gB;\sigma,\pi))\equiv g(\gB;\sigma,\pi)$, which can be computed using Euler's formula \eqref{eq:EulerFormulaEmbGraphs}. Jackets have the following properties.

\begin{proposition}\label{jacketprop}
i) A jacket $\gJ(\gB;\sigma,\pi)$ does not depend on the choice of the partition $\pi$. ii) $\gJ(\gB;\sigma)=\gJ(\gB;\sigma^{-1})$. We can thus associate $\smash{\frac{1}{2}(d-1)!}$ distinct jackets to a given $d$-bubble. iii) If one jacket $\gJ(\gB;\sigma)$ is orientable, then all the jackets $\gJ(\gB;\sigma')$, for any $\sigma'$, are orientable. We then say that the bubble $\gB$ itself is orientable. iv) A bubble is orientable if and only if the underlying graph is bipartite. In particular, planar jackets $\gJ(\gB;\sigma)$ can exist only if $\gB$ is bipartite.
\end{proposition} 

\proof i) If one changes the type of a given vertex, then, by definition, the new jacket is obtained from the old one by a local switch. We can thus denote the jackets simply by $\gJ(\gB;\sigma)$, even though, in practice, when one draws the graph, one chooses a partition $\pi$. ii) This is a consequence of i), because changing $\sigma$ to $\sigma^{-1}$ is equivalent to flipping the type of all the vertices of the graph. iii) If $\gJ(\gB;\sigma)$ is orientable, then it is well-known that, modulo local switches, it has a representation with untwisted ribbons only. The underlying graph is thus manifestly bipartite. We can then use the partition $\pi$ associated with the bipartite structure to find a manifestly orientable embedding for any jacket $\gJ(\gB;\sigma')$. iv) Immediately follows from the proof of iii).
\qed

\

At this stage, one has in principle two distinct notions of faces. On the one hand, we have the $(\alpha\beta)$-faces of the $d$-bubble $\gB$, as defined in Section \ref{sec:AppABubbles}. On the other hand, we have the usual faces associated with the ribbon graphs $\gJ(\gB;\sigma)$. There is a fundamental relation between the two notions given by the following proposition.
\begin{proposition}\label{faceprop} The faces of the ribbon graph $\gJ(\gB;\sigma)$ are in one-to-one correspondence with the subset of $(\alpha\beta)$-faces of $\gB$ satisfying $\beta=\sigma(\alpha)$ or $\alpha=\sigma(\beta)$.
\end{proposition}

\proof This is a direct consequence of the precise definition of the jackets. In particular, it relies on the link between the signed rotation system representation and the ribbon graph representation of embedded graphs, where an edge of signature $+1$ and $-1$ corresponds to an untwisted and twisted ribbon edge respectively (see Section \ref{sec:AppA23}).
\qed

\

The jackets of $d$-bubbles as defined above naturally generalize the definition of \cite{LargeNColored}. In particular, one can check that the two definitions match for bipartite $d$-bubbles, in which case they correspond to orientable embedded graphs by Proposition \ref{jacketprop}. In contrast, we emphasize that in the general case, jackets may correspond to non-orientable embedded graphs; hence, $g(\gB;\sigma,\pi)\in\frac{1}{2}\mathbb{N}$. 

The case of $3$-bubbles is particularly simple because there is only one jacket modulo the orientation of the cycle $\sigma$. Furthermore, the genus of the jacket, which follows from Eq.~\eqref{eq:EulerFormulaEmbGraphs}, coincides with the genus of the $3$-bubble defined in Eq.\ \eqref{EulerFormFeynGraphColor}. Indeed, given a $3$-bubble $\cB$ and its jacket $\gJ(\gB;\sigma)=\gJ(\gB;\sigma^{-1})$ with $\sigma = (123)$, we naturally have $c(\cB)=c(\gJ(\gB;\sigma))$, $V(\cB)=V(\gJ(\gB;\sigma))$, $E(\cB)=E(\gJ(\gB;\sigma))$. Then, Proposition \ref{faceprop} implies that $F(\cB)=F(\gJ(\gB;\sigma))$. As a result,
\be\label{genusJacket3bubble}
g(\cB)=g(\gJ(\gB;\sigma))\, \qquad \text{for a 3-bubble}\ \cB\, .
\ee
In this sense, $3$-bubbles are equivalent to embedded graphs.

Based on the notion of jackets, we then define the degree of a $d$-bubble $\gB$, for $d\geq 3$, as the sum of the genera of its jackets \cite{LargeNColored} 
\be\label{degreedef} \deg\gB = \frac{1}{2}\sum_{\text{cycles}\ \sigma\in S_{d}} g(\gB;\sigma)\, .\ee
The factor $\frac{1}{2}$ simply takes into account that $\gJ(\gB;\sigma)=\gJ(\gB;\sigma^{-1})$. By construction, the degree is a non-negative half-integer; a stronger result can actually be proven, see Eq.~\eqref{degreeN}. Note that the degree of a multiply-connected bubble is the sum of the degrees of its connected components. In the case of $3$-bubbles, since there is only one jacket modulo the orientation of the cycle, the degree coincides with the genus of the jacket, or equivalently using Eq.~\eqref{genusJacket3bubble}, with the genus of the $3$-bubble.

One can derive a useful generalization of Euler's formula for $d$-bubbles, which relates the degree to the number of faces, vertices and edges.
\begin{proposition}\label{degreeformprop} The degree satisfies
\be\label{degreeformula} (d-1)c(\gB) - \frac{2}{(d-2)!}\deg\gB = F(\gB) - \frac{1}{4}(d-1)(d-2) V(\gB)\, ,\ee
and in particular
\be\label{degreeN} \frac{2}{(d-2)!}\deg\gB\in\mathbb N\, .\ee
\end{proposition}

\proof Eq.\ \eqref{degreeformula} follows from a simple counting of the faces using Proposition \ref{faceprop} and Euler's formula \eqref{eq:EulerFormulaEmbGraphs} for the jackets. Taking into account the fact that $V(\gB)$ is always even (see Eq.\ \eqref{EVrel}), Eq.\ \eqref{degreeformula} implies \eqref{degreeN}.
\qed

\

One can check that in the case of a $3$-bubble, Eq.\ \eqref{degreeformula} reduces to Euler's formula \eqref{EulerFormFeynGraphColor} as expected. In this sense, the degree provides a generalization of the genus for $d$-bubbles. We will come back to this point later when we establish $1/N$ expansions for tensor models.

Finally, we shall need the following crucial lemma, which is a direct generalization of Lemma 7 in \cite{Vira}.
\begin{lemma}\label{degdeglem} For any $d\geq 4$ and any choice of color $\alpha$,
\be\label{degineq} \deg\gB \geq (d-1)\deg\gB^{(\alpha)}\, .\ee
\end{lemma}

\proof To any jacket $\gJ(\gB;\sigma)$ is associated the jacket $\gJ(\gB^{(\alpha)};\sigma^{(\alpha)})$, where $\sigma^{(\alpha)}$ is the cycle obtained from $\sigma$ by deleting the color $\alpha$. The jacket $\gJ(\gB^{(\alpha)};\sigma^{(\alpha)})$ is obtained from $\gJ(\gB;\sigma)$ by deleting the ribbon edges associated with the edges of color $\alpha$ in $\gB$. It is well-known that this operation cannot increase the genus of the ribbon graph; hence, $g(\gB;\sigma)\geq g(\gB^{(\alpha)};\sigma^{(\alpha)})$. Summing this inequality over all cycles $\sigma$ yields the inequality~\eqref{degineq}.
\qed

\subsection{$3$-bubble subgraphs and index}

We now move on to the definition of the index of a $d$-bubble, which plays a prominent role in this thesis and in \cite{Ref2,Ref1}. There are two equivalent ways of defining the index. The first one relies on the degree defined previously and Lemma \ref{degdeglem}; whereas the second one is based on identifying another class of embedded graphs associated with a $d$-bubble, namely its $3$-bubble subgraphs.\footnote{The idea of $3$-bubble subgraph of a $d$-bubble was introduced in \cite{LargeNColored} under the name of bubble in the context of colored tensor models.} To stick with the presentation of Ref.\ \cite{Ref2}, we use the first way to define the index and we deduce the second one as a proposition. This part corresponds to Sections 2.3.1 and 2.3.2 in Ref.\ \cite{Ref2}. We use the set of colors $\mathscr C=\{0,1,\ldots,d-1\}$, singling out the color 0 for convenience and using latin indices $i$, $j$, etc., to label the colors from $1$ to $d-1$, but not 0.

The index of a $d$-bubble $\gB$, $d\geq 4$, with respect to the color $0$ is defined by
\be\label{inddef} \ind_{0}\gB = \frac{1}{(d-3)!}\Bigl(\deg\gB - (d-1)\deg \gB^{(0)}\Bigr)\, .\ee
From \eqref{degreeN} and \eqref{degineq}, we deduce that the index is a non-negative half-integer,
\be\label{ind0N} \ind_{0}\gB\in\frac{1}{2}\,\mathbb N\, .\ee
If $\gB$ is bipartite, the decomposition formula \eqref{fundid0} below actually shows that $\ind_{0}\gB$ is an integer. If the bubble is multiply-connected, its index is the sum of the indices of the connected components. For $d=3$, the index of a $3$-bubble coincides with its degree and its genus.

Consider a $d$-bubble with $d\geq4$. As explained in Section \ref{sec:AppABubbles}, removing all the edges of some colors in $\cB$ yields another bubble. A particular case of interest is to consider the set of all $3$-bubbles $\cB_{(0ij)}$, $i,j\in \mathscr C=\{0,1,\ldots,d-1\}$, obtained from $\cB$ by keeping the edges of colors $0,i,j$ and deleting all the others. We call these $3$-bubbles the $3$-bubble subgraphs of $\cB$ (with respect to the color $0$). In particular, there are $\frac{1}{2}(d-1)(d-2)$ distinct $3$-bubble subgraphs associated with a $d$-bubble $\cB$. Note that the connectivity of the $3$-bubble subgraphs associated with $\cB$ is not necessarily the same as the connectivity of $\cB$ itself, since we removed edges.

The $3$-bubble subgraphs $\cB_{(0ij)}$ of a $d$-bubble $\cB$ are natural objects to consider because they are equivalent to ribbon graphs (in the sense of Eq.\ \eqref{genusJacket3bubble}). Hence, they provide a class of embedded graphs associated with any $d$-bubble, similarly to the jackets. Each $3$-bubble subgraph $\cB_{(0ij)}$ is characterized by a genus $g(\cB_{(0ij)})\in\frac{1}{2}\mathbb{N}$, given by a Euler's formula of the form \eqref{EulerFormFeynGraphColor}, 
\be\label{EulerForm3BubbleSub}
\begin{split}
2B_{(0ij)}-2g(\cB_{(0ij)})&= V(\cB_{(0ij)}) - E(\cB_{(0ij)}) + F(\cB_{(0ij)}) \\
	&= -\frac{1}{2}V(\cB_{(0ij)}) + F_{0i}(\cB_{(0ij)}) + F_{0j}(\cB_{(0ij)}) + F_{ij}(\cB_{(0ij)}) \, .
\end{split}
\ee
Additionally, there is yet another non-negative quantity associated with each $\cB_{(0ij)}$, given in Eq.~\eqref{del3}, which results from the connectivity inequalities, see Section \ref{sec:AppABubbles}. 

The following proposition, called the decomposition formula, relates the index of a $d$-bubble $\cB$ with the two above non-negative quantities associated with its $3$-bubble subgraphs $\cB_{(0ij)}$.

\begin{proposition} \label{fundidth} (First form of the decomposition formula) The index can be expressed as a sum of manifestly non-negative contributions as
\be\label{fundid0} \ind_{0}\gB = \sum_{i<j}\bigl(g(\gB_{(0ij)}) +  F_{ij}(\gB) -B_{(0ij)}-B^{(0)}+B\bigr)\, .\ee
\end{proposition}

Equivalently, using \eqref{del3}, the decomposition formula \eqref{fundid0} can be rewritten as
\be\label{fundid} \ind_{0}\gB = \sum_{i<j}g(\gB_{(0ij)}) + \delta_{0;d-3}(\gB)\, .\ee

For $d=3$, there is only one $3$-bubble subgraph of $\cB$, which is $\cB$ itself. In addition, the second term on the RHS of the above equation vanishes so that the index of $\cB$ coincides with its genus, as expected. For $d>3$, one can also use Eq.~\eqref{conid} to obtain an alternative expression. For example, for $d=4$, we get
\be\label{fid1} \ind_{0}\gB = \sum_{i=1}^{3}g(\gB^{(i)}) + \sum_{i=1}^{3}\bigl(
B^{(0i)}-B^{(i)}-B^{(0)}+B\bigr) \, , \ee
and for $d=5$,
\begin{multline}\label{fid2} \ind_{0}\gB = \sum_{1\leq i<j\leq 4}g(\gB^{(ij)}) + \frac{3}{2}\sum_{i=1}^{4}\bigl(
B^{(0i)}-B^{(i)}-B^{(0)}+B\bigr) \\+ \sum_{1\leq i < j\leq 4}\bigl(
B^{(0ij)}-B^{(ij)}-B^{(0i)}+B^{(i)}\bigr)\, .
\end{multline}

Proposition \ref{fundidth} provides a first form of the decomposition formula. In Section \ref{sec:TensLargeNExp2}, a second form will be derived (see Proposition \ref{fundidth2}), which has a natural physical interpretation. Remark that an equally valid presentation for the index of a $d$-bubble would consist in using the decomposition formula \eqref{fundid0} as a definition and then derive Eq.~\eqref{inddef}, instead of the other way around.

\proof The proof uses the following lemma, which is the analogue of Eq.\ \eqref{degreeformula} for the index.
\begin{lemma}\label{indexlem} The index can be expressed as
\begin{multline}\label{indexplicit} \ind_{0}\gB = \frac{1}{2}(d-1)(d-2) \bigl(B-B^{(0)}\bigr)+\frac{1}{8}(d-1)(d-2) V(\gB)\\
+ \frac{1}{2}\sum_{i<j}F_{ij}(\gB)-\frac{1}{2}(d-2)\sum_{i} F_{0i}(\gB)\, .
\end{multline}
\end{lemma}
The expression \eqref{indexplicit} is obtained by using Eq.~\eqref{degreeformula} for the $d$-bubble $\gB$ and the $(d-1)$-bubble $\gB^{(0)}$ and decomposing
\be\label{facedec} F(\gB) = \sum_{i=1}^{d-1}F_{0i}(\gB) + \sum_{i<j}F_{ij}(\gB)\, ,\quad F(\gB^{(0)}) = \sum_{i<j}F_{ij}(\gB)\, .\ee
To proceed further, we use Euler's formula \eqref{EulerForm3BubbleSub} for all the 3-bubble subgraphs $\gB_{(0ij)}$. Summing the resulting $\frac{1}{2}(d-1)(d-2)$ equations yields
\be\label{indstep1} \sum_{i<j} \bigl(2 B_{(0ij)} - 2 g(\gB_{(0ij)})\bigr) =(d-2)\sum_{i}F_{0i} (\gB) + \sum_{i<j}F_{ij}(\gB) -\frac{1}{4}(d-1)(d-2) V(\gB)\, .\ee
We then use this result to eliminate $V(\gB)$ from Eq.~\eqref{indexplicit} to get Eq.~\eqref{fundid0}.
\qed

\

One can check that Eq.\ \eqref{indexplicit} reduces to Euler's formula \eqref{EulerFormFeynGraphColor} in the case of a $3$-bubble, as expected. As a result, the index can also be regarded as a generalization of the genus for $d$-bubbles, at the same level as the degree. For a given $d$-bubble $\cB$, the two notions are related by Eq.\ \eqref{inddef}. 

\subsection{Addition formulas}
\label{sec:AddFormula}

We finish this section with useful formulas for the degree and the index of $d$-bubbles.
\begin{proposition}\label{addprop} (Addition formulas) Consider two $d$-bubbles $\gB_{1}$ and $\gB_{2}$ and the so-called $2$-point graphs $\tilde\gB_{1}$ and $\tilde\gB_{2}$ obtained by cutting open any edge of color 0 in $\gB_{1}$ and $\gB_{2}$ respectively. Build a new $d$-bubble $\gB$ by gluing the open edges of color 0 in $\tilde\gB_{1}$ and $\tilde\gB_{2}$ as depicted in Figure \ref{additionfig} (note that there are in general two inequivalent ways to perform this gluing). Then
\begin{figure}[]
\centerline{\includegraphics[width=6in]{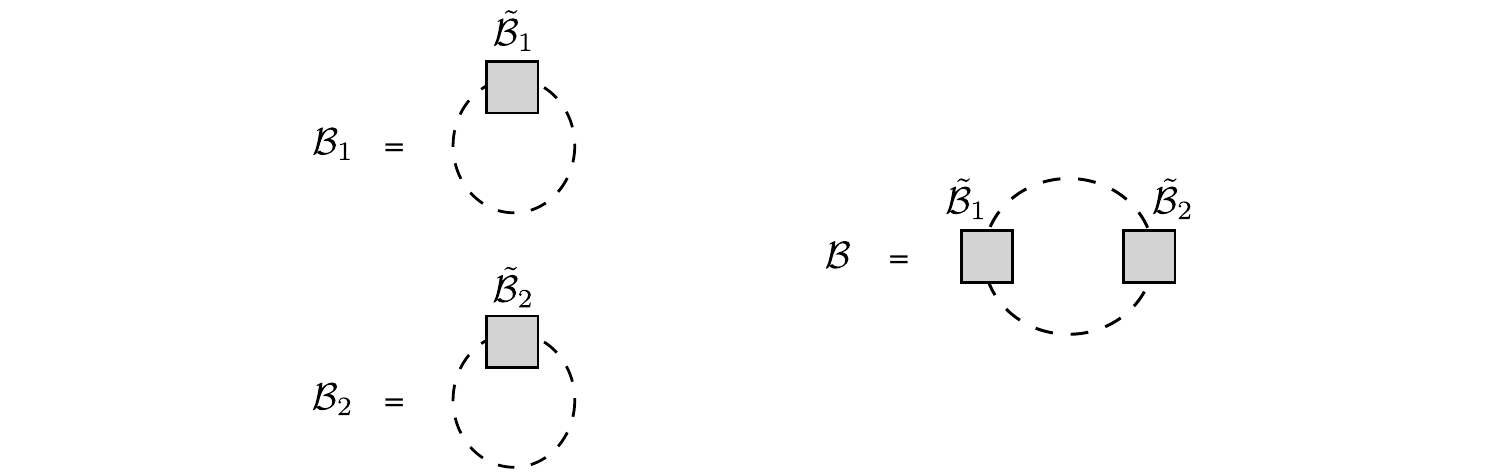}}
\caption{Construction of a new bubble $\gB$ from bubbles $\gB_{1}$ and $\gB_{2}$, such that $\deg \gB = \deg \gB_{1}+\deg\gB_{2}$ and $\ind_{0}\gB  = \ind_{0}\gB_{1}+\ind_{0}\gB_{2}$ (Proposition \ref{addprop}). Dashed edges represent edges of color 0.}\label{additionfig}
\end{figure}
\begin{align}\label{adddeg} \deg \gB &= \deg \gB_{1}+\deg\gB_{2}\\\label{addind}
\ind_{0}\gB & = \ind_{0}\gB_{1}+\ind_{0}\gB_{2}\, .
\end{align}
\end{proposition} 

\proof By construction, $c(\gB)=c(\gB_{1})+c(\gB_{2})-1$, $V(\gB)=V(\gB_{1})+V(\gB_{2})$ and $F_{ij}(\gB)=F_{ij}(\gB_{1})+F_{ij}(\gB_{2})$. Moreover, the two $(0i)$-faces, for some color $i$, in $\gB_{1}$ and $\gB_{2}$ going through the edges of color 0 that are cut open are joined in a unique $(0i)$-face in $\gB$, whereas the other $(0i)$-faces remain unchanged. This yields $F_{0i}(\gB)=F_{0i}(\gB_{1})+F_{0i}(\gB_{2})-1$. Overall, we thus get $F(\gB)=F(\gB_{1})+F(\gB_{2})-(d-1)$. Eq.\ \eqref{adddeg} then follows using Eq.~\eqref{degreeformula} for the bubbles $\gB$, $\gB_{1}$ and $\gB_{2}$. Eq.\ \eqref{addind} can be proven by a similar reasoning using Eq.~\eqref{indexplicit} or from the definition \eqref{inddef} using Eq.~\eqref{adddeg} and the trivial result $\deg\gB^{(0)}=\deg\gB_{1}^{(0)}+\deg\gB_{2}^{(0)}$.\qed

\

\begin{figure}[]
\centerline{\includegraphics[width=6in]{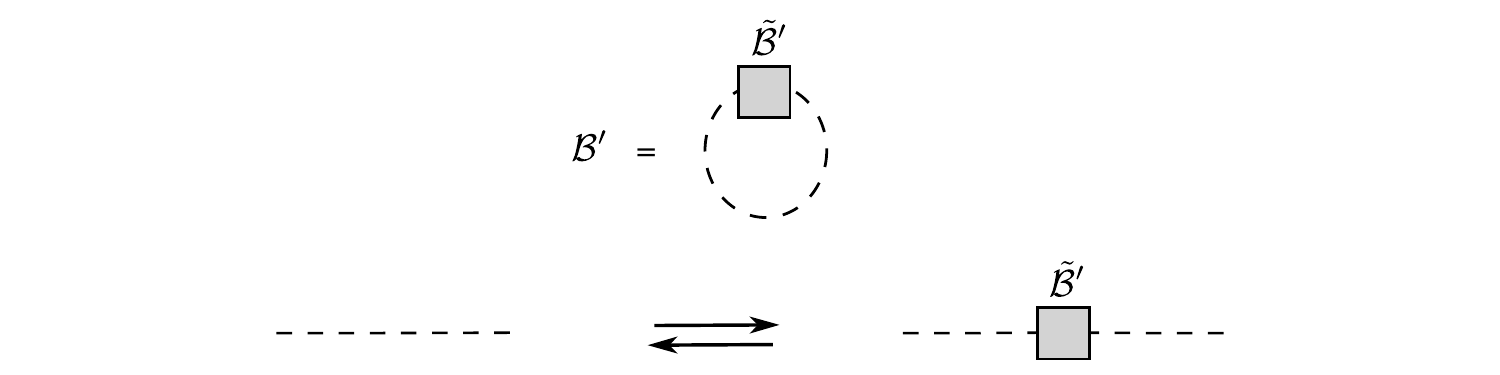}}
\caption{Bubble $\gB'$ insertion (from left to right) and contraction (from right to left) on an edge of color 0 in a bubble $\gB$. According to Proposition \ref{addprop}, the insertion (contraction) increases (decreases) the degree and the index of $\gB$ by $\deg\gB'$ and $\ind_{0}\gB'$ respectively.}\label{insertfig}
\end{figure}

We define the operation of bubble insertion of a $d$-bubble $\gB'$ as the replacement of an edge of color 0 in a given $d$-bubble $\gB$ by the $2$-point graph $\tilde\gB'$ obtained from $\gB'$, as depicted in Figure \ref{insertfig}. Proposition \ref{addprop} implies that $\deg\gB\mapsto\deg\gB + \deg\gB'$  and $\ind_{0}\gB\mapsto\ind_{0}\gB + \ind_{0}\gB'$ under this operation. The inverse operation is called a bubble contraction.

\section{Large $N$ expansion (1): BGR scaling}
\label{sec:TensLargeNExp1}

To define a large $N$ expansion for tensor models, we need to specify how the coupling constants $g_a$ in the action \eqref{TensAction} scale with $N$ when $N\rightarrow\infty$. The scaling must be such that the associated $1/N$ expansion is well-defined and non-trivial. We observed in Section \ref{sec:VecLargeNExp} for vector models and in Section \ref{sec:MatLargeNExp} for matrix models that the optimal scaling is known for the coupling constant associated with any interaction term that enters the action. As explained in the introduction of this chapter, the situation is completely different in the case of tensor models, as optimal scalings are understood only for a small subset of all possible interactions. 

In this section, we review the BGR scaling and the associated $1/N$ expansion, suitably generalized to $\oN^R$ tensor models for multiply-connected interaction bubbles. This part corresponds to bits of Sections 3.2 and 3.3.4 in Ref.\ \cite{Ref2}. 

\subsection{Large $N$ expansion}
\label{sec:TensLargeN1}

The BGR scaling introduced in \cite{BGR} can be generalized to the symmetry group $\oN^R$ instead of $\uN^R$ and to the case of multiply-connected interaction bubbles $\cB_a$. One defines new coupling parameters $\tau_a$ in terms of the couplings $g_a$ associated with the interaction terms $I_{\cB_a}$ in the action \eqref{TensAction} by
\be\label{BGRScaling}
g_a=N^{R-c(\cB_a) -\frac{2}{(R-2)!}\deg\cB_a}\tau_a\, ,
\ee
where $\cB_a$ is the $R$-bubble associated with $\cI_{\cB_a}$, which has $c(\cB_a)$ connected components and degree $\deg\cB_a$. To formulate a large $N$ limit, we keep the $\tau_a$'s fixed when $N\rightarrow\infty$. The action rewrites in terms of the BGR scaling as
\be\label{TensActionBGRScaling}
S(T)= N^{R-1}\biggl(\frac{1}{2} T\cdot T + \sum_{a\in\cS} N^{1-c(\cB_a)-\frac{2}{(R-2)!}\deg\cB_a}\tau_a \, \cI_{\cB_a}(T)\biggr) \, .
\ee
The BGR scaling scales the coupling constants according to the degree of the associated interaction bubbles. Since the degree is non-negative, it corresponds to a suppression of some interaction terms. As remarked in \cite{BGR}, the scaling according to the degree is however not required, the associated large $N$ expansion can also be derived in its absence. As in \cite{BGR}, we keep this scaling to simplify the computations. On the other hand, the scaling according to the number of connected components is required. The BGR scaling is illustrated in Figure \ref{TensFeynRulesColor} for some interaction bubbles.

The power of $N$ associated with a given connected Feynman graph $\cG$ in the perturbative expansion of the free energy can be deduced from the action \eqref{TensActionBGRScaling}: to each propagator is associated a factor $1/N^{R-1}$, to each interaction vertex of type $\cI_{\cB_a}$ a factor 
\begin{equation*}
N^{R-c(\cB_a)-\frac{2}{(R-2)!}\deg\cB_a}\, ,
\end{equation*}
and to each closed stranded loop a factor $N$. Overall, the amplitude $\cA(\cG)$ of $\cG$ thus scales with $N$ as
\be\label{TensFeynGraphBGRScaling}
\cA(\cG) \sim N^{-(R-1)p+\sum_{a\in\cS_\cG}\bigl(R-c(\cB_a)-\frac{2}{(R-2)!}\deg\cB_a\bigr)+f} \equiv N^{R-L} \, ,
\ee
where we introduced the parameter $L$ defined as
\be\label{TensLParam1}
\begin{split}
L & = R+(R-1)p-\sum_{a\in\cS_\cG}\Bigl(R-c(\cB_a)-\frac{2}{(R-2)!}\deg\cB_a\Bigr)-f \, ,\\
	& = R+(R-1)p - Rv +\sum_{a\in\cS_\cG}c(\cB_a)+\frac{2}{(R-2)!}\sum_{a\in\cS_\cG}\deg\cB_a-f  \, ,
\end{split}
\ee
where we used the relation $v=|\cS_\cG|$. The power of $N$ associated with the Feynman graphs of Figure \ref{TensFeynGraphsColor} is indicated below the graphs in terms of the coupling parameters $\tau_a$.

As emphasized earlier, it is more convenient to work with the $(R+1)$-bubble $\cB$ associated with $\cG$ in the colored representation. Using the identities \eqref{TensFeynGraphData}, the parameter $L$ can be rewritten as
\be\label{TensLParam2}
L = R+\frac{1}{2}(R-1)V(\cB) - Rv(\cB) +B^{(0)}+\frac{2}{(R-2)!}\deg\cB_0-\sum_{i=1}^R F_{0i}(\cB) \, ,
\ee
where we also used the definition of the degree for a multiply-connected bubble,
\be\label{IntBubbleDegree}
\sum_{a\in{\cS_\cG}}\deg\cB_a = \deg\cB^{(0)} \, .
\ee

The following theorem ensures that the BGR scaling \eqref{BGRScaling} yields a well-defined $1/N$ expansion.

\begin{theorem}\label{TensTheoremBGR}
Let $\cG$ be a connected Feynman graph of the $\oN^R$ tensor model with BGR scaling and let $\cB$ be the corresponding $(R+1)$-bubble in the colored representation. Then, $L\geq 0$. In other words, the power of $N$ associated with any connected Feynman graph $\cG$ is bounded above by $R$.
\end{theorem}

\proof We apply the formula \eqref{degreeformula} for the degree of the bubbles $\cB^{(0)}$ and $\cB$,
\begin{equation*}
(R-1)B^{(0)} - \frac{2}{(R-2)!}\deg\gB^{(0)} = F(\gB^{(0)}) - \frac{1}{4}(R-1)(R-2) V(\gB^{(0)})\, ,
\end{equation*}
\begin{equation*}
RB - \frac{2}{(R-1)!}\deg\gB = F(\gB) - \frac{1}{4}R(R-1) V(\gB) \, .
\end{equation*}
Now, we subtract these two equations, we plug the result into Eq.\ \eqref{TensLParam2} and we use the face decomposition formulas \eqref{facedec} as well as $V(\gB^{(0)})=V(\gB)$. This yields
\be\label{TensLParam3}
L = \frac{2}{(R-1)!}\deg\cB + R \Bigl[1+ \sum_{a\in\cS_\cG} (c(\cB_a)-1)-B\Bigr] \, .
\ee
The first term on the RHS of this expression is non-negative according to \eqref{degreeN}. As for the second term within brackets, it is also non-negative using the same argument as in the proofs for Theorem \ref{VecTheorem} or Theorem \ref{MatTheorem}. In details, if we break down the interaction vertices of type $\cB_a$ in $\cB$ into their $c(\cB_a)$ connected components, it increases the number of effective connected components of $\cG$ by at most $\sum_{a\in\cS_\cG}(c(\cB_a)-1)$. Since $\cG$ is connected, the total number of connected components generated in the process is $B-1$; hence, $B-1\leq\sum_{a\in\cS_\cG}(c(\cB-a)-1)$. As a result, $L$ is non-negative because it is a sum of two non-negative terms. Furthermore, \eqref{degreeN} implies that $L\in\mathbb{N}$. \qed

\

If we denote by $\{\cG_L\}$ the set of connected Feynman graph of fixed $L$, then the $1/N$ expansion of the $\oN^R$ tensor models with BGR scaling \eqref{BGRScaling} can be written as
\be \label{TensLargeNExpFeyn}
F=\sum_{L\in\mathbb{N}} N^{R-L} F_L\, ,
\ee
where $F_L$ corresponds to a sum over connected Feynman graphs $\cG$ in $\{\cG_L\}$ weighted by some amplitude $\tilde{\cA}(\cG)$,
\be \label{TensLargeNExpFeynCoeff}
F_L= \sum_{\cG\in\{\cG_L\}} \tilde{\cA}(\cG) \, .
\ee
In particular, the $1/N$ expansion is well-defined. Similar results also hold for expectation values of invariants, as usual.

\

\noindent\emph{Case of connected interaction bubbles}

In the special case of connected interaction bubbles, one has $c(\cB_a)=1 \ \forall a\in\cS$ and $B=1$, which implies that 
\be \label{ParamLConnBubb}
L= \frac{2}{(R-1)!}\deg\cB\, .
\ee
Thus, we obtain a $1/N$ expansion \eqref{TensLargeNExpFeyn} indexed by the degree of the Feynman graphs. In particular, it reduces to the result obtained in \cite{BGR} for $\uN^R$ tensor models if we further restrict the expansion to bipartite Feynman graphs only. 

Before moving on to the study of the leading sector, let us make a few remarks. First, the $1/N$ expansion \eqref{TensLargeNExpFeyn} reduces for $R=2$ to the one \eqref{MatLargeNExpFeyn} obtained in Section \ref{sec:MatLargeNExp} for $\oN^2$ matrix models. This follows from comparing the parameter $L$ defined in Eq.~\eqref{TensLParam3} for tensor models with the parameter $h$ defined in Eq.\ \eqref{MatHParam3} for matrix models and using $\deg\cB = g(\cB)$ and $\sum_{a\in\cS_\cG} c(\cB_a)=\sum_{a\in\cS_\cG} t(\cB_a) = F_{12}(\cB)$. 

Second, we note that in the case of a $3$-bubble $\cB$, we have two distinct notions of embedded graph: on the one hand, the ribbon graph $\tilde{\cG}$ introduced in Section \ref{sec:MatStrandRep} for matrix models and on the other hand, the unique (up to orientation of the cycle $\sigma$) jacket $\gJ(\gB;\sigma)$ defined in Section \ref{sec:DBubbles}. These two ribbon graphs have the same genus, which match the genus $g(\cB)$ of the $3$-bubble. In fact, they can be related from one to another by contracting/resolving the ribbon graph vertices associated with the interaction vertices. 

Finally, for $R\geq3$, the degree is not a topological invariant likewise the genus for $R=2$. It is in some sense expected because topology is complicated in dimensions $R\geq3$. Rather, the degree of a $R$-bubble combines topological and combinatorial information about the $R$-dimensional discretized geometry described by the bubble. Bubbles of fixed degree have been fully classified and enumerated \cite{GurauSchaeffer}. In particular, they yield convergent series, that is, $F_L$ is convergent for fixed $L$ for connected interaction bubbles.

\subsection{Leading sector: melonic graphs}
\label{sec:TensBGRLO}

We now study the leading order (LO) Feynman graphs for the $1/N$ expansion \eqref{TensLargeNExpFeyn} obtained with the BGR scaling. They satisfy $L=0$, which is equivalent, using Eq.\ \eqref{TensLParam3}, to the conditions 
\be \label{TensLOGraph1}
\sum_{a\in\cS_\cG} \bigl(c(\cB_a)-1\bigr) = B-1 \, ,
\ee
\be \label{TensLOGraph2}
\deg\cB=0 \, .
\ee
As usual, the first condition implies that a LO Feynman graph $\cB$ is effectively ``maximally disconnected" with respect to its interaction bubbles (see for instance Section \ref{sec:MatLO}) and it is automatically satisfied for connected interaction bubbles. As for the second condition, it means that each connected component of $\cB$ must have degree zero. One can verify that the Feynman graphs $(a)$, $(d)$ and $(e)$ in Figure \ref{TensFeynGraphsColor} are all LO while the others are not. In the following, we focus on the second condition, which concerns the degree. In other words, we look at each connected component of a LO Feynman graph or we restrict the action to connected interaction bubbles only. 

For $R=2$, Feynman graphs are $3$-bubbles and the degree coincide with the genus. Therefore, the LO Feynman graphs of degree zero are the planar graphs. In particular, they correspond to discretized two-dimensional spheres. For $R\geq3$, the LO Feynman graphs of degree zero have been classified in \cite{Melons}; they belong to a family called melons or melonic graphs. It was shown in \cite{LargeNColored} that they correspond to some discretized spheres in $R$ dimensions (more precisely, they correspond to some colored triangulations of the $R$-sphere $\mathbb{S}^R$). The goal of this section is to describe to some extent the family of melons and to present the large $N$ solution of the $\oN^R$ tensor models for specific interactions included in the action. More details and rigorous proofs can be found in \cite{LargeNColored,Melons,BGR}. 

\subsubsection{Melons}

A $(R+1)$-bubble $\cB$ of degree zero is called melon or melonic graph. Formally, melons form a larger family than the LO Feynman graphs described above because they may contain $R$-dipoles with external color $0$, which correspond to ``fictitious" interaction bubbles with two vertices. By Lemma \ref{degdeglem}, the fact that $\deg\cB=0$ has strong implications on the class of subgraphs that can appear in $\cB$. In particular, this lemma implies that the subgraphs of colors $1$ to $R$ in $\cB$, which correspond to interaction bubbles for Feynman graphs, must all have degree zero. In other words, they must also be melons as $R$-bubbles. As a result, in the large $N$ limit, only melonic interaction bubbles can contribute to the leading sector. This explains why the BGR scaling is optimal for melonic interactions but not optimal in general, as mentioned in the introduction of this chapter.

Interestingly, melonic graphs are more restricted than planar graphs. From the definition \eqref{degreedef} of the degree, all their jackets must be planar so that melons can be interpreted as being ``super planar". They also form a simple class of graphs that can be constructed in a recursive way and summed explicitly, similarly to the bubble graphs of the $\oN$ vector models. In the following, we describe the recursive structure of the melonic graphs in a combinatorial way. For more details, see \cite{Melons,BGR}.

\begin{figure}[]
\centerline{\includegraphics[scale=1]{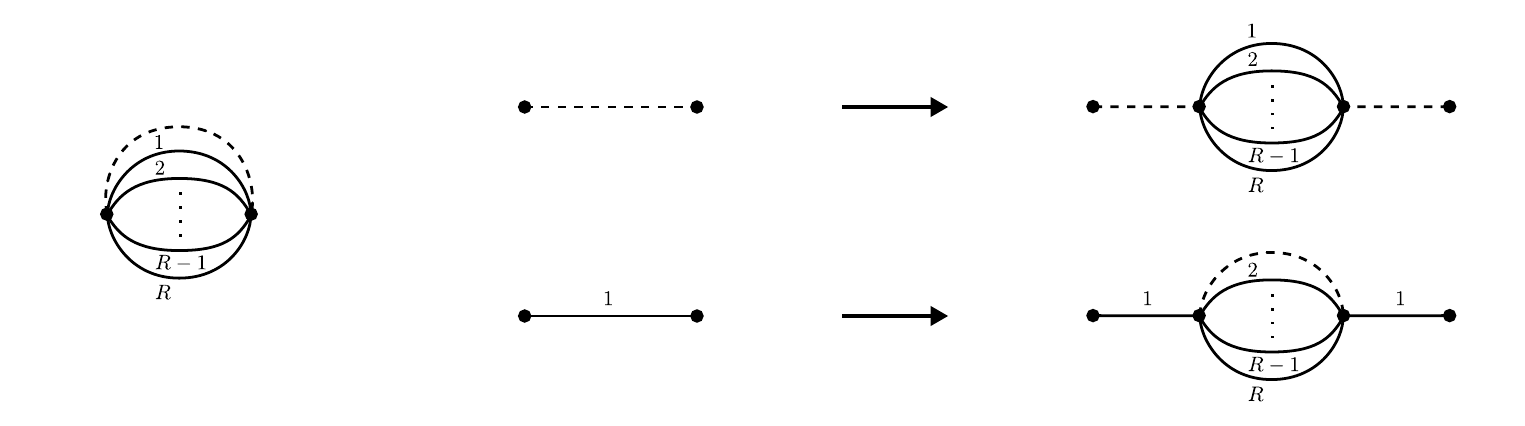}}
\caption{Melons are constructed by recursive melonic insertions. Left: elementary melon. Right: examples of melonic insertions, which consist in replacing an edge of color $0$ or $i=1,2,\ldots,R$ in a $(R+1)$-bubble by the elementary $2$-point melon with external color $0$ or $i$ respectively.}\label{TensMelonicMove1}
\end{figure}

We define the elementary melon to be the (unique) $(R+1)$-bubble made of two vertices. This bubble necessarily has $\frac{R(R+1)}{2}$ faces and a single connected component; hence, Eq.~\eqref{degreeformula} implies that the elementary melon has degree zero. If we cut one of the $R+1$ edges in the elementary melon, we obtain a graph with two external (half-)edges, which can have color $0$ to $R$. We call the $R+1$ possible resulting graphs elementary 2-point melons (with external color $0$ or $i=1,2,\ldots,R$). Next, we define a melonic insertion in a $(R+1)$-bubble $\cB$ as the replacement of an edge of color $0$ or $i$ in $\cB$ by the elementary 2-point melon with external color $0$ or $i$ respectively. This is illustrated in Figure \ref{TensMelonicMove1}. This operation leaves the degree of the bubble unchanged because the number $V$ of vertices increases by two and the number $F$ of faces increases by $\frac{R(R-1)}{2}$ (see Eq.~\eqref{degreeformula}). Then, one can show that any melon can be obtained by starting from the elementary melon and then applying recursively an arbitrary number of melonic insertions on any edge. For instance, in Figure \ref{TensFeynGraphsColor}, the melons $(a)$ and $(e)$ can both be obtained by this recursive procedure.

The connection between this combinatorial definition of melons and the LO Feynman graphs of the models is as follows. Any melonic insertion on an edge of color $i$ in a LO Feynman graph $\cB$ corresponds to a modification of one of its interaction bubbles (which remains melonic afterwards). As for a melonic insertion on an edge of color $0$ in $\cB$, it does not generate another Feynman graph strictly speaking because there is no interaction bubble with two vertices, as explained previously. However, the resulting graph can be regarded as a temporary step to create other LO Feynman graphs after further melonic insertions.

\begin{figure}[]
\centerline{\includegraphics[scale=1]{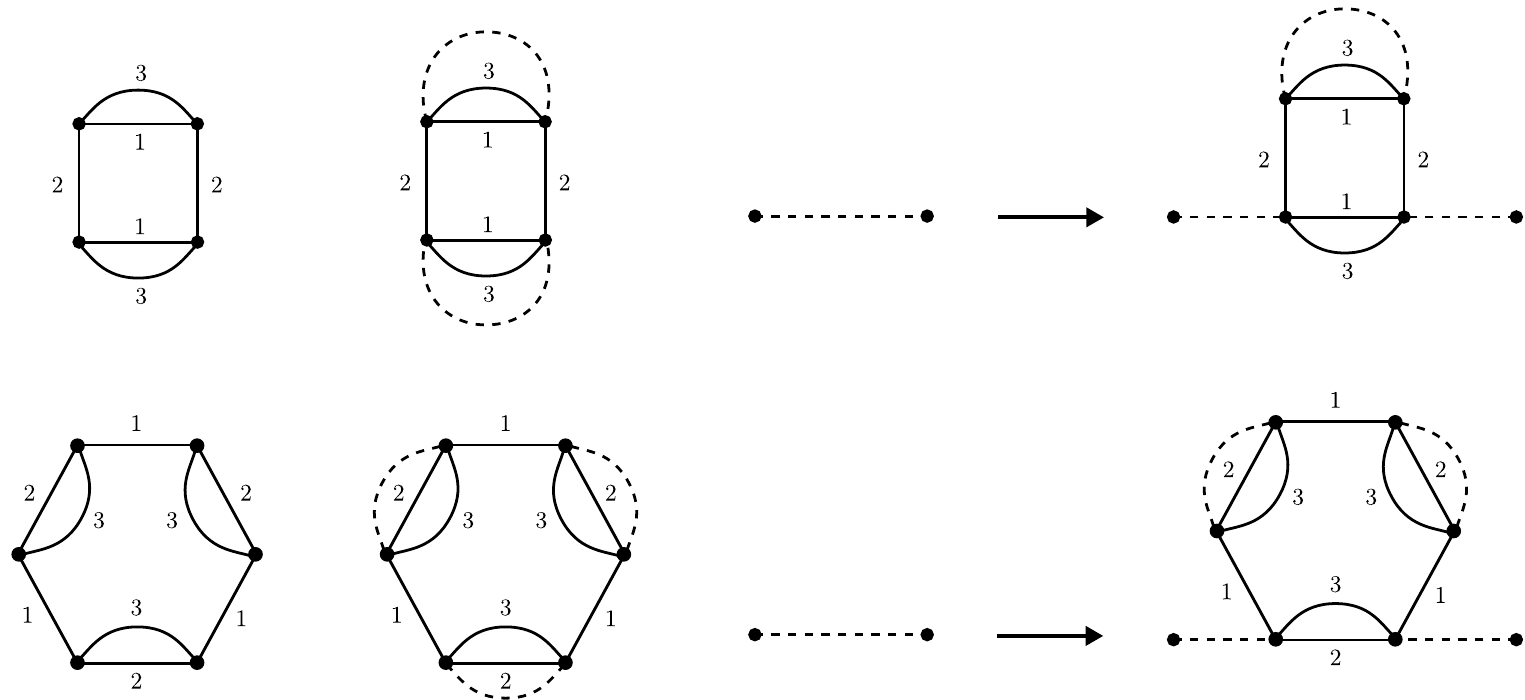}}
\caption{Illustration of the melonic structure of the LO Feynman graphs for the $\oN^R$ tensor models with BGR scaling. Left: melonic interaction bubbles $\cB_a^m$ with $R=3$. Middle: elementary melons of type $\cB_a^m$. Right: melonic insertion of type $\cB_a^m$.}\label{TensMelonicMove2}
\end{figure}

There is actually another way of understanding the melonic structure of the LO Feynman graphs, which is based on the addition formula in Section \ref{sec:AddFormula}. Consider the set of interaction terms $\cI_{\cB^m_a}$ that correspond to melonic interaction bubbles $\cB^m_a$. We know that these interaction bubbles are the only ones that can appear in a LO Feynman graph. For each interaction bubble $\cB^m_a$, we construct the LO Feynman graph that has $\cB^m_a$ as the only interaction bubble. Due to the melonic structure, one can check that this LO Feynman graph is uniquely obtained by appropriately adding edges of color $0$ to $\cB^m_a$. We call this unique Feynman graph associated with $\cB^m_a$ the elementary melon of type $\cB^m_a$, which has degree zero by construction. Then, we define a melonic insertion of type $\cB^m_a$ as the bubble insertion of the elementary melon of type $\cB^m_a$ in a Feynman graphs $\cB$. This is illustrated in Figure \ref{TensMelonicMove2} for two melonic interaction bubbles with $R=3$. From Proposition~\ref{addprop}, any such insertion leaves the degree unchanged. Finally, one can verify that any LO Feynman graph can be obtained by recursively inserting melons of any type starting from the elementary melon of a given type. We note that this recursive procedure can be naturally understood in terms of multiple successive melonic insertions, as defined above for melons.

\subsubsection{Large $N$ universality}

The melonic structure of the LO Feynman graphs has interesting consequences. Firstly, melons can be mapped to $(R+1)$-ary trees \cite{Melons}. Hence, they can be enumerated and their number at fixed number of vertices is exponentially bounded, which is necessary for $F_0$ to be a convergent series. Note that this is reminiscent of vector models, see Section \ref{sec:VecLO}. Secondly, the melonic structure implies that tensor models with BGR scaling become Gaussian in the large $N$ limit, which means that they are entirely determined by the full connected $2$-point function. This universality theorem, initially proven in \cite{GaussianBGR}, puts strong constraints on the type of random continuous geometries that can be generated from the leading sector in the continuum limit. On the other hand, it allows one to derive a closed equation for the full connected $2$-point function at large $N$, which can be explicitly solved in some particular cases. We first discuss the large $N$ universality.

Consider an invariant interaction $\cI_{\cB_a}$ in the action \eqref{TensActionBGRScaling}, with coupling constant $\tau_a$ and with connected interaction bubble $\cB_a$. The connected expectation value of this invariant, denoted as $\bigl\langle\cI_{\cB_a}\bigr\rangle_c$, can be obtained from the free energy by taking a partial derivative with respect to $\tau_a$. Since the free energy admits a $1/N$ expansion onto connected Feynman graphs, see Eq.\ \eqref{TensLargeNExpFeyn}, this connected expectation value admits a $1/N$ expansion onto connected Feynman graphs having $\cB_a$ as a marked subgraph. 

\begin{figure}[]
\centerline{\includegraphics[scale=1]{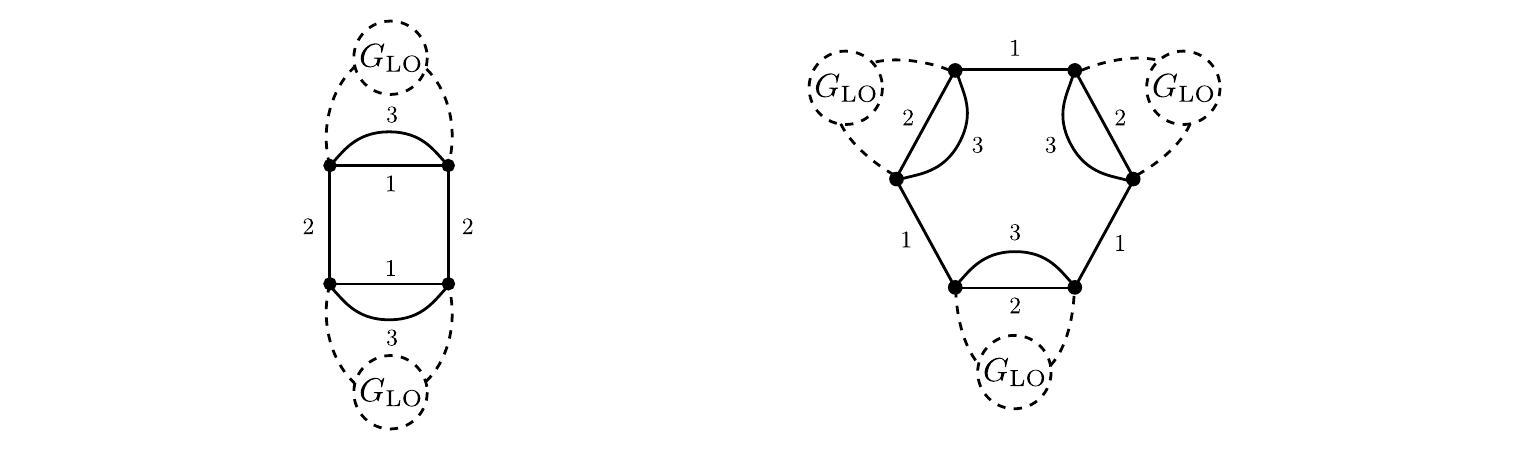}}
\caption{Large $N$ expectation values $\bigl\langle\cI_{\cB^m_a}\bigr\rangle_c$ for the melonic interaction bubbles $\cB^m_a$ of Figure \ref{TensMelonicMove2}. $G_{\text{LO}}$ corresponds to the LO full connected $2$-point function.}\label{TensConnExpVal}
\end{figure}

In the large $N$ limit, only melonic interaction bubbles can appear in LO Feynman graphs; hence, only expectation values $\bigl\langle\cI_{\cB^m_a}\bigr\rangle_c$ of melonic invariants $\cI_{\cB^m_a}$ survive. Furthermore, due to the melonic structure, there is only one way of building LO Feynman graphs that contain the interaction bubble $\cB^m_a$: they are all obtained by starting from the elementary melon of type $\cB^m_a$ and by recursively inserting melons of any type on the edges of color $0$. The large $N$ full expectation value $\bigl\langle\cI_{\cB^m_a}\bigr\rangle_c$ is therefore obtained by replacing the edges of color $0$ in the elementary melon of type $\cB^m_a$ by the LO full connected $2$-point function. Let us write the full connected $2$-point function as
\be \label{TensConn2PtFunction}
\bigl\langle T_{a_1a_2\cdots a_R}T_{b_1b_2\cdots b_R}\bigr\rangle_c \ = \frac{1}{N^{R-1}}\delta_{a_1b_1}\delta_{a_2b_2} \ldots \delta_{a_Rb_R} G(N,\tau_a)\, ,
\ee
and denote by $G_{\text{LO}}(\tau_a)$ the restriction of $G(N,\tau_a)$ to the leading sector. Then, the large $N$ expectation value of melonic invariants can be pictured as in Figure \ref{TensConnExpVal}.

The above large $N$ universal behavior is usually expressed as 
\be \label{TensLargeNUniv}
\frac{1}{N}\bigl\langle\cI_{\cB^m_a}\bigr\rangle_c \ = \bigl(G_{\text{LO}}\bigr)^{V(\cB^m_a)/2} \, ,
\ee
where $V(\cB^m_a)/2$ corresponds to the number of $G_{\text{LO}}$'s that can be attached to the interaction bubble $\cB^m_a$. This so-called universality theorem implies that in the large $N$ limit, the expectation value of any observable is fully determined by the LO full connected $2$-point function.\footnote{We remark that this universality theorem can also be obtained using Schwinger-Dyson equations \cite{BonzomSD}.} In other words, the models become Gaussian. Remark however that the large $N$ limit still depends on the details of the model, as it is the case for $G_{\text{LO}}(\tau_a)$.

The universality theorem for tensor models with BGR scaling is a strong outcome. In particular, there is no such behavior for matrix models, which have non-Gaussian large N limits. In this sense, Gaussian tensor models are closer to vector models, which also become Gaussian in the large $N$ limit. In the next subsection, we explain that the universal large $N$ behavior allows one to obtain the large $N$ solution for specific models and it typically leads to the susceptibility critical exponents of branched polymers.

\subsubsection{Large $N$ solution and critical behavior}

\begin{figure}[]
\centerline{\includegraphics[scale=1]{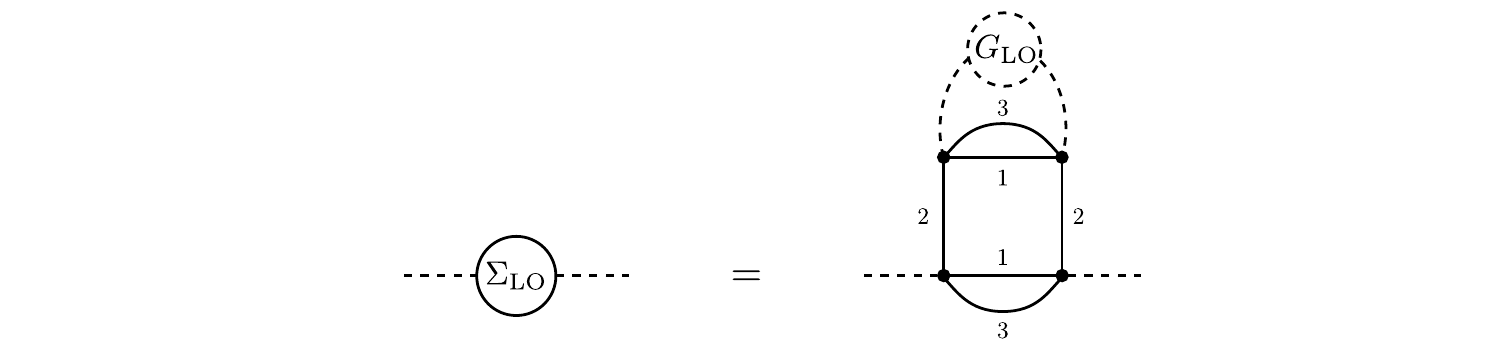}}
\caption{Diagrammatic equation for the LO self-energy $\Sigma_{\text{LO}}$ in terms of the LO connected 2-point function $G_{\text{LO}}$.}\label{TensFreeEnergy}
\end{figure}

Let us compute the large $N$ solution for a specific $\oN^3$ tensor model based on the pillow interaction $\cB_1$ (see Figure \ref{TensFeynRulesColor}). We denote by $\tau$ the corresponding coupling constant with BGR scaling. As usual, $G(N,\tau)$ is related to the connected 1PI $2$-point function $\Sigma(N,\tau)$ (or self-energy) by
\be \label{Tens2PtFunction}
G(N,\tau) = \frac{1}{1-\Sigma(N,\tau)}\, ,
\ee
where the free propagator is $1$ in the models at hand (see Eq.\ \eqref{TensFree2PtFunction}). Because of the melonic structure of the LO Feynman graphs, the LO self-energy $\Sigma_{\text{LO}}(\tau)$ necessarily has the structure represented in Figure \ref{TensFreeEnergy}, which is equivalent to the Schwinger-Dyson equation
\be \label{TensSDEqBGR}
\Sigma_{\text{LO}}(\tau) = -4 \tau G_{\text{LO}}(\tau)\, .
\ee

Then, Eq.\ \eqref{Tens2PtFunction} together with Eq.\ \eqref{TensSDEqBGR} provide the following closed equation for $G_{\text{LO}}(\tau)$ in the large $N$ limit,
\be \label{TensClosedEqBGR}
G_{\text{LO}}(\tau)=1-4\tau G_{\text{LO}}(\tau)^2\, .
\ee
This is in fact the same equation as the one obtained in Section \ref{sec:VecLO} for $\oN$ vector models with quartic interactions, see Eq.\ \eqref{VecClosedEq}. Hence, the solution \eqref{VecLOSolution} in terms of Catalan numbers or the explicit solution \eqref{VecLOSolutionExpl} are also valid for the tensor model at hand. This emphasizes once more the similarity between tensor models with BGR scaling and vector models.

One can generalize the above solution to any $\oN^R$ tensor models with BGR scaling based on a single melonic interaction bubble $\cB^m_a$ with coupling constant $\tau$ and with $V(\cB^m_a)=V$ vertices. The corresponding Schwinger-Dyson equation would write
\be \label{TensSDEqBGRGen}
\Sigma_{\text{LO}}(\tau) = - V \tau G_{\text{LO}}(\tau)^{V/2-1}\, , 
\ee
yielding the following large $N$ closed equation for $G_{\text{LO}}(\tau)$
\be \label{TensClosedEqBGRGen}
G_{\text{LO}}(\tau)=1-V\tau G_{\text{LO}}(\tau)^{V/2}\, .
\ee
This type of equation is known in the literature \cite{BookComb} and the ``physical" solution, which goes to one when $\tau$ goes to zero, can be written as a power series in $\tau$ with coefficients given by the order-$V/2$ Fuss-Catalan numbers \cite{FussCatalanNumber}:
\be \label{TensLOSolutionBGR}
G_{\text{LO}}(\tau)=\sum_{k=0}^\infty C^{(V/2)}_k (-V\tau)^k \quad \text{with} \quad C^{(V/2)}_k=\frac{1}{\frac{V}{2}k+1} \begin{pmatrix} 
\frac{V}{2}k+1 \\
k 
\end{pmatrix} \, .
\ee
For $V=4$, the solution coincides with the one in terms of the Catalan numbers. There are many combinatorial interpretations for the order-$V/2$ Fuss-Catalan numbers \cite{BookComb}. The one relevant in the present context is that they enumerate $V/2$-ary trees, which can be obtained by the recursive procedure of Figure \ref{TensMelonicMove2} to construct the LO Feynman graphs. We remark that Eq.\ \eqref{TensClosedEqBGRGen} can be solved explicitly for specific values of $V$ \cite{Melons}. 

Let us now work out the summability of $G_{\text{LO}}$ and the corresponding critical behavior. We can assume that $\tau<0$ as $G_{\text{LO}}$ is generically singular for negative values of $\tau$. In this case, the series \eqref{TensLOSolutionBGR} have positive coefficients and it can be shown that $G_{\text{LO}}$ is singular on the boundary of its domain of convergence \cite{CT}. It is useful to look at the asymptotic behavior $k\rightarrow\infty$ of the coefficients of $G_{\text{LO}}$ in the perturbative series \eqref{TensLOSolutionBGR}. Using the definition of the order-$V/2$ Fuss-Catalan numbers and Stirling's formula $k! \sim \sqrt{2 \pi} e^{-k}k^{k+\frac{1}{2}}$ for $k\rightarrow \infty$, one finds that 
\be \label{AsymptBehavGLOBGR}
G_{\text{LO}}(\tau) \underset{k\rightarrow\infty}{\sim} A \, |\tau_c|^{-k} \, k^{-\alpha}\, ,
\ee
where $A$ is a constant independent of $k$ and 
\be \label{AsymptBehavGLOBGRROC}
\tau_c=- \frac{1}{2}\frac{(\frac{V}{2}-1)^{(\frac{V}{2}-1)}}{\frac{V}{2}^{(\frac{V}{2}+1)}}\quad \text{and} \quad \alpha=\frac{3}{2} \, .
\ee
This proves the summability of $G_{\text{LO}}$, whose radius of convergence is given by $|\tau_c|$. Note that the critical value $\tau_c$ depends on $V$ and thus on the interaction term used in the model. The summability of $G_{\text{LO}}$ then implies the existence of a critical regime when $\tau\rightarrow \tau_c$. 

The associated critical behavior can also be deduced from the above asymptotic behavior. Using Appell's comparison theorem \cite{Appell}, one can show that the leading singular part of $G_{\text{LO}}$ behaves close to the critical point as
\be \label{TensCritBehavGLOBGR}
G_{\text{LO,sing}} \sim (\tau_c-\tau)^{\alpha-1}=(\tau_c-\tau)^{\frac{1}{2}} \, .
\ee
In other words, the critical behavior is controlled by the polynomial part of the asymptotics \eqref{AsymptBehavGLOBGR}. Remark that the leading singular part of $G_{\text{LO}}$ has a similar behavior as the one for vector models (see Eq.~\eqref{VecCritBehaviorG}). In fact, this can be traced back to the fact that in both cases, $G_{\text{LO}}$ is a generating function of tree-like objects, which generically leads to square-root singularities \cite{CT}. 

Finally, to compute the susceptibility critical exponent of the models, we need an expression that relates the full connected two-point function $G(N,\tau)$ to the free energy $F(N,\tau)$. This can be obtained in the same way as for Eq.\ \eqref{VecConn2PtFunctionFreeEnergy} for vector models. The final result is 
\be \label{TensConn2PtFunctionFreeEnergy}
G(N,\tau) = 1- V \tau \frac{d}{d\tau} \frac{F(N,\tau)}{N^{R}} \, .
\ee
Hence, we deduce that the most singular part of the LO free energy $F_0(\tau)$ when $\tau\rightarrow\tau_c$ is
\be \label{tensCritBehaviorBGR}
F_{0,\text{sing}}(\tau) \sim (\tau_c-\tau)^{2-\gamma} \quad \text{with} \ \gamma=2-\alpha=\frac{1}{2} \, .
\ee
It indicates that tensor models with BGR scaling fall into the universality class of branched polymers, likewise vector models. This was first observed in \cite{Melons} and the connection between tensor models and branched polymers was made more precise in \cite{TensorPolymer}. Besides, multi-critical behaviors can also be observed in tensor models with BGR scaling, leading to the universality class of multi-critical branched polymers (see \cite{BGR} for more details).

\subsection{Discussion}
\label{sec:TensDiscussionBGR}

In the previous sections, we derived the $1/N$ expansion of $\oN^R$ tensor models with BGR scaling, generalizing the one initially derived in \cite{BGR}. In the large $N$ limit, melonic graphs dominate the expansion and lead to interesting results, such as Gaussianity and the universality class of branched polymers. The study of tensor models described in this section is far from exhaustive. In parallel developments, tensor field theories, with Laplacian-based propagator, have been studied and renormalized \cite{TensorMore1}, and their renormalization group flows have been investigated \cite{TensorMore2}. Non-perturbative or constructive aspects are also actively studied \cite{TensorMore3}. 

Finally, we make a brief comment on the use of tensor models with BGR scaling in the context of quantum black holes in holography. There is another important similarity between the LO graphs of these models and the bubble graphs of vector models: they both generate a tadpole at leading order in the two-point function. This can be directly related to the melonic structure of the LO self-energy (see Figure \ref{TensFreeEnergy}), which involves a single interaction vertex at a given time insertion. Therefore, it is unlikely that tensor models with BGR scaling capture the physics associated with black holes. In some sense, the family of melons is not large enough; more Feynman graphs need to be kept at large $N$. This is the precise motivation for enhancing the scaling associated with non-melonic interaction bubbles.

\section{Large $N$ expansion (2): enhanced scaling}
\label{sec:TensLargeNExp2}

Now that we have reviewed the BGR scaling, we move on to describing the new large $N$ scaling of Ref.\ \cite{Ref2}. This scaling enhances all the non-melonic interactions compared to the BGR scaling. In spite of this non-trivial enhancement, we show that the models still admit a well-defined $1/N$ expansion, which is governed by a new non-negative integer related to the index of the Feynman graphs. In particular, the results coincide with \cite{CT} in the case of $R=3$ for the pillow and the tetrahedric interactions. We then analyse the leading order Feynman graphs, called generalized melons. Next, we prove that the enhanced scaling is optimal for a class of interactions larger than the melonic ones. Finally, we explain the classification of all leading order Feynman graphs for models based on the complete interaction. This section corresponds to Sections 3.3, 3.4 and 4 in Ref.\ \cite{Ref2}.

\subsection{Large $N$ expansion}
\label{sec:TensLargeNExpEn}

The enhanced scaling of Ref.\ \cite{Ref2} is defined as follows. We introduce new coupling parameters $\lambda_a$ in terms of the coupling constants $g_a$ in the action \eqref{TensAction},
\be\label{EnScaling}
g_a=N^{R-c(\cB_a) +\frac{2}{(R-1)!}\deg\cB_a}\lambda_a\, ,
\ee
and we define a new large $N$ limit by keeping the $\lambda_a$'s fixed when $N\rightarrow\infty$. The action rewrites in terms of the enhanced scaling as
\be\label{TensActionEnScaling}
S(T)= N^{R-1}\biggl(\frac{1}{2} T\cdot T + \sum_{a\in\cS} N^{1-c(\cB_a)+\frac{2}{(R-1)!}\deg\cB_a}\lambda_a \, \cI_{\cB_a}(T)\biggr) \, .
\ee
Similarly to the BGR scaling \eqref{BGRScaling}, the coupling constants are scaled with respect to $N$ according to the degree of the corresponding interaction bubbles. However, the crucial difference is the sign of the scaling: the enhanced scaling corresponds to an enhancement of all the non-melonic interactions ($\deg \cB_a>0$). In other words, the limit $N\rightarrow\infty$ at fixed $\lambda_a$ amounts to an infinite amplification of all the non-melonic interactions compared to the BGR scaling. For the melonic interactions ($\deg \cB_a=0$), the two scalings coincide, which is consistent with the fact that the BGR scaling is already optimal for these interactions and therefore cannot be further enhanced. Besides, the scaling according to the number of connected components is the same for the two scalings. 

The enhanced scaling associated with some interaction bubbles is given in Figure \ref{TensFeynRulesColor}. Remark that since $\cB_1$, $\cB_3$ and $\cB_4$ are melonic, the enhanced scaling matches the BGR scalings. On the other hand, the tetrahedric bubble $\cB_2$ is enhanced by a factor of $N^{3/2}$. The enhanced scaling for $\cB_2$ turns out to be the optimal one (see below), as first realized in \cite{CT}.

As usual, the power of $N$ associated with a given connected Feynman graph $\cG$ can be deduced from the above action. The only difference with the BGR scaling is the factor associated with the interaction vertices, which is now 
\begin{equation*}
N^{R-c(\cB_a)+\frac{2}{(R-1)!}\deg\cB_a}\, ,
\end{equation*}
for each interaction vertex of type $\cI_{\cB_a}$. Overall, the amplitude $\cA(\cG)$ of $\cG$ thus scales with $N$ as
\be\label{TensFeynGraphEnScaling}
\cA(\cG) \sim N^{-(R-1)p+\sum_{a\in\cS_\cG}\bigl(R-c(\cB_a)+\frac{2}{(R-1)!}\deg\cB_a\bigr)+f} \equiv N^{R-\frac{\ell}{R-1}} \, ,
\ee
where we introduced the parameter $\ell$ defined as
\be\label{TensEllParam1}
\frac{\ell}{R-1}  =  R+(R-1)p - Rv +\sum_{a\in\cS_\cG}c(\cB_a)-\frac{2}{(R-1)!}\sum_{a\in\cS_\cG}\deg\cB_a-f  \, .
\ee
The scaling in $N$ of the amplitude is illustrated in Figure \ref{TensFeynGraphsColor}, where the power of $N$ associated with the Feynman graphs is indicated below the graphs in terms of the coupling parameters $\lambda_a$. It is convenient to rewrite the parameter $\ell$ in terms of the Feynman data in the colored representation using the identities \eqref{TensFeynGraphData}, which yields
\be\label{TensEllParam2}
\frac{\ell}{R-1} = R+\frac{1}{2}(R-1)V(\cB) - Rv(\cB) +B^{(0)}-\frac{2}{(R-1)!}\deg\cB_0-\sum_{i=1}^R F_{0i}(\cB) \, ,
\ee
where $\cB$ is the $(R+1)$-bubble associated with $\cG$.

Importantly, the following theorem ensures that the $1/N$ expansion associated with the enhanced scaling \eqref{EnScaling} is well-defined.

\begin{theorem}\label{TensTheoremEn}
Let $\cG$ be a connected Feynman graph of the $\oN^R$ tensor models with enhanced scaling and let $\cB$ be the corresponding $(R+1)$-bubble in the colored representation. Then, $\ell\geq 0$. In other words, the power of $N$ associated with any connected Feynman graph $\cG$ is bounded above by $R$.
\end{theorem}

\proof We can rewrite the parameter $\ell$ in terms of the parameter $L$ in Eq.\ \eqref{TensLParam2} as 
\begin{equation*}
\ell = (R-1)L - \frac{2R}{(R-2)!} \deg\cB_0 \, .
\end{equation*}
Using Eq.\ \eqref{TensLParam3} and the definition \eqref{inddef} of the index of $\cB$, we then obtain
\be\label{TensEllParam3}
\ell = 2\ind_0\cB + R(R-1) \Bigl[1+ \sum_{a\in\cS_\cG} (c(\cB_a)-1)-B\Bigr] \, .
\ee
The first term on the RHS is non-negative according to \eqref{ind0N} and the same holds for the second term within brackets, as explained in the proof of Theorem \ref{TensTheoremBGR}. Hence, $\ell$ is also non-negative because it is a sum of two non-negative terms. In addition, \eqref{ind0N} implies that $\ell\in\mathbb{N}$. \qed

\

If we denote by $\{\cG_\ell\}$ the set of connected Feynman graph of fixed $\ell$, then the $1/N$ expansion of the $\oN^R$ tensor models with enhanced scaling \eqref{BGRScaling} can be written as
\be \label{TensLargeNExpFeynEn}
F=\sum_{\ell\in\mathbb{N}} N^{R-\frac{\ell}{R-1}} F_\ell\, ,
\ee
where $F_\ell$ corresponds to a sum over connected Feynman graphs $\cG$ in $\{\cG_\ell\}$ weighted by some amplitude $\tilde{\cA}(\cG)$,
\be \label{TensLargeNExpFeynCoeffEn}
F_\ell= \sum_{\cG\in\{\cG_\ell\}} \tilde{\cA}(\cG) \, .
\ee
The enhanced scaling therefore leads to a well-defined $1/N$ expansion. The expansion parameter is now $1/N^{\frac{1}{R-1}}$, in contrast with $1/N$ for the BGR scaling. Moreover, the expansion is governed by a new non-negative integer $\ell$. This parameter has a totally different combinatorial interpretation than the parameter $L$ for the BGR scaling because it depends on the index of the Feynman graphs. The corresponding $1/N$ expansion is new in the sense that the Feynman graphs are rearranged in a new way. As we explain in the next section, it has important consequences on the leading sector of the models.

One can check that the $1/N$ expansion \eqref{TensLargeNExpFeynEn} reduces for $R=2$ to the one \eqref{MatLargeNExpFeyn} obtained in Section \ref{sec:MatLargeNExp} for $\oN^2$ matrix models. Besides, likewise the degree, the index is not a topological invariant for $R\geq 3$.

\

\noindent\emph{Case of connected interaction bubbles}

In the special case of connected interaction bubbles, one has $c(\cB_a)=1 \ \forall a\in\cS$ and $B=1$, which implies that 
\be \label{ParamEllConnBubb}
\ell= 2\ind_0\cB\, .
\ee
Thus, we obtain a $1/N$ expansion \eqref{TensLargeNExpFeynEn} governed by the index of the Feynman graphs. 

\ 

\noindent\emph{Second form of the decomposition formula for the index}

The $1/N$ expansion for the enhanced scaling suggests an interesting rewriting of the important decomposition formula \eqref{fundid0} for the index. The basic idea is that any $3$-bubble subgraph of a Feynman graph can also be regarded as a Feynman graph of a matrix model. In particular, it is characterized by a parameter $h$ that governs the $1/N$ expansion of this matrix model, as explained in Section \ref{sec:MatLargeNExp}. Then, the following proposition relates the index of a Feynman graph $\cB$ to the parameters $h$ associated to all the possible $3$-bubble subgraphs of $\cB$. 

\begin{proposition} \label{fundidth2} (Second form of the decomposition formula) The index of a connected $d$-bubble $\gB$, interpreted as the Feynman graph of a tensor model with connected interaction bubbles, is the sum of the quantities $\frac{1}{2}h_{ij}$ that govern the $1/N$ expansions of all the possible $(ij)$ matrix models one can build from the tensor model by singling out any two distinct colors $1\leq i<j\leq d-1$, 
\be\label{fundid1} \ind_{0}\gB = \frac{1}{2}\sum_{i<j}h_{ij}\, .\ee
\end{proposition}
\proof Note that one can always interpret a connected $d$-bubble $\gB$ as being a connected Feynman graph in a tensor model with connected interaction bubbles, by setting the rank of the tensor to $R=d-1$ and considering that the connected components of $\gB^{(0)}$ are the interaction bubbles of the model. Consider the $3$-bubble subgraph $\cB_{(0ij)}$ of $\cB$. We can view this $3$-bubble as a Feynman graph of the $(ij)$ matrix model constructed from the tensor model by singling out the two indices $i$ and $j$ and deleting all the other indices. In this $(ij)$ matrix model, the amplitude of $\cB_{(0ij)}$ is proportional to $N^{2-h_{ij}}$ (see Eq.\ \eqref{MatLargeNExpFeyn}), where $h_{ij}$ is given by (see Eq.\ \eqref{MatHParam3})
\be\label{hfij} 
\frac{h_{ij}}{2} = g(\gB_{(0ij)})+ 1+F_{ij}(\gB_{(0ij)}) - v(\cB_{(0ij)})-B_{(0ij)} \, .
\ee
Since $\cB$ only has connected interaction bubbles, we have $B=1$ and $B^{(0)}=v(\cB)$. We further have the trivial equalities $F_{ij}(\cB)=F_{ij}(\gB_{(0ij)})$ and $v(\cB)= v(\cB_{(0ij)})$. As a result, Eq.\ \eqref{fundid1} follows from the decomposition formula \eqref{fundid0} and Eq.\ \eqref{hfij}.\qed

\

Recall that the term $\smash{1+F_{ij}(\gB_{(0ij)}) - v(\cB_{(0ij)})-B_{(0ij)}}$ in Eq.\ \eqref{hfij} is required because the $(ij)$ matrix model may be multi-trace. This happens when $\smash{F_{ij}(\cB_a)>1}$, where $\cB_a$ is an interaction bubble in $\cB$. Note that this is not in contradiction with the fact that the interaction bubbles $\cB_a$ of the tensor model are connected. 

These considerations suggest to consider an interesting class of $\oN^R$ tensor models, which contain only interaction terms $\gB_{a}$ such that $F_{ij}(\gB_{a})=1$ for all choices of colors $i$ and $j$. Interaction bubbles with this property are called maximally single-trace (MST).\footnote{In the mathematical literature, a MST $d$-bubble is a $d$-regular graph whose edge-coloring corresponds to a perfect one-factorization of the graph \cite{perfectone}.} The index of a Feynman graph $\cB$ of any model in this particular class is given by the simpler formula
\be \label{singfaceind}
\ind_{0}\gB = \sum_{i<j}g(\gB_{(0ij)}) \, . 
\ee
This family of interactions is further studied in Section \ref{sec:MSTInt} below and in Appendix \ref{app:AppD}.

Before moving on to the description of the leading sector, we point out that unlike the degree, the index has not been studied extensively in the literature. In particular, there is no existing classification of graphs of fixed index and the entire class of interactions for which the enhanced scaling is optimal is not known. It opens interesting research routes both for tensor models and physics.  

\subsection{Leading sector: generalized melonic graphs}
\label{sec:LOGenMel}

In this section, we describe the LO Feynman graphs in the $1/N$ expansion \eqref{TensLargeNExpFeynEn} for the enhanced scaling. They satisfy $\ell=0$ or equivalently 
\be \label{TensLOGraph1En}
\sum_{a\in\cS_\cG} \bigl(c(\cB_a)-1\bigr) = B-1 \, ,
\ee
\be \label{TensLOGraph2En}
\ind_0\cB=0 \, ,
\ee
using Eq.\ \eqref{TensEllParam3}. The first condition is the same as the one obtained for the LO Feynman graphs with the BGR scaling (see Eq.\ \eqref{TensLOGraph1}); it implies that $\cB$ is effectively ``maximally disconnected" with respect to its interaction bubbles. On the other hand, the second condition says that the connected components of the LO Feynman graphs must all have index zero. 

As we explain below, melons, i.e.\ graphs of degree zero, necessarily have index zero. Thus, the melonic Feynman graphs $(a)$, $(d)$ and $(e)$ in Figure \ref{TensFeynGraphsColor} are LO with the enhanced scaling, as expected. Importantly, the Feynman graph $(f)$ is also LO when one uses the enhanced scaling. This is a hint that the leading sector of tensor models with enhanced scaling is actually larger than the one for the BGR scaling. In particular, non-melonic interaction vertices can now contribute to the leading sector. As explained in the introduction of this thesis, the Feynman graph $(f)$ based on the tetrahedric interaction has played a crucial role in the recent developments of the SYK-like tensor models. It was first studied in \cite{CT} and then in \cite{KT} and many other works. In the following, we focus on Feynman graphs of index zero. 

For $R=2$, the story is the same as before: Feynman graphs are $3$-bubbles, their index matches their genus and thus the LO Feynman graphs are the planar ones. However, the story is completely different for $R\geq 3$, as the study of graphs of index zero remains an open problem. In Ref.\ \cite{Ref2}, they are called generalized melons. Up to now, there exists no general classification for this class of graphs. In particular, we do not know all the types of interaction bubble that can appear in generalized melons, in contrast with melons whose interaction bubbles are necessarily melonic. In this section, we describe several properties of generalized melons. It corresponds to Sections 2.3.4 and 3.3.3 in Ref.\ \cite{Ref2}. Then, we show in the next section that the enhanced scaling is optimal for all MST interactions. In other words, any MST interaction bubble can appear an arbitrary number of times in a generalized melon. Finally, we explain the full classification of the LO Feynman graphs for models based on the complete interaction on $R+1$ vertices, where $R\geq 3$ is a prime number. 

\subsubsection{Generalized melons and generalized melonic moves}
\label{gmSec}

A $(R+1)$-bubble of index zero is called generalized melon or generalized melonic graph. Since the index of a bubble is the sum of the indices of its connected components, a bubble is of index zero if and only if all its connected components have index zero. For a connected $(R+1)$-bubble $\gB$, the first form of the decomposition formula \eqref{fundid0} implies that $\ind_{0}\gB=0$ is equivalent to the conditions
\begin{align}\label{ind01} 
& g(\gB_{(0ij)}) = 0\, ,\\ \label{ind02}
& F_{ij}(\gB) - B_{(0ij)} = B^{(0)}-1\, ,
\end{align}
for all choices of pairs of colors $(i,j)$ with $i,j\in\{1,2,\ldots,R\}$. The first condition simply says that the $3$-bubble subgraphs $\cB_{(0ij)}$ of $\cB$ must all be planar. As for the second condition, it can also be written as
\be\label{zeroindex1} 
\Delta_{0}B^{(i_{1}\cdots i_{R-2})}=B^{(0i_{1}\cdots i_{R-2})}-B^{(i_{1}\cdots i_{R-2})}=\Delta_{0}B=B^{(0)}-1\, ,
\ee
for all choices of indices $i_{1},\ldots,i_{R-2}$, or, as explained in Appendix \ref{ineqSec}, as
\be\label{zeroindex}
\Delta_{0}B^{(i_{1}\cdots i_{p})}= B^{(0 i_{1}\cdots i_{p})}-B^{(i_{1}\cdots i_{p})}=\Delta_{0}B^{(i_{1}\cdots i_{p-1})}=B^{(0 i_{1}\cdots i_{p-1})}-B^{(i_{1}\cdots i_{p-1})}\, ,
\ee
for all choices of indices and $1\leq p\leq R-2$. A convenient way to understand these conditions is to consider a graph for which the set of $B^{(0)}$ connected components of $\gB^{(0)}$ are given. The edges of color 0 must then be adjusted in such a way that $B_{(0ij)}$ takes its maximum possible value, that is to say, the $3$-bubble subgraphs $\gB_{(0ij)}$ must all be ``maximally disconnected".

Generalized melons can also be understood from the point of view of the second form of the decomposition formula \eqref{fundid1}. For a connected $(R+1)$ bubble $\gB$, it implies that
\be\label{ind03} 
h_{ij}=0
\ee
for all choices of pairs of colors $(i,j)$. In other words, the $3$-bubble subgraphs $\cB_{(0ij)}$ of $\cB$ must all correspond to LO Feynman graphs in their respective $(ij)$ matrix model, i.e.\ they must be ``maximally disconnected" planar graphs.

Using the definition \eqref{inddef} of the index and the inequality \eqref{degineq}, it is straightforward to see that ordinary melons, i.e.\ graphs of degree zero, are also generalized melons. However, the converse is not true: the class of generalized melons is \textit{a priori} much larger. In particular, the connected components of $\gB^{(0)}$ can be non-melonic interaction bubbles. 

Finally, we mention an important property regarding the structure of the generalized melons. Recall that in the description of melons in Section \ref{sec:TensBGRLO}, the addition formula \eqref{adddeg} for the degree was used in combination with the melonic structure in order to construct recursively all the LO Feynman graphs of the models. In the case of the index, Proposition \ref{addprop} also provides an addition formula, which implies that inserting or contracting a generalized melon on an edge of color $0$ in an arbitrary Feynman graph $\gB$ does not change its index. We call such operations generalized melonic moves. A useful consequence is that, from any given generalized melon, one can immediately construct an infinite family of generalized melons by repeated generalized melonic insertions. 

\subsection{Optimal scalings, MST interactions and mirror melons}
\label{sec:MSTInt}

Proving that a scaling is optimal for the coupling associated with a given interaction term and identifying the LO Feynman graphs are two different and difficult problems in general. In this section, we give a simple criterion that allows one to conclude for the first problem in the context of tensor models with enhanced scaling. Then, we apply the criterion to MST interaction terms by constructing specific generalized melons based on MST interaction bubbles.

\begin{lemma}\label{optimallem} If a given interaction bubble $\cB_a$ can contribute to the leading sector, then the enhanced scaling \eqref{EnScaling} is optimal for the associated coupling constant.
\end{lemma}

\proof Suppose that an interaction bubble $\gB_{a}$, with associated coupling constant $\lambda_{a}$ and with enhanced scaling \eqref{EnScaling}, contributes to the leading sector. Then, it is necessarily part of a generalized melon. We can use this generalized melon and the corresponding generalized melonic moves to build new generalized melons containing an arbitrary number of interaction bubbles $\gB_{a}$. As a result, the enhanced scaling is necessarily optimal for the coupling constant associated with $\cB_a$.\qed

\

As a simple application, we study the case of MST interaction bubbles. As explained earlier, these interaction bubbles, denoted as $\cB_a^{\text{MST}}$, satisfy $F_{ij}(\cB_a^{\text{MST}})=1$ for all choices of colors $i,j\in\{1,2,\ldots,R\}$. Suppose that $\cB_a^{\text{MST}}$ has $V(\cB_a^{\text{MST}})=V$ vertices. Such a bubble is automatically connected and has $\frac{1}{2}R(R-1)$ faces. Its degree, which follows from Eq.\ \eqref{degreeformula} with $d=R$, is thus given by $\deg\gB_a^{\text{MST}} = \frac{1}{8}(R-2)(R-1)!(V-2)$. As for the associated enhanced scaling in the action, it is given by \eqref{EnScaling}, that is,
\be\label{scalingMST}
g_a^{\text{MST}}=N^{R-1+\frac{1}{4}(R-2)(V-2)}\lambda_a^{\text{MST}}\, .
\ee

\begin{proposition}\label{MSToptprop} The enhanced scaling \eqref{scalingMST} is optimal for all MST interaction terms. 
\end{proposition}

\

\begin{figure}
\begin{center}
{\includegraphics[width=6in]{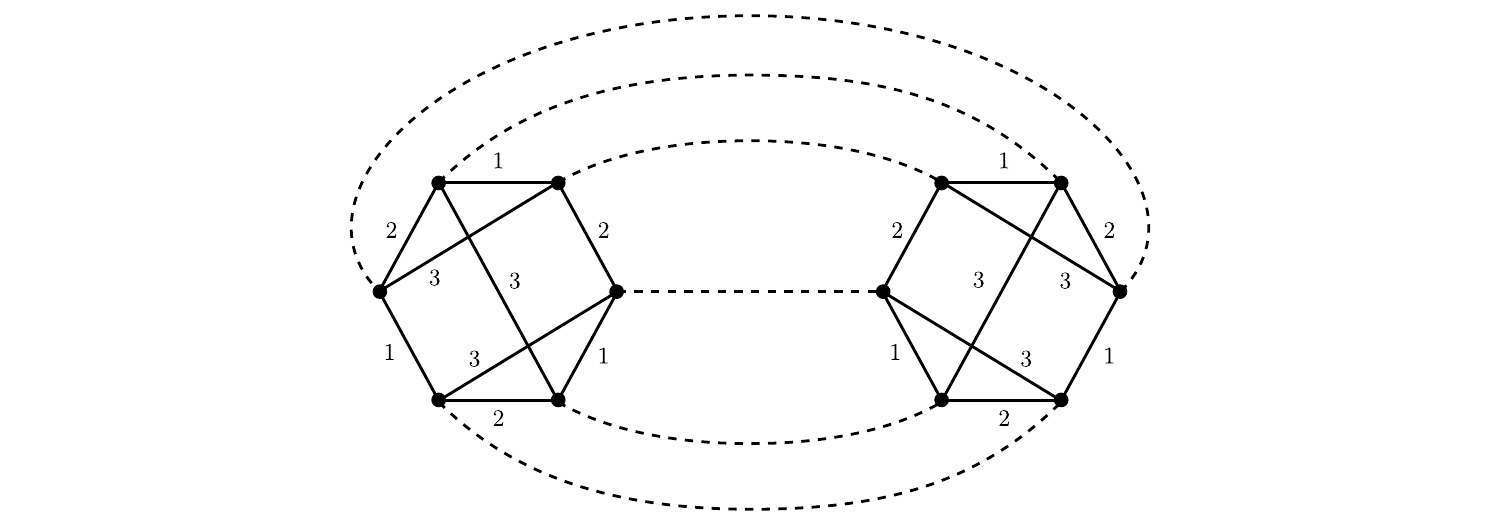}}
\end{center}
\caption{An elementary mirror melon built from two identical interaction bubbles.}
\label{figureGraphMirror}
\end{figure}

The proof of this result relies on a specific type of Feynman graphs, called mirror melons, which we define now. Let $\gB_a$ be any interaction bubble with labeled vertices. Consider the Feynman graph obtained by connecting with an edge of color $0$ vertices with identical labels in two copies of $\gB_a$. We call this unique Feynman graph associated with $\gB_a$ the elementary mirror melon of type $\gB_a$. This is illustrated in Figure \ref{figureGraphMirror} for a given interaction bubble. Note that it is convenient to picture the second interaction bubble as the mirror reverse image of the first one; hence the name. Then, we construct an infinite family of Feynman graphs by recursively inserting elementary mirror melons of type $\gB_a$, in the spirit of Section \ref{sec:AddFormula}, starting from the elementary mirror melon of type $\gB_a$. We call this infinite family of Feynman graphs the mirror melons associated with $\gB_a$. Proposition~\ref{MSToptprop} follows from the fact that mirror melons are generalized melons if $\gB_a$ is MST.

\proof Consider a MST interaction bubble $\cB_a^{\text{MST}}$ with $V$ vertices. In order to prove that the scaling \eqref{scalingMST} is optimal, it is sufficient to find, according to Lemma \ref{optimallem}, a generalized melon that contains $\cB_a^{\text{MST}}$. Let us show that the elementary mirror melon of type $\cB_a^{\text{MST}}$, which we denote by $\cB$ in the proof, is a generalized melon. By construction, all the $(0i)$-faces of $\cB$ contain one edge of color $i$ in each copy of $\cB_a^{\text{MST}}$. Their total number thus matches the number of edges in $\cB_a^{\text{MST}}$: $\sum_{i}F_{0i}(\gB)=E(\cB_a^{\text{MST}})=\frac{1}{2}RV$. On the other hand, using the MST property, we have $\smash{\sum_{i<j}F_{ij}(\gB)=2\sum_{i<j}F_{ij}(\gB_{\text{MST}}) = R(R-1)}$. Finally, since $B=1$ and $B^{(0)}=2$, Eq.\ \eqref{indexplicit} then yields $\ind_{0}\gB=0$. In other words, the elementary mirror melon of type $\cB_a^{\text{MST}}$ is a generalized melon. \qed

\

One can also show that the elementary mirror melon of type $\cB_a$ corresponds to a generalized melon if and only if $\cB_a$ is MST. A natural but much more difficult question to ask is whether the mirror melons yield all the possible generalized melons. In the next section, we explain that it is indeed the case for the complete interaction on $R+1$ vertices, when $R \ge 3$ is a prime number.

\subsection{Case of the complete interaction}
\label{sec:CompleteInt}

In this section, we give a few details on the $\oN^R$ tensor models based on the complete interaction on $\smash{R+1}$ vertices. This is an important result as it provides a full classification of the LO Feynman graphs when $R$ is a prime number. The proof, although interesting, is rather long and technical, the details of which can be found in Appendix \ref{app:AppC}. It corresponds to Section 4 in Ref.\ \cite{Ref2}. Here, we focus on describing the main results. 

The case $R=3$ was first solved in \cite{CT}, where the complete interaction on four vertices corresponds to the tetrahedric interaction. Interestingly, the cases $R>3$ turn out to be qualitatively different and their analysis requires to introduce several new ingredients. 

Models based on the complete interaction on $R+1$ vertices play an important role in the development of quantum models with the SYK/black hole physics, as they correspond to tensor versions of the SYK model with $q=(R+1)$-fold random interactions \cite{SYKTensorQ}. We defer the discussion of such models to Section \ref{sec:MatTensDiscussion} in the context of matrix-tensor models.

Finally, the recent paper \cite{NewKleba} also studies the case of the complete interaction on $R+1$ vertices, in the context of SYK-like tensor models. In particular, they provide a sketch of a proof for the classification of the LO Feynman graphs for any $R$ odd, based on the existence of MST colorings of the complete graph $\cK_{R+1}$ for any $R$ odd (see below).

\subsubsection{Definition of the complete interaction bubble}

\begin{figure}
\centerline{\includegraphics[width=6in]{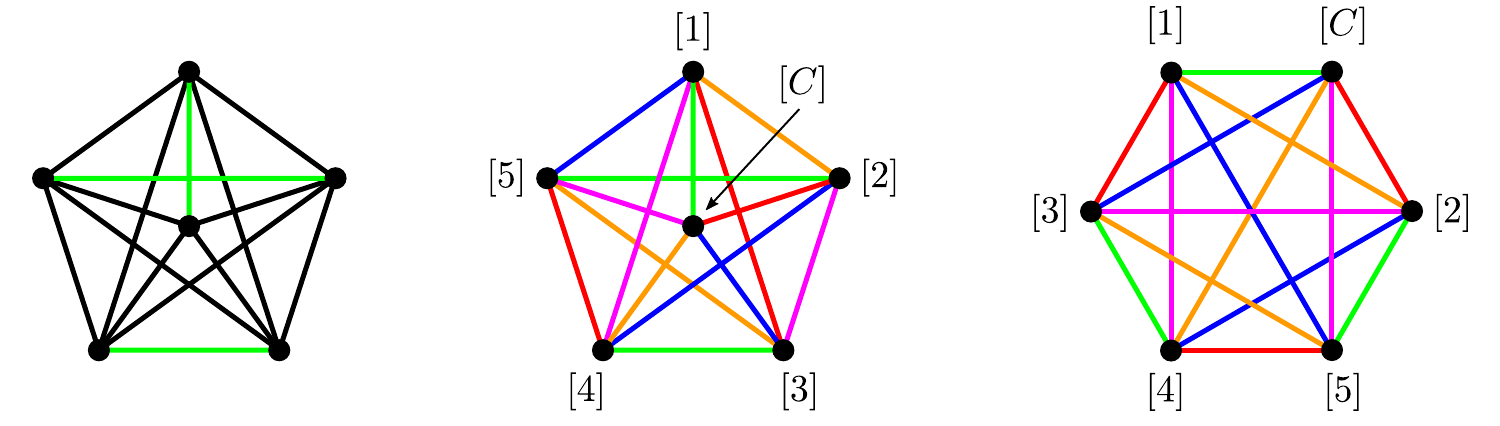}}
\caption{Edge-coloring for the complete graph $\cK_6$. Left: rule for the coloring of the edges of a particular color, here green. Center: full edge-coloring and standard vertex labeling, here $1=\text{green}$, $2=\text{red}$, $3=\text{blue}$, $4=\text{orange}$ and $5=\text{purple}$. Right: the equivalent $(\text{green},\text{red})$-polygonal representation in the shape of a six-sided polygon whose boundary is the face of colors green and red. This polygonal representation is natural when $R$ is prime.}\label{figureA}
\end{figure}

For $R$ odd, the complete graph $\cK_{R+1}$ with $R+1$ vertices and $\frac{1}{2}R(R+1)$ edges is edge-colorable with $R$ colors. The explicit $R$-regular edge-coloring that we shall use can be described as follows \cite{soifer}. We consider a regular $R$-sided polygon plus its center. The center is labeled as $[C]$ and the vertices of the polygon are cyclically numbered as $[1]$ to $[R]$. For each color $1 \le i \le R$, we draw an edge of color $i$ from the center $[C]$ to the vertex $[i]$ of the polygon. Then, we use the same color for all edges between polygon vertices that are perpendicular to the edge $[C][i]$. If we identify the polygon vertices with ${\mathbb Z } / R {\mathbb Z }$, it means that the polygon vertices $[n]$ and $[n'] \ne [n]$ are joined by an edge of color $i$ if and only if $n+n'= 2i \mod R$. Equivalently, the edges of color $i$ join the polygon vertices $[i+p]$ and $[i-p]$ for all $1\leq p\leq \frac{1}{2}(R-1)$. The $R$-bubble obtained in this way will also be denoted as $\gK_{R+1}$\footnote{Depending on the context, $\cK_{R+1}$ can thus mean either the complete graph on $R+1$ vertices or the corresponding $R$-bubble with standard coloring.} and the above color and vertex labeling will be called the ``standard'' coloring. The construction is illustrated in Figure \ref{figureA} for $\mathcal K_6$, where we distinguish the colored edges with real colors instead of positive integers.

\begin{figure}
\centerline{\includegraphics[width=2.8in]{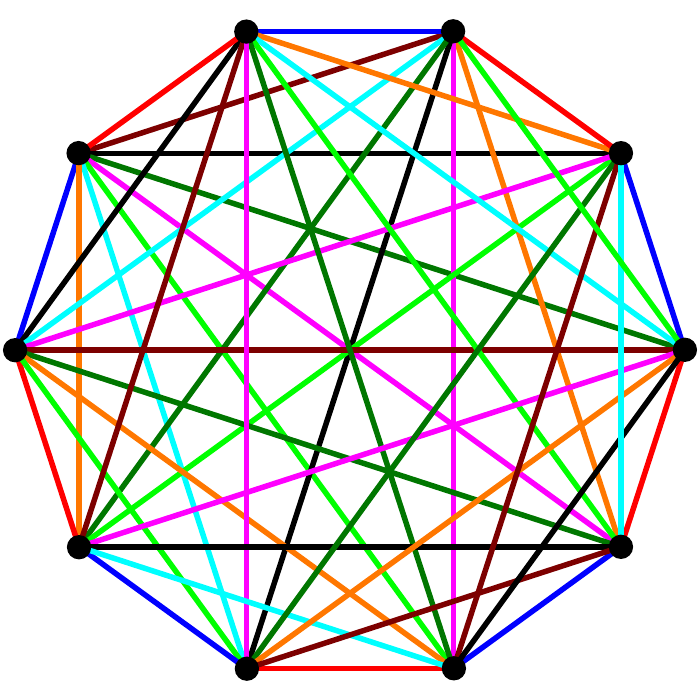}}
\caption{MST coloring of $\cK_{10}$.}\label{nonstandardK10}
\end{figure}

We shall say that two edge-colorings for the complete graph are equivalent if there exist a permutation $\sigma$ of the $R+1$ vertices and a permutation $\tau$ of the $R$ colors that change one edge-coloring into the other. It is easy to check directly that all the possible colorings for $R=3$ and $R=5$ are equivalent to the standard one. More generally, the number of non-equivalent edge-colorings for the complete graph is counted by the sequence A000474 in OEIS \cite{OEIS}. For example, there are $6$ non-equivalent edge-colorings for $\cK_8$, 396 for $\cK_{10}$, etc. If we also impose the MST condition, there remains only one possibility for $\cK_{8}$, which is the standard coloring, and also one possibility for $\cK_{10}$, which is not the standard coloring (as explained below, the standard coloring for $\cK_{R+1}$ is MST if and only if $R$ is a prime number; the standard coloring for $\mathcal K_{8}$ is thus MST but the one for $\mathcal K_{10}$ is not). The non-standard MST coloring for $\cK_{10}$ is depicted in Figure \ref{nonstandardK10}.

We focus on the complete interaction bubble with the standard coloring. All the results strongly depend on this choice and it is an open question as to whether similar results can be derived for edge-colorings that are not equivalent to the standard one. For instance, we do not know the classification of the generalized melons in the case of the non-standard MST coloring for $\cK_{10}$ depicted in Figure \ref{nonstandardK10} (see however \cite{NewKleba} for a sketch of a proof).

\subsubsection{The MST condition}

From the above discussion, the cases $R$ prime and $R$ not prime seem to be qualitatively different with respect to the standard coloring. This is confirmed by Proposition \ref{prime} in Appendix \ref{app:AppC}, which states that the $R$-bubble $\gK_{R+1}$ is MST if and only if $R$ is a prime number. The interaction bubbles $\gK_{R+1}$ with $R$ prime are called prime-complete. These bubbles have convenient $(ij)$-polygonal representations in the shape of a $(R+1)$-sided polygon whose boundary is the unique $(ij)$-face. This is illustrated on the right of Figure \ref{figureA} for $\mathcal K_6$.

A simple application of Proposition \ref{prime} is the computation of the degree of the prime-complete bubbles. Indeed, since we know that there is exactly one face per pair of colors, the total number of faces is simply $F(\gK_{R+1}) = \frac{1}{2}R(R-1)$. Together with $c(\gK_{R+1})=1$ and $V(\gK_{R+1}) = R+1$, Eq.\ \eqref{degreeformula} yields
\begin{equation}  
\frac{2}{(R-1)!}\deg \gK_{R+1} = \frac{1}{4} (R-1)(R-2)\, .
\label{degreeint}
\end{equation}
In contrast, for $R$ not a prime, we have by Proposition \ref{prime} that $F(\gK_{R+1}) > \frac{1}{2}R(R-1)$ so that
\begin{equation}  
\frac{2}{(R-1)!}\deg \gK_{R+1} < \frac{1}{4} (R-1)(R-2)\, .
\label{degreeintnotprime}
\end{equation}

\subsubsection{Distinguishing edges and vertices}

One can check that all the edges of a given color in $\gK_{4}$ are equivalent and that all the vertices of $\gK_{4}$ are equivalent too. The situation is drastically different at higher rank. It turns out that the $\frac{1}{2}(R+1)$ edges of any given color in the prime-complete bubble $\gK_{R+1}$ are all inequivalent, for all primes $R>3$. The same is true for the $R+1$ vertices of the prime-complete bubble as well. This major difference between $R=3$ and $R>3$ goes a long way in explaining why a new proof of the classification theorem for the leading graphs must be provided. We give here an intuitive understanding of equivalent edges in the present context. The full details can be found in Appendix \ref{app:AppC}. 

Consider an edge of a given color $k$ that joins two vertices in $\cK_{R+1}$, say $[\nu_{1}]$ and $[\nu_{2}]$. Then, for any pair of distinct colors $i$ and $j$, we follow the $(ij)$-path, i.e.\ the path of alternating colors $i$ and $j$, which starts at $[\nu_{1}]$ with an edge of color $i$ and ends at $[\nu_{2}]$. In the case of $\cK_{4}$, one can convince himself that the $(ij)$-paths obtained by the above procedure are essentially the same for two edges of the same color. However, for $R>3$, this is not the case anymore for all pairs of distinct colors $i$ and $j$. In other words, two edges of the same color can be unambiguously distinguished from one another using the coloring of the graph. 

\subsubsection{Action and index}

The interaction term associated with the complete interaction bubble $\gK_{R+1}$ is given explicitly by
\be\label{uncoloredint}
\cI_{\gK_{R+1}}(T) = \prod_{\nu=0}^{R} T_{a_{\nu,1}  \cdots \,  a_{\nu,R}} \, .
\ee
In this expression, we set $a_{0,i}=a_{i,i}$ and $a_{n,i}=a_{n',i}$ for $1\leq n,n'\leq R$, $n'\not = n$ and $n+n' = 2i \mod R$, so that we reproduce the standard edge-coloring of $\cK_{R+1}$ explained above. 

The action for models based on the complete interaction with enhanced scaling is given by (see Eq.\ \eqref{TensActionEnScaling} and Eq.\ \eqref{degreeint})
\be \label{actionGenCT} S=N^{R-1} \Bigl( \frac{1}{2} T\cdot T +  N^{\frac{1}{4}(R-1)(R-2)}  \lambda\, \cI_{\gK_{R+1}}(T) \Bigr) \, , \ee
for $R$ prime.\footnote{As usual, instead of a zero-dimensional action, we could consider quantum mechanical or field-theoretic generalizations; the discussion would remain unchanged.}

As explained in Section \ref{sec:TensLargeNExpEn}, when one takes the large $N$ limit at fixed $\lambda$, one gets a well-defined expansion in powers of $1/N^{\frac{1}{R-1}}$. The Eqs.~\eqref{TensLargeNExpFeynEn} and \eqref{ParamEllConnBubb} show that the connected Feynman graphs $\gB$ contributing at a given order $\smash{N^{R-\frac{\ell}{R-1}}}$ have a fixed index $\smash{\ind_{0}\gB=\frac{\ell}{2}}$. Since the prime-complete interaction bubble is MST, the index is given by formula \eqref{singfaceind}. A more explicit expression for the index can be obtained from Eq.\ \eqref{indexplicit}. Indeed, since the interaction bubbles of the Feynman graphs are all prime-complete, we have $B=1$, $B^{(0)}=v$, $V=(R+1)v$ and $\sum_{i<j}F_{ij}=\frac{1}{2}R(R-1)v$ so that Eq.\ \eqref{indexplicit} yields
\begin{equation}\label{indexmodel}
\ind_{0}\gB =\frac{1}{2}R(R-1) + \frac{1}{8}R(R-1)^2 v -\frac{1}{2}(R-1)\sum_i F_{0i} \, .
\end{equation}
The LO Feynman graphs that dominate the large $N$ expansion are the generalized melons, which are, by definition, of index zero. For the present models, we call them prime-complete generalized melons, or PCGMs for short. They maximize the total number of $(0i)$-faces for a fixed number of vertices. The condition for a Feynman graph to be a PCGM is equivalent to  
\begin{equation}
\sum_i F_{0i} = \frac{1}{4}R(R-1)v + R \, ,
\label{leadinggraphs}
\end{equation}
which, using formula \eqref{singfaceind}, is itself equivalent to
\be\label{leadg} g(\gB_{(0ij)}) = 0\ \text{for all pairs of distinct colors $(i,j)$.}\ee

\subsubsection{Description of the PCGMs}

\begin{figure}
\centerline{\includegraphics[scale=1]{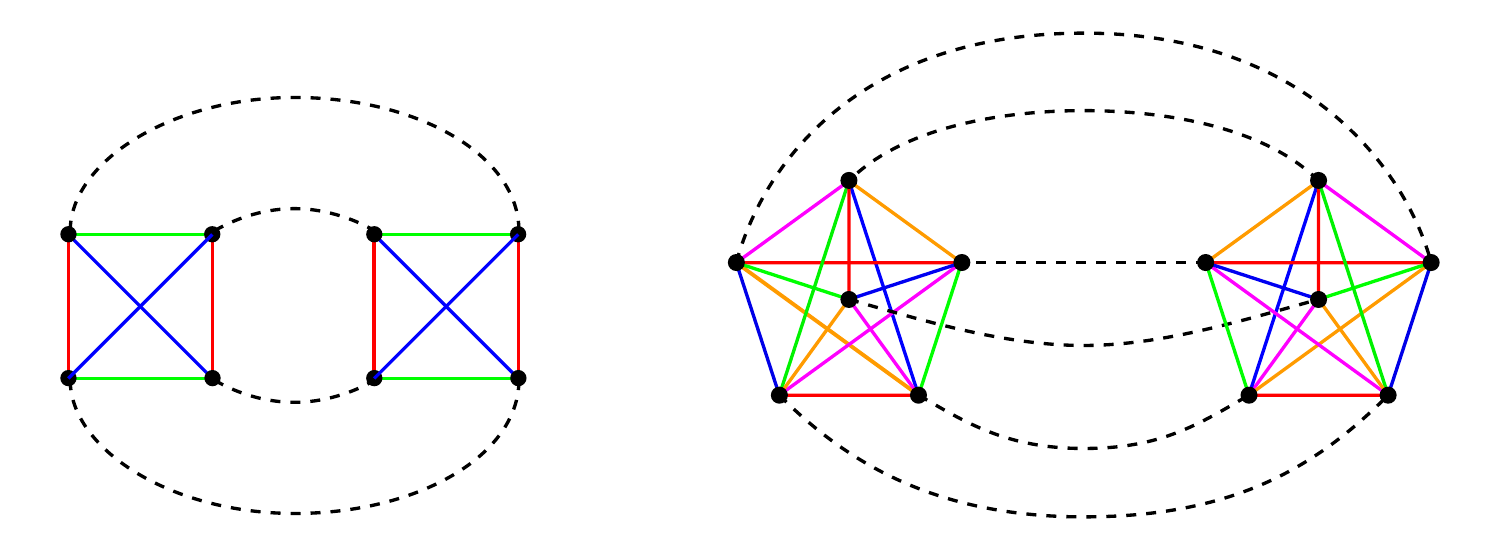}}
\caption{Elementary mirror melons of type $\cK_4$ (on the left) and $\cK_6$ (on the right). The PCGMs are constructed by recursively inserting these Feynman bubbles onto themselves. The colors are explained in Figure \ref{figureA}.}\label{figureB}
\end{figure}

In Appendix \ref{app:AppC} or in Ref.\ \cite{Ref2}, it is shown in details that for $R$ prime, the PCGMs exactly coincide with the mirror melons associated with $\cK_{R+1}$, as defined in Section \ref{sec:MSTInt}. Recall that these Feynman graphs are obtained by recursively inserting elementary mirror melons of type $\cK_{R+1}$ onto themselves. As an illustration, the elementary mirror melons of type $\cK_4$ and $\cK_6$ are shown in Figure \ref{figureB}. 

We now make a remark for $R$ not prime. In this case, the complete interaction bubble $\cK_{R+1}$ is no longer MST. As a result, the elementary mirror melon of type $\cK_{R+1}$ does not correspond to a generalized melon for the enhanced scaling. However, we can still construct a model with action \eqref{actionGenCT} where the large $N$ scaling is strictly enhanced compared to the enhanced scaling (see Eq.\ \eqref{degreeintnotprime}). One can check that it is the right scaling for the elementary mirror melon of type $\cK_{R+1}$ and the corresponding mirror melons to scale like $N^R$. However we do not know if the leading sector is made solely of the mirror melons in this case, or even if the large $N$ limit makes sense.

\subsubsection{Application: Schwinger-Dyson equations and large $N$ solution}

Finally, we can use the description of the leading sector for models with action \eqref{actionGenCT} and $R$ prime to derive a Schwinger-Dyson equation for the LO self-energy $\Sigma_{\text{LO}}$ and then a closed equation for the LO full connected $2$-point function $G_{\text{LO}}$, in the same spirit as in Section \ref{sec:TensBGRLO}. We first focus on $\cK_{4}$, which was initially treated in \cite{CT}, and then we generalize the result to $\cK_{R+1}$.

\begin{figure}[]
\centerline{\includegraphics[scale=1]{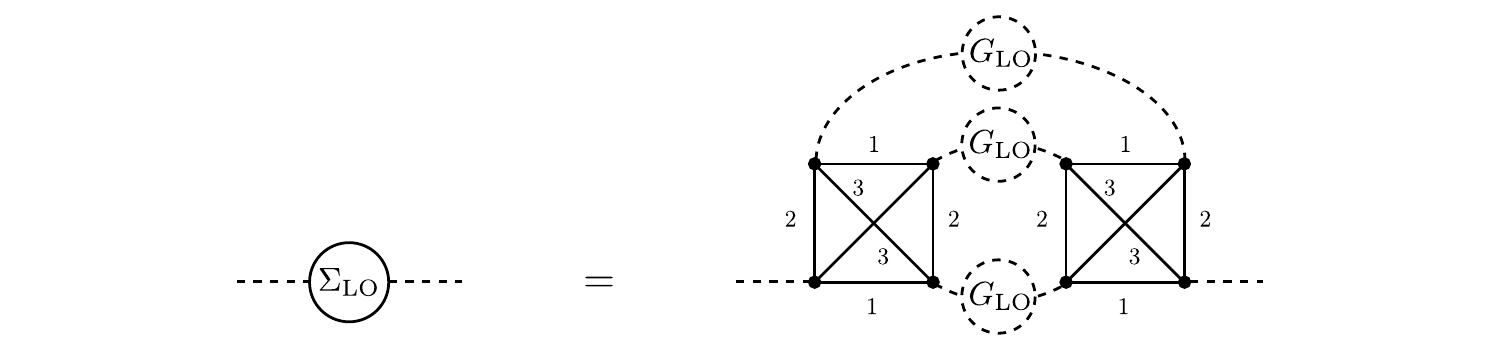}}
\caption{Diagrammatic equation for the LO self-energy $\Sigma_{\text{LO}}$ for models based the prime-complete interaction bubble $\cK_4$ (or tetrahedric interaction).}\label{TensFreeEnergyEn}
\end{figure}

Because of the recursive structure of the mirror melons, $\Sigma_{\text{LO}}(\lambda)$ necessarily has the structure represented in Figure \ref{TensFreeEnergyEn}, which is equivalent to the Schwinger-Dyson equation
\be \label{TensSDEqEn}
\Sigma_{\text{LO}}(\lambda) = 16 \lambda^2 G_{\text{LO}}(\lambda)^3 \, .
\ee
Remark the important factor $\lambda^2$, which results from having two distinct interaction bubbles in the elementary mirror melon. This is in contrast with melons, whose recursive structure is based on elementary melons containing a single interaction bubble (see Figure \ref{TensMelonicMove2} or Figure \ref{TensFreeEnergy}). This has important physical consequences (see below). As for the combinatorial factor, it follows from the fact that for $R=3$, the elementary mirror melon is obtained from four distinct sets of Wick contractions (see Appendix \ref{app:AppC}).

The Schwinger-Dyson equation \eqref{TensSDEqEn} for $\cK_4$ generalized to $\cK_{R+1}$ and $R$ prime as
\be \label{TensSDEqEnGen}
\Sigma_{\text{LO}}(\lambda) = (R+1) \lambda^2 G_{\text{LO}}(\lambda)^{R} \, .
\ee
The combinatorial factor $16$ when $R=3$ does not generalize to $(R+1)^2$ for larger values of $R$ because of the lack of symmetry between the edges and vertices of $\cK_{R+1}$ for $R>3$. In particular, the elementary mirror melon is obtained from a unique set of Wick contractions in this case.

The two above Schwinger-Dyson equations together with Eq.\ \eqref{Tens2PtFunction} yield the following two closed equations for $G_{\text{LO}}(\lambda)$ in the large $N$ limit
\begin{align} \label{TensClosedEqEn1}
G_{\text{LO}}(\lambda)&=1+16\lambda^2 G_{\text{LO}}(\lambda)^4 \qquad \quad \quad \, \, \text{if} \ R=3 \, , \\ \label{TensClosedEqEn2}
G_{\text{LO}}(\lambda)&=1+(R+1)\lambda^2 G_{\text{LO}}(\lambda)^{R+1} \quad \text{if} \ R\geq 3 \ \text{and} \ R \ \text{prime} \, .
\end{align}
These equations are polynomial in $G_{\text{LO}}$, as in the case of melons (see Eq.~\eqref{TensClosedEqBGR}). Similarly, the physical solution can be written as a power series in $\lambda$ with the order-$(R+1)$ Fuss-Catalan number as coefficients. In particular, it means that the PCGMs can be mapped to $(R+1)$-ary trees and therefore look similar to the LO Feynman graphs of colored tensor models \cite{LargeNColored, Melons}. 

One can compute the critical behaviors of $G_{\text{LO}}$ and the LO free energy $F_0$, and then deduce the susceptibility critical exponent, as in Section \ref{sec:TensBGRLO}. Since $G_{\text{LO}}(\lambda)$ corresponds to a generating function of $(R+1)$-ary trees, it exhibits a critical behavior with a square-root singularity, that is, if we set $\mu=\lambda^2$, 
\be \label{TensCritBehavGLOEn}
G_{\text{LO,sing}} \sim (\mu_c-\mu)^{\frac{1}{2}} \, .
\ee
for some critical value $\mu_c$. Then, the critical behavior of $F_0$ can be deduced from Eq.~\eqref{TensConn2PtFunctionFreeEnergy} with $\tau=\lambda$ and $V=R+1$. The most singular part of $F_0(\mu)$ when $\mu\rightarrow\mu_c$ thus writes
\be \label{tensCritBehaviorEn}
F_{0,\text{sing}}(\mu) \sim (\mu_c-\mu)^{2-\gamma} \quad \text{with} \ \gamma=\frac{1}{2} \, .
\ee
As a result, tensor models based on the prime-complete interaction $\cK_{R+1}$ with enhanced scaling are part of the universality class of branched polymers. This was first observed for $R=3$ in \cite{CT}. In fact, this is typical of Schwinger-Dyson equations which are polynomials in $G_{\text{LO}}$.

\subsection{Discussion and outlook}
\label{sec:TensDiscussionEn}

In the previous sections, we described the large $N$ expansion of $\oN^R$ tensor models with enhanced scaling. In spite of all non-melonic interactions being enhanced compared to the BGR scaling, the expansion is still well-defined. It is governed by the index of the Feynman graphs, which offers a new combinatorial interpretation of the perturbative expansion. In the large $N$ limit, generalized melonic graphs dominate and form a larger class of Feynman graphs than the standard melons. In particular, non-melonic interaction bubbles can contribute to the leading sector, such as the MST interaction bubbles. 

Unlike tensor models with BGR scaling, little is known about tensor models with enhanced scaling. The latter provide interesting research directions, from the point of view of both combinatorics and physics. We present some of them in the following. 

\subsubsection{Optimal scalings}

As explained in the introduction, random tensor models play a special role in two different approaches to quantum gravity: on the one hand in the discretized approach where the Feynman graphs generate, in the continuum limit, random higher dimensional geometries, and on the other hand in the holographic approach with the SYK-like tensor models. In both cases, finding optimal scalings for non-melonic interactions is an important issue because it allows one to extend the class of leading order Feynman graphs, which means that in principle, more types of random geometries could be generated and more types of physics could be described. 

With this motivation in mind, we observed in this chapter that for the known cases of interaction bubbles for which the enhanced scaling is optimal and for which the associated leading order Feynman graphs are classified, namely the melonic interactions and the complete interactions $\cK_{R+1}$ for $R$ prime, the models give rise, in the continuum limit, to a branched polymer phase. It would be interesting to investigate whether the enhanced scaling can allow one to escape this branched polymer phase and probe new universality classes with more attractive physical features. A first step in this direction would be to characterize the full set of interaction bubbles for which the enhanced scaling is optimal, then to classify the corresponding leading order Feynman graphs and finally to study the possible critical behaviors.

There exist many results in the literature regarding optimal scalings (see second to last reference in \cite{nonBGR2} for a good account). In particular, there are known examples of interactions bubbles whose optimal scaling has been found and which reproduce the universality class of two-dimensional pure gravity, like random matrix models. In fact, these interaction bubbles often correspond to modified versions of matrix models interactions, the necklaces being an example \cite{nonBGR2}. A straightforward check shows that the enhanced scaling introduced in this chapter is not optimal for these interaction bubbles. Hence the question: can we also achieve the universality class of two-dimensional pure gravity with the enhanced scaling for particular interaction bubbles? Within the framework provided by the index, one should be in principle able to check whether this is true or not. 

Of course, the questions of optimal scalings and new universality classes extend far beyond the present work. A main motivation is quantum gravity in four dimensions; however, being able to generate random four-dimensional geometries in the continuum limit has not been possible in the realm of random tensor models up to now. It remains an exciting research avenue, which may require new, unknown inputs (see \cite{LucaBM} for a recent development, though unrelated to random tensor models).

On the other hand, the enhanced scaling has been shown to be crucial for constructing SYK-like tensor models in the context of holography. An important example is the CTKT model with action \eqref{CTKT}, which is based on the tetrahedric interaction $\cK_4$ \cite{CT}. This model can be generalized to the complete interaction $\cK_{R+1}$ for $R$ prime. In the large $N$ limit, the leading order Feynman graphs then correspond to the PCGMs and the resulting Schwinger-Dyson equations are similar to the ones in the SYK model with $q=(R+1)$-fold random interactions \cite{SYKq}. As a result, the physics described in both models is the same at leading order. It is worth emphasizing why the structure of the PCGMs is important in comparison with the structure of standard melons. The latter generate a tadpole at leading order in the two-point-function, likewise bubbles graphs in vector models. On the other hand, the PCGMs allow for a leading order two-point function which is bi-local in time, like in the SYK model. This can be seen from the Schwinger-Dyson equation in Figure \ref{TensFreeEnergyEn}, which involves two interactions bubbles at two distinct time insertions. This bi-local structure is a key element for the SYK physics \cite{EffectiveSYK,PaoloThesis}. In this context, characterizing the full set of interaction bubbles for which the enhanced scaling is optimal and classifying the corresponding leading order Feynman graphs is interesting as it may enlarge the class of known SYK-like tensor models.

\subsubsection{Beyond leading order}

In random tensor models with enhanced scaling, the large $N$ expansion is governed by the index of the Feynman graphs. Up to now, only the leading sector of this expansion, corresponding to generalized melonic graphs which have index zero, is relatively well understood for some particular interaction bubbles. However, it is important to have a better analytical control over the perturbative expansion. This requires a detailed study of the subleading contributions in $1/N^{\frac{1}{R-1}}$, which is equivalent to studying the sets of Feynman graphs of fixed index. This program is similar to the one initiated in \cite{GurauSchaeffer} for $\uN^R$ random tensor models with BGR scaling, where the structure of the Feynman graphs of fixed degree is analyzed in detail. In particular, it is shown using enumeration techniques that the generating series of the family of Feynman graphs of fixed degree has a finite radius of convergence. The work of \cite{GurauSchaeffer} was then extended to MO tensor models in \cite{FusyTanasa}, whose perturbative expansion includes more Feynman graphs compared to their $\uN^3$ tensor model counterparts. In the same line, a similar analysis could be performed for $\oN^R$ tensor models with enhanced scaling, at least for some interaction bubbles such as the tetrahedric interaction $\cK_4$. We note that such an analysis might be more involved than the one in \cite{GurauSchaeffer,FusyTanasa} because the perturbative expansion includes more Feynman graphs than both MO tensor models and $\uN^R$ tensor models with BGR scaling. Besides, we remark that the subleading contributions for the $\oN^3$ tensor models with tetrahedric interaction has already been investigated in \cite{BonzomON} up to next-to-next-to-next-to-leading order. 

An important motivation for the analysis of the subleading contributions is the implementation of a double scaling limit for $\oN^R$ tensor models with enhanced scaling, in the same spirit as in matrix models \cite{DiFran}, $\uN^R$ tensor models with BGR scaling \cite{GurauSchaeffer,DSTM} and MO tensor models \cite{DSMOTM}. The idea of the double scaling limit is to take, in a correlated way, both the large $N$ limit and the limit to criticality so that Feynman graphs of arbitrary order in the large $N$ expansion can contribute. One can then study the summability of the double scaling series, check whether new universality classes can be achieved and possibly reiterate the double scaling procedure. In the case of matrix models, the double scaling series corresponds to a sum over all genera, which is divergent. In contrast, the double scaling series of $\uN^R$ tensor models with BGR scaling is convergent for $R\leq5$ while it becomes divergent for $R\geq6$, as in matrix models \cite{GurauSchaeffer,DSTM}. Finally, the case of MO orientable tensor models is similar to the previous one for $R=3$. In this context, an interesting question for the doubling scaling limit of $\oN^R$ tensor models with enhanced scaling is whether the double scaling series has a behavior similar to $\uN^R$ tensor models with BGR scaling, and whether genuinely new universality classes can be achieved. 

Another important motivation for understanding the structure of the subleading contributions is the connection between random tensor models and the SYK physics. As explained in the introduction, even though the leading order Feynman graphs of the SYK model and the ones of the SYK-like tensor models are similar, there is a clear distinction between the subleading contributions. Indeed, the SYK model remains a vector model in disguise whereas the SYK-like tensor models correspond to genuine tensor models. In particular, the subleading contributions are dual to random discretized geometries in dimension $R$. The difference between the subleading contributions was first pointed out in \cite{SYKCorr}; however, the corresponding physics still needs further investigations.

\subsubsection{Case of reduced symmetry}

The results of this chapter and most of the literature on tensor models rely on the fact that each index of the tensors is associated with a distinct symmetry group and thus a distinct color. This property is lost if some symmetry properties on the indices are imposed, as explained in Section \ref{sec:MatRedSym} for matrix models. In particular, the colored representation is not available anymore, as well as the results on the degree and the index. The situation for tensor models with reduced symmetry is more complicated than the one for matrix models. It has recently attracted some particular attention.

To our knowledge, the first instance of reduced symmetry was considered in \cite{FrankLargeD} for the planar limit of matrix-tensor models based on Hermitian matrices, for which the $\text{U}(N)^{2}$ symmetry associated with the two indices of the matrices is broken down to a single $\text{U}(N)$. In particular, the result holds for any interaction term included in the action.

It was later conjectured in \cite{KlebaReduced}, based on numerical computations, that tensor models with tetrahedric interaction and with totally symmetric tensors of rank three also admit a well-defined $1/N$ expansion, with the important additional requirement that the tensors must be traceless. Without the tracelessness condition, it is in fact straightforward to construct Feynman graphs whose power of $N$ is not bounded from above. They are obtained by recursively inserting an arbitrary number of tadpoles (see Ref.~\cite{Ref1}). Interestingly, these problematic graphs all disappear when one imposes the tracelessness condition, because of non-trivial cancellations between Feynman graphs. We remark that the conjecture of \cite{KlebaReduced} also applies to tensor models with totally antisymmetric tensors, in which case the undesired tadpoles do not appear in the Feynman perturbative expansion.

Another way of eliminating the undesired tadpoles was then put forward in \cite{GurauReducedBip}, who considers a bipartite model with the complete interaction $\cK_{R+1}$ and two symmetric tensors of rank $R$. In this case, the Feynman graphs are bipartite and therefore do not allow for self-contractions. It was then proved that such models support a well-defined $1/N$ expansion. The combinatorial approach of \cite{GurauReducedBip}, based on stranded Feynman graphs, then allowed for a full proof in \cite{ReducedRank3} of the above conjecture for both totally symmetric traceless and totally antisymmetric tensors of rank three. This proof relies on a detailed but tedious case-by-case analysis of the Feynman graphs. One of the key ingredients of \cite{ReducedRank3} is that the tensors are in an irreducible representation of the symmetry group $\oN$. Interestingly, the generalized melonic graphs still dominate the expansion at large $N$. Finally, another irreducible representation of $\oN$ at rank three, based on tensors with mixed permutation symmetry, was solved in \cite{ReducedRank3Mixed}.

The extension of the above results to tensors of rank $R>3$ is an exciting research avenue. A full understanding of a method that avoids a case-by-case analysis is still missing, partly because of the cancellations that result from the tracelessness condition, which do not have a straightforward interpretation in terms of graphs.

%%%%%%%%%%%%%%%%%%%%%%%%%%%%%%%%%%%%%%%%%%%%%%%%
%%%%%%%%%%%%%%%%%%%  Chapter 5  %%%%%%%%%%%%%%%%%%%%%%%%%
%%%%%%%%%%%%%%%%%%%%%%%%%%%%%%%%%%%%%%%%%%%%%%%%

\chapter{Large $N$ and large $D$ matrix-tensor models}
\label{chap:MatrixTensor}

In the matrix models of Chapter \ref{chap:Matrix} and in the tensor models of Chapter \ref{chap:Tensor}, we assumed that the independent indices of the matrix and the tensor have the same size $N$, that is, their indices correspond to spaces with the same dimension $N$. Of course, this is not required in general. In fact, the study of random matrices started with Wishart's theory of random rectangular matrices. On the other hand, it was also noticed in \cite{BGR} that the indices of the tensor can have different sizes. In the case of tensor models, the symmetry group $\oN^R$ becomes the direct product of $R$ orthogonal groups of different sizes, $\text{O}(N_i)$ with $i=1,2,\ldots, R$, and the field corresponds to a ``rectangular" real tensor of rank $R$, $T_{a_1a_2\cdots a_R}$, of size $N_1\times N_2\times \cdots \times N_R$, where $N_i$ is the range of the index $a_i$. Many ingredients of the previous chapter for ``hypercubic" tensors, i.e.\ tensors with $N_i=N$ for all $i$, can be applied to the case of rectangular ones, such as the transformation law and the bijection between trace invariants and $R$-bubbles. A new important feature is however the presence of $R$ parameters $N_i$ instead of a single one $N$. In particular, it means that one can take several large $N_i$ limits and check if they lead to well-defined expansions. 

The first instance of interesting expansions for rectangular tensors was in \cite{BonzomNewLargeExp}, where the author scales the coupling constants differently with respect to the parameters $N_i$. It is then shown that the models admit well-defined large $N_i$ expansions, whose leading sector is larger than the class of melons.

An interesting application of rectangular tensor models consists in rewriting the tensor $T$ in terms of a $N\times N$ matrix $X$ with additional indices having the same size $D$. In other words, we single out two indices of $T$ out of the $R=r+2$, $(a_1a_2\cdots a_R)=(a_1a_2\mu_1\mu_2\cdots\mu_r)$, and we rewrite $T$ as
\be \label{TensorMatrixTensor}
T_{a_1a_2\cdots a_R} = (X_{a_1a_2})_{\mu_1\mu_2\cdots\mu_r} \, ,
\ee
with $a_1,a_2=1,2,\ldots, N$ and $\mu_i=1,2\ldots, D$ for $i=1,2,\ldots, r$. The field $X$ can then be interpreted as a special case of rectangular tensor of rank $R=r+2$ or as a set of $D^r$ matrices. This simple rewriting builds an appealing connection between matrix models and tensor models. Results from one type of models may in principle be translated into the other type. The natural symmetry group associated with \eqref{TensorMatrixTensor} is $\smash{\oN^2\times\oD^r}$, or similar groups (see below). This class of models were studied in Ref.\ \cite{Ref2} under the name of matrix-tensor models. Of course, one can always consider the case $D=N$, for which the theory reduces to that of a hypercubic tensor of rank $R=r+2$. However, as explained below, some non-trivial aspects of matrix-tensor models require a hierarchy between the two parameters $N$ and $D$. 

The program of connecting matrix models and tensor models was first initiated in \cite{TensorMatrix1,TensorMatrix2}. In \cite{TensorMatrix1}, models similar to matrix-tensor models with $R=3$ and unitary groups are studied. From the point of view of matrix models, the perturbative expansion of these models can be nicely interpreted in terms of loops drawn on random surfaces. More precisely, when $N\rightarrow\infty$, the perturbative expansion can be rearranged as a $1/N$ expansion onto orientable discretized surfaces, indexed by the genus (see Chapter \ref{chap:Matrix}). Then, when $D\rightarrow\infty$, the perturbative expansion further admits a $1/D$ expansion governed by the number of oriented loops drawn on these discretized surfaces. From the point of view of tensor models, the perturbative expansion corresponds to a sum over $4$-bubbles, as explained in Chapter \ref{chap:Tensor}. To define the large $D$ limit, the authors use an equivalent of the BGR scaling. Then, they show using the combinatorial techniques of tensor models that the number of loops is directly related to the degree of the $4$-bubbles, which provides a new combinatorial interpretation of the degree. In particular, the leading sector is given by the usual melonic graphs, which correspond to the configurations that maximize the number of loops.

One of the main motivations for matrix-tensor models is to find new interesting limits for matrix models using the technology of tensor models. Indeed, by taking the large $N$ limit at fixed $D$, one obtains the usual sum over planar graphs of matrix models. In the most interesting cases for physics, including matrix QFTs in dimension $d>0$, computing the sum over planar graphs remains a difficult technical challenge. But with the additional parameter $D$ at hand, one can now study the large $D$ limit of the planar graphs. 

This program was initiated in \cite{FrankLargeD} in the case of $r=1$ and was extended in Ref.~\cite{Ref1}. The models studied in these references are based on a set of $D$ matrices of size $N\times N$. These matrices can be complex, in which case the symmetry group is $\uN^2\times\oD$, or Hermitian, in which case the symmetry group reduces to $\uN\times\oD$, where the Hermitian matrices transform in the adjoint representation of $\uN$. As explained in the introduction of this thesis, such models have a natural interpretation in the context of $D$-branes and string theory. An important feature of these models is that the large $D$ limit is of different nature whether one uses an equivalent of the BGR scaling, as in \cite{TensorMatrix1}, or an equivalent of the enhanced scaling. With the BGR scaling, one can show that the large $N$ and large $D$ limits commute, whereas with the enhanced scaling, it is essential to take $N\rightarrow\infty$ first and $D\rightarrow\infty$ second for the limits to make sense. Thus, in the case of the enhanced scaling, there is a natural hierarchy between the two parameters $N$ and $D$ and one typically assumes $N>>D$. This point also applies in the general case of matrix-tensors of rank $R=r+2$ and turns out to have important physical consequences.

A simple intuitive explanation of why the difference between the two scalings is crucial for physics can be given in the case of $r=1$, i.e.\ for vectors of matrices \cite{FrankLargeD,Ref1}. If the large $N$ and large $D$ limits commute, one can take the limit $D\rightarrow\infty$ at fixed $N$ first, which corresponds to a standard large $D$ limit of vector models (see Chapter \ref{chap:Vector}). This limit selects a very restrictive class of Feynman graphs--that is, the bubble graphs--and the resulting physics is the same as in standard vector models. This is true already at fixed $N$ and remains of course true when $N$ goes to infinity, which eliminates even more Feynman graphs. However, vector models are much simpler than matrix models. The large $D$ approximation of the planar graphs obtained in this way is thus bound to be a very poor approximation and it does not reproduce the most crucial physical properties of the full sum over planar graphs. 

On the other hand, the situation is very different with the enhanced scaling, for which large $N$ and large $D$ do not commute \cite{FrankLargeD}. The large $D$ expansion with enhanced scaling is totally different from the large $D$ expansion of vector models, because it includes a much wider class of Feynman graphs. The remarkable point, emphasized in \cite{FrankLargeD}, is that the main physical properties expected for the full sum over planar graphs seem to be captured already at leading order. This property highlights the importance of the enhanced scaling. It also provides another perspective on the deep relationship between matrix and tensor models and a new and reliable way to study physically relevant planar matrix models which were thought to be intractable before.

In this context, the objective of this chapter is to describe the large $N$ and large $D$ limits of general matrix-tensor models with enhanced scaling, as derived in Ref.\ \cite{Ref2}. We use real matrix-tensors with symmetry group $\text{O}(N)^{2}\times\text{O}(D)^{r}$, since this is the most general case. Indeed, as explained in the previous chapters, the use of other symmetry groups, like unitary groups, amounts to consider a complex matrix-tensor $X$ and its conjugate $X^\dagger$. The corresponding study is then typically easier since the Feynman graphs in the complex case are bipartite and therefore consist in a strict subset of the ones in the real case. Besides, as usual, the large $N$ and large $D$ expansions do not depend on the dimension $d$ of spacetime nor on the statistics of the fields. Unless otherwise stated, we thus set ourselves in $d=0$ with a real, bosonic matrix-tensor. 

Many parts of this chapter correspond to (slightly modified) parts of Ref.\ \cite{Ref2}. The results are directly based on the graph-theoretic tools developed in Chapter \ref{chap:Tensor} for tensor models, which also apply to matrix-tensor models. We emphasize that there are important differences, both conceptual and technical, between matrix-tensor models and tensor models. First, the presence of two parameters $N$ and $D$ in matrix-tensor models makes the connection with higher dimensional discretized geometries less direct. Second, matrix-tensor models with $r=1$ have a natural interpretation in the context of $D$-branes and string theory, whereas it is difficult to find a similar interpretation for tensors of rank three or higher. Third, the large $D$ expansion of matrix-tensor models does not coincide with the large $N$ expansion of tensor models because it is made at fixed genus. This being said, the results for matrix-tensor models naturally reduce to the ones for tensor models by setting $N=D$.

The results presented in this chapter connects with the ongoing effort to understand quantum models of black holes via holography, following ideas first put forwards in the SYK model. This model is non-standard because it uses quenched disorder, but it was pointed out that an ordinary quantum mechanics based on tensor models shares the same basic properties. It was then realized in \cite{FrankLargeD} that the basic structure of the Feynman graphs responsible for the properties of the SYK model is also relevant to planar matrix quantum mechanics, through the large $D$ limit for $r=1$ explained above. This makes the link with holography and string theory clearer, since planar matrix models are singled out in this framework, the two indices of the matrices being the Chan-Paton factors associated with the two end points of open strings. The presence of the additional $\text{O}(D)$ index on the matrices corresponds to the rotation group transverse to $D$-branes. The limit $D\rightarrow\infty$ is then physically similar to the large dimension limit of gravity. The results of this chapter provide a general framework to build a large class of solvable models with relevant properties to describe quantum black holes which have many interesting properties yet to be discovered.

We start this chapter by describing the $\oN^2\times\oD^r$ matrix-tensor models and the corresponding invariant interactions. We then explain the stranded and colored representations for the Feynman graphs. Next, we define a large $N$ and two large $D$ scalings for the coupling constants, which allow for well-defined large $N$ and large $D$ expansions. The two large $D$ scalings are the equivalents of the BGR and the enhanced scalings for tensor models. We then focus on the large $D$ expansion for the enhanced scaling and we discuss its properties. As an application, we study two types of quantum mechanical matrix-tensor models based on the complete interaction, which reproduce in the large $N$ and large $D$ limit the SYK/black hole physics. Finally, we conclude the chapter with open questions and perspectives.

\section{Definition of the models}
\label{sec:MatTensDefModels}

The basic degrees of freedom of matrix-tensor models are a set of $D^r$ real matrices $X_{\mu_1\mu_2\cdots\mu_r}=(X_{a_1a_2})_{\mu_1\mu_2\cdots\mu_r}$ of size $N\times N$, where $a_1,a_2=1,2,\ldots,N$ and $\mu_i=1,2,\ldots,D$ for $i=1,2\ldots,r$. The symmetry group is $\oN^2\times\oD^r$, whose action is defined by the following transformation law
\be \label{MatTensTransfLaw}
(X_{a_1a_2})_{\mu_1\mu_2\cdots \mu_r} \rightarrow (X'_{a_1a_2})_{\mu_1\mu_2\cdots \mu_r} = O^{(1)}_{a_1a_1'}O^{(2)}_{a_2a_2'}O^{(3)}_{\mu_1\mu_1'}\ldots O^{(r+2)}_{\mu_r\mu_r'}(X_{a_1'a_2'})_{\mu_1'\mu_2'\cdots \mu_r'} \, ,
\ee
where $O^{(1)}, O^{(2)}$ and $O^{(i)}, i=1,2,\ldots,r$, are independent $N\times N$ and $D\times D$ orthogonal matrices respectively. The number of degrees of freedom is $\cN=N^2D^r$. Similarly to the tensor models of Chapter \ref{chap:Tensor}, each orthogonal group acts independently on its corresponding index; hence, the $R$ indices are distinguishable.

The invariants of matrix-tensor models can be constructed in a similar way as in tensor models. Their interpretation is however different because we now have matrix indices; therefore, there is an associated notion of trace. An invariant thus corresponds to a product of traces with respect to the matrix indices $a_1, a_2$ such that the $\oD$ indices transforming under the same group are all contracted two by two. We denote by $\cI_a(X)$ an invariant, where $a$ is some label in a discrete set $\cS$, by $t_a$ the number of traces in $\cI_a(X)$ and by $s_a$ the number of entries $X$ in $\cI_a(X)$. An invariant of matrix-tensor models can thus be formally written as
\be \label{MatTensInteraction}
\cI_a(X) = \prod_{i=1}^{t_a} \tr (X_{\mu_{1,i}\cdots \mu_{r,i}}\tra{X}_{\nu_{1,i}\cdots \nu_{r,i}} \cdots X_{\rho_{1,i}\cdots \rho_{r,i}}\tra{X}_{\sigma_{1,i}\cdots \sigma_{r,i}})\, ,
\ee
where the $\oD$ indices are properly contracted pairwise and summed over. Note that $\oD$ indices belonging to distinct traces can be contracted together. Also, the number $s_a$ of entries is always even, in agreement with the transformation law \eqref{MatTensTransfLaw}. Like in tensor models, the invariants are in bijection with $R$-bubbles $\cB_a$, with $R=r+2$. We thus write $\cI_a(X)\equiv\cI_{\cB_a}(X)$ and $t_a\equiv t(\cB_a)$. Examples of invariants include:
\begin{itemize}
\item $X\cdot X \equiv \tr (X_{\mu_1\mu_2\cdots\mu_r}\tra{X}_{\mu_1\mu_2\cdots\mu_r})=(X_{a_1a_2})_{\mu_1\mu_2\cdots \mu_r}(X_{a_1a_2})_{\mu_1\mu_2\cdots \mu_r}$ ($t=1$, $s=2$);
\item $\cI_{\cB_1} = \tr (X_\mu \tra{X}_\mu X_\nu \tra{X}_\nu) = (X_{a_1a_2})_\mu (X_{b_1a_2})_\mu (X_{b_1b_2})_\nu (X_{a_1b_2})_\nu$ ($r=1$, $t_1=1$, $s_1=4$);
\item $\cI_{\cB_2} = \tr (X_\mu \tra{X}_\nu X_\mu \tra{X}_\nu) = (X_{a_1a_2})_\mu (X_{b_1a_2})_\nu (X_{b_1b_2})_\mu (X_{a_1b_2})_\nu$ ($r=1$, $t_2=1$, $s_2=4$);
\item \small{ $\cI_{\cB_3} = \tr (X_\mu \tra{X}_\mu) \tr (X_\nu\tra{X}_\nu) = (X_{a_1a_2})_\mu (X_{a_1a_2})_\mu (X_{b_1b_2})_\nu (X_{b_1b_2})_\nu$ ($r=1$, $t_3=2$, $s_3=4$);}
\item \small{$\cI_{\cB_4} = \tr (X_{\mu_1\mu_2} \tra{X}_{\mu_1\mu_2} X_{\nu_1\nu_2} \tra{X}_{\nu_1\nu_2}) = (X_{a_1a_2})_{\mu_1\mu_2} (X_{b_1a_2})_{\mu_1\mu_2} (X_{b_1b_2})_{\nu_1\nu_2} (X_{a_1b_2})_{\nu_1\nu_2}$ ($r=2$, $t_4=1$, $s_4=4$);}
\item etc.
\end{itemize}
In these examples, the first invariant corresponds to the mass term in the action whereas the other invariants correspond to the ones in Section \ref{sec:TensDefModels} translated into the language of matrix-tensors.

An invariant action $S(X)$ for the matrix-tensor models can be written as
\be\label{MatTensAction}
S(X)= \frac{ND^r}{2} X \cdot X + \sum_{a\in\cS} g_a \, \cI_{\cB_a}(X) = \frac{ND^r}{2} X \cdot X + V(X) \, ,
\ee
where $g_a$ is the coupling constant associated with $\cI_{\cB_a}(X)$. The factor of $ND^r$ in front of the mass term is set for further convenience. Note that in the case $N=D$, the action reduces to the one \eqref{TensAction} of tensor models. 

The partition function $Z$ and the free energy $F$ associated with the action \eqref{MatTensAction} are given by the usual expressions
\be \label{MatTensPartitionFunctionFull}
Z = \exp (-F) = \int [dX] \, e^{-S(X)}=\int [dX] \, e^{-\frac{ND^r}{2} X \cdot X - \sum_{a\in\cS} g_a \, \cI_{\cB_a}(X)} \, ,
\ee
where the path integral measure is given in $d=0$ by the product of $N^2D^r$ simple integral measures on $\mathbb{R}$. As usual, similar expressions hold for general $n$-point functions and the free $2$-point function or free propagator is given by
\be \label{MatTensFree2PtFunction}
\bigl\langle(X_{a_1a_2})_{\mu_1\mu_2\cdots \mu_r}(X_{b_1b_2})_{\nu_1\nu_2\cdots \nu_r}\bigr\rangle_0\,=\frac{1}{ND^r}\delta_{a_1b_1}\delta_{a_2b_2}\delta_{\mu_1\nu_1}\cdots \delta_{\mu_r \nu_r} \, .
\ee

We now define a large $N$ and a large $D$ expansions for matrix-tensor models. To do so, we go to perturbation theory and study the resulting Feynman graphs.

\section{Feynman graphs}
\label{sec:MatTensFeynmanGraphs}

We focus on the perturbative expansion of the free energy $F$ onto connected Feynman graphs. The Feynman rules for matrix-tensor models are found in the usual way. We denote by $\cG$ a connected Feynman graph appearing in the perturbative expansion of $F$ and we write its amplitude $\cA(\cG)$, which depends on $N$ and $D$. Like in the previous chapter, it is useful to describe the Feynman graphs using the stranded or the colored representations. These representations for matrix-tensor models are the same as for tensor models. The only difference is their interpretation, which we highlight in this section. 

\subsection{Stranded representation}
\label{sec:MatTensStrandRep}

\begin{figure}[]
\centerline{\includegraphics[scale=1]{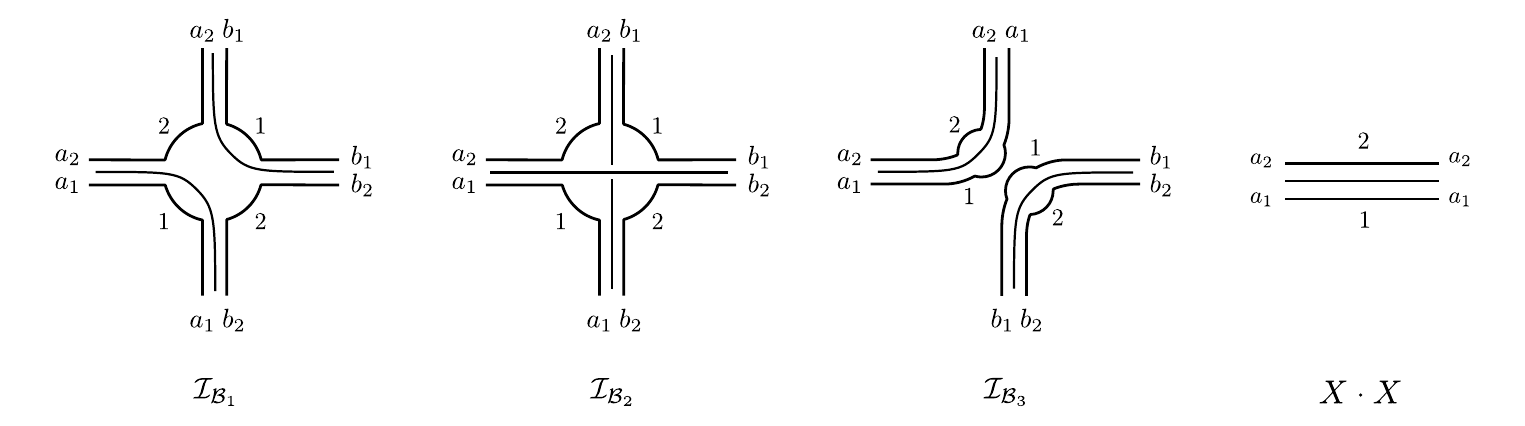}}
\caption{Stranded representation for some vertices and the propagator of matrix-tensor models in the special case $r=1$. They can be interpreted as being ribbon graph vertices and edge with an additional internal line corresponding to the $\oD$ index. The resulting Feynman graph thus describe loops, made by connecting internal lines together, drawn on discretized surfaces.}
\label{MatTensFeynRulesStrand}
\end{figure}

Like in tensor models, the field components $(X_{a_1a_2})_{\mu_1\mu_2\cdots \mu_r}$ have $R=r+2$ distinguished indices, which we label with $R$ distinct colors. We use again the color $1$ for the first index, the color $2$ for the second index, etc. In the case of matrix-tensor models, the colors $1$ and $2$ have a ``special" status because they correspond to the matrix indices of $X$. To emphasize this, we draw the interaction vertices and the propagator like in a usual ribbon graph (see Section \ref{sec:MatStrandRep}), whose strands correspond to the matrix indices, supplemented by additional ``internal" strands for the additional $\oD$ indices.\footnote{Note that these ``internal" strands may connect distinct ribbon vertices at a given interaction vertex.} This is illustrated in Figure \ref{MatTensFeynRulesStrand} for the interaction terms $\cI_{\cB_1}$, $\cI_{\cB_2}$ and $\cI_{\cB_3}$ defined above and for the propagator. The resulting Feynman graphs can then be interpreted as ribbon graphs with internal loops made by connecting the internal strands together. This makes the link between matrix-tensor models and loops drawn on discretized surfaces. 

Given a connected Feynman graph $\cG$ in the stranded representation, we denote by $p$ its number of propagators, by $v$ its number of interaction vertices, by $f$ its number of closed stranded loops made of $\uN$ matrix indices (which correspond to the number of faces of the matrix ribbon graph), by $\varphi$ its number of closed stranded loops made of $\oD$ indices and by $\cS_\cG$ the discrete set that accounts for the different types of interaction vertices in $\cG$. As usual, $\cG$ may effectively correspond to a disconnected stranded graph, denoted as $\tilde{\cG}$, because nothing restricts an interaction vertex to be effectively disconnected (see $\cI_{\cB_3}$ in Figure \ref{TensFeynRulesStrand}). We denote by $\tilde{v}$ the number of effective ``connected'' interaction vertices in $\tilde{\cG}$.

Even though the stranded representation of matrix-tensor models makes an interesting connection with matrix models ribbon graphs, it is more practical to work with the colored representation and the corresponding graph-theoretical tools.

\subsection{Colored representation}
\label{sec:MatTensColorRep}

\begin{figure}[]
\centerline{\includegraphics[scale=1]{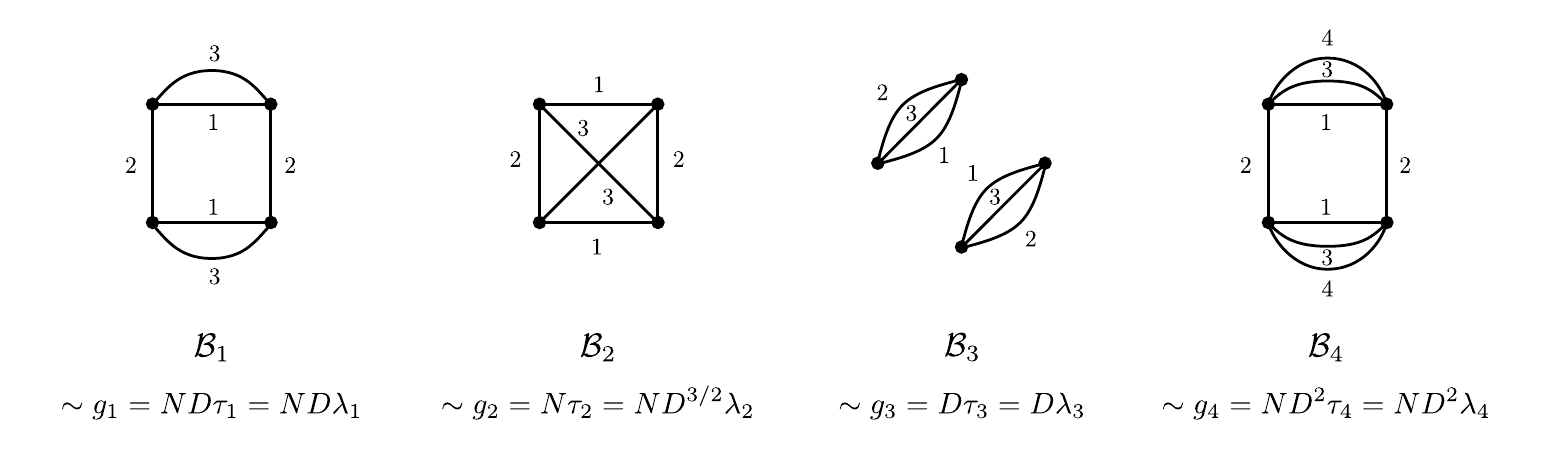}}
\caption{Interaction bubbles of Figure \ref{TensFeynRulesColor} for matrix-tensor models. The coupling parameters $\tau_a$ for the BGR scaling are defined in Eq.\ \eqref{MatTensBGRScaling} and the coupling parameters $\lambda_a$ for the enhanced scaling in Eq.\ \eqref{MatTensEnScaling}.}\label{MatTensFeynRulesColor}
\end{figure}

The colored representation of matrix-tensor models is constructed in the same way as for tensor models. The interaction terms $\cI_{\cB_a}$ entering the action \eqref{MatTensAction} are in bijection with $R$-bubbles $\cB_a$. The $R$-bubbles associated with $\cI_{\cB_1}$, $\cI_{\cB_2}$, $\cI_{\cB_3}$ and $\cI_{\cB_4}$ are presented in Figure \ref{MatTensFeynRulesColor}. As expected, they look the same as in Figure \ref{TensFeynRulesColor} but the scalings of the coupling constants are different (see below). Remark that if we restrict the interaction bubbles to the edges of colors $1$ and $2$ only, we obtain interaction bubbles for the associated $\oN^2$ matrix models. In particular, the number $t(\cB_a)$ of traces in $\cI_{\cB_a}$ corresponds to the number of $(12)$-faces in the $R$-bubble $\cB_a$. 

Each interaction $R$-bubble $\cB_a$ is thus characterized by:
\begin{itemize}
\item its number of vertices $V(\cB_a)=s_a$;
\item its number of traces $t(\cB_a)=F_{12}(\cB_a)$;
\item its number of connected components $c(\cB_a)$, which satisfies $t(\cB_a)\geq c(\cB_a)$;
\item its degree $\deg \cB_a$.
\end{itemize}

A particular case of interest is given by models with only single-trace interactions, for which $t(\gB_{a})=1$ and thus $c(\gB_{a})=1$. In the following, we denote the colors $1$ to $R$ by lower case latin indices $i$, $j$, etc., and the $r$ colors $3$ to $R$ by upper case latin indices $I$, $J$, etc. 

Regarding the Feynman graphs, like in tensor models, they are mapped to $(R+1)$-bubbles, the color 0 being associated with the propagators. In particular, the Feynman graphs of Figure \ref{TensFeynGraphsColor} remain valid in the context of matrix-tensor models. We present them again in Figure \ref{MatTensFeynGraphsColor} with the correct scalings with respect to $N$ and $D$ (see below). 
\begin{figure}[]
\centerline{\includegraphics[scale=1]{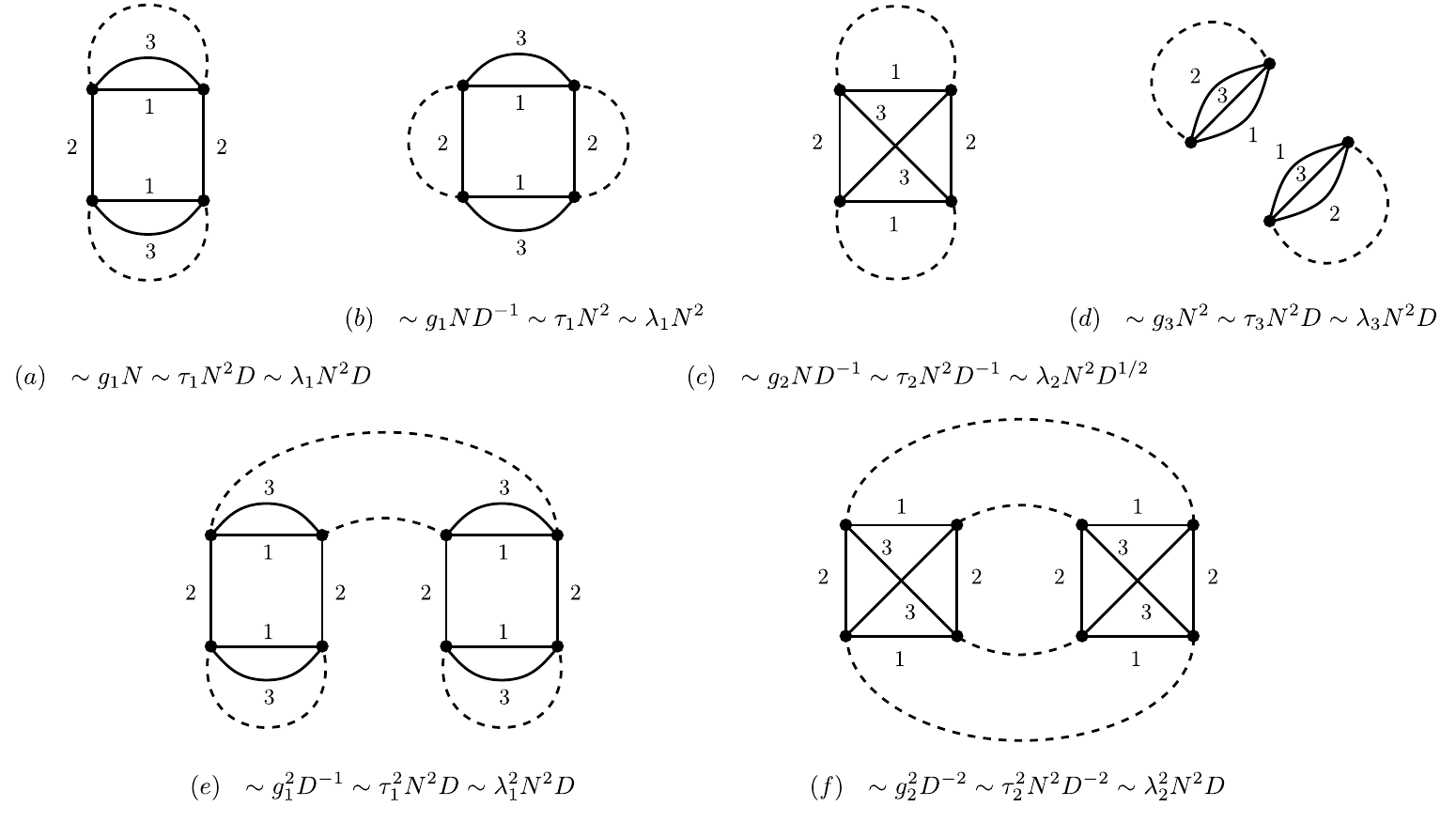}}
\caption{Feynman graphs of Figure \ref{TensFeynGraphsColor} for matrix-tensor models. The coupling parameters $\tau_a$ (BGR scaling) are defined in Eq.\ \eqref{MatTensBGRScaling} and the coupling parameters $\lambda_a$ (enhanced scaling) in Eq.\ \eqref{MatTensEnScaling}. LO Feynman graphs scale like $N^2D$.}\label{MatTensFeynGraphsColor}
\end{figure}

We use the notation $\cB$ to denote a given connected Feynman graph $\cG$ in the colored representation. Most of the relations \eqref{TensFeynGraphData} between the data associated with the stranded and the colored representations remain valid for matrix-tensor models. Only the ones associated with the faces of the bubbles must be adapted:
\be\label{MatTensFeynGraphData}
f=F_{01}(\cB)+F_{02}(\cB)\, , \quad \varphi=\sum_{I} F_{0I}(\cB)\, , \quad  \sum_{a\in\cS_\cG}t(\cB_a) = F_{12}(\cB) \, .
\ee

\section{Large $N$ and large $D$ expansions}
\label{sec:MatTensLargeNDExp}

In the context of matrix-tensor models, we need to prescribe how the coupling constants $g_a$ in the action \eqref{MatTensAction} scale with both $N$ and $D$ to define a large $N$ and a large $D$ limits. In particular, we want to obtain well-defined and non-trivial expansions when both $N\rightarrow\infty$ and $D\rightarrow\infty$. In this section, we first review the large $N$ scaling directly imported from matrix models. Then, we introduce the equivalents of the BGR scaling (see Section \ref{sec:TensLargeNExp1}) and the enhanced scaling (see Section \ref{sec:TensLargeNExp2}) for the large $D$ scaling. Finally, we focus on the enhanced scaling and we derive the associated large $N$ and large $D$ expansions. This part corresponds to Sections 3.2 and 3.3 of Ref.\ \cite{Ref2}. 

\subsection{Large $N$ and large $D$ scalings}
\label{sec:MatTensScalings}

\subsubsection{Large $N$ 't Hooft scaling}

Because the large $N$ limit of matrix-tensor models corresponds to the large $N$ limit of $\oN^2$ matrix models, we use 't Hooft scaling for the large $N$ scaling of the coupling constants (see Eq.\ \eqref{MatOptScaling}). In other words, we introduce new coupling parameters $\mu_a$ related to the couplings $g_a$ as
\be\label{MatTensLargeNScaling} 
g_a=N^{2-t(\cB_a)}\mu_{a} \, ,
\ee
and we keep the $\mu_a$'s fixed when $N\rightarrow\infty$. In this way, we reproduce the usual large $N$ limit of matrix models.

\subsubsection{Large $D$ BGR scaling}

The BGR scaling of tensor models can be generalized to matrix-tensor models as follows. We define new coupling parameters $\tau_a$ in terms of the couplings $\mu_a$ by
\be\label{MatTensBGRScaling} 
\mu_{a}=D^{r+t(\gB_{a})-c(\gB_{a})-\frac{2}{r!}\deg \gB_{a}}\tau_{a}\, .
\ee
Then, the limits $N\rightarrow\infty$ and $D\rightarrow\infty$ are formulated by keeping the $\tau_{a}$'s fixed. The large $D$ BGR scaling is illustrated in Figure \ref{MatTensFeynRulesColor}. The action \eqref{MatTensAction} rewrites in terms of the couplings $\tau_a$ as 
\be\label{MatTensActionBGR}
S(X)= ND^r \biggl(\frac{1}{2} X \cdot X + \sum_{a\in\cS} N^{1-t(\cB_a)}D^{t(\gB_{a})-c(\gB_{a})-\frac{2}{r!}\deg \gB_{a}} \tau_a \, \cI_{\cB_a}(X) \biggr) \, .
\ee
One can show (see below) that the limits $N\rightarrow\infty$ and $D\rightarrow\infty$ commute in this case. When $D=N$, the large $N$ and large $D$ scalings reduce together to the BGR scaling \eqref{BGRScaling}. One then obtains the $1/N$ expansion described in Section \ref{sec:TensLargeN1}, which is organized according to the degree of the Feynman graphs. The leading order Feynman graphs are the melons, which have degree zero. 

As emphasized in the previous chapter, the class of melons is quite restricted. In particular, only melonic interaction bubbles can contribute to the leading sector. From the point of view of matrix-tensor models, the scarcity of the melonic family implies that the BGR scaling \eqref{MatTensBGRScaling} yields a physics similar to the large $D$ limit of vector models. This is why enhancing non-melonic interactions is important, so that the large $D$ limit includes many more Feynman graphs and hopefully captures the essential aspects of the full sum over planar graphs. A natural candidate is the enhanced scaling introduced in the previous chapter. Indeed, it is optimal for a large class of interactions and in particular for the tetrahedric interaction, which plays a crucial role in reproducing the SYK/black hole physics.

\subsubsection{Large $D$ enhanced scaling}

The enhanced scaling of tensor models can be generalized to matrix-tensor models in a similar way. We introduce new couplings parameters $\lambda_a$ as 
\be\label{MatTensEnScaling} 
\mu_{a}= D^{r+t(\gB_{a})-c(\gB_{a})+\frac{2}{(r+1)!}\deg \gB_{a}}\lambda_{a}\, ,
\ee
and we define the associated large $N$ and large $D$ limits by keeping the $\lambda_a$'s fixed. Examples of enhanced scalings are given in Figure \ref{MatTensFeynRulesColor}. In terms of the enhanced scaling, the action \eqref{MatTensAction} rewrites as
\be\label{MatTensActionEn}
S(X)= ND^r \biggl(\frac{1}{2} X \cdot X + \sum_{a\in\cS} N^{1-t(\cB_a)}D^{t(\gB_{a})-c(\gB_{a})+\frac{2}{(r+1)!}\deg \gB_{a}} \lambda_a \, \cI_{\cB_a}(X) \biggr) \, .
\ee
Note that
\be\label{BGREnScalings} 
\tau_{a} = D^{\frac{2(r+2)}{(r+1)!}\deg\gB_{a}}\lambda_{a}\, ,
\ee
which means that similarly to the case of tensor models, taking $D\rightarrow\infty$ at fixed $\lambda_{a}$ amounts to an infinite amplification of all the non-melonic ($\deg\gB_{a}>0$) interactions with respect to the BGR scaling. In spite of this enhancement, the large $N$ and large $D$ expansions are both still well-defined, with the important feature that the two limits no longer commute (see below). One should first take $N\rightarrow\infty$ and then $D\rightarrow\infty$ at each order in the $1/N$ expansion. Moreover, the large $D$ expansion is governed by the index of the Feynman graphs, as expected from the results for tensor models.

\subsection{Case of the enhanced scaling}
\label{sec:MatTensScalingsExp}

The powers of $N$ and $D$ associated with a given connected Feynman graphs $\cG$ follows from the action \eqref{MatTensActionEn}. To each propagator is associated a factor $1/(ND^{r})$, to each interaction bubble of type $\cB_a$ a factor 
\begin{equation*}
N^{2-t(\gB_{a})}D^{r+t(\gB_{a})-c(\gB_{a})+\frac{2}{(r+1)!}\deg\gB_{a}} \, ,
\end{equation*}
to each closed stranded loop made of $\oN$ indices a factor $N$ and to each closed stranded loop made $\oD$ indices a factor $D$. Overall, the amplitude $\cA(\cG)$ of $\cG$ thus scales with $N$ and $D$ as
\be\label{MatTensFeynGraphScaling}
\cA(\cG) \sim N^{-p+\sum_{a\in\cS_\cG}\bigl(2-t(\cB_a)\bigr) +f} D^{-rp +\sum_{a\in\cS_\cG}\bigl(r+t(\cB_a)-c(\cB_a)+\frac{2}{(r+1)!}\deg\cB_a\bigr)+\varphi} \equiv N^{2-h}D^{r+h-\frac{\ell}{r+1}} \, ,
\ee
where we introduced the parameters $h$ and $\ell$ defined as
\be\label{MatTensHParam1}
\begin{split}
h & = 2+p-\sum_{a\in\cS_\cG}\Bigl(2-t(\cB_a)\Bigr)-f \, ,\\
	& = 2+p-2v+\sum_{a\in\cS_\cG}t(\cB_a) -f \, ,
\end{split}
\ee
and
\be\label{MatTensEllParam1}
\begin{split}
\frac{\ell}{r+1} & = r+h+rp-\sum_{a\in\cS_\cG}\Bigl(r+t(\cB_a)-c(\cB_a)+\frac{2}{(r+1)!}\deg\cB_a\Bigr)-\varphi \, ,\\
	& = r+2+(r+1)p-(r+2)v+\sum_{a\in\cS_\cG}\Bigl(c(\cB_a)-\frac{2}{(r+1)!}\deg\cB_a\Bigr)-f-\varphi \, .
\end{split}
\ee
The powers of $N$ and $D$ associated with the Feynman graphs of Figure \ref{MatTensFeynGraphsColor} are indicated in terms of the couplings $\lambda_a$ for the enhanced scaling. 

It is convenient to rewrite the parameters $h$ and $\ell$ with the colored representation data. Using the identities \eqref{TensFeynGraphData}, \eqref{IntBubbleDegree} and \eqref{MatTensFeynGraphData}, this yields
\be\label{MatTensHParam2}
h = 2+\frac{1}{2}V(\cB)-2v(\cB) + F_{12}(\cB) - F_{01}(\cB) - F_{02}(\cB) \, ,
\ee
and
\be\label{MatTensEllParam2}
\frac{\ell}{r+1} = r+2 +\frac{1}{2}(r+1)V(\cB)-(r+2)v(\cB)+B^{(0)} - \frac{2}{(r+1)!}\deg \cB^{(0)}- \sum_{i=1}^{r+2} F_{0i}   \, .
\ee

\subsubsection{Factor of $N$} 
\label{MatTensFactorN}

The factor of $N$ in Eq.\ \eqref{MatTensFeynGraphScaling} is similar to the one for $\oN^2$ matrix models, as expected. We can actually use the results of Chapter \ref{chap:Matrix} to prove that the large $N$ limit of matrix-tensor models is well-defined.

\begin{theorem}\label{MatTensTheoremLargeN}
Let $\cG$ be a connected Feynman graph of the matrix-tensor models with enhanced scaling and let $\cB$ be the corresponding $(R+1)$-bubble in the colored representation, with $R=r+2$. Then, $h\geq 0$. In other words, the power of $N$ associated with any connected Feynman graph $\cG$ is bounded above by $2$.
\end{theorem}

\proof Consider the $3$-bubble subgraph $\cB_{(012)}$ obtained from $\cB$ by deleting all the edges of colors $3$ to $r+2$. This $3$-bubble corresponds to the matrix part of $\cB$ and can be equivalently described as a ribbon graph. The expression \eqref{MatTensHParam2} for the parameter $h$ straightforwardly holds if we replace $\cB$ by the $3$-bubble $\cB_{(012)}$. Since $\cB_{(012)}$ is a $3$-bubble, we can then use the Euler's formula \eqref{EulerFormFeynGraphColor} to rewrite $h$ as
\be\label{MatTensHParam3}
h = 2g(\cB_{(012)}) + 2 \Bigl[1+ F_{12}(\cB_{(012)}) - v(\cB_{(012)}) - B_{(012)} \Bigr] \, ,
\ee
which is similar to the expression \eqref{MatHParam3} for the parameter $h$ of matrix models. Finally, by following the proof of Theorem \ref{MatTheorem}, we deduce that $h$ is non-negative and also $h\in\mathbb{N}$.\qed

\

\noindent\emph{Case of connected interaction bubbles}

A first particular case of interest is obtained by including in the models only connected interaction bubbles, that is, $c(\cB_a)=1 \ \forall a\in\cS$. Then, $B=1$ and $B^{(0)}=v(\cB)$ so that Eq.~\eqref{MatTensHParam3} takes the form
\be\label{MatTensHParam4}
h = 2g(\cB_{(012)}) + 2 \Bigl[1+ F_{12}(\cB_{(012)}) - B^{(0)} - B_{(012)} \Bigr] \, ,
\ee
Note that the second term on the RHS of this expression is of the form \eqref{Deltadec} for $\gG=\gB$, $p=r$ and the colors $(i_{1}\ldots i_{p})=(3\ldots r+2)$.

\

\noindent\emph{Case of single-trace interaction bubbles}

A second interesting case corresponds to models with only single-trace interactions, i.e.\ $t(\cB_a)=1 \ \forall a\in\cS$. Note that this case implies the previous one because $t(\cB_a)\geq c(\cB_a)$. It further implies $B_{(012)}=1$ so that Eq.\ \eqref{MatTensHParam3} reduces to
\be\label{MatTensHParam5}
h = 2g(\cB_{(012)}) \, .
\ee
This is consistent with the case of matrix models with single-trace interactions, see Eq.\ \eqref{MatHSingleTrace}. 

\subsubsection{Factor of $D$} 
\label{MatTensFactorD}

We now discuss the factor of $D$ in Eq.~\eqref{MatTensFeynGraphScaling} and prove that the large $D$ of matrix-tensor models is well-defined. We remark that the arguments are similar to the ones in the proofs for Theorems \ref{TensTheoremBGR} and \ref{TensTheoremEn}.

\begin{theorem}\label{MatTensTheoremLargeD}
Let $\cG$ be a connected Feynman graph of the matrix-tensor models with enhanced scaling and let $\cB$ be the corresponding $(R+1)$-bubble in the colored representation, with $R=r+2$. Then, $\ell\geq 0$. In other words, the power of $D$ associated with any connected Feynman graph $\cG$ is bounded above by $r+h$.
\end{theorem}

\proof We apply the formula \eqref{degreeformula} for the degree of the bubbles $\cB^{(0)}$ and $\cB$,
\begin{equation*}
(r+1)B^{(0)} - \frac{2}{r!}\deg\gB^{(0)} = F(\gB^{(0)}) - \frac{1}{4}r(r+1) V(\gB^{(0)})\, ,
\end{equation*}
\begin{equation*}
(r+2)B - \frac{2}{(r+1)!}\deg\gB = F(\gB) - \frac{1}{4}(r+1)(r+2) V(\gB) \, .
\end{equation*}
Subtracting these two equations, plugging the result into Eq.\ \eqref{MatTensEllParam2} and using the face decomposition formulas \eqref{facedec} as well as $V(\gB^{(0)})=V(\gB)$ then yields
\be\label{MatTensEllParam3}
\ell = 2\ind_0\cB + (r+1)(r+2) \Bigl[1+ \sum_{a\in\cS_\cG} (c(\cB_a)-1)-B\Bigr] \, .
\ee
This expression is similar to the one \eqref{TensEllParam3} for the parameter $\ell$ of tensor models with enhanced scaling, as expected. Then, by following the proof of Theorem \ref{TensTheoremEn}, we conclude that $\ell$ is non-negative and $\ell\in\mathbb{N}$.\qed

\

\noindent\emph{Case of connected interaction bubbles}

For connected interaction bubbles, the parameter $\ell$ reduces to twice the index of $\cB$:
\be\label{MatTensEllParam4}
\ell = 2\ind_0\cB \, .
\ee
In this case, the second form of the decomposition formula applies (see Proposition \ref{fundidth2}) so that the index of $\cB$ can be interpreted as the sum of the parameters $\frac{1}{2}h_{ij}$ that govern the $1/N$ expansions of all the possible $(ij)$ matrix models embedded in the matrix-tensor model. This interpretation is particularly relevant in the present context as singling out two distinct colors is equivalent to singling out two indices of a tensor to rewrite it in terms of matrices. 

Besides, the notion of maximally single-trace (MST) interaction bubbles, introduced at the end of Section \ref{sec:TensLargeNExpEn}, is also perfectly defined for matrix-tensor models. For this particular class of interactions, the index of a Feynman graph $\cB$ reduces to Eq.~\eqref{singfaceind}.

\subsubsection{Form of the expansions} 
\label{MatTensExp}

Theorems \ref{MatTensTheoremLargeN} and \ref{MatTensTheoremLargeD} imply that the free energy $F$ admits well-defined large $N$ and large $D$ expansions for the enhanced scaling \eqref{MatTensEnScaling}. More precisely, if we denote by $\{\cG_h\}$ the set of connected Feynman graphs of fixed $h$ and by $\{\cG_{h,\ell}\}$ the one for fixed $h$ and $\ell$, then we first expand $F$ at large $N$,
\be\label{MatTensFexp} 
F = \sum_{h\in\mathbb N} N^{2-h} F_h \, ,
\ee
where $F_h$ corresponds to a $N$-independent weighted sum over connected Feynman graphs $\cG$ in $\{\cG_h\}$. Then, we expand each $F_h$ at large $D$,
\be\label{MatTensFhexp} 
F_{h}=\sum_{\ell\in\mathbb N} D^{r+h-\frac{\ell}{r+1}} F_{h,\ell} \, ,
\ee
where $F_{h,\ell}$ corresponds to a $N$- and $D$-independent weighted sum over connected Feynman graphs $\cG$ in $\{\cG_{h,\ell}\}$.

One can observe that both the large $N$ and the large $D$ expansions are well-defined. For the latter, it follows from the fact that at fixed $h$, the power of $D$ associated with any Feynman graph is bounded above. Note that it crucially implies that the limits $N\rightarrow\infty$ and $D\rightarrow\infty$ do not commute, as first realized in \cite{FrankLargeD}. Indeed, at fixed $N$, the power of $D$ grows linearly with $h$ so that the limit $D\rightarrow\infty$ does not exist. One must always consider the limit $N\rightarrow\infty$ first and then the limit $D\rightarrow\infty$ next, at each order in the $1/N$ expansion.

On the other hand, the large $D$ expansion is governed by the integer $\ell$, which is directly related to the index of the Feynman graphs, and the expansion parameter is $1/D^{\frac{1}{r+1}}$. This is very similar to the large $N$ expansion of tensor models with enhanced scaling. In fact, by setting $N=D$ and taking the limit $N\rightarrow\infty$, the above expansions reduce to the large $N$ expansion \eqref{TensLargeNExpFeynEn}. 

\subsubsection{Leading sector} 
\label{sec:MatTensLO}

Let us briefly study the leading order (LO) Feynman graphs of matrix-tensor models with enhanced scaling. According to Eqs.\ \eqref{MatTensFexp} and \eqref{MatTensFhexp}, they satisfy $h=0$ and $\ell=0$ and they scale like $N^2D^r$. Since the parameters $h$ and $\ell$ only depend on the structure of the Feynman bubbles, they have the same combinatorial interpretation as the parameters $h$ in matrix models and $\ell$ in tensor models with enhanced scaling, respectively. It means that the description of the leading sector in Section \ref{sec:MatLO} for matrix models and in Section \ref{sec:LOGenMel} for tensor models hold for matrix-tensor models.

The condition $h=0$ means that the LO Feynman graphs $\cB$ are such that their $3$-bubble subgraph $\cB_{(012)}$ is maximally disconnected and its connected components are all planar. As for the condition $\ell=0$, it says that $\cB$ is itself maximally disconnected and all of its connected components have index zero, that is, they are generalized melons. 

As mentioned in the previous chapter, the class of generalized melons obtained with the enhanced scaling is much larger than the class of ordinary melons obtained with the BGR scaling. In particular, non-melonic interaction bubbles can contribute at leading order. 

Finally, to emphasize the point of view of the large $D$ limit of matrix models, let us restrict the models to single-trace interaction bubbles. Then, the usual sum over planar graphs truncates to a sum over generalized melons. In the previous chapter, we explained that the classification of generalized melons is not known in general but it is the case for the physically relevant complete interaction $\cK_{R+1}$ for $R$ prime. In this particular case, the truncation of the full sum over planar graphs is tractable and yet non-trivial. This result is central in the context of the SYK/black hole physics; we come back to this point in Section \ref{sec:MatTensDiscussion}. 

\subsubsection{Upper bound on the power of $D$ at fixed $h$} 
\label{sec:MatTensUpBound}

From the large $D$ expansion \eqref{MatTensFhexp}, it is clear that, at fixed $h$, the highest possible power of $D$ in a Feynman graph is $r+h$. For models with single-trace interactions, this is $r+2g$ where $g\equiv g(\cB_{(012)})$. This upper bound can actually be improved as follows. From the decomposition formula \eqref{fundid0}, we have
\be\label{indgt} \ind_{0}\gB\geq g + F_{12}-B_{(012)}-B^{(0)}+B\, .\ee
Using Eq.\ \eqref{MatTensHParam3} and the identities \eqref{TensFeynGraphData} and \eqref{MatTensFeynGraphData}, we obtain
\be\label{indgt2} \ind_{0}\gB\geq \frac{h}{2}-\sum_{a\in\cS_\cG}\bigl(c(\gB_{a})-1\bigr) + B -1\, .\ee
Together with Eq.\ \eqref{MatTensEllParam3}, this yields
\be\label{ellgt} \ell\geq h+r(r+3)\Bigl[\sum_{a\in\cS_\cG}\bigl( c(\gB_{a})-1\bigr) - B + 1\Bigr]\, .\ee
Finally, using the large $D$ expansion \eqref{MatTensFhexp}, we see that the highest possible power of $D$ is actually
\be\label{Dpowermax} r+\frac{r}{r+1}h - \frac{r(r+3)}{r+1}\Bigl[\sum_{a\in\cS_\cG}\bigl( c(\gB_{a})-1\bigr) - B + 1\Bigr]\, .\ee
For models with single-trace interactions, this reduces to $r+\frac{2r}{r+1}g$, generalizing the bound $1+g$ obtained in the case $r=1$ in \cite{FrankLargeD}.

\subsubsection{Other scalings and expansions}
\label{sec:MatTensOtherScalings}

We finally mention two other natural scalings that yield non-trivial large $D$ and/or large $N$ expansions, albeit keeping less Feynman graphs at leading order.

\

\noindent\emph{Large $D$ BGR scaling}

This scaling is defined by \eqref{MatTensBGRScaling}. Straightforward modifications of the above derivations show that a Feynman bubble $\cB$ is proportional to $N^{2-h}D^{r+h-L}$, where the parameter $h$ is given by Eq.~\eqref{MatTensHParam3} and the parameter $L$ by
\be\label{Lform} L = \frac{2}{(r+1)!}\deg\gB + (r+2)\Bigl[1+\sum_{a}\bigl( c(\gB_{a})-1\bigr) - B \Bigr]\, .\ee
Note that this expression is similar to the one \eqref{TensLParam3} obtained for tensor models with BGR scaling, as expected. In particular, $L\in\mathbb N$. We thus get an expansion in powers of $1/D$ governed by the degree. 

Using the bubble inequalities \eqref{degineq} successively for the colors 3 to $R$, together with $\smash{\deg\gB_{(012)}=g}$, one finds that
\be\label{deggt} \deg\gB\geq\frac{1}{2}(r+2)!\, g\, .\ee
Then, one can show using this bound that the combination $h-L$ is non-positive, so that the power of $D$ of any Feynman graph is bounded above by $r$, independently of $h$. As a result, the large $N$ and large $D$ limits commute with the BGR scaling.

\ 

\

\

\noindent\emph{Splitted scaling}

Another natural procedure is to define a scaling by treating the matrix and tensor parts of the matrix-tensor separately.\footnote{Even more generally, we could consider splitted scalings for which we separate the $R$ indices of the tensor into several groups, $R=r_{1}+\cdots +r_{s}$. This approach was first proposed in \cite{BonzomNewLargeExp} under the name of color slice scaling.} For the matrix part, we use the standard 't Hooft scaling \eqref{MatTensLargeNScaling}. For $r=1$, we further scale the corresponding couplings as $\mu_a\rightarrow D\mu_a$, so that they correspond to the large $D$ scaling of vector models. For $r=2$, which is a bi-matrix model, we use an action of the form
\be\label{act2} 
S = ND\Bigl( \frac{1}{2} X\cdot X + \sum_{a\in\cS}N^{1-t(\gB_{a})}D^{1-\tilde t(\gB_{a})}\kappa_{a}I_{\gB_{a}}(X)\Bigr)\, ,
\ee
where we have defined $\tilde t(\gB_{a})=F_{34}(\gB_{a})$ and we keep the couplings $\kappa_{a}$ fixed. For $r\geq 3$, we choose to use the enhanced scaling for the tensor part of the matrix-tensor, which amounts to keeping the $\kappa_{a}$ defined by
\be\label{act3} 
S = ND^{r-1}\Bigl( \frac{1}{2} X\cdot X + \sum_{a\in\cS}N^{1-t(\gB_{a})}D^{1-c(\gB_{a}^{(12)})+\frac{2}{(r-1)!}\deg\gB_{a}^{(12)}}\kappa_{a}I_{\gB_{a}}(X)\Bigr) \, ,
\ee
fixed. One can check that these scalings are less optimal than \eqref{MatTensEnScaling}, in the sense that the ratios $\kappa_{a}/\lambda_{a}$ are always proportional to a positive power of $D$. The large $N$ and large $D$ limits always commute in the splitted scalings, because the highest power of $D$ of any Feynman diagram is $D^{r}$.

\section{Application: quantum models and SYK physics}
\label{sec:QuantumModels}

Matrix-tensor models can be used to build interesting quantum mechanical (and field theoretic) models based on the prime-complete interaction $\cK_{R+1}$, which reproduce the important features of the SYK model with $q=(R+1)$-fold random interactions \cite{SYKq}. Such models are natural candidates to describe quantum black holes in the context of holography and string theory. This part corresponds to Section 4.4 in Ref.~\cite{Ref2}.

\subsubsection{Prime-complete Majorana fermion model}

We consider real fermionic matrix-tensor operators $$(\psi_{a_1a_2})_{\mu_{1}\cdots\mu_{r}}=(\psi_{a_1a_2})^\dagger_{\mu_{1}\cdots\mu_{r}}\, ,\quad 1\leq a,b\leq N\, ,\ 1\leq\mu_{i}\leq D\, ,$$ satisfying the quantization conditions
\be\label{qc1}\bigl\{(\psi_{a_1a_2})_{\mu_{1}\cdots\mu_{r}},(\psi_{b_1b_2})_{\nu_{1}\cdots\nu_{r}}\bigr\} = \frac{1}{ND^{r}} \delta_{a_1b_1}\delta_{a_2b_2}\delta_{\mu_{1}\nu_{1}}\cdots
\delta_{\mu_{r}\nu_{r}}\, .\ee
The $\text{O}(N)^{2}\times\text{O}(D)^{r}$ symmetric Hamiltonian is
\be\label{Hdef} H=-\frac{1}{2} i^{\frac{1}{2}(r+2)(r+3)}ND^{r+\frac{1}{4}r(r+1)}\lambda\tr\Bigl(\psi_{[C]}\psi^{T}_{[2]}\prod_{p=1}^{\frac{r+1}{2}}\psi_{[2-2p]}\psi^{T}_{[2+2p]}\Bigr) + \text{H. c.}
\ee
We use here the matrix notation associated with the colors $1$ and $2$, the trace in the Hamiltonian being associated with the $(12)$-face of the prime-complete interaction bubble $\gK_{r+3}$. We have indexed the variables according to which vertex they are associated to in $\gK_{r+3}$. The appropriate contractions of the $\text{O}(D)$ indices are assumed. For example,
\begin{align}\label{H1ex} H &= ND^{\frac{3}{2}}\lambda_{1}\tr\bigl(\psi_{\mu}\psi_{\nu}^{T}\psi_{\mu}\psi_{\nu}^{T}\bigr)\quad\text{for $r=1$}\, ,\\\label{H3ex}
H&= \frac{i}{2}ND^{6}\lambda\tr\bigl(\psi_{\alpha\mu\theta}\psi^{T}_{\beta\nu\phi}\psi_{\gamma\rho\theta}\psi^{T}_{\beta\mu\xi}\psi_{\alpha\rho\phi}\psi^{T}_{\gamma\nu\xi}\bigr) + \text{H. c.}\quad\text{for $r=3$}\, .
\end{align}
Note that for $r=1$, only $\lambda_{1}=\re\lambda$ contributes, whereas for $r\geq 3$, it is essential to add the Hermitian conjugate term to get a unitary theory. This is related to the fact that the vertices are all inequivalent in $\gK_{r+3}$ for $r\geq 3$, as discussed in Section \ref{disbubbleSec}. However, the difference between the two terms in the Hamiltonian is a subleading effect at large $N$ and large $D$. The factor of $i$ in Eq.\ \eqref{H3ex} has been chosen so that only $\lambda_{1}=\re\lambda$ contributes at leading order, for any $r$.

The basic quantity in the model is the Euclidean two-point function at finite temperature
\be\label{Gdef1Test} G(t) = \frac{1}{N}\bigl\langle\tr\bigl(\psi_{\mu_{1}\cdots\mu_{r}}(t)\psi^{T}_{\mu_{1}\cdots\mu_{r}}\bigr)\bigr\rangle_{\beta}\, .\ee
It is a real and odd function of the Euclidean time $t$ satisfying $G(t+\beta)=-G(t)$. From the classification Theorem \ref{theorem1} in Appendix \ref{app:AppC}, we can write down a Schwinger-Dyson equation that determines $G$ at leading order. We introduce the Fourier transform
\be\label{GFourier} G(t) = \frac{1}{\beta}\sum_{k\in\mathbb Z+\frac{1}{2}}G_{k}\,e^{-\frac{2i\pi k t}{\beta}}\ee
and the self-energy $\Sigma$,
\be\label{Sigmadef} \frac{1}{G_{k}}=-\frac{2i\pi k}{\beta} + \Sigma_{k}\, ,\quad \Sigma(t) = \frac{1}{\beta}\sum_{k\in\mathbb Z+\frac{1}{2}}\Sigma_{k}\,e^{-\frac{2i\pi k t}{\beta}}\, .\ee
The Schwinger-Dyson equation then reads
\be\label{SDMajorana} \Sigma_{\text{LO}} (t) = 
\begin{cases} -16 \lambda_{1}^{2}G_{\text{LO}}(t)^{3} & \text{if $r=1$,}\\
-(r+3)\lambda_{1}^{2}G_{\text{LO}}(t)^{r+2} & \text{if $r\geq 3$ and $r+2$ is prime.}
\end{cases}\ee
Remark that it looks similar to the Schwinger-Dyson equation obtained for tensor models (cf.\ Eqs.\ \eqref{TensSDEqEn} and \eqref{TensSDEqEnGen}), but with the crucial difference that we now work in one dimension with fermionic variables. Again, the fact that the combinatorial factor 16 when $r=1$ does not generalize to $(r+3)^{2}$ for larger values of $r$ is an effect of the lack of symmetry between the vertices in the prime-complete interaction bubble for $r\geq 3$.

\vspace{0.3cm}

We now make several comments:

\vspace{0.2cm}

\begin{itemize}

\item We have obtained a matrix-tensor version of the SYK model with $q$-fold interactions \cite{SYKq} and we can in particular study the limit $q=r+3\rightarrow\infty$. Note, however, that the result holds in principle only when $r+2=q-1$ is a prime number. As mentioned in Section \ref{sec:CompleteInt}, we do not know the form of the leading order graphs and thus we cannot write down the Schwinger-Dyson equation if $r+2$ is not prime. It is a logical possibility that \eqref{SDMajorana} is still correct, but the proof would have to be generalized non-trivially, since the relation with the index is then lost. 

\vspace{0.2cm}

\item In the case of the SYK model, a straightforward and purely algebraic derivation of the leading order Schwinger-Dyson equation from standard manipulations of the path integral with an auxiliary field is possible, without any study of the Feynman diagrams themselves \cite{EffectiveSYK}. An analogue of such a derivation is not known for matrix-tensor models and most probably does not exist. The reason is that when there is a formulation in terms of an auxiliary field, the large $N$ expansion at all orders and for all the correlation functions can be obtained straightforwardly from the loop expansion generated by the effective action for the auxiliary field. In matrix-tensor models such as \eqref{Hdef}, the study of subleading contributions is typically much more difficult and cannot be reduced to a simple loopwise expansion. For a discussion of subleading contributions in related models, see e.g.\ \cite{GW,SYKCorr}.

\vspace{0.2cm}

\item In spite of the lack of an auxiliary field formulation, all the thermodynamical functions (free energy, etc.) in our model can be expressed at leading order in terms of the two-point function \eqref{Gdef1Test} only, see \cite{FrankPlus}.

\vspace{0.2cm}

\item The fact that a tensor model mimicking the $q$-fold random SYK interaction can be built has been mentioned in the literature \cite{SYKTensorQ}. These papers seem to have assumed that the analysis of the simplest case in \cite{CT} could be immediately generalized. As we explained, this is incorrect. Still, it might be that the Schwinger-Dyson equation \eqref{SDMajorana} remains valid (plausibly modulo appropriate combinatorial factors) beyond the case $R$ prime which is fully solved in the present paper. For $R$ prime, the particular combinatorial factors in \eqref{SDMajorana} are related to the fact that the vertices of the prime-complete bubble are distinguishable when $r\geq 3$. These factors are not correctly written down in \cite{SYKTensorQ}. This is however irrelevant for the bulk of the results in these references.

\end{itemize}

\subsubsection{Prime-complete Dirac fermion model}

The SYK model has an interesting complex version \cite{ComplexSYK}; hence, it is natural to also consider a complex version of the matrix-tensor model \eqref{Hdef}. An important physical motivation to do so is to be able to add a non-trivial mass term. The resulting phase diagram when $r=1$ has been shown to display many surprising features, the most notable being the existence of a non-trivial critical point \cite{FrankPlus}. The generalization of the model to any prime $R=r+2$ opens the way to a detailed analytical study of this critical point in the large $r$ limit.

The model is built from complex matrix-tensor operators satisfying the quantization conditions
\be\label{qc2}\bigl\{(\psi_{a_1a_2})_{\mu_{1}\cdots\mu_{r}},(\psi^{\dagger}_{b_2b_1})_{\nu_{1}\cdots\nu_{r}}\bigr\}= \frac{1}{ND^{r}} \delta_{a_1b_1}\delta_{a_2b_2}\delta_{\mu_{1}\nu_{1}}\cdots
\delta_{\mu_{r}\nu_{r}}\, ,\ee
where we use the convention $$(\psi_{a_1a_2})^\dagger_{\mu_{1}\cdots\mu_{r}}=(\psi^{\dagger}_{a_2a_1})_{\mu_{1}\cdots\mu_{r}}\, ,\quad 1\leq a,b\leq N\, ,\ 1\leq\mu_{i}\leq D.$$ The $\text{U}(N)^{2}\times\text{O}(D)^{r}$ symmetric Hamiltonian is
\be\label{Hdef2} H= ND^{r}\tr\biggl[ m \psi^{\dagger}_{\mu_{1}\cdots\mu_{r}}\psi_{\mu_{1}\cdots\mu_{r}} + D^{\frac{1}{4}r(r+1)}\Bigl(\lambda
\psi_{[C]}\psi^{\dagger}_{[2]}\prod_{p=1}^{\frac{r+1}{2}}\psi_{[2-2p]}\psi^{\dagger}_{[2+2p]}
+ \text{H. c.}\Bigr)\biggr]\, ,\ee
with notations similar to what we used in the model \eqref{Hdef}. When $r=1$, the model reduces to the one studied in the first reference in \cite{FrankPlus}. When $r\geq 3$, the addition of the Hermitian conjugate term is essential for unitarity. Unlike in the model \eqref{Hdef}, this is crucial even at leading order. Actually, with only one term in the Hamiltonian, no generalized melon respecting the $\text{U}(N)^{2}$ symmetry could be built.

We introduce the Euclidean two-point function at finite temperature
\be\label{Gdef2} G(t) = \frac{1}{N}\bigl\langle\tr\psi_{\mu_{1}\cdots\mu_{r}}(t)\psi^{\dagger}_{\mu_{1}\cdots\mu_{r}}\bigr\rangle_{\beta}\, .\ee
It is real, satisfies $G(t+\beta)=-G(t)$, but it is not odd as in the Majorana case, except when $m=0$. In terms of the Fourier transform defined as in Eq.~\eqref{GFourier} and the self-energy 
\be\label{Sigmadef2} \frac{1}{G_{k}}=m-\frac{2i\pi k}{\beta} + \Sigma_{k}\, ,\quad \Sigma(t) = \frac{1}{\beta}\sum_{k\in\mathbb Z+\frac{1}{2}}\Sigma_{k}\,e^{-\frac{2i\pi k t}{\beta}}\, ,\ee
the Schwinger-Dyson equation reads
\be\label{SDDirac} \Sigma_{\text{LO}} (t) = 
\begin{cases} 16 |\lambda|^{2} G_{\text{LO}}(t)^{2}G_{\text{LO}}(-t) & \text{if $r=1$,}\\
(-1)^{\frac{r+3}{2}}(r+3)|\lambda|^{2}G_{\text{LO}}(t)^{\frac{r+3}{2}}G_{\text{LO}}(-t)^{\frac{r+1}{2}} & \text{if $r\geq 3$, $r+2$ prime.}
\end{cases}\ee
Following \cite{FrankPlus}, a rich physics for this model is expected to be found in the $(T,m)$-plane, with small and large black hole phases, a line of first order phase transition between them, terminating at a non-trivial critical point.

The physics of the quantum mechanical versions of matrix-tensor models in the large $N$ and large $D$ limits remains, to a large extent, to be uncovered. The above results, in particular the possibility to study the large $r$ limit, might open the way to a better analytical understanding of the phase transition and the non-trivial critical point discussed in \cite{FrankPlus}. More generally, matrix-tensor models seem to capture basic qualitative properties associated with quantum black holes (emergence of a reparameterization symmetry in the IR, quasi-normal behavior, non-zero zero-temperature entropy and maximal chaos), but a detailed and satisfactory picture of the relationship with black holes has not emerged yet. In particular, a model with a genuine gravity-like holographic dual has not been constructed and it is unclear how the black hole geometry can be seen directly from the quantum models. These are interesting research directions for the future.

\section{Discussion and outlook}
\label{sec:MatTensDiscussion}

In this chapter, we described the large $N$ and large $D$ expansions for general matrix-tensor models with enhanced scaling. In particular, we stressed that the large $N$ limit must always be taken first, yielding a matrix model-like large $N$ expansion governed by the genus for single-trace interactions. Then, the large $D$ limit is taken at each order in the $1/N$ expansion, that is, at fixed genus. The resulting large $D$ expansion is governed by the index of the Feynman graphs, similarly to the case of $\oN^R$ tensor models with enhanced scaling introduced in the previous chapter. Finally, in the large $N$ and large $D$ limits, the leading order Feynman graphs again correspond to the generalized melonic graphs of index zero. 

Matrix-tensor models provide an interesting connection between matrix models and tensor models. Indeed, from the point of view of large $N$ matrix models, the large $D$ expansion provides a new way of understanding the sum over planar graphs, which is in many cases too difficult to compute. In particular, in the large $D$ limit, the sum over planar graphs truncates to a sum over generalized melons, which becomes tractable and yet non-trivial for some particular interaction bubbles such as the complete interaction $\cK_{R+1}$ for $R$ prime. This result is key in the context of holography where string-inspired matrix quantum mechanical models based on $\cK_{R+1}$ can be constructed with all the relevant features of the SYK physics. 

Some interesting questions however remain open within the framework of matrix-tensor models, in particular with regards to their connections with matrix and tensor models and with quantum mechanical models in holography. They provide several research directions, which we describe in the following. 

\subsubsection{Optimal scalings}

The problem of optimal scaling in matrix-tensor models is equivalently formulated as in usual tensor models, whose case is presented in Section \ref{sec:TensDiscussionEn}, because it boils down to studying Feynman graphs of index zero. Since the connection between matrix-tensor models and random higher dimensional geometries is not direct, a main motivation for finding optimal scalings in this framework is to enlarge the set of interesting quantum mechanical models for holography. The reader may refer to Section \ref{sec:QuantumModels} for a discussion of some remaining open questions regarding the physics associated with such models in holography.

\subsubsection{Beyond leading order}

Likewise tensor models, whose case is described in Section \ref{sec:TensDiscussionEn}, the analysis of the subleading contributions in the large $N$ and large $D$ expansions of matrix-tensor models is particularly interesting. In these models, there are two parameters $N$ and $D$; hence, there are two types of subleading contributions: the ones with respect to the large $N$ expansion and the ones with respect to the large $D$ expansion at fixed order in the $1/N$ expansion. This stands in contrast with usual tensor models where there is only one type of subleading contributions. In particular, a given subleading Feynman graphs in tensor models can correspond to one of the two types of subleading contributions in matrix-tensor models. 

Beyond gaining analytical control over the perturbative expansion of matrix-tensor models, the description of the subleading contributions is also appealing from the point of view of matrix models via double scaling limits. As an illustration, we rewrite below the large $N$ and large $D$ expansions of matrix-tensor models for $r=1$ and for single-trace interaction bubbles (cf.\ Eqs.\ \eqref{MatTensFexp} and \eqref{MatTensFhexp} together with Eq.\ \eqref{MatTensHParam5})
\be\label{MatTensFexpDisc} 
F = \sum_{g\in\frac{1}{2}\mathbb N} N^{2-2g} \sum_{\ell\in\mathbb N} D^{1+2g-\frac{\ell}{2}} F_{g,\ell}\, ,
\ee
where it is understood that the limit $N\rightarrow\infty$ is taken first and the limit $D\rightarrow\infty$ next. At large $N$ and large $D$, the sum over planar graphs ($g=0$) truncates to a sum over generalized melonic graphs ($\ell=0$). A natural follow-up would be to study the subleading contributions in $1/\sqrt{D}$ to the sum over planar graphs. More precisely, one could analyse the set of Feynman graphs characterized by $g=0$ and fixed $\ell$ and evaluate their generating series. We note that such an analysis should be in principle simpler that studying the set Feynman graphs of fixed index, as proposed in Section \ref{sec:TensDiscussionEn}, due to the reduced number of Feynman graphs. As a second step, one could then work out a double scaling analysis by sending both $D\rightarrow\infty$ and the coupling constant to criticality in a correlated way.

On the other hand, another interesting question is whether the framework of matrix-tensor models can provide a simplification of the sum over all graphs of arbitrary genus, in the same way as it does for the sum over planar graphs. Such a sum corresponds to the full perturbative expansion of matrix models, which is known to be divergent. In the present framework, the additional parameter $D$ might allow, in some limit, to reduce the number of graphs at fixed genus and then study the reduced sum over all genera. However, a direct answer can not be obtained from \eqref{MatTensFexpDisc} because the large $N$ limit is taken first and suppresses the higher genus contributions. Fortunately, there is a way out which consists in two steps.

The first step consists in defining a new parameter $\alpha$ for the large $D$ expansion, based on the lower bound \eqref{ellgt} for the parameter $\ell$, as follows: $\alpha=\ell-2g\in\mathbb{N}$ (recall that $h=2g$ for single-trace interactions). The reason behind this redefinition is that we want to look for a limit that selects the Feynman graphs of arbitrary genus $g$, but with $\ell=0$. However, the second form of the decomposition formula \eqref{fundid1} implies that $\ell=0$ constrains $g=0$. In contrast, the new parameter $\alpha$ can be set to zero without constraining the genus $g$.

Then, the second step consists in reorganizing the resulting perturbative expansion as
\be\label{MatTensFexpDisc3} 
F = \sum_{g\in\frac{1}{2}\mathbb N} \biggl(\frac{N}{\sqrt{D}}\biggr)^{2-2g} \sum_{\alpha\in\mathbb N} D^{2-\frac{\alpha}{2}} \tilde{F}_{g,\alpha}\, ,
\ee
where $\tilde{F}_{g,\alpha}\equiv F_{g,\ell-2g}$. This expression suggests that, by taking the limits $N\rightarrow\infty$ and $D\rightarrow\infty$ while keeping the ratio $M=N/\sqrt{D}$ fixed, one obtains a reduced sum over all genera,
\be\label{MatTensFexpDisc4} 
\lim_{\substack{{D\rightarrow\infty}\\ {M<\infty}}} \frac{1}{D^{2}}F = \sum_{g\in\frac{1}{2}\mathbb N} M^{2-2g} \tilde{F}_{g,0}\, ,
\ee
in the sense that only Feynman graphs of arbitrarily genus $g$ and with $\alpha=0$ are kept in this limit. Because the number of such Feynman graphs is much lower that all the possible graphs, one may hope that the reduced sum over all genera is convergent. This could be checked, for instance, by first analyzing the set of Feynman graphs characterized by $\alpha=0$ and fixed $g$ and evaluating their generating series, and then by initiating a double scaling limit by sending $M\rightarrow\infty$ and the coupling constant to criticality in a correlated way. 

It is worth emphasizing that the two double scaling limits introduced above are based on the study of strict subsets of the set of all Feynman graphs of fixed index. In particular, their analysis may be considerably simpler than the general case and new structures may emerge. These two limits offer interesting research directions that we hope to address in future work.

\subsubsection{Case of reduced symmetry}

Finally, we mention a possible research direction concerning reduced symmetry in matrix-tensor models, in relation with the discussion of Section \ref{sec:TensDiscussionEn} for usual tensor models. In the present framework, one can impose symmetry or antisymmetry on the $\oN$ matrix indices or on the additional $\oD$ indices.

As already mentioned in Section \ref{sec:TensDiscussionEn}, it was proved in \cite{FrankLargeD} that in the case of $\uN^2\times\oD$ matrix-tensor models based on complex matrices, the symmetry can be broken down to $\uN\times\oD$ at the planar level by imposing Hermiticity on the matrices. More precisely, it is shown that at leading order in the $1/N$ expansion, which corresponds to the planar limit, the large $D$ expansion remains well-defined. The argument of \cite{FrankLargeD} can be extended to general matrix-tensor models: the large $D$ limit of matrix-tensor models for which the $\text{O}(N)^{2}\times\text{O}(D)^{r}$ symmetry is broken down to $\text{O}(N)\times\text{O}(D)^{r}$ by imposing symmetry or antisymmetry on the real matrices still makes sense at the planar level. 

An interesting open problem is to generalize this result beyond the planar limit. As explained in Ref.\ \cite{Ref1} (see next chapter), the consistency of the large $D$ limit is lost at higher genera if no further constraint is imposed on the Hermitian or symmetric matrices. The problematic Feynman graphs displayed in Ref.~\cite{Ref1} however disappear in the case of traceless matrices, yielding to the natural conjecture that the large $D$ limit is consistent at all genera in this case. 

The question of reduced symmetry in matrix-tensor models is of course directly related to the one in usual tensor models. In particular, it can be seen as a special case because the symmetry or antisymmetry is imposed on a subset of the indices of the matrix-tensor. As a result, the arguments of \cite{ReducedRank3} may be useful for proving the conjecture, though they would need some adaptation. Another important tool might be the existing results in matrix models and topology because of their natural connection with matrix-tensor models.

%%%%%%%%%%%%%%%%%%%%%%%%%%%%%%%%%%%%%%%%%%%%%%%%
%%%%%%%%%%%%%%%%%%%  Conclusion  %%%%%%%%%%%%%%%%%%%%%%%%%
%%%%%%%%%%%%%%%%%%%%%%%%%%%%%%%%%%%%%%%%%%%%%%%%

%
%
%\chapter*{Conclusion}
%\addcontentsline{toc}{chapter}{Conclusion}
%\label{chap:Conclu}
%
%
%
%%%%%%%%%%%%%%%%%%%%%%%%%%%%%%%%%%%%%%%%%%%%%%%%%
%%%%%%%%%%%%%%%%%%%  APPENDICES  %%%%%%%%%%%%%%%%%%%%%
%%%%%%%%%%%%%%%%%%%%%%%%%%%%%%%%%%%%%%%%%%%%%%%%

% Appendices
\newpage
\chapter*{Appendices}
\addcontentsline{toc}{chapter}{Appendices}

\begin{appendix}

%%%%%%%%%%%%%%%%%%%%%%%%%%%%%%%%%%%%%%%%%%%%%%%%
%%%%%%%%%%%%%%%%%%%  Appendix A  %%%%%%%%%%%%%%%%%%%%%%%%%
%%%%%%%%%%%%%%%%%%%%%%%%%%%%%%%%%%%%%%%%%%%%%%%%

\chapter{Graphs embedded on surfaces}
\label{app:AppA}

Graphs are mathematical structures that appear in many areas of sciences. They are the theoretical backbone of the work realized in this thesis, as Feynman diagrams obtained in perturbative QFT are in fact graphs. 

The aim of this section is to provide a short introduction to various elements of graph theory that are used in this thesis and in the related Refs.\ \cite{Ref2,Ref1,Ref3}. This section is far for being an exhaustive exposition of graph theory; it rather focuses on giving the reader who is unfamiliar with this broad topic the necessary tools to appreciate the combinatorial aspects of this thesis. For a complete and rigorous treatment of graph theory, the interested reader is referred to, for example, \cite{Bollo,MohThom,EllMof}. Most of the following is inspired from \cite{MohThom,EllMof}.

In Section \ref{sec:App1}, we introduce useful notions on abstract graphs, where abstract is used to emphasize that the graph is not embedded \cite{EllMof}. We then describe abstract graphs whose edges are assigned a color and we discuss the related concept of bubble. In Section \ref{sec:App2}, we move on to defining graphs embedded on surfaces. In particular, we detail three possible representations for embedded graphs and we explain the notion of genus.

\section{Abstract graphs}
\label{sec:App1}

\subsection{General definitions}
\label{sec:AppAGeneralDef}

A \emph{graph} $\cG=(\cV(\cG), \cE(\cG))$ is a set $\cV(\cG)$ of \emph{vertices} together with a set $\cE(\cG)$ of \emph{edges}, each given by the data of two (not necessarily distinct) elements in $\cV(\cG)$. An edge $e\in\cE(\cG)$ is denoted as $\nu_1\nu_2$ with $\nu_1, \nu_2 \in \cV(\cG)$. We say that it \emph{joins} or \emph{connects} the vertices $\nu_1$ and $\nu_2$, which are called the \emph{endvertices} of $e$. Two vertices $\nu_1$ and $\nu_2$ of $\cG$ are \emph{adjacent} if there exists an edge $e=\nu_1\nu_2\in\cE(\cG)$ that connects them. An edge $e$ of $\cG$ is \emph{incident} to a vertex $\nu$ of $\cG$ if $\nu$ is an endvertex of this edge. Finally, the number of vertices and the number of edges of $\cG$ are denoted by $V(\cG)= | \cV(\cG) |$ and $E(\cG)= | \cE(\cG) |$ respectively. 

In the context of QFT, we allow for graphs with \emph{multiple edges}, that is, distinct elements in $\cE(\cG)$ having the same pair of endvertices, as well as \emph{self-loops}, that is, edges with only one endvertex. In the terminology of QFT, edges are often called \emph{propagators} and self-loops \emph{tadpoles}. We use these names interchangeably in this thesis.

The \emph{degree} of a vertex $\nu\in\cV(\cG)$ is defined as the number of incident edges to this vertex (with each self-loop counted twice). If all the vertices of $\cG$ have the same degree $k$, the graph is \emph{k-regular}.

In some instances, we may want to describe graphs contained within other graphs. A graph $\cH$ is a \emph{subgraph} of $\cG$, written as $\cH\subseteq\cG$, if $\cV(\cH)\subseteq\cV(\cG)$ and $\cE(\cH)\subseteq\cE(\cG)$. A class of subgraphs of particular interest in this thesis are \emph{spanning subgraphs}, which are subgraphs $\cH$ such that $\cV(\cH)=\cV(\cG)$. Note that spanning subgraphs of $\cG$ can always be obtained from $\cG$ by deleting edges in $\cE(\cG)$.

A sequence $W=\nu_1e_1\nu_2e_2\ldots e_{k-1}\nu_k$ $(k\geq1)$ of vertices and edges of $\cG$, such that for $1\leq i\leq k-1$, the edge $e_i$ connects the vertices $\nu_i$ and $\nu_{i+1}$, is called a \emph{walk} in $\cG$. The vertices $\nu_1$ and $\nu_k$ are the \emph{endvertices} of the walk. If $\nu_1=\nu_k$, the walk $W$ is \emph{closed}. When there is no confusion about the edges used in the sequence, they can be omitted and the walk through the consecutive vertices $\nu_1, \nu_2, \ldots, \nu_k$ is written as $W=\nu_1\nu_2\ldots \nu_k$. Remark that a walk allows for repeated vertices and edges. A walk together with the constraint that it contains no repeated vertices and edges is called a \emph{path} $P$ in $\cG$. If we add to $P$ the edge $\nu_k\nu_1$, the resulting sequence is called a \emph{cycle} $C$ in $\cG$. A path or a cycle passing through $k$ vertices is said to be of \emph{length} $k$.

We can now define precisely what we mean by a graph being connected. A graph $\cG$ is \emph{connected} if any pair of vertices of $\cG$ are connected by a path in $\cG$. Naturally, a \emph{connected component} of $\cG$ is defined to be a maximal connected subgraph of $\cG$. A general graph $\cG$ can thus be seen as the disjoint union of its connected components. We denote the number of connected components of $\cG$ by $G$ or more explicitly by $c(\cG)$. 

We now introduce two classes of graphs used in this thesis. First, the \emph{complete graph} $\cK_n$ on $n$ vertices is the graph with $n$ vertices and no self-loops such that any two vertices is joined by exactly one edge. It is straightforward to check that $E(\cK_n)=n(n-1)/2$. Second, a \emph{tree} $\cT_n$ on $n$ vertices is a connected graph with $n$ vertices and no cycles. One can check that $E(\cT_n)=n-1$ and for $n\geq2$, trees necessarily have at least two vertices of degree one.

\begin{figure}[]
\centerline{\includegraphics[width=6in]{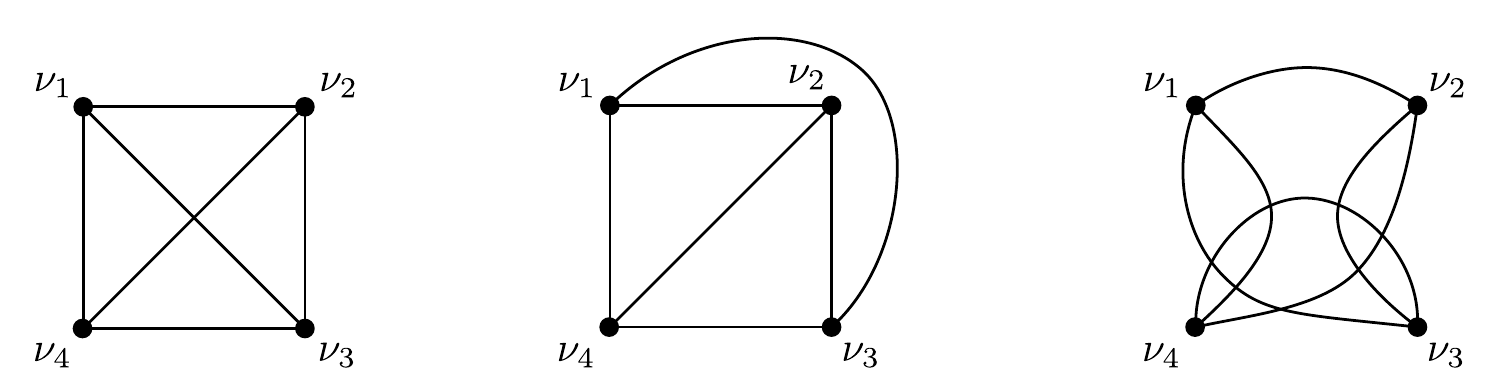}}
\caption{Different drawings of the graph $\cK_4$, with vertex set $\cV(\cK_4)=\{\nu_1, \nu_2, \nu_3, \nu_4\}$ and edge set $\cE(\cK_4)=\{\nu_1\nu_2, \nu_1\nu_3, \nu_1\nu_4, \nu_2\nu_3, \nu_2\nu_4, \nu_3\nu_4\}$, in the plane.}\label{ExampleK4}
\end{figure}

By definition, graphs are abstract mathematical structures made of vertices and edges. At some point, one may want to represent them, say in the plane. To do so, we represent vertices as points (or dots) in the plane and edges as simple curves joining the vertices that correspond to their endvertices. Edges are allowed to cross each other, but they should touch no other vertices of the graph than their endvertices. This representation of a graph $\cG$ is called a \emph{drawing} of $\cG$ in the plane. We emphasize that in general, there are many ways of drawing a given graph $\cG$ in the plane. This is illustrated in Figure \ref{ExampleK4} for $\cK_4$.

Finally, the following notion is also useful in the context of this thesis. A graph $\cG$ is \emph{bipartite} if we can partition the set of vertices into two sets $\cV_+(\cG)$ and $\cV_-(\cG)$ such that edges of $\cG$ connect vertices in $\cV_+(\cG)$ to vertices in $\cV_-(\cG)$ only. An important theorem that characterizes bipartite graphs was proved by K\"onig \cite{Konig}: a graph is bipartite if and only if it contains no cycle of odd length.

\subsection{Edge-colored graphs and bubbles}
\label{sec:AppABubbles}

From Chapter \ref{chap:Matrix} onwards, we use the notion of edge-colored graphs, that is, graphs whose edges are assigned a given color. This part corresponds to Sections 2.1.1, 2.1.2 and 2.2.1 of Ref.\ \cite{Ref2}. Formally, an edge-coloring of a graph $\cG$ with $d$ colors is a surjective map $\cE(\cG)\rightarrow\mathscr C$ where the set of colors $\mathscr C$ is isomorphic to $\{1,\ldots, d\}$. Unless explicitly stated otherwise, we assume that $\mathscr C =  \{1,\ldots,d\}$, in which case the colors are typically denoted by greek letters $\alpha$, $\beta$, etc., or that $\mathscr C =  \{0,\ldots,d-1\}$, in which case the color 0 is singled out and the colors $1,2,\ldots,d-1$ are denoted by latin indices $i$, $j$, etc. The number of edges of color $\alpha$ is $E_{\alpha}(\gG)$. The graph $\gG^{(\alpha_{1}\cdots \alpha_{p})}$ is obtained from $\gG$ by deleting all the edges of colors $\alpha_{1},\ldots ,\alpha_{p}$, whereas the graph $\gG_{(\alpha_{1}\cdots \alpha_{p})}$ is obtained from $\gG$ by keeping the edges of colors $\alpha_{1},\ldots ,\alpha_{p}$ and deleting all the others. The number of connected components of a graph $\gG$ is denoted by $G$ or more explicitly by $c(\gG)$. Similarly, we write $c(\gG^{(\alpha_{1}\cdots \alpha_{p})}) = G^{(\alpha_{1}\cdots \alpha_{p})}$ and  $c(\gG_{(\alpha_{1}\cdots \alpha_{p})}) = G_{(\alpha_{1}\cdots \alpha_{p})}$. The number of loops (more precisely independent cycles) of a graph $\gG$ is given by $L(\gG)=E(\gG)-V(\gG)+G$.

\subsubsection{\label{ineqSec}Connectivity inequalities and identities}

The following inequality is useful.
\begin{lemma}\label{ineqlem}
\be\label{inequal} G^{(\alpha\beta)}-G^{(\alpha)}-G^{(\beta)}+ G \geq 0\, .\ee
\end{lemma}
\noindent A straightforward generalization is obtained by replacing the colors $\alpha$ and $\beta$ by many colors $\alpha_{1},\ldots,\alpha_{p}$ and $\beta_{1},\ldots,\beta_{q}$ and by substituting $G^{(\gamma_{1}\cdots \gamma_{r})}$ to $G$ in \eqref{inequal},
\be\label{inequal2} G^{(\alpha_{1}\cdots\alpha_{p}\beta_{1}\cdots\beta_{q} \gamma_{1}\cdots \gamma_{r})}-G^{(\alpha_{1}\cdots\alpha_{p}\gamma_{1}\cdots \gamma_{r})}-G^{(\beta_{1}\cdots\beta_{q}\gamma_{1}\cdots \gamma_{r})}+ G^{(\gamma_{1}\cdots \gamma_{r})} \geq 0\, .\ee
These inequalities are valid in full generality, without putting any constraint on the coloring of the graph. 

\proof When one removes the edges of color $\alpha$ from $\gG$ and $\gG^{(\beta)}$, one creates $G^{(\alpha)}-G$ and $G^{(\alpha\beta)}-G^{(\beta)}$ new connected components, respectively. But a graph that splits when the edges of colors $\beta$ are not taken into account may remain connected otherwise. This implies that $G^{(\alpha)}-G\leq G^{(\alpha\beta)}-G^{(\beta)}$, which is \eqref{inequal}. Another  argument amounts to noting that the left-hand side of \eqref{inequal} matches with the number of loops in the abstract bipartite graph $\gG_{\alpha,\beta}$ built as follows: the $+$ and $-$ vertices of $\gG_{\alpha,\beta}$ are the connected components of $\gG^{(\alpha)}$ and $\gG^{(\beta)}$ respectively and an edge joins a $+$ to a $-$ vertex for each connected component of $\smash{\gG^{(\alpha\beta)}}$ included in both vertices. One can check that $\gG_{\alpha,\beta}$ has $G$ connected components. Moreover, by construction, $V(\gG_{\alpha,\beta})=G^{(\alpha)}+G^{(\beta)}$ and $E(\gG_{\alpha,\beta})=G^{(\alpha\beta)}$; thus $L(\gG_{\alpha,\beta})=G^{(\alpha\beta)}-G^{(\alpha)}-G^{(\beta)}+G\geq0$.\qed

\

In some instances, it is convenient to use the above inequalities in the following form. We single out the color 0 and label the other colors with latin indices. The quantity
\be\label{Deltadef} \Delta_{0} G^{(i_{1}\cdots i_{p})} = G^{(0i_{1}\cdots i_{p})}-G^{(i_{1}\cdots i_{p})}\ee
represents the number of new connected components that are created when one removes the lines of color 0 from $\gG^{(i_{1}\cdots i_{p})}$. One can decompose
\be\label{Deltadec}\begin{split}
\Delta_{0} G^{(i_{1}\cdots i_{p})} - \Delta_{0} G & = G^{(0i_{1}\cdots i_{p})}-G^{(i_{1}\cdots i_{p})} - G^{(0)}+ G\\
& = \bigl(\Delta_{0} G^{(i_{1}\cdots i_{p})}-\Delta_{0} G^{(i_{1}\cdots i_{p-1})}\bigr) \\
& \hskip .4cm  + \bigl(\Delta_{0} G^{(i_{1}\cdots i_{p-1})}-\Delta_{0} G^{(i_{1}\cdots i_{p-2})}\bigr)
 + \cdots + \bigl(\Delta_{0} G^{(i_{1})}-\Delta_{0} G\bigr)
\end{split}\ee
as a sum of terms that are all positive according to \eqref{inequal2}. In particular, the condition 
\be\label{vcond1} \Delta_{0} G^{(i_{1}\cdots i_{p})} = \Delta_{0} G\ee
is equivalent to the conditions 
\be\label{vcond2}\Delta_{0} G^{(i_{1}\cdots i_{k})}=\Delta_{0} G^{(i_{1}\cdots i_{k-1})}\ee
for all $1\leq k\leq p$. It is useful to sum the decomposition \eqref{Deltadec} over all the possible indices $i_{1},\ldots,i_{p}$. If we introduce the positive integers
\be\label{ncon1} \delta_{0;p}(\gG)  =\sum_{i_{1}<i_{2}< \cdots< i_{p}} \bigl(\Delta_{0} G^{(i_{1}\cdots i_{p})} - \Delta_{0} G\bigr)=
\sum_{i_{1}<i_{2}< \cdots< i_{p}}\bigl( G^{(0 i_{1}\cdots i_{p})}-G^{(i_{1}\cdots i_{p})}-G^{(0)}+G\bigr)\ee
and
\begin{multline}\label{nconnex1} \tilde\delta_{0;p} (\gG)  = \sum_{i_{1}< i_{2}< \cdots < i_{p}}
\bigl(\Delta_{0} G^{(i_{1}\cdots i_{p})}-\Delta_{0} G^{(i_{1}\cdots i_{p-1})}\bigr)\\=
\sum_{i_{1}< i_{2}< \cdots < i_{p}}\bigl( G^{(0 i_{1}\cdots i_{p})}-G^{(i_{1}\cdots i_{p})}-G^{(0 i_{1}\cdots i_{p-1})}+G^{(i_{1}\cdots i_{p-1})}\bigr) \, ,
\end{multline}
then \eqref{Deltadec} yields
\be\label{conid} \delta_{0;p}(\gG)  = \frac{1}{p!(d-p-1)!}\sum_{k=1}^{p}k!(d-k-1)!\, \tilde\delta_{0;k}(\gG)\, .\ee
The condition $\delta_{0;p}=0$ is equivalent to \eqref{vcond1} for all possible indices $i_{1},\ldots,i_{p}$, which is also equivalent to $\tilde\delta_{0;k}=0$ and to \eqref{vcond2} for all possible indices $i_{1},\ldots,i_{k}$ and $1\leq k\leq p$.

Finally, we note that the formula \eqref{ncon1} for $\delta_{0;p}$ can also be written as
\be\label{ncon2} \delta_{0;d-1-p}(\gG) =
\sum_{i_{1}<i_{2}< \cdots< i_{p}}\bigl( G_{(i_{1}\cdots i_{p})}-G_{(0i_{1}\cdots i_{p})}-G^{(0)}+G\bigr)\, .\ee
For example, we obtain
\begin{align}\label{del1}\delta_{0;d-1}(\gG) & = V(\gG) - E_{0}(\gG) - G^{(0)} + G\, ,\\
\label{del2}
\delta_{0;d-2}(\gG) & =\sum_{i}\bigl( E_{i}(\gG) - G_{(0i)} - G^{(0)} + G\bigr)\, ,\\
\label{del3}
\delta_{0;d-3}(\gG) & =\sum_{i<j}\bigl(G_{(ij)} - G_{(0ij)}-G^{(0)}+G\bigr)\, .
\end{align}

\subsubsection{Bubbles}

We define a $d$-bubble $\cB$ as a $d$-regular graph $\cB=(\cV(\cB),\cE(\cB))$ which is edge-colored with $d$ colors such that each color is incident exactly once at each vertex of $\cB$. In particular, one can check that
\be\label{EVrel} 2 E(\gB) = d\, V(\gB) = 2 d\, E_{\alpha}(\gB)\, .\ee
Note that, if $\gB$ is a $d$-bubble, $\gB^{(\alpha_{1}\cdots \alpha_{p})}$ and $\gB_{(\alpha_{1}\cdots \alpha_{p})}$ are $(d-p)$ and $p$-bubbles, respectively, for any $p\leq d$. Given two distinct colors $\alpha$ and $\beta$, a \emph{face} of colors $\alpha$ and $\beta$, also called an $(\alpha\beta)$-face, is defined to be a cycle of $\gB$ made of edges of alternating colors $\alpha$ and $\beta$. Equivalently, the $(\alpha\beta)$-faces are the connected components of $\gB_{(\alpha\beta)}$. The total number of $(\alpha\beta)$-faces is $F_{\alpha\beta}(\gB)=B_{(\alpha\beta)}$. The total number of faces is $F(\gB) = \sum_{\alpha<\beta}F_{\alpha\beta}(\gB)$.

For the matrix models studied in Chapter \ref{chap:Matrix}, we deal with $2$-bubbles and $3$-bubbles whereas in Chapter \ref{chap:Tensor} for tensor models and Chapter \ref{chap:MatrixTensor} for matrix-tensor models, we use general $d$-bubbles.

\section{Embedded graphs}
\label{sec:App2}

As explained in the previous section, graphs are abstract mathematical objects made of vertices and edges. When one is lead to draw a given graph on the plane or on a general surface, we pointed out that there are in general many ways to do so. Thus, additional data is required to distinguish the drawings of a graph on a surface. This is part of the subject of graphs embedded on surfaces. This broad and rich topic provides an important interplay between discrete mathematics, which includes graph theory and combinatorics, and continuous mathematics, which includes the theory of Riemann surfaces. In the context of this thesis, graphs embedded on surfaces play an important role, especially in matrix models and in (matrix-)tensor models.

The aim of this section is to provide an overview of graphs embedded on surfaces. Many results and equivalences are stated with intuitive explanations and pedagogical examples rather than with rigorous proofs, which can be found in, for example, \cite{MohThom,EllMof}. The important concepts that are used throughout this thesis are clearly explained. The emphasis is on the combinatorial properties of surfaces through the use of embedded graphs. The section is divided in four parts. The first three outline three equivalent methods for describing graphs embedded on surfaces. The fourth part defines the notion of genus for graphs embedded on surfaces.

For the sake of clarity, we deal in this section with compact, closed 2-dimensional surfaces.\footnote{The results can be extended to surfaces with boundary, see \cite{MohThom,EllMof}} Surfaces are considered up to homeomorphism, which is why we can say ``the sphere" instead of ``a sphere". Unless otherwise stated, the surfaces are connected and can be \emph{orientable} or \emph{non-orientable}. An important topological invariant of a surface $S$ is its \emph{genus} $g(S)$, which is in $\mathbb{N}$ for orientable surfaces and in $\frac{1}{2}\mathbb{N}$ for non-orientable surfaces.\footnote{In this thesis, we use a different convention for the definition of the genus of non-orientable surfaces than most references on the subject. Usually, non-orientable surfaces are given a non-orientable genus which is in $\mathbb{N}$. However, the factor two between the two conventions is not relevant.} Genus together with orientability provides a complete classification of closed compact surfaces. Examples of such surfaces include the sphere $\mathbb{S}^2$ (genus $0$) and the torus $\mathbb{T}^2$ (genus $1$), which are orientable, as well as the real projective plane $\mathbb{R}\text{P}^2$ (genus $\frac{1}{2}$) and the Klein bottle $\mathbb{K}^2$ (genus $1$), which are non-orientable.

\subsection{Cellularly embedded graphs}
\label{sec:AppA21}

The main idea of graphs embedded on surfaces is that we want to decompose a surface into fundamental pieces that can be seen as the faces of a graph drawn in a given way on the surface. A \emph{cellularly embedded graph} $\cG=(\cV(\cG),\cE(\cG))\subset S$ is a graph drawn on a surface $S$ such that the vertices correspond to distinct points in $S$, the edges are simple curves in $S$ joining their endvertices, the edges can only cross each other at vertices and each connected component of $S\setminus \cG$ is homeomorphic to an open disk. The connected components of $S\setminus \cG$ are called the \emph{faces} of the cellularly embedded graph $\cG$.\footnote{In this thesis, we follow the terminology of \cite{EllMof} and we will loosely use the term embedded graph to mean cellularly embedded graph, whereas the former formally does not require the faces to be disks.} 

An illustration of an embedded graph on the torus $\mathbb{T}^2$ is given in Figure \ref{fig:EmbGraphsa}. The complement of the graph in the surface corresponds to a disjoint union of disks. Thus, it means that embedding a graph on a surface amounts to describing a combinatorial way to obtain this surface by gluing disks together. In other words, it provides a notion of discretized surface. We emphasize that the additional data associated with an embedded graph compared to an abstract graph is the notion of faces.

If a graph is embedded on an orientable (resp.\ non-orientable) surface, we refer to it as an orientable (resp.\ non-orientable) embedded graph.

\subsection{Ribbon graphs}
\label{sec:AppA22}

An equivalent description of graphs embedded on surfaces is given by \emph{ribbon graphs}. This is the representation that is mostly used in the theoretical physics literature.

\begin{figure}
    \centering
    \hspace{1cm}
    \begin{subfigure}[b]{0.4\textwidth}
        \includegraphics[]{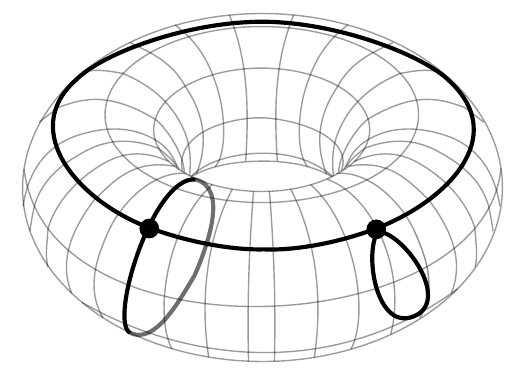}
        \caption{Graph $\cG$ embedded on the torus $\mathbb{T}^2$ in the cellularly embedded graph representation. The two faces of $\cG$ are open disks.}
        \label{fig:EmbGraphsa}
    \end{subfigure}
    \hspace{1cm} %add desired spacing between images, e. g. ~, \quad, \qquad, \hfill etc. 
    \begin{subfigure}[b]{0.45\textwidth}
        \includegraphics[]{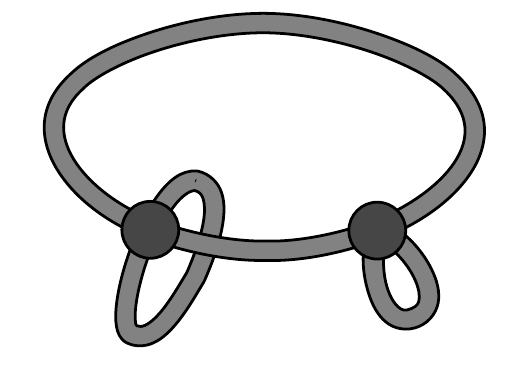}
        \caption{Same embedded graph $\cG$ in the ribbon graph representation, made of vertex disks and edge ribbons. The two faces of $\cG$ correspond to its two boundary components.}
        \label{fig:EmbGraphsb}
    \end{subfigure}
    \\
    \vspace{1cm}
    \hspace{1cm}
    \begin{subfigure}[b]{0.4\textwidth}
        \includegraphics[]{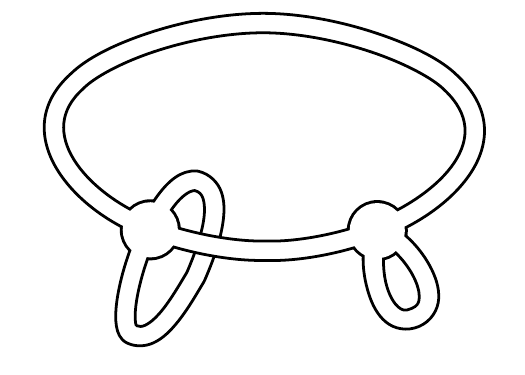}
        \caption{Embedded graph $\cG$ in the ribbon graph representation without the internal structure. The faces of $\cG$ can be viewed as closed curves.}
        \label{fig:EmbGraphsc}
    \end{subfigure}
    \hspace{1cm} %add desired spacing between images, e. g. ~, \quad, \qquad, \hfill etc. 
    \begin{subfigure}[b]{0.45\textwidth}
        \includegraphics[]{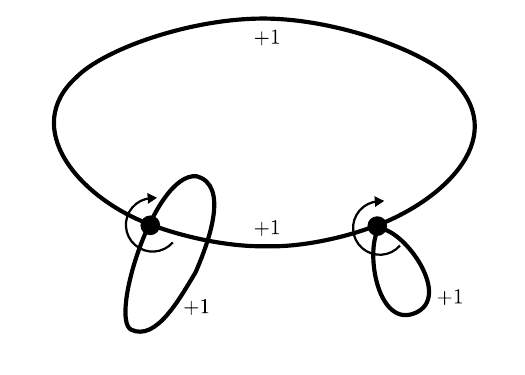}
        \caption{Embedded graph $\cG$ in the signed rotation system representation. To each vertex is assigned a cyclic ordering and to each edge a signature.}
        \label{fig:EmbGraphsd}
    \end{subfigure}
    \\
    \vspace{1cm}
    \hspace{1cm}
    \begin{subfigure}[b]{0.4\textwidth}
        \includegraphics[]{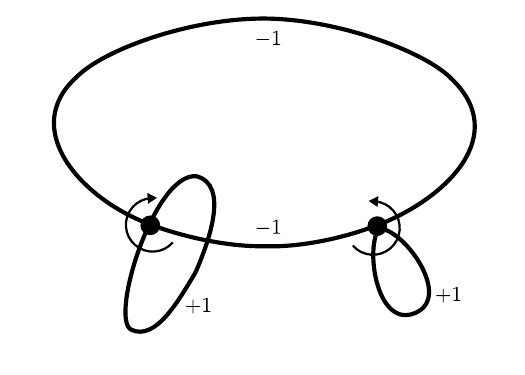}
        \caption{Illustration of a local switch on the vertex on the right. This corresponds to reversing the cyclic ordering around the vertex and flipping the signature of the incident edges.}
        \label{fig:EmbGraphse}
    \end{subfigure}
    \hspace{1cm} %add desired spacing between images, e. g. ~, \quad, \qquad, \hfill etc. 
    \begin{subfigure}[b]{0.45\textwidth}
        \includegraphics[]{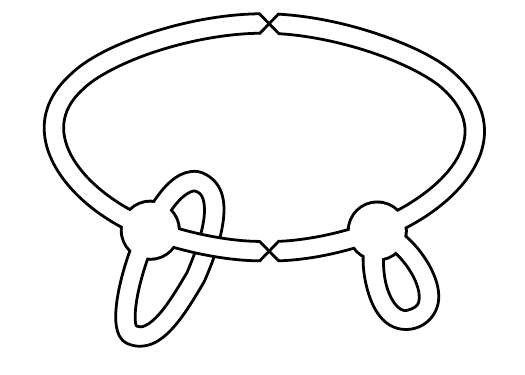}
        \caption{Ribbon graph associated to the embedded graph in (e) (which is equivalent to $\cG$). Edges with signature $-1$ correspond to twisted ribbons in the ribbon graph representation.}
        \label{fig:EmbGraphsf}
    \end{subfigure}
    \caption{Illustration of a graph $\cG$ embedded on the torus $\mathbb{T}^2$ in different representations.}\label{fig:EmbGraphs}
\end{figure}

A straightforward way to describe a ribbon graph is by starting with a cellularly embedded graph $\cG\subset S$ and taking a small neighbourhood of $\cG$ in $S$. More explicitly, we attach disjoint disks on the vertices of $\cG\subset S$, one for each vertex, then we attach rectangular strips (or ribbons) to the disks, with one strip for each edge. Finally, we cut off the remaining of $S$. This is illustrated in Figure \ref{fig:EmbGraphsb}.

Hence, a ribbon graph $\cG=(\cV(\cG),\cE(\cG))$ is a surface with boundary, seen as the union of disks representing the vertices, connected together with ribbons representing the edges. Note that ribbons only intersect the disks that correspond to their endvertices. In particular, they do not intersect each other (i.e.\ they should be viewed in $\mathbb{R}^3$). Finally, nothing restricts ribbons to be twisted, see for instance Figure \ref{fig:EmbGraphsf}.

To obtain a cellularly embedded graph from a ribbon graph, we perform the reverse operation. We first sew disks into each boundary component of the ribbon graph and then we contract the disks into points, which correspond to the vertices, and the ribbons into simple curves, which correspond to the edges. 

From the above, it is clear that the faces of a cellularly embedded graph correspond to the boundary components of the ribbon graph, which are therefore also called the \emph{faces} of the ribbon graph. If we forget the internal structure of the disks and the ribbons, the faces can also be viewed as closed curves, see Figure \ref{fig:EmbGraphsc}.

Ribbon graphs constitute a useful representation of graphs embedded on surfaces in some instances. In particular, it is easy to visualize the faces of a ribbon graph. Further, the connection between ribbon graphs and the Feynman graphs of matrix models in the stranded representation is straightforward (see Section \ref{sec:MatStrandRep}). It is one of the reasons why the ribbon graph representation is often used in theoretical physics.

\subsection{Signed rotation system}
\label{sec:AppA23}

A third equivalent description of graphs embedded on surfaces involves purely combinatorial data without direct reference to the surfaces. It is based on \emph{signed rotation systems}, which can be traced back to works of Heffter \cite{Heffter} and Edmonds \cite{Edmonds}. This description is useful in some instances such as counting the number of graphs that can be embedded on a given surface.

We focus for a moment on graphs embedded on orientable surfaces. In this case, one can convince himself that the combinatorial data required to distinguish between distinct embedded graphs correspond to the \emph{cyclic ordering} of the edges incident to each vertex. In other words, an orientable embedded graph is equivalent to an abstract graph $\cG$ together with a cyclic permutation $\sigma_\nu$ of the edges incident to each vertex $\nu \in \cV(\cG)$. In particular, if an edge $e$ of $\cG$ is incident to $\nu$, $\sigma_\nu(e)$ is the successor of $e$ in the clockwise ordering around $\nu$. The set $\sigma=\{ \sigma_\nu | \nu\in\cV(\cG)\}$ is called the \emph{rotation system} of the embedded graph $\cG$.

Now, if we return to graphs embedded on general surfaces, the distinct embedded graphs are labeled by a choice of cyclic ordering of the edges incident to each vertex together with a choice of \emph{signature} $+1$ or $-1$ on each edge. The signature defines a map $\lambda: \cE(\cG) \rightarrow \{+1,-1\}$ and the pair $\Pi=(\sigma,\lambda)$ is called the \emph{signed rotation system} of the embedded graph. 

The notion of signed rotation system is best understood by making the link with the ribbon graph representation. Given a ribbon graph, we choose an orientation for each vertex disk. This defines a cyclic ordering of the edges incident to each vertex for the underlying abstract graph (obtained by contracting the vertex disks and the edge ribbons). Then, we assign a signature $+1$ to each edge that corresponds to an untwisted edge ribbon and $-1$ to each edge corresponding to a twisted edge ribbon. This is illustrated in Figure \ref{fig:EmbGraphsd}.

Two signed rotation systems are \emph{equivalent}, in the sense that they describe the same embedded graph, if they can be obtained from one another by reversing the cyclic ordering of the incident edges at a given vertex and at the same time by flipping the signatures of all these incident edges. This operation is called a \emph{local switch}. Signed rotation systems are defined up to a sequence of local switches. This is illustrated in Figure \ref{fig:EmbGraphse}, where a local switch has been implemented on the vertex on the right. We also give the corresponding ribbon graph in Figure \ref{fig:EmbGraphsf} to make explicit the equivalence between the two representations.

Finally, we state two important results that make a distinction between orientable and non-orientable embedded graphs in the signed rotation system representation. These results can be easily translated into the ribbon graph representation. First, an orientable embedded graph can always be represented, using a sequence of local switches, with a signed rotation system which gives a signature $+1$ on all the edges. In the ribbon graph language, it means with untwisted edge ribbons only. Second, if an embedded graph has at least one cycle with an odd number of edges with signature $-1$, then it is non-orientable.

\subsection{Genus of embedded graphs}
\label{sec:AppA24}

In the previous parts, we defined three equivalent ways of describing graphs embedded on surfaces. We also pointed out how to go from one representation to another. Hence, in the following and throughout this thesis, we use the three representations interchangeably and we refer to them collectively as embedded graphs. 

The genus $g(\cG)$ of an embedded graph $\cG\subset S$ with $V$ vertices, $E$ edges and $F$ faces corresponds to the genus $g(S)$ of the surface $S$ on which it is embedded. For instance, an orientable embedded graph $\cG\subset S$ has genus zero if $S=\mathbb{S}^2$ (in this case, we say that $\cG$ is a planar graph), genus one if $S=\mathbb{T}^2$, etc. The same holds for non-orientable surfaces. The genus of an embedded graph $\cG$ can be readily computed using the well-known Euler's formula
\be\label{eq:EulerFormulaConEmbGraphs}
2-2g(\cG)=V(\cG) - E(\cG) + F(\cG) \, .
\ee

Up to now, we only dealt with connected surfaces and connected graphs (connected graphs are embedded on connected surfaces). All the results generalize straightforwardly to graphs and surfaces with several connected components, by studying each connected component separately. In this case, we define the genus (of an embedded graph or of a surface) to be the sum of the genera of the connected components. If we use this definition for the genus, Euler's formula must be modified to
\be\label{eq:EulerFormulaEmbGraphs}
2c(\cG)-2g(\cG)=V(\cG) - E(\cG) + F(\cG) \, ,
\ee
where $c(\cG)$ is the number of connected components of $\cG$.

%%%%%%%%%%%%%%%%%%%%%%%%%%%%%%%%%%%%%%%%%%%%%%%%
%%%%%%%%%%%%%%%%%%%%%%%%%%%%%%%%%%%%%%%%%%%%%%%%

\chapter{Classification of PCGMs}
\label{app:AppC}

In this appendix, we provide the proof of the classification of the prime-complete generalized melons (PCGMs), which are the generalized melons for the tensor model based on the prime-complete interaction bubble $\cK_{R+1}$ with $R$ prime. The model is described in Section \ref{sec:CompleteInt} and this part corresponds to Section 4 in Ref.\ \cite{Ref2}. 

\

\noindent\emph{Notations}\\ Labeled vertices belonging to an interaction bubble are denoted in brackets, like $[\nu_{1}], [\nu_{2}]$, etc. When several interaction bubbles are present, an upper index may be added to distinguish between the bubbles, like $[\nu_{1}]^{I}$, $[\nu_{1}]^{II}$, etc. A path going successively through the vertices $[\nu_{1}],[\nu_{2}],\ldots,[\nu_{q}]$ is denoted as $[\nu_{1}][\nu_{2}]\cdots[\nu_{q}]$. This is unambiguous inside interaction bubbles, which have at most one edge joining two given vertices. In other cases, a possible ambiguity is waived by specifying the edge colors. The path is oriented if we distinguish between $[\nu_{1}][\nu_{2}]\cdots[\nu_{q}]$ and $[\nu_{q}][\nu_{q-1}]\cdots [\nu_{1}]$. A path $[\nu_{1}][\nu_{2}]$ is an edge. 

Equality in $\mathbb Z/R\mathbb Z$, i.e.\ equality modulo $R$, is denoted as $\equiv$. When $R$ is prime, $\mathbb Z/R\mathbb Z$ is a field in the algebraic sense and the inverse of an element $x$ is denoted by $x^{-1}$; for example, $2^{-1}\equiv\frac{1}{2}(R+1)$.

\section{Properties of the complete graph and its edge-coloring}

\subsection{The MST condition}
The complete interaction bubble $\cK_{R+1}$ is obtained from the complete graph on $\smash{R+1}$ vertices together with the standard coloring described in Section \ref{sec:CompleteInt}. The following proposition shows that the cases $R$ prime and $R$ not prime are qualitatively different. 

\begin{proposition} \label{prime}
The $R$-bubble $\gK_{R+1}$ is maximally single-trace (MST) if and only if $R$ is a prime number.
\end{proposition}

\proof
For any pair of two colors $(i,j)$, let us consider the $(ij)$-face that goes through the center vertex $[C]$. Let us denote its length as $2q$, which is the even integer defined to be the number of its edges of colors $i$ and $j$.  Using the rule for the edge-coloring of the complete bubble, we can explicitly write this face, starting from the center vertex $[C]$ with the edge of color $j$, as $[C][k_1\equiv j][k_2]\cdots [k_{2q-1}\equiv i][C]$ and check inductively that $k_{2p}\equiv 2pi-(2p-1)j$ and $k_{2p+1}\equiv (2p+1)j - 2pi$ for $1\leq p\leq q-1$. Therefore, $k_{2q-1}\equiv i$ is equivalent to $(2q-1)(j-i) \equiv 0$. 

If $R$ is a prime number, this implies $2q-1 \equiv 0$ because $j-i \not\equiv 0$. The length is the smallest possible solution, that is, $2q=R+1$. Our $(ij)$-face must then visit all the vertices of $\gK_{R+1}$ and is thus unique: $F_{ij}=1$ for all pairs $(i,j)$ and the bubble is MST. 

If $R$ is not prime, write $R=R_{1}R_{2}$ where $R_{1}$ and $R_{2}$ are odd integers with $1<R_{1}<R$ and $1<R_{2}<R$. Moreover, set $j-i=R_{2}$. The smallest solution to $(2q-1)(j-i) \equiv 0$ is then $2q=R_{1}+1<R+1$. This implies that there are vertices in $\gK_{R+1}$ that are not visited by our $(ij)$-face, which therefore cannot be unique: $F_{ij}>1$ and the bubble is not MST.\qed

\

For instance, when $R=9$, there are two $(14)$-faces, namely $[C][1][7][4][C]$ of length four going through the center and $[2][6][5][3][8][9][2]$ of length six.

The bubbles $\gK_{R+1}$ with $R$ prime are called \emph{prime-complete}. These bubbles have convenient \emph{$(ij)$-polygonal} representations in the shape of an $(R+1)$-sided polygon whose boundary is the unique $(ij)$-face. This is illustrated on the right of Figure \ref{figureA} for $\mathcal K_6$.

\subsection{\label{disbubbleSec}Distinguishing edges and vertices}

As explained in Section \ref{sec:CompleteInt}, when $R>3$, the edges of a given color in the prime-complete interaction bubble $\cK_{R+1}$ are all inequivalent and the same holds for the vertices. In this section, we introduce an elegant way to distinguish between these edges and vertices.

Consider an oriented edge $[\nu_{1}][\nu_{2}]$ of color $k$. For any ordered pair of distinct colors $(i,j)$, there exists a unique $(ij)$-path, i.e.\ a path of alternating colors $i$ and $j$, that starts at $[\nu_{1}]$ with an edge of color $i$ and ends at $[\nu_{2}]$. The existence of this path is ensured by the fact that the unique $(ij)$-face visits all the vertices of the prime-complete bubble. If $\ell$ is the length of the path, defined to be the number of its edges of colors $i$ and $j$, we then say that the ordered pair of colors $(i,j)$ \emph{indexes the oriented edge $[\nu_{1}][\nu_{2}]$ at length $\ell$}. Unoriented edges can also be indexed by unordered pairs in an obvious way. The indexing enjoys the following properties.

\begin{lemma}\label{simpleindex} The edge $[\nu_{1}][\nu_{2}]$ is indexed by $(i,j)$ at length $\ell$ if and only if is it indexed by $(j,i)$ at length $\ell'=R+1-\ell$. The edge $[\nu_{1}][\nu_{2}]$ is indexed by $(i,j)$ at even length $\ell$ if and only if the edge $[\nu_{2}][\nu_{1}]$ is indexed by $(j,i)$ at even length $\ell$. The edge $[\nu_{1}][\nu_{2}]$ is indexed by $(i,j)$ at odd length $\ell$ if and only if the edge $[\nu_{2}][\nu_{1}]$ is indexed by $(i,j)$ at odd length $\ell$.\end{lemma}

\proof Trivial by following the $(ij)$-face.\qed

\

We say that two oriented edges of the same color are \emph{weakly equivalent} if they are indexed by the same set of pairs of colors at length two. Otherwise, they are \emph{strongly inequivalent}. 

At length one, any edge of color $k$ is obviously indexed by the $R-1$ pairs $(k,i)$ for all $i\not = k$. For $R=3$, any edge of color $k$ is indexed at length two by the two ordered pairs $(i,j)$ and $(j,i)$ of complementary colors. For $R>3$, the computation of the pairs of colors indexing an arbitrary edge at any length $\ell\geq 2$ is a straightforward exercise. The results at lengths two and three are summarized by the following lemma.

\begin{lemma} \label{lengthtwolemma} Consider the bubble $\mathcal K_{R+1}$ for $R$ prime and $R>3$. We use the standard vertex labeling and we name the colors by integers modulo $R$.

The edge $[C][k]$, of color $k$, is indexed at length two by the pairs of colors $(i,j)$ with $j\equiv \frac{1}{2}(R+1)(k+i)$, for all $i\not\equiv k$. This yields $R-1$ distinct pairs. It is indexed at length three by the pairs of colors $(i,j)$ with $j\equiv \frac{1}{2}(R+1)(3i-k)$, for all $i\not\equiv k$. This yields $R-1$ distinct ordered pairs.

The edge $[k-p][k+p]$, for any $1\leq p\leq \frac{1}{2}(R-1)$, of color $k$, is indexed at length two by the pairs of colors $(k-p,k+p)$ and $(i,i+p)$ for all $i$ different from $k$ and $k-p$. This yields a total of $R-1$ distinct pairs. It is indexed at length three by the pairs of colors $(k-p,k-3p)$, $(k+p, k+3p)$ and $(i,2i-k)$ for all $i$ different from $k$, $k-p$ and $k+p$. This yields $R-1$ distinct ordered pairs.
\end{lemma} 

\proof
The results follow from a direct computation using the rule of the edge-coloring. For example, the edge of color $i\not\equiv k$ attached to $[C]$ is $[C][i]$. The edge $[i][k]$ is then of color $j$ with $i+k\equiv 2j$, which proves the first statement since $2^{-1}\equiv\frac{1}{2}(R+1)$ in $\mathbb Z/R\mathbb Z$. At length three, one again considers the edge $[C][i]$ of any color $i\not\equiv k$ and the edge of color $i$ starting from $[k]$, which is $[k][2i-k]$. The color of the edge $[i][2i-k]$ is then $j\equiv 2^{-1}(3i-k)$, which proves the second statement. The results for the edges $[k-p][k+p]$ follow from a similar analysis.\qed

\

From Lemma \ref{lengthtwolemma}, we can prove two useful results.

\begin{proposition} \label{inequivalent}
For $R$ prime and $R>3$, two distinct oriented edges of the same color in $\gK_{R+1}$ are always strongly inequivalent. Equivalently, two weakly equivalent edges necessarily coincide. 
\end{proposition} 

\proof Colors are defined modulo $R$. Lemma \ref{simpleindex} implies that if $(i,j)$ indexes an edge $[\nu_{1}][\nu_{2}]$ at length two, then $(j,i)$ indexes $[\nu_{2}][\nu_{1}]$ at lenght two but does not index $[\nu_{1}][\nu_{2}]$ at length two as soon as $R>3$. Thus $[\nu_{1}][\nu_{2}]$ and $[\nu_{2}][\nu_{1}]$ are strongly inequivalent. Moreover, using Lemma \ref{lengthtwolemma}, one can check that, for any $k$, the pair of colors $(k-p,k+p)$ indexes the edge $[k-p][k+p]$ at length two for all $p\not\equiv 0$ but does not index the edge $[C][k]$ at length two; and for all $p'\not\equiv p$, $p\not\equiv 0$, $p'\not\equiv 0$, the pair of colors $(k+\frac{1}{2}(R+1)(p'-p),k+\frac{1}{2}(R+1)(p'+p))$ indexes the edge $[k-p][k+p]$ at length two but does not index the edge $[k-p'][k+p']$ at length two.\qed

\

We thus see that two distinct oriented edges of a given color in the prime-complete bubble can be unambiguously distinguished from one another using the coloring of the graph. The same is then automatically true for the vertices, since two distinct vertices $[\nu_{1}]$ and $[\nu_{2}]$ are the endpoints of two distinct oriented edges $[\nu_{1}][\nu_{2}]$ and $[\nu_{2}][\nu_{1}]$.

\begin{lemma} \label{lengththree}
For $R$ prime and $R>3$, choose two distinct unoriented edges $[\nu_{1}][\nu_{2}]$ and $[\nu_{3}][\nu_{4}]$ (i.e.\ such that $[\nu_{1}][\nu_{2}]\not = [\nu_{3}][\nu_{4}]$ and $[\nu_{1}][\nu_{2}]\not = [\nu_{4}][\nu_{3}]$) of the same color $k$. One can then always find a pair of colors $(i,j)$ indexing one of the edge at length two and the other at length three. Note that, of course, $i\not = k$ and $j\not = k$.
\end{lemma} 

\proof
If the two distinct edges of color $k$ are $[C][k]$ and $[k-p][k+p]$ for some $1\leq p\leq \frac{1}{2}(R-1)$, one considers the pair of colors $(k+2p,k+3p)$. Using Lemmas \ref{simpleindex} and \ref{lengthtwolemma}, it is straightforward to check that it indexes $[k-p][k+p]$ at length two and both $[C][k]$ and $[k][C]$ at length three. Similarly, $(k-2p,k-3p)$ indexes $[k+p][k-p]$ at length two and both $[C][k]$ and $[k][C]$ at length three. If the two distinct edges of color $k$ are $[k-p][k+p]$ and $[k-p'][k+p']$ for some $1\leq p,p'\leq \frac{1}{2}(R-1)$, $p\not = p'$, one considers the pair of colors $(k+p',k+2p')$. Using again Lemmas \ref{simpleindex} and \ref{lengthtwolemma}, we see that it indexes $[k-p'][k+p']$ at length two and both $[k-p][k+p]$ and $[k+p][k-p]$ at length three. Similarly, $(k-p',k-2p')$ indexes $[k+p'][k-p']$ at length two and both $[k-p][k+p]$ and $[k+p][k-p]$ at length three.\qed

\section{Action and index}

See Section \ref{sec:CompleteInt}. 

\section{The classification theorem}
In this section, we explicitly solve the condition \eqref{leadinggraphs}, or equivalently \eqref{leadg}, for a Feynman graph to be a PCGM and we provide a full description of all the PCGMs.

\subsection{Useful tools}

The lemma below is used repeatedly in the following.
\begin{lemma}\label{cyclelemma} A planar 3-bubble cannot have cycles of odd length. 
\end{lemma}

\proof This is a direct consequence of a more general result, explained in Section \ref{sec:DBubbles}, which states that the underlying graph of an orientable bubble is bipartite, together with the well-known facts that a planar graph is orientable and that a graph is bipartite if and only if it does not contain cycles of odd length.\qed

\

We shall also need standard results on the deletion of edges and vertices from a planar ribbon graph. The edge deletion is defined in the trivial way, maintaining the cyclic ordering of the remaining edges around vertices. The vertex deletion is defined only for vertices of valency two. It simply amounts to replacing the two ribbons attached to the vertex by a unique ribbon, twisted if precisely one of the two original ribbons is twisted or untwisted otherwise. It is useful to introduce the following terminology \cite{MohThom}: an edge of a ribbon graph is called \emph{regular} if it belongs to two distinct faces and it is called \emph{singular} otherwise; in other words, the borders of the ribbon associated with a regular edge are on two distinct faces, whereas they are on the same face in the case of a singular edge.
\begin{lemma}\label{oplemma} i) (Edge deletion) If one deletes a regular edge from a connected planar ribbon graph, one gets another connected planar ribbon graph. If one deletes a singular edge from a connected planar ribbon graph, one gets a ribbon graph with two planar connected components.\\
ii) (Vertex deletion) If one deletes a vertex of valency two from a connected planar ribbon graph, one gets a connected planar ribbon graph.
\end{lemma}
The claims i) are special cases of standard results on edge deletions for ribbon graphs of arbitrary genus. The proof, which is elementary, will not be included here, see e.g.\ \cite{MohThom}. The claim ii) is trivial to check.

\subsection{Results on faces}
\begin{lemma}\label{revisitlemma} A $(0i)$-face in a PCGM cannot visit a given interaction bubble more than once. Equivalently, two distinct edges of color $i$ of a $(0i)$-face in a PCGM belong to two distinct interaction bubbles. 
\end{lemma}
\proof Let $\gB$ be a PCGM and assume that there exists a $(0i)$-face $$[\nu_{1}][\nu_{2}][\nu_{3}]\cdots [\nu_{2p-1}][\nu_{2p}]\cdots [\nu_{2q}][\nu_{1}]$$ such that $[\nu_{1}][\nu_{2}]$ and $[\nu_{2p-1}][\nu_{2p}]$ are two edges of color $i$ belonging to the same interaction bubble. Since this bubble is a complete graph, there exists an edge $[\nu_{2p-1}][\nu_{1}]$ of color $j\not = i$. The path $[\nu_{1}][\nu_2]\cdots [\nu_{2p-1}][\nu_{1}]$ is then a cycle of odd length in $\gB_{(0ij)}$. Using Lemma \ref{cyclelemma}, this contradicts the PCGM condition \eqref{leadg}.\qed

\

We denote by $F_{\ell/2}$ the number of $(0i)$-faces of length $\ell$, for any color $i$, where the length of a $(0i)$-face is defined as usual to be the total number of its edges, which is twice the number of edges of color 0. A \emph{self-contraction} is an edge of color 0 attached to two vertices of the same interaction bubble.
\begin{lemma}\label{onefacelemma} A PCGM does not have self-contractions. In particular, $F_{1}=0$.\end{lemma}
\proof Let us assume that the PCGM $\gB$ has a self-contraction and denote by $[\nu_{1}]$ and $[\nu_{2}]$ the two endpoints of the corresponding edge of color 0. Since $[\nu_{1}]$ and $[\nu_{2}]$ belong to the same interaction bubble, there is an edge $[\nu_{1}][\nu_{2}]$ of some color $k$ in this interaction bubble. Choose a pair of colors $(i,j)$ that indexes $[\nu_{1}][\nu_{2}]$ at length two and thus forms a triangle with that edge and a third vertex, say $[\nu_{3}]$. The three-bubble $\gB_{(0ij)}$ then has a cycle of length three, namely $[\nu_{1}][\nu_{3}][\nu_{2}][\nu_{1}]$, which, using Lemma \ref{cyclelemma}, contradicts \eqref{leadg}. The result $F_{1}=0$ immediately follows because the edge of color $0$ in a $(0i)$-face of length two is automatically a self-contraction.

Note that the result can also be obtained immediately from Lemma \ref{revisitlemma}, by considering, for any color $i\not = k$, the $(0i)$-face passing through the vertices $[\nu_{1}]$ and $[\nu_{2}]$.\qed

\

\begin{lemma} \label{lengthface} A PCGM has $F_{2}\geq\frac{1}{2}R(R+1)$.\end{lemma}
\proof By definition, we have
\begin{equation}
\sum_i F_{0i} = \sum_{n\geq 1} F_{n} \, .
\label{Fldef}
\end{equation}
Moreover, each edge of color $0$ belongs to $R$ different $(0i)$-faces, one for each color $i$. The number of edges of color $0$ is the number of propagators $p$ and thus we have
\begin{equation}
\sum_{n\geq 1} n F_{n} = Rp = \frac{R(R+1)}{2} v \, ,
\label{Flrelation}
\end{equation}
where we have used $2p=(R+1)v$ since our Feynman graphs are $(R+1)$-regular. Using  \eqref{leadinggraphs}, \eqref{Fldef} and \eqref{Flrelation} combined with $F_{1}=0$ from Lemma \ref{onefacelemma}, we get
\begin{equation} \label{F2bound}
F_2  = \frac{1}{2}R(R+1) + \frac{1}{4} \sum_{n\geq 3}\Bigl[(R-1)n - 2R-2 \Bigr] F_{n} \, .
\end{equation}
If $R\geq 5$, the second term on the right-hand side is non-negative, which yields the inequality $\smash{F_2 \geq \frac{1}{2} R(R+1)}$. If $R=3$ we further show that $F_{3}=0$ (see also \cite{CT} for this case). Indeed, assume that one has a face of length six, say of colors $0$ and $3$. Since there is no self-contraction, this face must visit three distinct interaction bubbles and is thus of the form $[\nu_{1}]^{I}[\nu_{2}]^{I}[\nu_{3}]^{II}[\nu_{4}]^{II}[\nu_{5}]^{III}][\nu_{6}]^{III}[\nu_{1}]^{I}$, where the edges of color 3, namely $[\nu_{1}]^{I}][\nu_{2}]^{I}$, $[\nu_{3}]^{II}[\nu_{4}]^{II}$ and $[\nu_{5}]^{III} [\nu_{6}]^{III}$, belong to three distinct interaction bubbles. Since $R=3$, these three edges are all indexed by the same pair of colors $(1,2)$ at length two, which forms a triangle with each of the edges and third vertices that we call $[\nu_{3/2}]^{I}$, $[\nu_{7/2}]^{II}$ and $[\nu_{11/2}]^{III}$ respectively. The three-colored graph $\gB_{(012)}$ then has a cycle of length nine $[\nu_{1}]^{I}][\nu_{3/2}]^{I}[\nu_{2}]^{I}[\nu_{3}]^{II}[\nu_{7/2}]^{II}[\nu_{4}]^{II}[\nu_{5}]^{III}[\nu_{11/2}]^{III}[\nu_{6}]^{III}[\nu_{1}]^{I}$, contradicting \eqref{leadg} by using Lemma \ref{cyclelemma}.\qed

\

As a result, our generalized melons always have several $(0i)$-faces of length four. The following fundamental lemma fixes the structure of these faces.
\begin{lemma}\label{fourfaces} The $(0k)$-faces of length four in a PCGM are of the form $$[\nu_{1}]^{I}[\nu_{2}]^{I}[\nu_{2}]^{II}[\nu_{1}]^{II}[\nu_{1}]^{I}\, ,$$ where $[\nu_{1}]^{I}[\nu_{2}]^{I}$ and $[\nu_{1}]^{II}[\nu_{2}]^{II}$ are two equivalent oriented edges of color $k$ in two distinct interaction bubbles.
\end{lemma}

\proof In the proof, we use the standard labels for the vertices and in particular, vertices $[\nu]^{I}$ and $[\nu]^{II}$ that have the same label $\nu$ are equivalent vertices in two distinct interaction bubbles $I$ and $II$.

There is nothing to prove if $R=3$. We thus assume that $R>3$ and we consider a $(0k)$-face of length four $[\nu_{1}][\nu_{2}][\nu_{3}][\nu_{4}][\nu_{1}]$ in a PCGM, choosing the edges $[\nu_{1}][\nu_{2}]$ and $[\nu_{3}][\nu_{4}]$ to be of color $k$ and thus the edges $[\nu_{2}][\nu_{3}]$ and $[\nu_{4}][\nu_{1}]$ to be of color 0. From Lemma \ref{revisitlemma}, we know that the edges of color $k$ are in two distinct interaction bubbles; we can thus write $[\nu_{1}]=[\nu_{1}]^{I}$, $[\nu_{2}]=[\nu_{2}]^{I}$, $[\nu_{3}]=[\nu_{3}]^{II}$ and $[\nu_{4}]=[\nu_{4}]^{II}$.

Let us first assume that $[\nu_{1}][\nu_{2}]$ and $[\nu_{3}][\nu_{4}]$ are inequivalent unoriented edges. Using Lemma \ref{lengththree}, we can find $(i,j)$ indexing, e.g., $[\nu_{1}][\nu_{2}]$ at length two and $[\nu_{3}][\nu_{4}]$ at length three. Following the $(ij)$-path of length two joining $[\nu_{1}]^{I}$ to $[\nu_{2}]^{I}$, then the edge of color 0 joining $[\nu_{2}]^{I}$ to $[\nu_{3}]^{II}$, then the $(ij)$-path of length three joining $[\nu_{3}]^{II}$ to $[\nu_{4}]^{II}$ and finally the edge of color 0 joining $[\nu_{4}]^{II}$ to $[\nu_{1}]^{I}$, we get a cycle of length seven in $\gB_{(0ij)}$, contradicting planarity by Lemma \ref{cyclelemma} and thus the PCGM condition \eqref{leadg}.

\begin{figure}
\centerline{\includegraphics[width=6in]{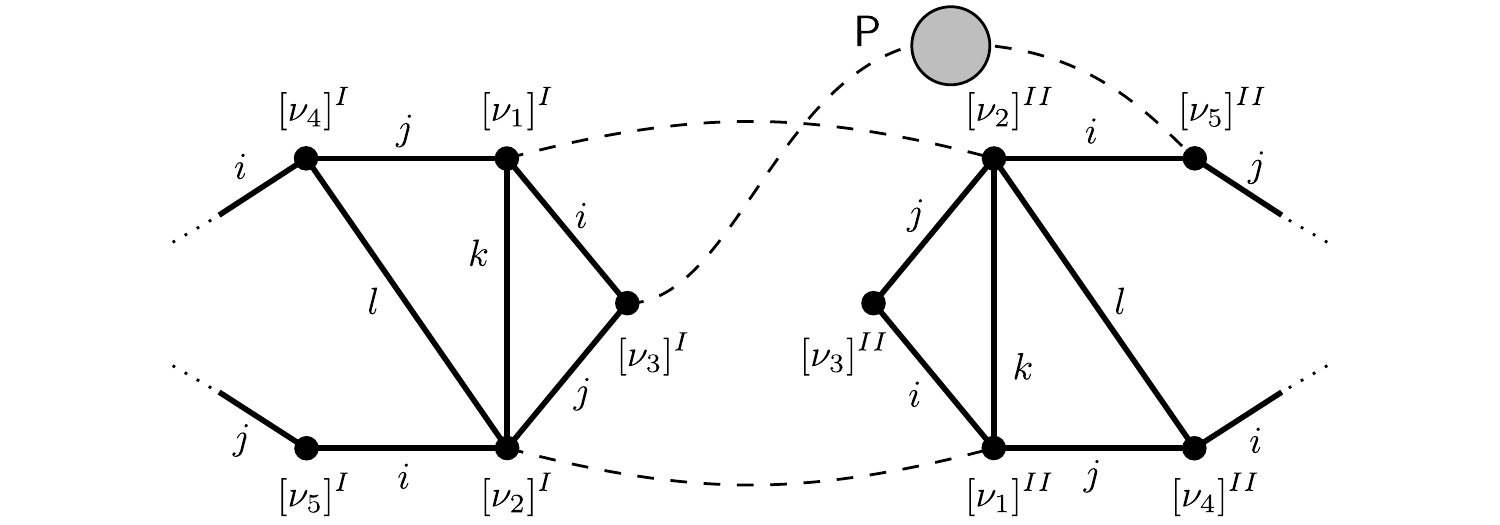}}
\caption{Configuration for which $[\nu_{1}][\nu_{2}]=[\nu_{3}][\nu_{4}]$ and thus $[\nu_{1}]^{I}$ and $[\nu_{2}]^{I}$ are contracted with $[\nu_{2}]^{II}$ and $[\nu_{1}]^{II}$ respectively. Dashed lines represent edges of color 0. Only the relevant parts of the graph are depicted. The $(0i)$-path $\mathsf P$ joining $[\nu_{3}]^{I}$ and $[\nu_{5}]^{II}$ is stylized as a grey disk attached to two edges of color 0. This path is represented in more details in Figure \ref{figproofb}.}\label{figproofa}
\end{figure}
\begin{figure}
\centerline{\includegraphics[width=6in]{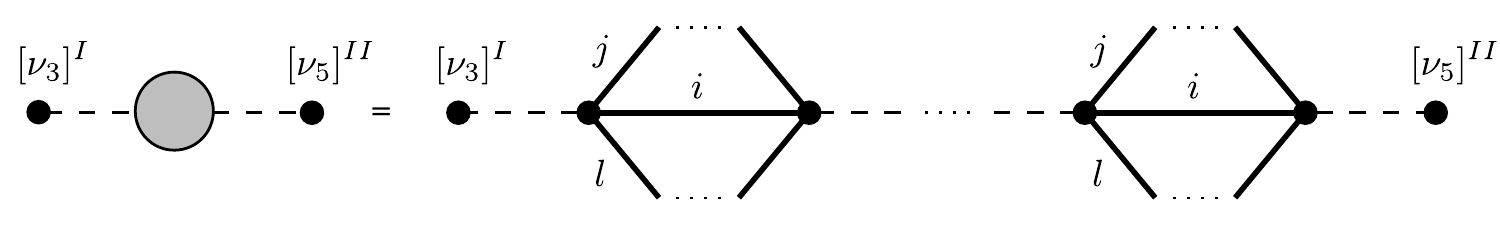}}
\caption{$(0i)$-path $\mathsf P$ joining $[\nu_{3}]^{I}$ to $[\nu_{5}]^{II}$, together with the $(jl)$-faces in the various distinct interaction bubbles visited by the path.}\label{figproofb}
\end{figure}

There remains two possibilities: $[\nu_{1}][\nu_{2}]=[\nu_{4}][\nu_{3}]$, which is what we want to prove, or $[\nu_{1}][\nu_{2}]=[\nu_{3}][\nu_{4}]$. Let us assume that the second possibility is realized, which means that $[\nu_{1}]^{I}$ and $[\nu_{2}]^{I}$ are contracted with $[\nu_{2}]^{II}$ and $[\nu_{1}]^{II}$ respectively. The resulting configuration is depicted in Figure \ref{figproofa}. The important features are as follows. We have indexed the edge $[\nu_{1}][\nu_{2}]$ at length two by the pair of colors $(i,j)$, with associated triangles $[\nu_{1}][\nu_{2}][\nu_{3}]$ in the two interaction bubbles. We have depicted part of the $(ij)$-face in both interaction bubbles, introducing in particular the vertices $[\nu_{4}]$ and $[\nu_{5}]$ and the edge $[\nu_{2}][\nu_{4}]$. The color of this edge is denoted by $l$. Note that $i$, $j$, $k$ and $l$ must be four distinct colors. We have also explicitly depicted the $(0i)$-face that contains the edges $[\nu_{1}]^{I}[\nu_{3}]^{I}$ and $[\nu_{2}]^{II}[\nu_{5}]^{II}$. From Lemma \ref{revisitlemma}, we know that the $(0i)$-path joining $[\nu_{3}]^{I}$ to $[\nu_{5}]^{II}$ along this face, which we call $\mathsf P$, visits distinct interaction bubbles at each intermediate edge of color $i$ (so, for example, the edge $[\nu_{2}]^{I}[\nu_{1}]^{II}$ of color 0 cannot belong to this path). We have represented in more details the path $\mathsf P$ in Figure \ref{figproofb}, indicating as well the $(jl)$-faces in each intermediate interaction bubble visited by the path. 

\begin{figure}
\centerline{\includegraphics[width=6in]{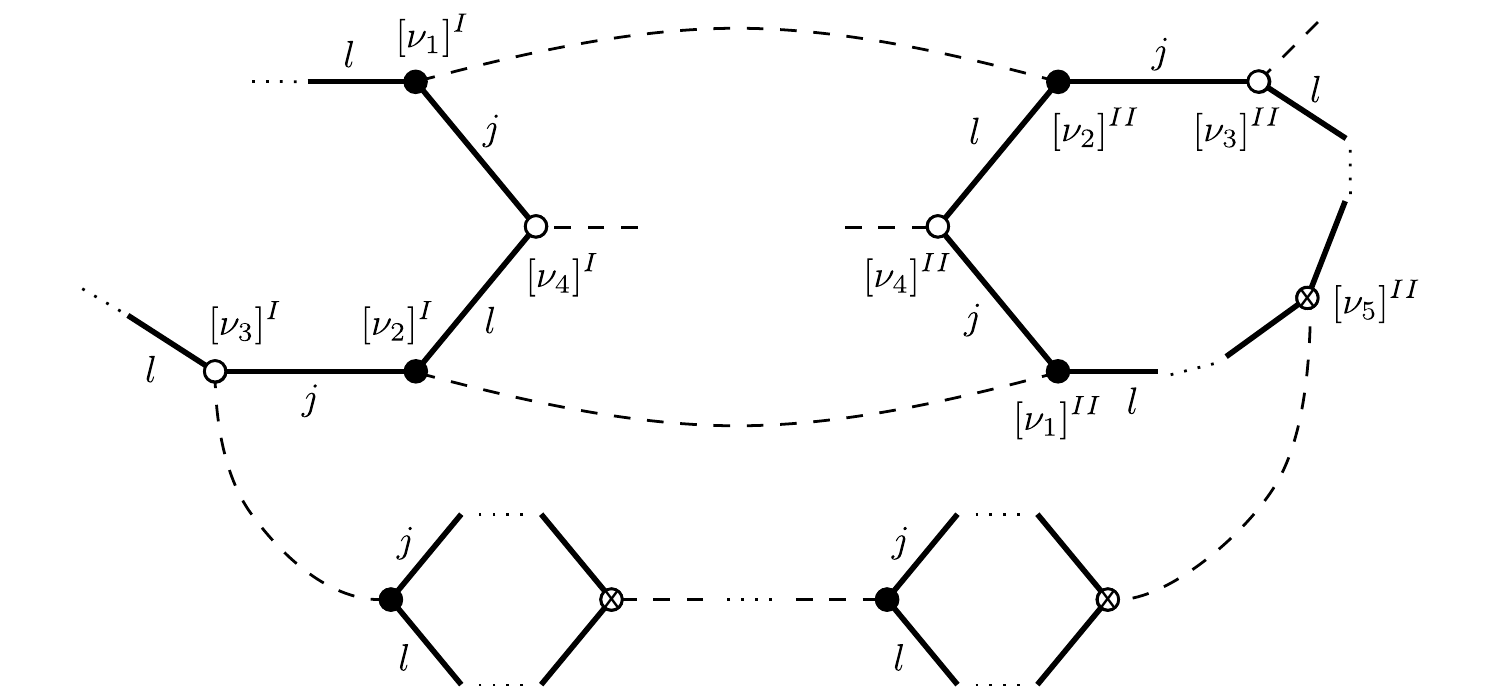}}
\caption{Three-bubble $\gB_{(0jl)}$, deduced from the graphs depicted in Figures \ref{figproofa} and \ref{figproofb}, drawn with a convenient choice of embedding. The type (filled or unfilled) of the vertices indicated with a cross is a priori unknown.}\label{figproofc}
\end{figure}

\begin{figure}
\centerline{\includegraphics[width=6in]{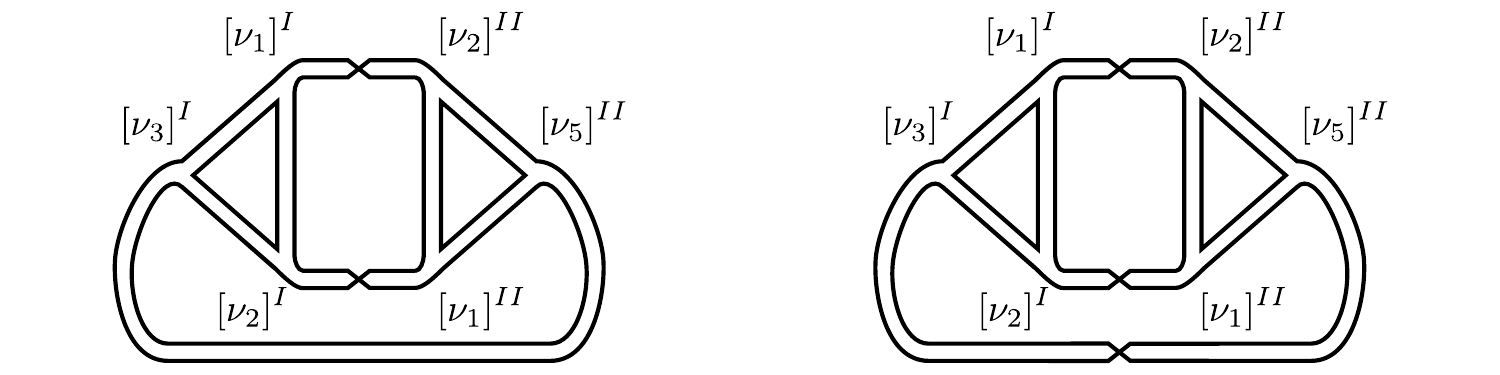}}
\caption{The two possible ribbon graphs obtained from the three-bubble $\gB_{(0jl)}$ depicted in Figure \ref{figproofc} after performing the edge and vertex deletion operations described in the main text. Both have genus one, contradicting planarity.}\label{figproofd}
\end{figure}

Consider now the three-bubble $\gB_{(0jl)}$. A convenient representation, obtained starting from the graphs in Figures \ref{figproofa} and \ref{figproofb}, is given in Figure \ref{figproofc}. Without loss of generality, the embedding is chosen such that the edges of color 0 attached to the polygonal $(jl)$-faces always point outwards. This implies that the ribbons $[\nu_{1}]^{I}[\nu_{2}]^{II}$ and $[\nu_{2}]^{I}[\nu_{1}]^{II}$ are twisted. Because we are in a PCGM, we know that $\gB_{(0jl)}$ must be planar. Let us then use the edge and vertex deletion operations described in Lemma \ref{oplemma} in the following way: we first delete all the edges of color 0, except for $[\nu_{1}]^{I}[\nu_{2}]^{II}$, $[\nu_{2}]^{I}[\nu_{1}]^{II}$ and the edges in the path $\mathsf P$ that join the distinct interaction bubbles within this path; we then delete pieces of each $(jl)$-faces in $\mathsf P$ (the upper or lower parts in each interaction bubble when the path is depicted as in Figure \ref{figproofc}) so that the path is reduced to a succession of ribbons attached to vertices of valency two; finally, we delete all the vertices of valency two. Taking into account the fact that the type of some vertices is a priori unknown in the embedding, these operations can produce one of the two ribbon graphs depicted in Figure \ref{figproofd}. From Lemma \ref{oplemma}, at least one of these graphs must be planar. However, it is easy to check that they both have genus one. Our initial hypothesis, that $[\nu_{1}][\nu_{2}]=[\nu_{3}][\nu_{4}]$, is thus impossible. This concludes the proof.\qed

\subsection{The PCGM with two bubbles} \label{subsectiontwovertices}

A PCGM with $v=1$ cannot exist, because it would necessarily have self-contractions. The simplest PCGMs thus have $v=2$, which is the case considered in the present subsection. We use the standard vertex labeling, with vertices $[C]^{I}$, $[k]^{I}$ and $[C]^{II}$, $[k]^{II}$, $1\leq k\leq R$, for the interaction bubbles number one and two respectively.

\begin{lemma}\label{twoverticeslemma} There exists a unique PCGM with $v=2$, called the elementary generalized melon, corresponding to the symmetric configuration (i.e.\ elementary mirror melon, see Section \ref{sec:MSTInt}) where edges of color 0 join $[C]^{I}$ to $[C]^{II}$ and $[k]^{I}$ to $[k]^{II}$ for all $1\leq k\leq R$. In the case $R=3$, this elementary generalized melon is obtained from four distinct Wick contractions between the two prime-complete interaction bubbles, whereas for $R>3$ it is obtained from a unique Wick contraction.
\end{lemma}

\proof Lemmas \ref{revisitlemma}, \ref{onefacelemma} and \ref{fourfaces} immediately fix the PCGM with $v=2$ to the symmetric configuration. Note that this is of course consistent with \eqref{leadinggraphs} and the inequality $F_{2}\geq \frac{1}{2}R(R+1)$, which predict that a PCGM with $v=2$ has precisely $F_{2}=\frac{1}{2}R(R+1)$ and $F_{n}=0$ if $n\not =2$.

When $R>3$, since all the vertices of the prime-complete bubble are inequivalent, there is clearly a unique Wick-contraction that yields this symmetric configuration. When $R=3$, since all the vertices are equivalent, one can start by Wick-contracting any given vertex of the first bubble with any vertex of the second bubble, yielding four possibilities. One can check that, after this initial choice is made, the other contractions are automatically fixed by the requirement that all the $(0i)$-faces have length four. Because of the special symmetry properties of $\mathcal K_{4}$, the four graphs we get in this way are actually four copies of the same elementary generalized melon.\qed

\

The resulting elementary generalized melon is depicted in Figure \ref{figureB} in the cases $R=3$ (on the left) and $R=5$ (on the right).

\begin{figure}
\centerline{\includegraphics[width=6in]{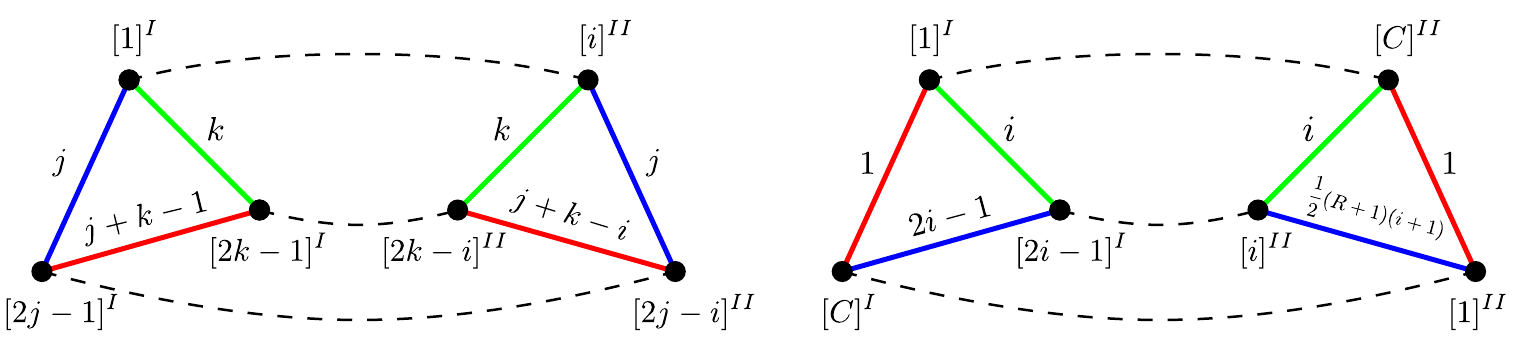}}
\caption{\label{proofLemFig1} Configurations used in the elementary proof of Lemma \ref{twoverticeslemma}. Consistency of the left picture requires $i=1$, whereas the right picture is impossible.}
\end{figure}

For completeness, let us also mention a completely elementary proof of Lemma \ref{twoverticeslemma} that does not use the non-trivial Lemma \ref{fourfaces} but only the fact that all the $(0i)$-faces must be of length four. It goes as follows (see Figure \ref{proofLemFig1}).

Let us first assume that $[1]^{I}$ is Wick-contracted with $[i]^{II}$. We then consider two distinct edges $[1]^{I}[2j-1]^{I}$ and $[1]^{I}[2k-1]^{I}$, which have respectively colors $j$ and $k$ both different from $i$ and 1. This is always possible because $R>3$. The $(0j)$-face containing $[1]^{I}[2j-1]^{I}$ is then of length four if and only if $[2j-1]^{I}$ is contracted with the vertex $[2j-i]^{II}$, so that $[1]^{I}[2j-1]^{I}$ and $[i]^{II}[2j-i]^{II}$ have the same color $j$ (note that $[2j-1]^{I}$ cannot be contracted with the vertex $[C]^{II}$, because by choice $j\not = i$). Similarly, $[2k-1]^{I}$ must be contracted with $[2k-i]^{II}$. The face of colors 0 and $j+k-1$ containing the edge $[2j-1]^{I}[2k-1]^{I}$ is then of length four if and only if the edge $[2j-i]^{II}[2k-i]^{II}$ is of color $j+k-1$, which yields $i\equiv 1$.

Let us second assume that $[1]^{I}$ is Wick-contracted with $[C]^{II}$. The face containing $[C]^{I}[1]^{I}$ is of length four if and only if $[C]^{I}$ is contracted with $[1]^{II}$. Let us also consider the $(0i)$-face, $i\not = 1$, containing the edge $[1]^{I}[2i-1]^{I}$. It is of length four if and only if $[2i-1]^{I}$ is Wick-contracted with $[i]^{II}$. But then, the face of colors $0$ and $2i-1$ containing the edge $[C]^{I}[2i-1]^{I}$ is of length four if and only if the color of the edge $[1]^{II}[i]^{II}$ is $2i-1$, which yields $3i\equiv 3$. For $R>3$, this implies $i\equiv 1$, which is impossible.

We thus conclude that, for $R>3$, $[1]^{I}$ must be contracted with $[1]^{II}$. Exactly the same reasoning shows that $[k]^{I}$ must be contracted with $[k]^{II}$ for all $1\leq k\leq R$. The two center vertices $[C]^{I}$ and $[C]^{II}$ are then also automatically contracted.\qed

\

In the following, it will also be useful to consider ``elementary generalized two-point melons,'' which are obtained from the elementary generalized melon by cutting open an edge of color 0. Note that, since such an edge belongs to exactly one face of colors 0 and $i$, for a given $i$, and the original elementary generalized melon contains $\frac{1}{2}(R+1)$ faces of colors 0 and $i$, an elementary generalized two-point melon itself contains $\frac{1}{2}(R+1) - 1 = \frac{1}{2}(R-1)$ faces of colors 0 and $i$, for any given $i$.

\subsection{The most general PCGMs}
\begin{figure}
\centerline{\includegraphics[width=6in]{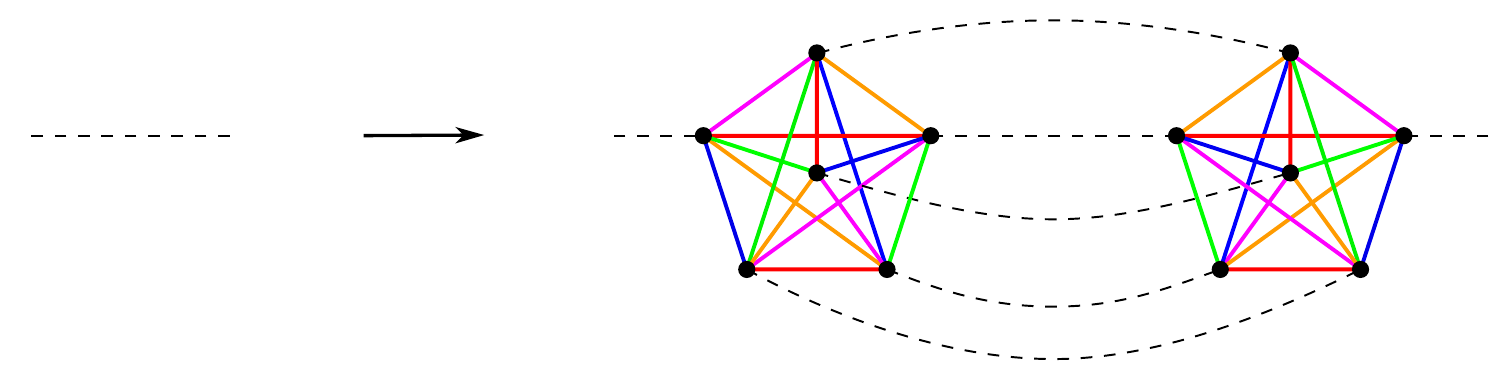}}
\caption{Example of a generalized melonic insertion, in the case $R=5$. This operation does not change the index of a Feynman graph, see the discussion in Section \ref{gmSec}.}
\label{figureC}
\end{figure}

By combining the results of Lemmas \ref{twoverticeslemma} and the generalized melonic moves depicted in Figure \ref{figureC} (see Section \ref{gmSec}), we can build an infinite family of PCGMs, starting from the elementary generalized melon and using an arbitrary number of melonic insertions. We are now going to prove that the most general PCGMs can be obtained in this way.

A prime-complete generalized two-point melon, or PCG2M for short, is defined to be the graph obtained from a PCGM by cutting open any edge of color 0. The trivial PCG2M is simply a single edge of color 0.

\begin{figure}
\centerline{\includegraphics[width=6in]{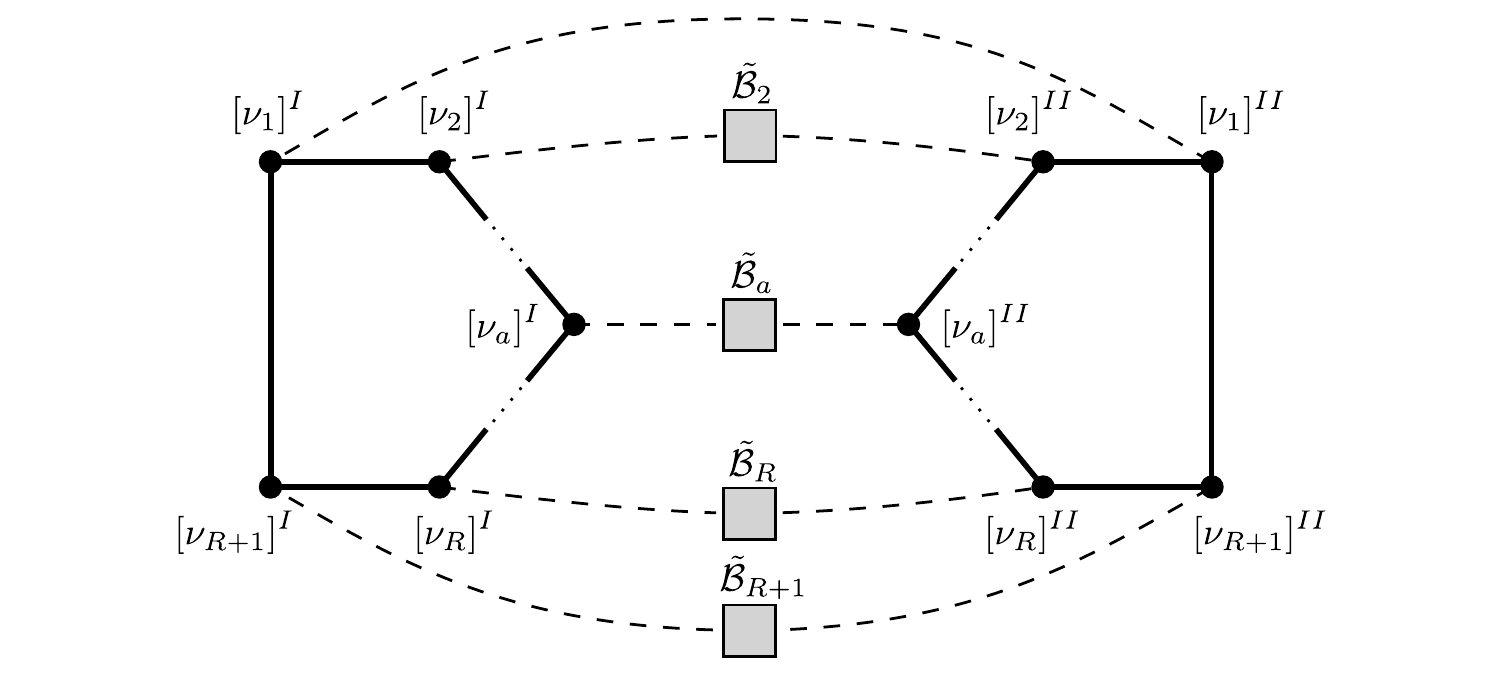}}
\caption{General structure of a PCGM in the form of $R+1$ PCG2Ms $\tilde\gB_{a}$, $1\leq a\leq R+1$, each attached to equivalent vertices $[\nu_{a}]^{I}$ and $[\nu_{a}]^{II}$ in two distinct interaction bubbles. The bubbles $\gK_{R+1}$ are depicted sketchily in an arbitrary $(ij)$-polygonal representation for which only the boundary of the polygon is drawn. The PCG2Ms are stylized as light grey squares attached to two edges of color 0. At least one of the $\tilde\gB_{a}$, $2\leq a\leq R+1$, is trivial.}
\label{fig2M}
\end{figure}
\begin{lemma}\label{thlemma1} Any PCGM can be represented in the form of $R+1$ PCG2Ms, with at least two of them being trivial, each attached to equivalent vertices in two distinct interaction bubbles $\gK_{R+1}$, see Figure \ref{fig2M}.
\end{lemma}

\begin{figure}
\centerline{\includegraphics[width=6in]{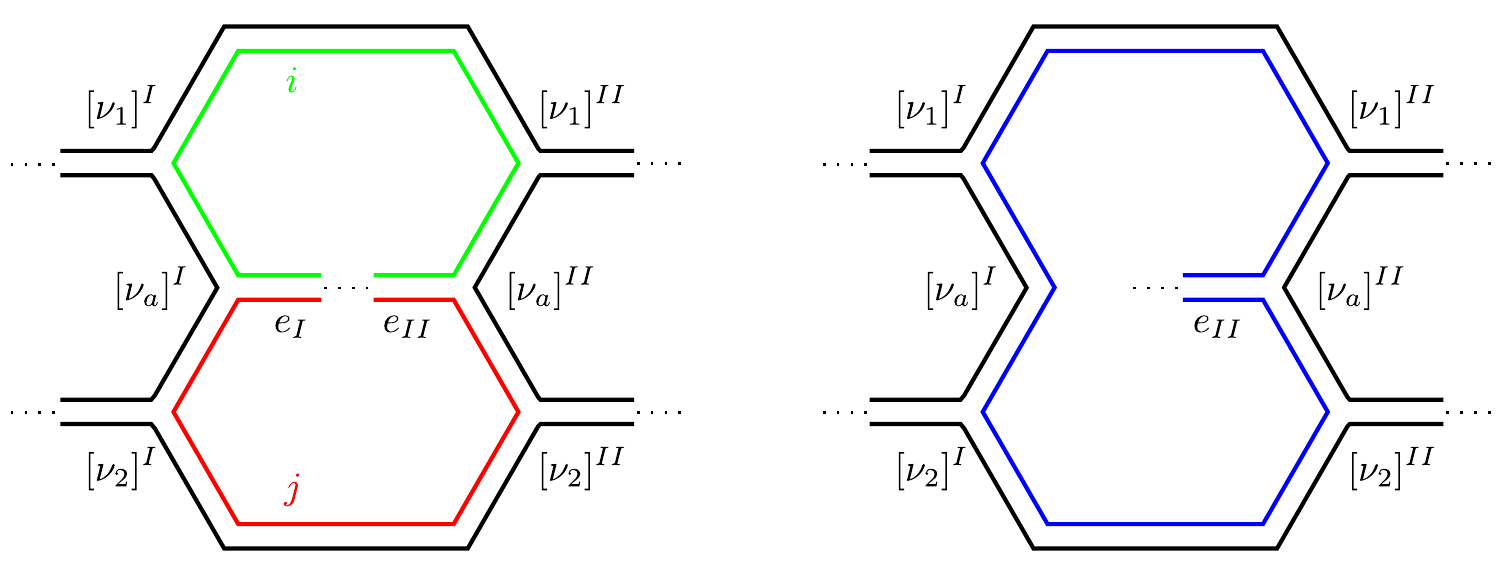}}
\caption{Ribbon graph for the planar three-bubble $\gB_{(0ij)}$ used in the proof of Lemma \ref{thlemma1} (left inset) and connected planar ribbon graph obtained after deletion of the regular edge $e_{I}$ (right inset). The green and red faces merge into a unique blue face and the edge $e_{II}$ becomes singular.}
\label{fig3M}
\end{figure}

\proof Let $\gB$ be a PCGM. We start by using a $(0k)$-face of length four, whose existence is ensured by Lemma \ref{lengthface}. The structure of this face is given by Lemma \ref{fourfaces}. This provides our two interaction bubbles $I$ and $II$. We then pick any vertex $[\nu_{a}]$ in $\gK_{R+1}$ different from $[\nu_{1}]$ and $[\nu_{2}]$. The edges $[\nu_{1}][\nu_{a}]$ and $[\nu_{a}][\nu_{2}]$ have colors $i$ and $j$ respectively. We call $e_{I}$ and $e_{II}$ the edges of color 0 attached to the equivalent vertices $[\nu_{a}]^{I}$ and $[\nu_{a}]^{II}$. As usual, the PCGM condition \eqref{leadg} implies that the three-bubble $\gB_{(0ij)}$ must be planar. The associated ribbon graph, in a convenient embedding, is depicted on the left of Figure \ref{fig3M}. We have outlined in green and red the $(0i)$- and $(0j)$-faces containing $[\nu_{1}][\nu_{a}]$ and $[\nu_{a}][\nu_{2}]$ respectively.

\begin{figure}
\centerline{\includegraphics[width=6in]{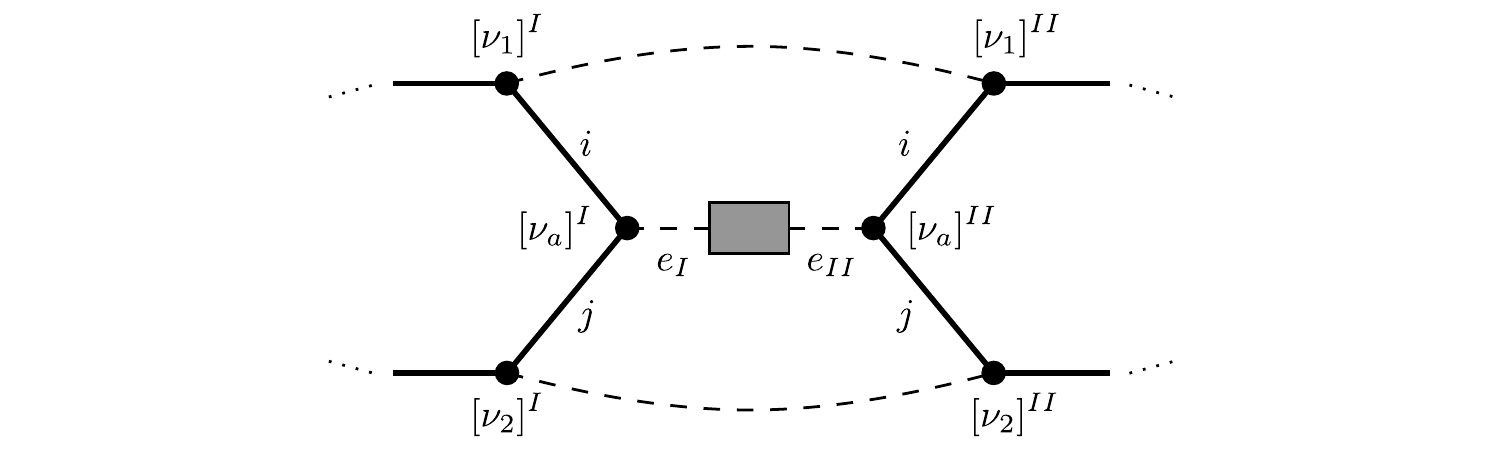}}
\caption{Structure of $\gB_{(0ij)}$, as implied by its two-particle reducibility with respect to the edge $e_{I}$ and $e_{II}$. The graph $\gB$ itself must have a similar structure.}
\label{fig4M}
\end{figure}

Let us now delete the regular edge $e_{I}$. This yields the ribbon graph depicted on the right of Figure \ref{fig3M}. From Lemma \ref{oplemma}, we know that this graph must be connected and planar. In this new graph, the edge $e_{II}$ is singular, because the original green and red faces have merged together. Using again Lemma \ref{oplemma}, we conclude that the deletion of the edge $e_{II}$ produces two connected planar components. In other words, $\gB_{(0ij)}$ is two-particle reducible with respect to the edges $e_{I}$ and $e_{II}$. The structure of $\gB_{(0ij)}$ must then be as illustrated in Figure \ref{fig4M}. The dark grey rectangular region in this figure represents one of the connected planar components obtained after deleting $e_{I}$ and $e_{II}$.

Because the interaction bubbles $\gK_{R+1}$ are MST, it is obvious that the connectivity properties of $\gB$ and $\gB_{(0ij)}$ are the same. In particular, one can find a path that joins two vertices in $\gB$ and that does not contain the edges $e_{I}$ and $e_{II}$ if and only if the same is true in $\gB_{(0ij)}$. Therefore, after the deletion of the edges $e_{I}$ and $e_{II}$, the graph $\gB$ itself splits into two connected components. A picture similar to the one for $\gB_{(0ij)}$ on Figure \ref{fig4M} is thus valid for $\gB$ as well. Moreover, since the equivalent vertices $[\nu_{a}]^{I}$ and $[\nu_{a}]^{II}$ were chosen arbitrarily, we can repeat the argument for all the pairs of equivalent vertices in the interaction bubbles $I$ and $II$. Eventually, we obtain the picture of Figure \ref{fig2M}. We also know that one of the $\tilde\gB_{a}$, for some $2\leq a\leq R+1$, is trivial.\footnote{In Figure \ref{fig2M}, this trivial $\tilde\gB_{a}$ is not necessarily $\tilde\gB_{2}$, because the polygon boundaries used in this figure do not necessarily correspond to the $(ij)$-faces used in the proof.}

\begin{figure}
\centerline{\includegraphics[width=6in]{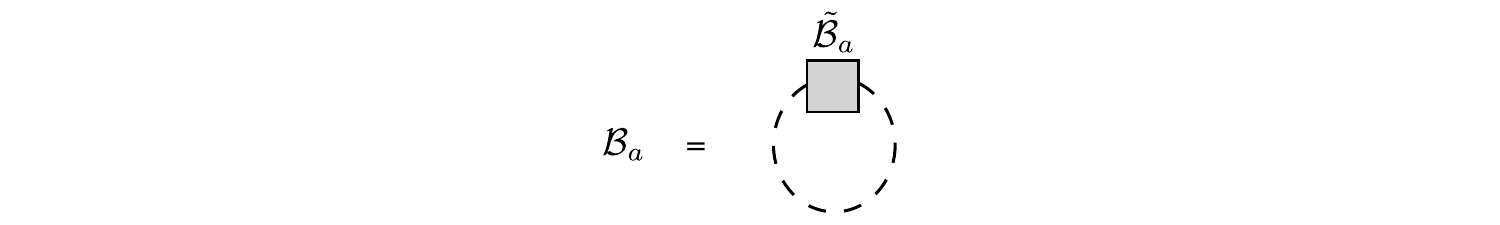}}
\caption{Bubbles $\gB_{a}$ constructed from the two-point graphs $\tilde\gB_{a}$.}
\label{fig5M}
\end{figure}

It remains to prove that the $\tilde\gB_{a}$ are all PCG2Ms. This is equivalent to the fact that the bubbles $\gB_{a}$ depicted in Figure \ref{fig5M} are PCGMs, which is itself equivalent to the planarity of the three-colored graphs $(\gB_{a})_{(0ij)}$ for all pairs of colors $(i,j)$. Then, let us pick two colors $i$ and $j$ and consider $\smash{\gB_{(0ij)}}$. The latter corresponds to a planar graph that looks like the one depicted in Figure \ref{fig2M}, the two polygon boundaries being the two $(ij)$-faces and the $\tilde\gB_{a}$ being replaced by the graphs $\smash{(\tilde\gB_{a})_{(0ij)}}$ obtained from $\tilde\gB_{a}$ by keeping the edges of colors 0, $i$ and $j$ only. If we delete all the edges of color 0 except $[\nu_{1}]^{I}[\nu_{1}]^{II}$ and the two attached to $[\nu_{a}]^{I}$ and $[\nu_{a}]^{II}$, then all the vertices of valency two and then two more edges, one joining $[\nu_{1}]^{I}$ to $[\nu_{a}]^{I}$ and the other $[\nu_{1}]^{II}$ to $[\nu_{a}]^{II}$, we get precisely the graph $(\gB_{a})_{(0ij)}$. Since we started from the planar graph $\gB_{(0ij)}$, Lemma \ref{oplemma} implies that $(\gB_{a})_{(0ij)}$ must be planar too and we conclude.\qed

\

We can now state and easily prove the main result of the present section.

\begin{theorem} \label{theorem1}
The most general PCGMs, which are the leading order vacuum graphs of the model defined by the action \eqref{actionGenCT} with \eqref{uncoloredint} and $R$ prime, are obtained by performing an arbitrary number of generalized melonic insertions starting from the elementary generalized melon.\end{theorem}

\proof Let $v$ be the total number of interaction bubbles in a PCGM. From Lemmas \ref{onefacelemma} and \ref{twoverticeslemma}, we know that the theorem is true for $v\leq 2$. We then proceed recursively. Assume that it is true for all $v<v_{0}$ and consider a PCGM $\gB$ having $v_{0}$ interaction bubbles. We use Lemma \ref{thlemma1} to put it in the form of Figure \ref{fig2M}. The PCGMs $\gB_{a}$, built from the $\tilde\gB_{a}$ as shown in Figure \ref{fig5M}, have  at most $v_{0}-2$ interaction bubbles. The theorem follows by using the recursion hypothesis on these PCGMs.\qed

\

In conclusion, the PCGMs coincide exactly with the mirror melons defined in Section \ref{sec:MSTInt}.

%%%%%%%%%%%%%%%%%%%%%%%%%%%%%%%%%%%%%%%%%%%%%%%%
%%%%%%%%%%%%%%%%%%%%%%%%%%%%%%%%%%%%%%%%%%%%%%%%

\chapter{A few remarks on MST interactions}
\label{app:AppD}

In this appendix, we provide some additional details on maximally single-trace (MST) interactions, which are introduced below Proposition \ref{fundidth2} in Section \ref{sec:TensLargeNExp2}. This part corresponds to Appendix A in Ref.\ \cite{Ref2}.

\section{The complete bipartite interaction}

\begin{figure}
\begin{center}
{\includegraphics[width=6in]{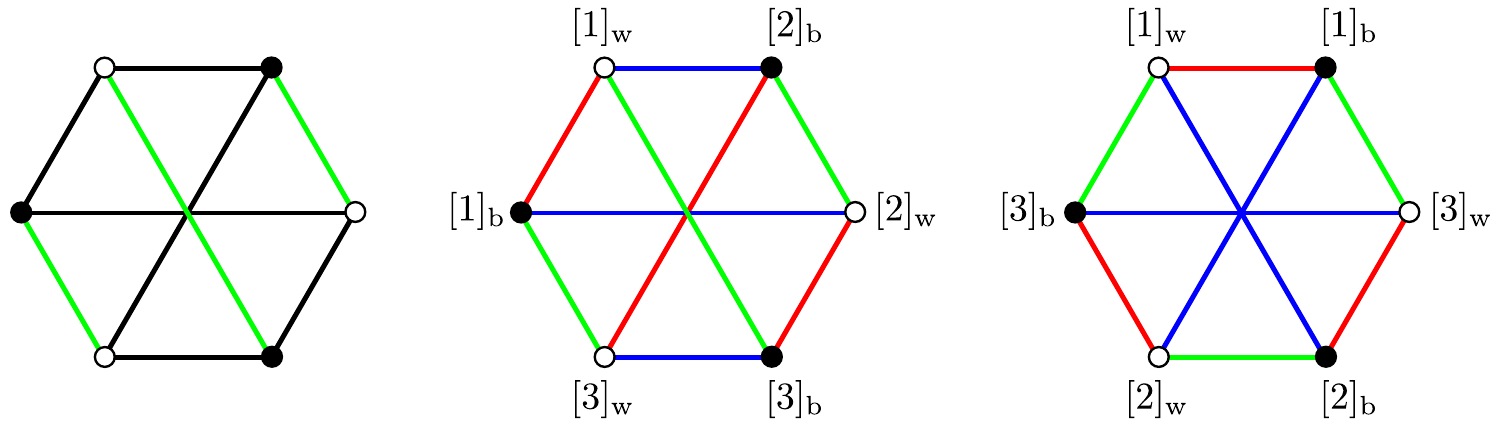}}
\end{center}
\caption{Edge-coloring for the complete bipartite graph $\cK_{3,3}$. Left: rule for the coloring of the edges of a given color, here green. Center: full edge-coloring and vertex labeling, here $1=\text{green}$, $2=\text{red}$ and $3=\text{blue}$. Right: equivalent (green,red)-polygonal representation; this type of representation is natural when $R$ is prime.}
\label{figureK33}
\end{figure}

The complete bipartite graph $\cK_{R,S}$ with $R$ black vertices and $S$ white vertices is edge-colorable with $\max{(R,S)}$ colors. In particular, $\cK_{R,R}$ is edge-colorable with $R$ colors. An explicit $R$-regular edge-coloring of $\cK_{R,R}$ is provided as follows. We first arrange the $2R$ vertices in the shape of a $2R$-sided polygon with alternating black and white vertices. We then label the black and white vertices cyclically as $[n]_{\text b}$ and $[n]_{\text w}$, where $n\in {\mathbb Z } / R {\mathbb Z }$, such that $[n]_\text{b}$ and $[n]_\text{w}$ follow each other. The vertices $[n]_\text{b}$ and $[m]_\text{w}$ are connected by an edge of color $i$, $1\leq i \leq R$, if and only if $n+m \equiv i$ where $\equiv$ denotes equality modulo $R$ as usual. The $R$-bubble obtained in this way is denoted by $\gK_{R,R}$. The above construction is illustrated for $\gK_{3,3}$ in Figure \ref{figureK33}, which also contains a convenient $(ij)$-polygonal representation.

\vspace{1cm}

We have the following result.
\begin{lemma} \label{primebiplemma}
The $R$-bubble $\gK_{R,R}$ is maximally single-trace (MST) if and only if $R$ is a prime number.
\end{lemma}

\proof
The proof follows the same strategy as for Proposition \ref{prime}. Consider a face of colors $i$ and $j$, $1 \le i <j \le R$, whose length is denoted by $2q$. Using the edge-coloring explained above, we can write the $(ij)$-face starting from the vertex $[1]_\text{b}$ with the edge of color $i$ as $[1]_\text{b}[k_1\equiv i-1]_\text{w}[k_2]_\text{b} \cdots [k_{2q-1}\equiv j-1]_\text{w}[1]_\text{b}$. One can check inductively that $k_{2p}\equiv p(j-i)+1$ and $k_{2p+1}\equiv (p+1)i - pj - 1$ for $1\leq p\leq q-1$. As a result, $k_{2q-1}\equiv j-1$ is equivalent to $q(i-j)\equiv 0$. 

If $R$ is a prime number, this implies $q\equiv 0$ because $i-j\not\equiv 0$. The smallest possible solution is given by $q=R$. Therefore, the $(ij)$-face has length $2R$ and it visits all the vertices in $\gK_{R,R}$; hence the bubble is MST.

If $R$ is not prime, write $R=R_1R_2$, where $R_1$ and $R_2$ are integers with $1<R_1<R$ and $1<R_2<R$, and set $i-j=R_2$. The smallest possible solution to $q(i-j)\equiv 0$ is then $q=R_1<R$. This implies that the $(ij)$-face has length $2R_1<2R$ and therefore there are vertices in $\gK_{R,R}$ that are not visited by this face. As a result, the bubble is not MST. \qed

Note that, by Proposition \ref{MSToptprop}, our enhanced scaling is thus optimal for the bubbles $\gK_{R,R}$ when $R$ is prime. In the case of $\gK_{3,3}$, this was already noticed in the second to last reference in \cite{nonBGR2}, together with the full characterization of the leading graphs.

\section{Building MSTs from MSTs}

In this section, we further study the family of MST interactions. As explained previously, tensor models that contain only MST interactions are of special interest in our construction because the index of a vacuum Feynman graph is then given elegantly by Eq.\ \eqref{singfaceind}.
 
\begin{figure}
\begin{center}
{\includegraphics[width=6in]{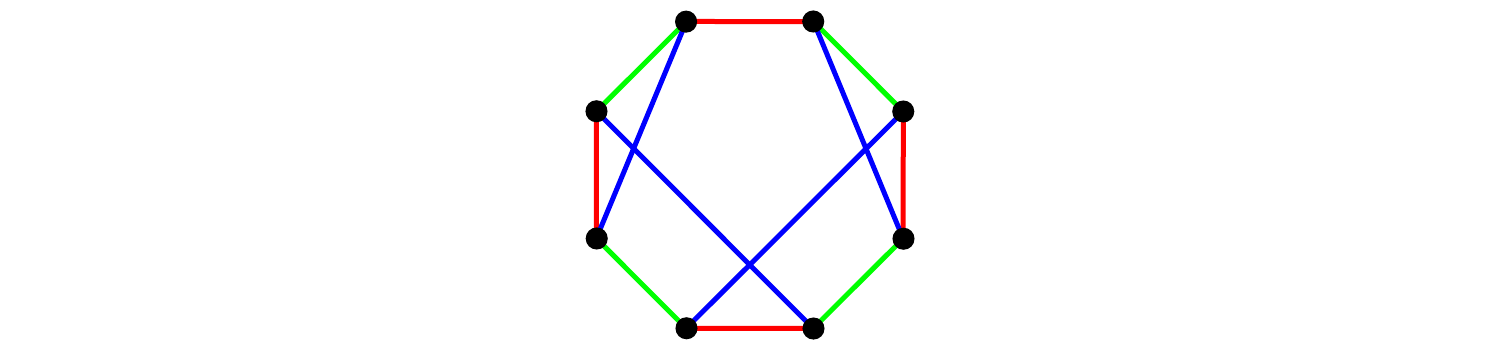}}
\end{center}
\caption{Example of an MST interaction that corresponds to a three-regular edge-colored graph with eight vertices.}
\label{figureMSTEx}
\end{figure}

Interesting examples of MST interactions include the complete interaction $\gK_{R+1}$ for $R$ a prime number and the complete bipartite interaction $\gK_{R,R}$ for $R$ a prime number. Of course, there are many more possibilities, see Figure \ref{figureMSTEx} for an example. To the best of our knowledge, a full classification of MST interactions is not known, but we gather a few remarks here.

No melonic graphs at rank greater than or equal to three with more than two vertices can be MST because melonic graphs have at least one face of length two \cite{Melons}. 

Given two disjoint MST $R$-bubbles $\gB_1$ and $\gB_2$, each with $V \ge 3$ 
vertices, and two distinguished vertices $[\nu_1]$ in $\gB_1$ and $[\nu_2]$ in $\gB_2$, we can build another MST $R$-bubble $\gB_1 \cup_{[\nu_1][\nu_2]} \gB_2$ by removing the vertices $[\nu_{1}]$ and $[\nu_{2}]$ and gluing together the edges incident to $[\nu_1]$ with those incident to $[\nu_2]$, respecting the colors.

Conversely, if there exists a $R$-reducible cut in an MST $R$-bubble $\gB$, that is, a set of $R$ edges of different colors whose removal partitions the bubble into two disjoint connected components, we can perform the reverse operation, 
that is, cut the set of $R$ edges and glue vertices $[\nu_1]$ and $[\nu_2]$ at both ends of the cut. This rewrites $\gB$ as $\gB_1 \cup_{[\nu_1][\nu_2]} \gB_2$.

Any MST $R$-bubble with no such $R$-reducible cut is said to be irreducible. 
The prime-complete interaction $\gK_{R+1}$ and the prime-complete bipartite interaction $\gK_{R,R}$ are both irreducible.

\begin{figure}
\begin{center}
{\includegraphics[width=6in]{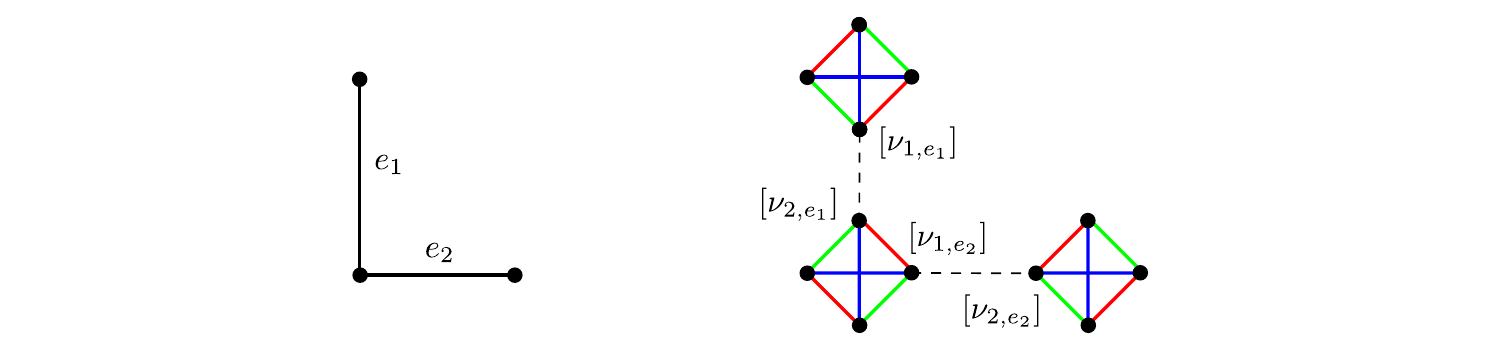}}
\end{center}
\caption{Example of a tetraedric colored rosette. Left: tree associated with the tetraedric colored rosette. Right: gluing of complete interactions $\gK_4$ according to the tree pattern. The resulting tetraedric colored rosette is the MST bubble depicted in Figure \ref{figureMSTEx}.}
\label{figureTreeMST}
\end{figure}

Now, consider an abstract tree $\cT$ with $V$ vertices, associate to each vertex of $\cT$ 
an MST $R$-bubble and for each edge $e \in \cT$ a pair of vertices $[\nu_{1,e}]$ and $[\nu_{2,e}]$ in the MST bubbles at both ends of the edge such that the $2(V-1)$ vertices $[\nu_{1,e}]$, $[\nu_{2,e}]$ are all distinct. Then, we can glue together the $V$ MST bubbles according to the tree pattern, that is, we glue the edges incident to $[\nu_{1,e}]$ with those incident to $[\nu_{2,e}]$ for all $e \in \cT$, and get another MST $R$-bubble. For instance, if prime-complete interactions with $R+1$ vertices are glued together along a tree, we obtain a family we could call the ``$(R+1)$-edric colored rosettes.'' If $R=3$, we obtain ``tetraedric colored rosettes.'' This family is a kind of $\text{O}(N)$ non-bipartite tetraedric generalization of the melonic family. The construction is illustrated in Figure \ref{figureTreeMST}.

\end{appendix}

%%%%%%%%%%%%%%%%%%%%%%%%%%%%%%%%%%%%%%%%%%%%%%%%
%%%%%%%%%%%%%%%%%%%  REFERENCES  %%%%%%%%%%%%%%%%%%%%
%%%%%%%%%%%%%%%%%%%%%%%%%%%%%%%%%%%%%%%%%%%%%%%%

% References
%\bibliographystyle{unsrt}
%\bibliography{references}
%\addcontentsline{toc}{chapter}{Bibliographie}

%

%

\end{document}